%% file: 27x27.tex
\documentclass[11pt,a4paper]{article}
\pdfoutput=1
\usepackage[dvipsnames]{xcolor}
\usepackage{graphicx,caption,array}
\usepackage{ifpdf}
\usepackage{xspace}
\usepackage{jheppub}
\usepackage{dsfont}
\usepackage{amsfonts}
\usepackage{amsmath}
\usepackage{euscript}
\usepackage{amssymb}
\usepackage{bbold}
\usepackage{xurl}
\usepackage{axodraw2}
\usepackage{multirow}
\usepackage{pstricks}
\usepackage{relsize,exscale,scalefnt,anyfontsize,cases}
\usepackage[utf8]{inputenc}
\usepackage{relsize} 

\usepackage{soul}

\usepackage{genyoungtabtikz}
\usepackage{tikz}
\usetikzlibrary{calc}

\usetikzlibrary{patterns}
\usetikzlibrary{decorations.pathmorphing,decorations.markings,trees,calc,shapes.geometric}
\tikzset{
  qua/.style={postaction={decorate},decoration={markings,mark=at position .55 with {\large\arrow{>}}}},
  anti/.style={postaction={decorate},decoration={markings,mark=at position .55 with {\large\arrow{<}}}},
  glu/.style={draw = black, decorate,decoration={coil,amplitude=1.4pt, segment length=2.6pt}} ,
  glu2/.style={draw = blue} ,
  vertical align/.style={baseline=-.5*(height("$+$")-depth("$+$"))},
  every picture/.style={vertical align}
}
 
\newcommand*{\boxcolor}{orange}
\makeatletter
\renewcommand{\boxed}[1]{\textcolor{\boxcolor}{%
\tikz[baseline={([yshift=-1ex]current bounding box.center)}] \node [rectangle, minimum width=1ex,rounded corners,draw] {\normalcolor\m@th$\displaystyle#1$};}}
 \makeatother

\newcommand{\setscaletikz}{0.22}

\newcommand{\qcolor}{Orange}
\newcommand{\qbcolor}{Green}
\include{macro}
\include{macroFIG}

\title{$\Irrep{27}{} \otimes \Irrep{27}{}$} 

\author[a,b]{Florian Cougoulic,}
\author[c]{St\'ephane Peign\'e,}

\affiliation[a]{Institute of Theoretical Physics, Jagiellonian University, ul. \L ojasiewicza 11, 30-348 Krak\'ow, Poland}

\affiliation[b]{Instituto Galego de F\'isica de Altas Enerx\'ias IGFAE, Universidade de Santiago de Compostela,
E-15782 Galicia-Spain}

\affiliation[c]{SUBATECH UMR 6457 (IMT Atlantique, Universit\'e de Nantes, IN2P3/CNRS), 4 rue Alfred Kastler, 44307 Nantes, France}

\emailAdd{florian.cougoulic@uj.edu.pl}
\emailAdd{stephane.peigne@subatech.in2p3.fr}

\abstract{We study the decomposition into $\sun$ irreducible representations (irreps) of the tensor product $\Irrep{27}{} \otimes \Irrep{27}{}$, where $\Irrep{27}{}$ is the highest-dimensional $\sun$ irrep present in a two-gluon system, and explicitly construct all Hermitian projectors on these irreps, as well as transition operators between them. This yields an explicit basis of the complete $\Irrep{27}{} \otimes \Irrep{27}{}$ color space (defined as the space of $\Irrep{27}{} \otimes \Irrep{27}{} \to \Irrep{27}{} \otimes \Irrep{27}{}$ color maps) in terms of orthogonal multiplets. 
This study shows that even complex color structures can be addressed, with the help of the birdtrack pictorial technique, using only elementary tools. In particular, we highlight the usefulness of the quadratic Casimir operator, whose eigenspaces allow efficient filtering of all projectors and transition operators, and of the permutation operators that further improve this filtering.
The product $\Irrep{27}{} \otimes \Irrep{27}{}$ also has an interesting feature: three equivalent irreps $\Irrep{27}{}$ appear in the decomposition, two of which are symmetric and can therefore be distinguished neither by the quadratic Casimir operator nor by their symmetry under permutation. In this case, it is convenient to use Clebsch-Gordan coefficients to derive the two associated, symmetric projectors. The latter are not uniquely determined (only their sum is), and we give the set of all solutions.
Finally, we explicitly derive the soft anomalous dimension matrix associated with $\Irrep{27}{} \otimes \Irrep{27}{} \to \Irrep{27}{} \otimes \Irrep{27}{}$, whose block-diagonal main structure is easy to understand, but whose detailed spectrum properties remain intriguing.
The approach presented for $\Irrep{27}{} \otimes \Irrep{27}{} \to \Irrep{27}{} \otimes \Irrep{27}{}$ could in principle be applied to any product of $\sun$ irreps, and eventually automated.
}


\begin{document} 


\maketitle
\setcounter{footnote}{0}
\renewcommand{\thefootnote}{\arabic{footnote}}
\interfootnotelinepenalty=10000

\newpage

\section{Introduction}
\label{sec:intro}

The improved precision of experimental data from hadron colliders, such as the Large Hadron Collider (LHC) or the future Electron Ion Collider (EIC), calls for theoretical developments that match the observed accuracy.
In perturbative Quantum Chromodynamics (pQCD), improving the accuracy of the theory requires higher-order calculations in the strong coupling constant $\alpha_s$.
These calculations involve an increasing number of loops and/or final state partons. This makes the color structure more complex, and even more so for QCD observables that require explicit color decomposition. 
Efficient approaches to dealing with the color structure of high-order pQCD calculations are therefore needed. 

The birdtrack technique~\cite{Cvitanovic:1976am,Cvitanovic:2008zz} is a visual approach, reminiscent of Feynman diagrams, used to compute group theory factors. In the case of QCD and the $\sun$ color symmetry (for $\Nc \geq 3$ quark colors), it has been the subject of growing interest over the past decades.\footnote{Several basic introductions to the birdtrack technique for $\sun$ can be found in the literature, see\ \eg~\cite{Dokshitzer:1995fv,Sjodahl:2015,Keppeler:2017kwt,Peigne:2023iwm,Peigne:2024srm}.} 
An important application of the birdtrack technique is the calculation of the Hermitian projectors associated to a given parton system, allowing its decomposition into a sum of  $\sun$ irreducible representations (irreps), or {\it color states}. Due to their orthogonality and completeness properties, these projectors are helpful for organizing perturbative computations, and are in fact needed to handle QCD observables that distinguish, at a certain stage of the calculation, the color states which are available in the underlying pQCD process. 
To give just a few examples of observables for which color decomposition is necessary (or at least beneficial), let's mention soft anomalous dimension matrices~\cite{Botts:1989kf,Sotiropoulos:1993rd,Contopanagos:1996nh,Kidonakis:1998nf,Oderda:1999kr,Kyrieleis:2005dt,Dokshitzer:2005ig}, high-energy evolution of cross-sections \cite{Kovner:2001vi}, high-energy scattering off an external color field~\cite{Angelopoulou:2023qdm}, transverse momentum broadening in nuclear media~\cite{Nikolaev:2003zf,Nikolaev:2005dd,Nikolaev:2005zj,Cougoulic:2017ust,Li:2023qkg}, quark-gluon plasma probes~\cite{Aurenche:2011rd,Zakharov:2018hfz,Zakharov:2019fov}, and medium-induced radiation~\cite{Arleo:2010rb,Arleo:2012rs,Liou:2014rha,Peigne:2014rka,Peigne:2014uha,Munier:2016oih,Arleo:2020hat,Arleo:2021bpv,Arleo:2021krm,Jackson:2023adv}.

Over the last few decades, considerable progress has been made in the construction of color projectors using birdtracks, alongside other developments of the birdtrack technique. The choice of color bases has been refined by the introduction of orthogonal bases such as the multiplet bases~\cite{Keppeler:2012ih} constructed from irreps and thus directly related to Hermitian projectors on color states. The corresponding algorithm has recently been automated~\cite{Chargeishvili:2024pnq}. 
Various tools have been developed for $\sun$ color calculations. These include the Mathematica \cite{Mathematica} package ``ColorMath''~\cite{Sjodahl:2012nk}, and the C++ library ``ColorFull''~\cite{Sjodahl:2014opa}. 
The construction of Hermitian Young operators~\cite{Keppeler:2013yla} improves on the construction of Young operators using birdtracks~\cite{Cvitanovic:2008zz}. 
In Refs.~\cite{Alcock-Zeilinger:2016sxc,Alcock-Zeilinger:2016cva,Alcock-Zeilinger:2016xgf,Alcock-Zeilinger:2017ija,Alcock-Zeilinger:2018fha}, simplification rules for birdtrack operators have been developed, and the MOLD algorithm \cite{Alcock-Zeilinger:2018fha} makes it possible to express projectors in a more compact form. Color flow decomposition~\cite{Maltoni:2002mq,Sjodahl:2009wx,Kilian:2012pz,Platzer:2013fha,AngelesMartinez:2018cfz,DeAngelis:2020rvq,Platzer:2020lbr} has also been developed, with a direct link to the large-$\Nc$ limit of QCD.
Furthermore, in recent years it has been suggested to use recoupling formulas~\cite{Cvitanovic:2008zz} to express some class of QCD color structures in terms of a finite set of Wigner 6j symbols~\cite{Alcock-Zeilinger:2022hrk,Keppeler:2023msu}. 

This manuscript studies the decomposition into irreps of the product $\Irrep{27}{} \otimes \Irrep{27}{}$ (where $\Irrep{27}{}$ is an $\sun$ irrep defined in section~\ref{sec:27def}), and presents the calculation, using the birdtrack technique, of the Hermitian projectors on the irreps appearing in that decomposition, as well as of the transition operators between equivalent irreps. In other words, we find an explicit basis of the space of $\Irrep{27}{} \otimes \Irrep{27}{} \to \Irrep{27}{} \otimes \Irrep{27}{}$ color maps (hereafter referred to as the `color space' of $\Irrep{27}{} \otimes \Irrep{27}{}$), in terms of orthogonal multiplets and transition operators between them. 

Our main aim is to illustrate that the most elementary birdtrack tools can be sufficient to study quite large color spaces, without necessarily requiring advanced techniques. In particular, in addition to the basic birdtrack tools (color conservation, Fierz identity, Schur's lemma), our study highlights the effective role of the quadratic Casimir operator $\hat{C}_2$ (a linear map of the color space on itself). Indeed, $\hat{C}_2$ divides the color space into a fairly large number of eigenspaces. (As we will see, the color space of $\Irrep{27}{} \otimes \Irrep{27}{}$ is of dimension 42, and $\hat{C}_2$ has 14 eigenspaces.) Since Hermitian projectors and transition operators are eigenvectors of $\hat{C}_2$, we can see that this operator acts as a powerful first filter for the entire color space. The quadratic Casimir operator has been previously used to handle spin and flavor projection operators~\cite{BandaGuzman:2020wrz,Guzman:2023vpq}, but to our knowledge it has not been applied to the derivation of color projectors of multiparton systems. Of course, when the color space becomes very large, systematic algorithms and automated calculation codes become indispensable. But the approach presented here, using the quadratic Casimir operator as a helpful tool for deriving multiplet bases, could undoubtedly be automated and possibly be an alternative to other systematic methods available in the literature.

Note that considering the specific `process' $\Irrep{27}{} \otimes \Irrep{27}{} \to \Irrep{27}{} \otimes \Irrep{27}{}$, chosen as a generic case to address the main purpose described above, is also motivated by its possible role in QCD phenomenology. For example, consider the production of four jets from two simultaneous $gg \to gg$ scatterings. This type of process can be important in proton-proton collisions~\cite{Blok:2010ge}, but also in proton-nucleus (pA) collisions.  In pA collisions, the prediction of an observable such as  four-jet transverse momentum nuclear broadening would require knowledge of the production probabilities of the four-gluon system in a given color state,\footnote{At least in some kinematical domain. See chapter 6 of Ref.~\cite{Peigne:2024srm} for a heuristic discussion of four-jet production in pA collisions from two simultaneous $2 \to 2$ scatterings.} and thus of the Hermitian projectors on the available irreps of that system. Finding the irreps appearing in the product $\Irrep{27}{} \otimes \Irrep{27}{}$ (corresponding to the case where each of the two final gluon pairs is produced in the $\Irrep{27}{}$) and their associated projectors is an important step towards the complete calculation of this observable. 
As another motivation, the $\Irrep{27}{} \otimes \Irrep{27}{} \to \Irrep{27}{} \otimes \Irrep{27}{}$ process might also provide clues to the general structure of the soft anomalous dimension matrix (sADM) arising in the context of the QCD resummation of soft gluons \cite{Botts:1989kf,Sotiropoulos:1993rd,Contopanagos:1996nh,Kidonakis:1998nf,Oderda:1999kr,Kyrieleis:2005dt,Dokshitzer:2005ig}. The sADM of a given $2 \to 2$ scattering (where any of the external partons could be a `generalized parton', defined as a compact multiparton system in a given irrep) depends on the irreps available in the different channels ($s$, $t$, $u$) of the $2 \to 2$ process, and therefore has a quite rich color structure. For $gg \to gg$, it was observed in Ref.~\cite{Dokshitzer:2005ig} that some of the sADM eigenvalues exhibit a mysterious symmetry under the exchange of internal and external variables (namely, the number $\Nc$ of colors and a kinematic variable), which is hard to imagine as accidental~\cite{Dokshitzer:2005ig}. 
For the process $\Irrep{27}{} \otimes \Irrep{27}{} \to \Irrep{27}{} \otimes \Irrep{27}{}$, the associated sADM can be directly obtained from our results (see section~\ref{sec:sADM}). The mysterious symmetry does not appear in that case, but we find that the sADM characteristic polynomial (of degree 42) factorizes into polynomials of degree $1$, $4$ and $6$, with a curious property: all these polynomials appear with multiplicity $> 1$, except the one of degree 6. This might suggest some peculiarity of the corresponding eigenvalues of the sADM, and it would be interesting to understand the fundamental origin (if any) of this structure.

The manuscript is organized as follows.
In {section~\ref{sec:27times27}}, we decompose $\Irrep{27}{} \otimes \Irrep{27}{}$ into a sum of $\sun$ irreps. We choose to represent the irrep $\Irrep{27}{}$ by that arising in the tensor product of two symmetrized quarks and two symmetrized antiquarks, see~\eq{6times6bar-deco}-\eq{eq:proj27-def}. We proceed with the full derivation of $\Irrep{27}{} \otimes \Irrep{27}{}$ using the `index method' (based on the characterization of $\sun$ irreps by traceless tensors having specific symmetries in quark and antiquark indices), and list all Young diagrams associated with the irreps of the decomposition. 
In {section~\ref{sec:ActionCasimir}}, we first define some birdtrack notations for the projection and transition operators. 
We then introduce the tensor basis ${\cal T}$ of the $\Irrep{27}{} \otimes \Irrep{27}{}$ color space we'll be using, and calculate the action of the quadratic Casimir operator $\hat{C}_2$ in this basis. 
The {section~\ref{sec:filtering}} contains the main results of our study. We show how the $\hat{C}_2$ operator, supplemented by permutation and complex conjugation operators, can be used to efficiently obtain nearly all Hermitian projectors and transition operators. In the slightly more delicate `sector of the $\Irrep{27}{}$' of the decomposition, which contains three irreps $\Irrep{27}{}$, additional help is provided by the use of Clebsch-Gordan operators (see subsection~\ref{sec:eigenspaces-dim9}). In section~\ref{sec:sADM}, the soft anomalous dimension matrix associated to $\Irrep{27}{} \otimes \Irrep{27}{} \to \Irrep{27}{} \otimes \Irrep{27}{}$ is derived explicitly. 
We conclude in section~\ref{sec:conclusion} by a summary of our main results and a brief discussion.

\section{Decomposition of $\Irrep{27}{} \otimes \Irrep{27}{}$ into irreps}
\label{sec:27times27}

An $\sun$ irrep can be characterized by a tensor carrying $n_{q}$ upper (quark) indices and $n_{\bar q}$ lower (antiquark) indices, having specific symmetry properties under the permutation of the quark indices and separately of the antiquark indices, and being moreover {\it traceless} when contracting one pair (or more) of quark and antiquark indices. 

In what follows, for general $\Nc$ an $\sun$ irrep will be named by the value of its dimension when $\Nc =3$. When different $\sun$ irreps have the same dimension for $\Nc =3$, each irrep can be unambiguously identified by its associated Young diagram. 

\subsection{Model for the irrep $\Irrep{27}{}$} 
\label{sec:27def}

The $\sun$ irrep $\Irrep{27}{}$ appears for instance in the decomposition of a gluon pair, which reads
\begin{align}
\Irrep{8}{} \otimes \Irrep{8}{}  = 
\Irrep{1}{} \oplus \Irrep{8}{+} \oplus \Irrep{8}{-} \oplus \Irrep{10}{} \oplus \overline{\Irrep{10}{}} \oplus \Irrep{27}{}  \oplus \Irrep{0}{} \ ,
\end{align}
where the equivalent, adjoint representations $\Irrep{8}{+}$ and $\Irrep{8}{-}$ can be distinguished by their symmetry (denoted by the subscript $\pm$) under permutation of the two gluons. 

The irrep $\Irrep{27}{}$ is characterized by two upper and two lower indices, and is symmetric under permutation of both types of indices. It it thus characterized by a tensor (implicitly traceless) 
\be
\psi^{ \left\{ i j \right\} }_{ \left\{ k l \right\} } \ , 
\ee
where the shorthand notation $\left\{ i \ldots j \right\}$ indicates symmetrization on indices $i \ldots j$. 
Thus, the irrep $\Irrep{27}{}$ must also appear in the product of a symmetric quark pair $\Pair{qq}{\bm 6}$ and a symmetric antiquark pair $\Pair{\bar{q}\bar{q}}{\bm{\bar 6}}$, which decomposes into a sum of $\sun$ irreps as 
\be
\label{6times6bar-deco}
\Irrep{6}{} \otimes \Irrep{\bar 6}{} \ = \ \Irrep{1}{} \, \oplus \, \Irrep{8}{} \, \oplus \, \Irrep{27}{} \ .
\ee
In our study, we choose to represent the irrep $\Irrep{27}{}$ as the color state of highest dimension found in the system $\Pair{qq}{\bm 6} \otimes \Pair{\bar{q}\bar{q}}{\bm{\bar 6}} \sim \Irrep{6}{} \otimes \Irrep{\bar 6}{}\,$. 

The decomposition~\eq{6times6bar-deco} can be expressed by the completeness relation satisfied by the Hermitian projectors associated to the irreps, 
\be
\label{6times6bar-comp}
\Id{0.25} \ = \ \Proj{\bm 1} + \Proj{\bm 8} + \Proj{\bm{27}} \ ,
\ee
where the l.h.s.~is the birdtrack for the identity operator of the vector space $\Irrep{6}{} \otimes \Irrep{\bar 6}{}$ and, following a standard notation, a white rectangle represents a symmetrizer on quark (or antiquark) indices (see for example~\cite{Keppeler:2017kwt}).

It can be shown (for instance using the `tensor method', see Refs.~\cite{Cvitanovic:2008zz,Peigne:2024srm}) that the projector on the irrep $\Irrep{27}{}$ reads\footnote{\label{foot:birdtrack-conv}In the present study, all birdtracks are defined using the conventions of Refs.~\cite{Dokshitzer:1995fv,Peigne:2023iwm,Peigne:2024srm}. The $\sun$ generators $t_F^a$ in the fundamental (quark) representation are defined by 
\begin{align}
t_F^a \ \equiv \ \TfaBT{0.15} \ ,  \nn 
\end{align}
and normalized as
\begin{align}
\tr t_F^a t_F^b = \frac{1}{2} \delta^{ab} \ \ \Longleftrightarrow \ \ \trtt{0.15} \ = \ \frac{1}{2}\ \DeltaAdj \ . \nn 
\end{align}
The $\sun$ generators $t^a_{\overline{F}}$ in the complex conjugate (antiquark) representation, $t^a_{\overline{F}} \equiv -(t^a_F)^\intercal$~\cite{Dokshitzer:1995fv}, include a minus sign which is made explicit by the pictorial rule: 
\begin{align} 
\BTtf{0.15} \ \equiv \ -\ \BTtfb{0.15}  \ . \nn 
\end{align}
This allows one to write color conservation in the form
\begin{align}
\ColorConservationA{0.2} \ + \ \ColorConservationB{0.2} \ + \ \ColorConservationC{0.2} \ + \ \cdots \ + \ \ColorConservationD{0.2} \ = \ 0 \ , \nn 
\end{align}
independently of the color singlet system of partons exiting the ellipse blob.}
\be
\label{eq:proj27-def}
\Proj{\bm{27}} \ \equiv \  \ProjB{0.25} \ = \ \Id{0.25} \ -\frac{4}{\Nc+2}\ \Q{0.25} \ + \frac{2}{(\Nc+1)(\Nc+2)}\ \S{0.25} \ \ , 
\ee
which uniquely defines the irrep $\Irrep{27}{}$. In the r.h.s.~of~\eq{eq:proj27-def}, the first term (the identity of $\Irrep{6}{} \otimes \Irrep{\bar 6}{}$) specifies the symmetries of the irrep $\Irrep{27}{}$, whereas the last two terms ensure that the multiplet associated with $\Proj{\bm{27}}$ is traceless (which is equivalent to subtracting from the first term the unwanted irreps $\Irrep{1}{}$ and $\Irrep{8}{}$). 
Thus, an economical way to prove that $\Proj{\bm{27}}$ is indeed given by~\eq{eq:proj27-def} is to check that it vanishes when a quark index is contracted with an antiquark index (on either the right or left of the birdtrack defining $\Proj{\bm{27}}$). 
In certain cases, we will use the following compact notation for the 
projector~\eq{eq:proj27-def}:\footnote{We will later attribute an upper label $\al$ to the dashed line representing the projector on an irrep $\al$, see~\eq{eq:proj-alpha-G}, but for simplicity this label will be omitted for the particular case of the irrep $\Irrep{27}{}$ which is omnipresent in our study.}
\be
\label{dashproj-def}
\Proj{\Irrep{27}{}} \ \equiv \  \ProjB{0.25} \  \equiv \ \dashproj{0.26} \ .
\ee

Let us mention that in birdtrack notation, the trace of an operator mapping a vector space to itself is simply obtained by connecting all incoming and outgoing parton lines.\footnote{\label{foot:trace}This trace should not be confused with the contractions, also called `traces', over pairs of quark and antiquark indices used in the next section to characterize an irrep.}  For the projector $\Proj{\bm{27}}$ mapping $\Irrep{27}{} \to \Irrep{27}{}$, this operation gives the dimension $\K{\Irrep{27}{}}$ of the $\sun$ irrep $\Irrep{27}{}$: 
\begin{align}
\label{27-dim}
\tr{\Proj{\bm{27}}} = \K{\Irrep{27}{}} = \frac{1}{4} (\Nc-1)\Nc^2(\Nc+3)  \ . 
\end{align}

\subsection{Decomposition of $\Irrep{27}{} \otimes \Irrep{27}{}$ using the index method}
\label{sec:index-method}

The decomposition of $\Irrep{27}{} \otimes \Irrep{27}{}$ into irreps can be obtained using the `index method' (see \eg\ Refs.~\cite{Dokshitzer:1995fv,Peigne:2024srm} for a simple presentation), considering a system of two `generalized partons' $\psi$ and $\chi$ (each being in the color state $\Irrep{27}{}$) represented by the tensor product  
\begin{align}
\label{psi-chi}
\psi^{ \left\{ i j \right\} }_{ \left\{ k l \right\} } \, \chi^{ \left\{ m n \right\} }_{ \left\{ p q \right\} } \ , 
\end{align}
and constructing all possible contractions (or `traces') satisfying the criteria recalled in the beginning of this section for characterizing an $\sun$ irrep. In the following, the symmetry in upper or lower indices of both $\psi$ and $\chi$ will be implicit.

\subsubsection{Irrep corresponding to four traces}

Since the tensors $\psi$ and $\chi$ are both traceless (by definition of an irrep), there is only one way to fully contract the indices in \eq{psi-chi}, corresponding to four `traces' between quark and antiquark indices: 
\be
\label{irreps-4traces}
\psi^{ij}_{kl} \, \chi^{kl}_{ij} \ . 
\ee

The above tensor must generate an irrep carrying no free index, which is the singlet (or trivial) representation of $\sun$, denoted by $\Irrep{1}{}$.

\subsubsection{Irreps corresponding to three traces}

There are two independent ways of contracting three pairs of upper and lower indices in \eq{psi-chi}, giving two tensor structures carrying one free upper index and one free lower index, 
\be
\psi^{ij}_{kl} \, \chi^{kn}_{ij} \ \ ; \ \ \ \ \psi^{ij}_{kl} \, \chi^{kl}_{iq} \ \ ,
\ee
yielding two $\sun$ adjoint representations. Since the above tensor structures are related by the exchange $\psi \leftrightarrow \chi$, these irreps can be traded for two adjoint representations being either symmetric or antisymmetric under $\psi \leftrightarrow \chi$, denoted respectively by $\Irrep{8}{+}$ and $\Irrep{8}{-}\,$: 
\begin{align}
\label{irreps-3traces}
\ba{c} 
\Irrep{8}{+}  \\ \tikeq{\bs[scale=0.7]\tyng(0cm,0cm,2,1^6)\es} 
\ea
\ \ ; \ \ \ \ 
\ba{c} \Irrep{8}{-}  \\ \tikeq{\bs[scale=0.7]\tyng(0cm,0cm,2,1^6)\es} 
\ea \ \ . 
\end{align}

Here and from now on, for each irrep of the product $\Irrep{27}{} \otimes \Irrep{27}{}$ we display its associated Young diagram. For simplicity the Young diagrams are drawn for the case $\Nc=8$, but this is sufficient to infer the exact shape of the Young diagrams for $\sun$. (For instance, the height of the adjoint representation Young diagram above is $\Nc-1$.) 

{Throughout this study, we will only consider symmetric or antisymmetric irreps under the permutation $\psi \leftrightarrow \chi$. Indeed, $\psi$ and $\chi$ are indistinguishable, and the complete wave function of the $\psi \chi$ pair must be either symmetric or antisymmetric (\ie\ it must satisfy Bose-Einstein or Fermi-Dirac statistics). This implies that the color part of the wave function, on which we are focusing, has a well-defined symmetry under $\psi \leftrightarrow \chi$.}

\subsubsection{Two traces}

There are three ways of contracting two pairs of upper and lower indices in \eq{psi-chi}: 
\be
\label{2traces}
\psi^{ij}_{kl} \, \chi^{mn}_{ij} \ \ ; \ \ \ \ \psi^{ij}_{kl} \, \chi^{kl}_{pq} \ \ ; \ \ \ \ 
\psi^{ij}_{kl} \, \chi^{kn}_{iq} \ . 
\ee
The first two tensors already have a specified symmetry under permutation of the free upper, and separately lower, indices, which is characteristic of the irrep $\Irrep{27}{}$. These two tensors are moreover related by the exchange $\psi \leftrightarrow \chi$. We thus infer the presence of two irreps $\Irrep{27}{}$ associated with some linear combinations (properly made traceless) of these tensors, one being symmetric and the other antisymmetric under $\psi \leftrightarrow \chi$.

The third tensor of \eq{2traces} is of the form $T^{jn}_{lq}$, whose symmetry properties under permutation of the upper and lower indices are not yet specified. A pair of quark (and similarly antiquark) indices can be either symmetrized or antisymmetrized, and  $T^{jn}_{lq}$ thus gives rise to the four independent tensors: 
\be
\label{Tjnlq}
T^{ \left\{ jn \right\} }_{ \left\{ lq \right\} } \ \ ; \ \ \ \ 
T^{ \left[ jn \right] }_{ \left[ lq \right] } \ \ ; \ \ \ \ 
T^{ \left\{ jn \right\} }_{ \left[ lq \right] } \ \ ; \ \ \ \ 
T^{ \left[ jn \right] }_{ \left\{ lq \right\} } \ \ , 
\ee 
where $\left[ i \ldots j \right]$ denotes antisymmetrization on indices $i \ldots j$. The four tensors \eq{Tjnlq} have fully specified symmetries and thus generate (after being made traceless) the $\sun$ irreps $\Irrep{27}{}$, $\Irrep{0}{a}$, $\Irrep{10}{}$ and $\Irrep{\overline{10}}{}$, respectively.    

Note that the irrep $\Irrep{27}{}$ (arising from the first tensor of~\eq{Tjnlq}) is symmetric in $\psi \leftrightarrow \chi$, thus providing, with the two other $\Irrep{27}{}$'s found previously, two symmetric and one antisymmetric irreps. These irreps will be denoted as $\Irrep{27}{+}$, $\Irrep{27'}{\!\!+}$, and $\Irrep{27}{-}$. The irrep $\Irrep{0}{a}$ (arising from the second tensor of~\eq{Tjnlq}) turns out to be absent when $\Nc=3$, and is so named to distinguish it from other $\Irrep{0}{}$'s appearing in the product $\Irrep{27}{} \otimes \Irrep{27}{}$ (see below). 

In summary, taking two traces in the product \eq{psi-chi} yields six $\sun$ irreps:  
\begin{align}
\label{irreps-2traces}
\begin{array}{c} 
\Irrep{0}{a} \\  
\tikeq{\bs[scale=0.7]
\draw[transparent] (0,0) -- (.45,-2.745);
\tyng(0cm,0cm,2^2,1^4)
\es}
\end{array}
\ \ ; \ \ \ \ 
\begin{array}{c} 
\Irrep{10}{} \\  
\tikeq{\bs[scale=0.7]
\draw[transparent] (0,0) -- (.45,-2.745);
\tyng(0cm,0cm,3,1^5)
\es}
\end{array}
\ \ ; \ \ \ \ 
\ba{c}  \Irrep{\overline{10}}{} \\ \tikeq{\bs[scale=0.7]\tyng(0cm,0cm,3^2,2^5)\es} \ea
\ \ ; \ \ \ \ 
\ba{c}  \Irrep{27}{+} \\ \tikeq{\bs[scale=0.7]\tyng(0cm,0cm,4,2^6)\es} \ea
\ \ ; \ \ \ \ 
\ba{c} \Irrep{27'}{\!\!+} \\ \tikeq{\bs[scale=0.7]\tyng(0cm,0cm,4,2^6)\es}  \ea
\ \ ; \ \ \ \ 
\ba{c} \Irrep{27}{-} \\ \tikeq{\bs[scale=0.7]\tyng(0cm,0cm,4,2^6)\es} \ea
\ . 
\end{align}

Note that for $\Nc=3$, two antisymmetrized antiquark indices can be traded for one quark index, and an irrep can be characterized by a tensor being {\it totally symmetric} in both quark and antiquark indices (see \eg\ Ref.~\cite{Dokshitzer:1995fv}). Thus, for $\Nc=3$ the irrep arising from $T^{ \left\{ jn \right\} }_{ \left[ lq \right] }$ (third tensor of~\eq{Tjnlq}) is equivalent to that carrying three (totally symmetrized) quark indices, of Young diagram $\,\tikeq{\bs[scale=0.7]\tyng(0cm,-0.1cm,3)\es}$\,. This ${\rm SU}(3)$ irrep is named $\Irrep{10}{}$, using the standard convention that only those ${\rm SU}(3)$ irreps containing more (totally symmetrized) antiquarks than quarks are denoted with a bar.  
Hence our convention to denote, for any $\Nc \geq 3$, by $\Irrep{10}{}$ the $\sun$ irrep associated with $T^{ \left\{ jn \right\} }_{ \left[ lq \right] }$, and by $\Irrep{\overline{10}}{}$ the $\sun$ irrep associated with $T^{ \left[ jn \right] }_{ \left\{ lq \right\} }$ (of associated Young diagrams given in~\eq{irreps-2traces}). 

Let us mention that for general $\Nc$, the Young diagram associated to a given irrep appearing in $\Irrep{27}{} \otimes \Irrep{27}{}$ is quickly obtained from the irrep symmetry properties. As an illustration, consider the $\sun$ irrep $\Irrep{\overline{10}}{}$. Since it is characterized by a (traceless) tensor $\sim T^{ \left[ jn \right] }_{ \left\{ lq \right\} }$, this irrep must also appear in the product of a symmetric antiquark pair 
$\sim q_{ \left\{ l \right. } q_{ \left. q \right\} }$  
and an antisymmetric quark pair 
$\sim q^{ \left[ j \! \right. } q^{ \left. n \right] }\,$: 
\be
\label{example}
q_{ \left\{ l \right. } q_{ \left. q \right\} } \ q^{ \left[ j \! \right. } q^{ \left. n \right] } \ \sim \ 
\tikeq{\bs[scale=0.7]\tyng(0cm,0cm,2^7)\es}  \otimes 
\tikeq{\bs[scale=0.7]\tyng(0cm,0cm,1^2)\es}  \ = \ \Irrep{\overline{10}}{} + \ldots \ , 
\ee
where the Young diagram for the symmetric antiquark pair is simply obtained from the Young diagram 
$\,\tikeq{\bs[scale=0.7]\tyng(0cm,-0.1cm,2)\es}\,$ of a symmetric quark pair. Clearly, the $\sun$ irrep $\Irrep{\overline{10}}{}$ appears only once in the decomposition of the product \eq{example}, and corresponds to taking {\it zero} trace between the sets of antiquark indices $lq$ and quark indices $jn$. From the multiplication rules of Young diagrams, it follows that the Young diagram of $\Irrep{\overline{10}}{}$ is simply obtained by keeping the two Young diagrams in \eq{example} unchanged, and juxtaposing them. This simple rule applies to all $\sun$ irreps (derived up to now and below) appearing in $\Irrep{27}{} \otimes \Irrep{27}{}$. 
 
\subsubsection{One trace}
\label{sec:one-trace}

Contracting over one pair of quark and antiquark indices in \eq{psi-chi} we obtain: 
\be
\label{1trace}
\psi^{ij}_{ k l  } \, \chi^{ m n  }_{iq} 
\ \ ; \ \ \ \ 
\psi^{  i j  }_{kl} \, \chi^{kn}_{  p q }  \ \ .
\ee

Let's focus on the first tensor, of the form $\sim T^{j \left\{ m n \right\}}_{ \left\{ k l \right\} q }$ (showing only free indices), whose symmetries under permutation of quark and antiquark indices are not yet fully specified. The possible symmetries in quark indices are the same as for the system
\be
\label{3times6}
q^{j} \ q^{ \left\{ m \right. } \! q^{ \left. n \right\} }  \ \sim \ 
\tikeq{\bs[scale=0.7]\tyng(0cm,0cm,1)\es}  \otimes 
\tikeq{\bs[scale=0.7]\tyng(0cm,0cm,2)\es} 
\ = \ \tikeq{\bs[scale=0.7]\tyng(0cm,0cm,3)\es}  \oplus
\tikeq{\bs[scale=0.7]\tyng(0cm,0cm,2,1)\es} 
\ \equiv \ \Irrep{10}{} \oplus \Irrep{8}{} \ . 
\ee
Likewise, the possible symmetries in antiquark indices follow from 
\be
\label{6bartimes3bar}
q_{ \left\{ k \right. } q_{ \left. l \right\} } \ q_{q}  \ \sim \ 
\tikeq{\bs[scale=0.7]\tyng(0cm,0cm,2^7)\es}  \otimes 
\tikeq{\bs[scale=0.7]\tyng(0cm,0cm,1^7)\es} 
\ = \ \tikeq{\bs[scale=0.7]\tyng(0cm,0cm,3^7)\es}  \oplus
\tikeq{\bs[scale=0.7]\tyng(0cm,0cm,2^6,1)\es} 
\ \equiv \ \Irrep{\overline{10}}{} \oplus \Irrep{8}{} \ . 
\ee
We note that for $\Nc >3$, the $\sun$ irreps named $(\Irrep{8}{}, \Irrep{10}{})$ and $(\Irrep{8}{}, \Irrep{\overline{10}}{})$ appearing respectively in \eq{3times6} and \eq{6bartimes3bar} are not the same as the $\sun$ irreps named similarly encountered so far. These names (which refer to the irrep dimensions in the case $\Nc =3$) are just used in the present section to label the different symmetries in quark and antiquark indices. In case of ambiguity, recall that an $\sun$ irrep is uniquely defined by its Young diagram. 
Thus, the first tensor of \eq{1trace} leads to four independent tensors with fully specified symmetries, represented by 
\be
\label{1trace-a}
T^{\left( j \left\{ m n \right\} \right)_{\mathsize{6}{\Irrep{8}{}}}}_{\left( \left\{ k l \right\} q \right)_{\mathsize{6}{\Irrep{8}{}}}}
\ \ ; \ \ \ \ 
T^{\left( j \left\{ m n \right\} \right)_{\mathsize{6}{\Irrep{10}{}}}}_{\left( \left\{ k l \right\} q \right)_{\mathsize{6}{\Irrep{8}{}}}}
\ \ ; \ \ \ \ 
T^{\left( j \left\{ m n \right\} \right)_{\mathsize{6}{\Irrep{8}{}}}}_{\left( \left\{ k l \right\} q \right)_{\mathsize{6}{\Irrep{\overline{10}}{}}}}
\ \ ; \ \ \ \ 
T^{\left( j \left\{ m n \right\} \right)_{\mathsize{6}{\Irrep{10}{}}}}_{\left( \left\{ k l \right\} q \right)_{\mathsize{6}{\Irrep{\overline{10}}{}}}} \ \ , 
\ee
where the subscripts $\Irrep{8}{}$, $\Irrep{10}{}$ and $\Irrep{\overline{10}}{}$ specify the color state of the corresponding quark or antiquark system, of indices $jmn$ and $klq$, respectively.  After being rendered traceless, the structures \eq{1trace-a} arising from the first tensor of \eq{1trace} thus provide four different irreps. 

Similarly, the second tensor of \eq{1trace} leads to the same four irreps, up to $\psi \leftrightarrow \chi$. Overall, one can organize the eight obtained irreps in four pairs of equivalent irreps, where the irreps within a pair are distinguished by their symmetry (even or odd) under $\psi \leftrightarrow \chi\,$:
\be
\label{irreps-1trace}
\ba{c} \left( \Irrep{0}{b+} , \Irrep{0}{b-} \right) \\ \tikeq{\bs[scale=0.7]\tyng(0cm,0cm,4,3,2^4,1)\es}   \ea
\ \ ; \ \ \ \ 
\ba{c}  \left( \Irrep{35}{+} , \Irrep{35}{-} \right) \\  \tikeq{\bs[scale=0.7]\tyng(0cm,0cm,5,2^5,1)\es} \ea
\ \ ; \ \ \ \ 
\ba{c} \left( \Irrep{\overline{35}}{+} , \Irrep{\overline{35}}{-} \right) \\ \tikeq{\bs[scale=0.7]\tyng(0cm,0cm,5,4,3^5)\es}\ea
\ \ ; \ \ \ \ 
\ba{c} \left( \Irrep{64}{+} , \Irrep{64}{-} \right) \\ \tikeq{\bs[scale=0.7]\tyng(0cm,0cm,6,3^6)\es}  \ea
\ \ . 
\ee
Using the same line of arguments as after \eq{example}, the Young diagrams of the latter $\sun$ irreps directly follow from juxtaposition of the Young diagrams of the three-quark sub-system (displayed in \eq{3times6}) and of the three-antiquark sub-system (displayed in \eq{6bartimes3bar}). 

\subsubsection{Zero trace}

In the absence of contraction between upper and lower indices in \eq{psi-chi}, we just need to specify all possible symmetries in quark and antiquark indices in the product $\psi^{ij}_{kl} \, \chi^{mn}_{pq}$. 

The possible symmetries in quark indices follow from 
\be
\label{6times6}
q^{ \left\{ i \right. } \! q^{ \left. j \right\} } \ q^{ \left\{ m \right. } \! q^{ \left. n \right\} }  \ \sim \ 
\tikeq{\bs[scale=0.7]\tyng(0cm,0cm,2)\es}  \otimes 
\tikeq{\bs[scale=0.7]\tyng(0cm,0cm,2)\es} 
\ = \ \tikeq{\bs[scale=0.7]\tyng(0cm,0cm,2^2)\es}  \oplus
\tikeq{\bs[scale=0.7]\tyng(0cm,0cm,3,1)\es} \oplus
\tikeq{\bs[scale=0.7]\tyng(0cm,0cm,4)\es} 
\ \equiv \ \Irrep{\bar{6}}{} \oplus \Irrep{15}{} \oplus \Irrep{15'}{} \ ,
\ee
and those in antiquark indices from:
\be
\label{6bartimes6bar}
q_{ \left\{ k \right. } q_{ \left. l \right\} } \ q_{ \left\{ p \right. } q_{ \left. q \right\} }  \ \sim \ 
\tikeq{\bs[scale=0.7]\tyng(0cm,0cm,2^7)\es}  \otimes 
\tikeq{\bs[scale=0.7]\tyng(0cm,0cm,2^7)\es} 
\ = \ \tikeq{\bs[scale=0.7]\tyng(0cm,0cm,2^6)\es}  \oplus
\tikeq{\bs[scale=0.7]\tyng(0cm,0cm,3^6,2)\es} \oplus
\tikeq{\bs[scale=0.7]\tyng(0cm,0cm,4^7)\es} 
\ \equiv \ \Irrep{6}{} \oplus \Irrep{\overline{15}}{} \oplus \Irrep{\overline{15'}}{} \ .
\ee
Again, any ambiguity in the naming of irreps (\eg\ $\Irrep{6}{}$ and $\Irrep{\bar{6}}{}$) is lifted by the corresponding Young diagrams.
We thus have three possible symmetries in quark indices (labelled by $\Irrep{\bar{6}}{}$, $\Irrep{15}{}$, $\Irrep{15'}{}$) and three in antiquark indices (labelled by $\Irrep{6}{}$, $\Irrep{\overline{15}}{}$, $\Irrep{\overline{15'}}{}$), leading to nine different irreps characterized by the tensors (duly made traceless) 
\begin{align}
\label{eq:0trace-indices}
\psi^{\left( ij \right.}_{ \left( kl \right. } \, \chi^{\left. mn \right)_{\mathsize{6}{\Irrep{\bar{6}}{}}}}_{ \left. pq \right)_{\mathsize{6}{\Irrep{6}{}}}}
\quad &; \quad
\psi^{\left( ij \right.}_{ \left( kl \right. } \, \chi^{\left. mn \right)_{\mathsize{6}{\Irrep{15}{}}}}_{ \left. pq \right)_{\mathsize{6}{\Irrep{6}{}}}}
\quad ; \quad
\psi^{\left( ij \right.}_{ \left( kl \right. } \, \chi^{\left. mn \right)_{\mathsize{6}{\Irrep{15'}{}}}}_{ \left. pq \right)_{\mathsize{6}{\Irrep{6}{}}}}
\quad ; \nn \\ 
\psi^{\left( ij \right.}_{ \left( kl \right. } \, \chi^{\left. mn \right)_{\mathsize{6}{\Irrep{\bar{6}}{}}}}_{ \left. pq \right)_{\mathsize{6}{\Irrep{\overline{15}}{}}}}
\quad &; \quad
\psi^{\left( ij \right.}_{ \left( kl \right. } \, \chi^{\left. mn \right)_{\mathsize{6}{\Irrep{15}{}}}}_{ \left. pq \right)_{\mathsize{6}{\Irrep{\overline{15}}{}}}}
\quad ; \quad
\psi^{\left( ij \right.}_{ \left( kl \right. } \, \chi^{\left. mn \right)_{\mathsize{6}{\Irrep{15'}{}}}}_{ \left. pq \right)_{\mathsize{6}{\Irrep{\overline{15}}{}}}}
\quad ; \\
\psi^{\left( ij \right.}_{ \left( kl \right. } \, \chi^{\left. mn \right)_{\mathsize{6}{\Irrep{\bar{6}}{}}}}_{ \left. pq \right)_{\mathsize{6}{\Irrep{\overline{15'}}{}}}}
\quad &; \quad
\psi^{\left( ij \right.}_{ \left( kl \right. } \, \chi^{\left. mn \right)_{\mathsize{6}{\Irrep{15}{}}}}_{ \left. pq \right)_{\mathsize{6}{\Irrep{\overline{15'}}{}}}}
\quad ; \quad 
\psi^{\left( ij \right.}_{ \left( kl \right. } \, \chi^{\left. mn \right)_{\mathsize{6}{\Irrep{15'}{}}}}_{ \left. pq \right)_{\mathsize{6}{\Irrep{\overline{15'}}{}}}}
\quad . \nn 
\end{align}

The Young diagrams of these $\sun$ irreps are obtained by juxtaposing the Young diagrams of the four-quark system displayed in \eq{6times6} and of the four-antiquark system displayed in \eq{6bartimes6bar}, and the irrep names are as usual given by their dimensions for $\Nc=3$:  
\begin{align}
\label{irreps-0trace}
&\ba{c}  \Irrep{0}{c} \\ \tikeq{\bs[scale=0.7]\tyng(0cm,0cm,4,4,2^4)\es} \ea  \ \ \ \ \ \ ; \ \ \ \,
\ba{c} \Irrep{0}{d} \\ \tikeq{\bs[scale=0.7]\tyng(0cm,0cm,5,3,2^4)\es} \ea \ \ \ \ \ \ ; \ \ \ \,
\ba{c}  \Irrep{{28}}{} \\ \tikeq{\bs[scale=0.7]\tyng(0cm,0cm,6,2^5)\es} \ea  \ \ \ \ \ \ ; 
\nn \\[1mm]
&\ba{c} \Irrep{\bar{0}}{d} \\ \tikeq{\bs[scale=0.7]\tyng(0cm,0cm,5,5,3^4,2)\es} \ea \ \ \ \ ; \ \ \
\ba{c} \Irrep{0}{e} \\ \tikeq{\bs[scale=0.7]\tyng(0cm,0cm,6,4,3^4,2)\es}   \ea  \ \ \ \ ; \ \ \ 
\ba{c}  \Irrep{81}{} \\ \tikeq{\bs[scale=0.7]\tyng(0cm,0cm,7,3^5,2)\es} \ea \ \ \ \ ; 
\\[1mm]
&\ba{c} \Irrep{\overline{28}}{} \\  \tikeq{\bs[scale=0.7]\tyng(0cm,0cm,6^2,4^5)\es} \ea \ \ ; \ \ \ \!
\ba{c} \Irrep{\overline{81}}{}  \\ \tikeq{\bs[scale=0.7]\tyng(0cm,0cm,7,5,4^5)\es} \ea  \ \ ; \ \ \,
\ba{c} \Irrep{125}{} \\ \tikeq{\bs[scale=0.7]\tyng(0cm,0cm,8,4^6)\es} \ea \ \ . \nn \\ \nn 
\end{align}

The naming of the irreps $\Irrep{28}{}$ and $\Irrep{\overline{28}}{}$ (as well as $\Irrep{81}{}$ and $\Irrep{\overline{81}}{}$) is in line with the convention mentioned earlier for ${\rm SU}(3)$ irreps (see paragraph after \eq{irreps-2traces}). Note, however, that this convention does not apply to the $\sun$ irreps $\Irrep{0}{d}$ and $\Irrep{\bar{0}}{d}$, which are absent when $\Nc=3$. For $\Nc > 3$ these two irreps are uniquely defined by their Young diagram, and their names is a matter of choice.

\subsection{Sum up} 
\label{sec:sumup}

Collecting the irreps~\eq{irreps-4traces}, \eq{irreps-3traces}, \eq{irreps-2traces}, \eq{irreps-1trace} and \eq{irreps-0trace} arising from taking in~\eq{psi-chi} a decreasing number of traces (from four to zero, respectively), we obtain the decomposition of $\Irrep{27}{} \otimes \Irrep{27}{}\,$ into the sum of 26 irreps: 
\begin{align}
\label{27times27-deco}
\Irrep{27}{} \otimes \Irrep{27}{} &= \Irrep{1}{} \oplus  \Irrep{8}{+} \oplus  \Irrep{8}{-} \oplus  \Irrep{0}{a} \oplus  \Irrep{10}{} \oplus  \Irrep{\overline{10}}{} \oplus  \Irrep{27}{+} \oplus  \Irrep{27'}{\!\! +} \oplus  \Irrep{27}{-} \nn \\ 
&\oplus   \ \Irrep{0}{b+} \oplus  \Irrep{0}{b-} \oplus  \Irrep{35}{+} \oplus  \Irrep{35}{-} \oplus  \Irrep{\overline{35}}{+} \oplus  \Irrep{\overline{35}}{-} \oplus  \Irrep{64}{+} \oplus  \Irrep{64}{-} \nn \\  
&\oplus \  \Irrep{0}{c} \oplus  \Irrep{0}{d} \oplus  \Irrep{\bar{0}}{d} \oplus  \Irrep{0}{e} \oplus  \Irrep{28}{} \oplus  \Irrep{\overline{28}}{} \oplus   \Irrep{81}{} \oplus  \Irrep{\overline{81}}{} \oplus  \Irrep{125}{}  \ . && 
\end{align}
These irreps have been named for convenience after their dimension for $\Nc=3$, but we recall that each $\sun$ irrep is uniquely defined by its associated Young diagram. 
The seven irreps denoted as $\Irrep{0}{}$ (or $\Irrep{\overline{0}}{}$) are absent for $\Nc=3$.
As should be clear from our derivation, the Young diagram of each irrep in \eq{27times27-deco} can be obtained by juxtaposing a Young diagram containing columns with 1 or 2 boxes (corresponding to the irrep of the quark sub-system remaining after taking a given number of traces in~\eq{psi-chi}), and a Young diagram containing columns with $\Nc-2$ or $\Nc-1$ boxes (corresponding to the irrep of the antiquark sub-system). Such a joined Young diagram can thus be represented by a list of four numbers, giving the number of columns with 1, 2, $\Nc-2$ and $\Nc-1$ boxes, respectively. For $\Nc =5$, these four numbers coincide with the Dynkin indices of the irrep, and each $\sun$ irrep $R$ in~\eq{27times27-deco} can thus be labelled by its ``${\rm SU}(5)$ Dynkin name'', $R \equiv [abcd]$.\footnote{As a consequence, for an irrep $R \equiv [abcd]$ in $\Irrep{27}{} \otimes \Irrep{27}{}$, we obviously have $\overline{R} \equiv [dcba]$.}

The $\sun$ irreps of the decomposition~\eq{27times27-deco} are listed in Table~\ref{table:irreps} of Appendix~\ref{app:A}, with their ``${\rm SU}(3)$ names'' and ``${\rm SU}(5)$ Dynkin names'', the latter defining unambiguously the Young diagram of each irrep. We also list the dimensions and quadratic Casimirs of the irreps, which directly follow from the Young diagrams.\footnote{The dimensions are given by the well-known ``factor over hooks rule''~\cite{Georgi:1999wka}, and the quadratic Casimir of an $\sun$ irrep of Dynkin indices $[a_1 a_2 \ldots a_{\Nc-1}]$ by (see \eg \ \cite{White:1992aa}) 
\be
C_2(R) = \frac{T_F}{\Nc} \sum_{m=1}^{\Nc-1} (\Nc-m) a_m  \left[ m(\Nc+a_m) + 2 \sum_{n < m} n a_n \right] \ , \nn
\ee
where $T_F$ is the normalization factor in $\tr t_F^a t_F^b = T_F \delta^{ab}$ ($T_F = \frac{1}{2}$ in our convention, see footnote~\ref{foot:birdtrack-conv}). 
The latter expression can also presented in the form (see \eg\ \cite{Gross:1993hu} or \cite{Dietrich:2006cm}) 
\begin{align*}
C_2(R) = T_F \left\{ b \frac{\Nc^2-b}{\Nc} + \sum_i r^2_i - \sum_i c^2_i \right\} \ , 
\end{align*}
where $b$ is the number of boxes of the irrep's Young diagram, $r_i$ the number of boxes in the $i^{th}$ row, and $c_i$ the number of boxes in the $i^{th}$ column. 
}
In Table~\ref{table:irreps}, the irreps are ordered with decreasing number of traces, and increasing Casimir (for sufficiently large $\Nc$).

\section{Action of the Casimir operator in the color space of $\Irrep{27}{} \otimes \Irrep{27}{}$}
\label{sec:ActionCasimir}

In this section we derive the action of the quadratic Casimir operator $\hat{C}_2$ in the full color space of $\Irrep{27}{} \otimes \Irrep{27}{}$. This will enable us to obtain, in section~\ref{sec:filtering}, most of the projectors on the irreps of the decomposition~\eq{27times27-deco}, as well as transition operators between equivalent irreps, almost directly from the eigenvectors of $\hat{C}_2$. 

We begin with a few general remarks and introduce some pictorial notations (section~\ref{sec:generalities}). 
We then choose a convenient basis of the full color space of $\Irrep{27}{} \otimes \Irrep{27}{}$ (section~\ref{sec:Basis_of_tensors}), and determine the action of $\hat{C}_2$ in this basis (section~\ref{sec:Casimir-action}). 

\subsection{Generalities}
\label{sec:generalities}

Each projector on an irrep $\al$ of the decomposition~\eq{27times27-deco} is a map of the vector space ${\cal E} \equiv \psi \otimes \chi \sim \Irrep{27}{} \otimes \Irrep{27}{}$ to itself, represented pictorially as 
\be
\label{eq:proj-alpha}
\Proj{\al} \ = \ \tensors{\Proj{\al}}{0.25} \ \equiv \ \projdash{\al}{0.26} \ \ ,
\ee
where we used the shorthand notation \eq{dashproj-def}. 

The projectors will obviously satisfy the completeness relation 
\be
\label{comp-rel}
\sum_{\al} \Proj{\al} \ = \ \sum_{\al}  \ \projdash{\al}{0.26}  \ = \  \identitydash{0.26}  \ , 
\ee
where $\,\identitydash{0.15}\,$ denotes the identity of ${\cal E}$:
\be
\identitydash{0.26} \ \equiv \ \fullidentity{0.25} \ \ \ . 
\ee

If $\al$ and $\beta$ are equivalent irreps, the $\al \to \beta$ transition operator can be defined as (see \eg\ Ref.~\cite{Peigne:2024srm}) 
\be
\label{trans-def}
\Trans{\al \to \beta} \ \equiv \ \transdash{\al}{\beta}{0.26} \ \ , 
\ee
where the middle blob stands for any map of ${\cal E} \to {\cal E}$ such that the result is non-zero. It follows from this definition that a transition operator is uniquely defined to within one global factor.
From Table~\ref{table:irreps}, we see that among the 26 irreps, there are eight pairs of equivalent irreps, and thus 16 transition operators. However, for simplicity the transition operators $\Trans{\al \to \beta}$ and $\Trans{\beta \to \al}$ associated with a given pair $(\al,\beta)$ of equivalent irreps can obviously be chosen such that $\Trans{\beta \to \al} =\Trans{\al \to \beta}^\dagger\,$.

We stress that the set of the 42 operators $\Proj{\al}$ and $\Trans{\al \to \beta}$ constitutes a basis of the full color space under consideration (\ie\ the space of all color maps from ${\cal E} \to {\cal E}$). Indeed, considering any color map $\cal M$, 
\be
\label{eq:color-map}
\tensors{\cal M}{0.25} \ \equiv \ \projdash{\cal M}{0.26} \ \ ,
\ee
and inserting to the right and to the left the completeness relation \eq{comp-rel}, we readily see using Schur's lemma that $\cal M$ must be a linear combination of the set $\{ \Proj{\al}, \Trans{\al \to \beta} \}$. 

\subsection{Basis of tensors ${\cal T}$} 
\label{sec:Basis_of_tensors}

The aim of section~\ref{sec:filtering} will be to explicitly determine each element of the basis $\{ \Proj{\al}, \Trans{\al \to \beta} \}$. This will follow from the knowledge of the Casimir operator $\hat{C}_2$ in another, simpler basis of the full color space, which we now describe. 

First, let us remind that an irrep is obtained by taking a given number of traces in \eq{psi-chi}, then specifying the symmetry properties of the resulting tensor (w.r.t.~the remaining free indices) and making it traceless. Note also that as a consequence of Schur's lemma, a transition operator $\Trans{\al \to \beta}$ must be characterized by the same number of traces as the irreps $\al$ and $\beta$. 
It is thus natural to organize the basis $\{ \Proj{\al}, \Trans{\al \to \beta} \}$ in subsets corresponding to the same number of traces. From section~\ref{sec:index-method}, the subsets characterized by a decreasing number of traces are $\{ \Proj{\Irrep{1}{}} \}$ (4 traces), $\{ \Proj{\Irrep{8}{+}}, \Proj{\Irrep{8}{-}}, \Trans{\Irrep{8}{+}\!\to\!\Irrep{8}{-}}, \Trans{\Irrep{8}{-}\!\to\!\Irrep{8}{+}} \}$ (3 traces), the set characterized by 2 traces containing 6 Hermitian projectors (on the irreps~\eq{irreps-2traces}) and 6 transition operators (between the equivalent irreps $\Irrep{27}{+}$, $\Irrep{27'}{\!\!+}$ and $\Irrep{27}{-}$), etc\ldots 

From the above observations, we easily infer that a simple basis of 42~\footnote{We thank P. Cvitanovi{\'c} for pointing out that this is indeed the expected solution~\cite{Adams1986-ea}.} tensors allowing to construct the set $\{ \Proj{\al}, \Trans{\al \to \beta} \}$ is the basis $\mathcal{T}$ given in Appendix~\ref{app:B1} (see Table~\ref{table:TensorBasis}), where for convenience the pictorial expression of a tensor $\tau\in\mathcal{T}$ is abbreviated as 
\be
\label{eq:tensor-abbrev}
\simpletensors{\cal T} \ \equiv \ \tensors{\cal T}{0.25}  \ \ , 
\ee
the ellipse blob of the l.h.s.~implicitly including the projectors $\Proj{27}$.

Indeed, let us consider the subsets of $\{ \Proj{\al}, \Trans{\al \to \beta} \}$ with a decreasing number of traces, starting with the subset $\{ \Proj{\Irrep{1}{}} \}$ characterized by 4 traces. Since the tensor $S$ (see Table~\ref{table:TensorBasis}) is the only map of ${\cal E} \to {\cal E}$ (up to an overall factor) satisfying this criterion, we clearly have $\Proj{\Irrep{1}{}} \propto S$, \ie\ $\{ S \}$ spans the same space as $\{ \Proj{\Irrep{1}{}} \}$. 
Then, the vector space spanned by $\{ \Proj{\Irrep{8}{+}}, \Proj{\Irrep{8}{-}}, \Trans{\Irrep{8}{+}\!\to\!\Irrep{8}{-}}, \Trans{\Irrep{8}{-}\!\to\!\Irrep{8}{+}} \}$ is characterized by 3 traces and orthogonal to $\Proj{\Irrep{1}{}}$. To construct a basis of the vector space spanned by $\{ \Proj{\Irrep{1}{}} \} \cup \{ \Proj{\Irrep{8}{+}}, \Proj{\Irrep{8}{-}}, \Trans{\Irrep{8}{+}\!\to\!\Irrep{8}{-}}, \Trans{\Irrep{8}{-}\!\to\!\Irrep{8}{+}} \}$, we thus need, in addition to the tensor $S$, four linearly independent tensors having each a non-zero overlap with $\{ \Proj{\Irrep{8}{+}}, \Proj{\Irrep{8}{-}}, \Trans{\Irrep{8}{+}\!\to\!\Irrep{8}{-}}, \Trans{\Irrep{8}{-}\!\to\!\Irrep{8}{+}} \}$ (and which do not need to be orthogonal to $\Proj{\Irrep{1}{}}$). A simple choice for these four tensors is $\{ B_1, B_2, B_3, B_4 \}$ (see Table~\ref{table:TensorBasis}). Indeed, by using the Fierz identity in the $q \bar{q}$ intermediate state of each tensor $B_i$, it is clear that these tensors have a non-zero overlap with the irreps $\Irrep{8}{+}$ and $\Irrep{8}{-}$ characterized by 3 traces (as well as with the singlet $\Irrep{1}{}$ characterized by 4 traces).  
Thus, the operators $\{ \Proj{\Irrep{1}{}}, \Proj{\Irrep{8}{+}}, \Proj{\Irrep{8}{-}}, \Trans{\Irrep{8}{+}\!\to\!\Irrep{8}{-}}, \Trans{\Irrep{8}{-}\!\to\!\Irrep{8}{+}} \}$ are linear combinations of $\{ S,B_1,B_2,B_3,B_4 \}$. 

The above can be directly generalized by joining subsets characterized by a smaller number of traces. 
The basis $\mathcal{T}$ of tensors (see Table~\ref{table:TensorBasis}) is thus obtained by joining the subsets $\{ S \}$, $\{ B_1 \ldots B_4 \}$,  $\{ C_1 \ldots C_4, Z_1 \ldots  Z_4, H_1\ldots  H_4 \}$, $\{ J_1 \ldots J_8, K_1 \ldots K_8 \}$, and $\{ L_1 \ldots L_9 \}$, classified by the number $m$.  
A tensor in `class $m$' has $m$ `traces' (\ie, contractions between a quark and an antiquark index) on each side of its defining birdtrack, and $4-m$ $q \bar{q}$ pairs in its middle intermediate state. Conversely, a projector or transition operator characterized by $n$ traces must be a linear combination of tensors $\tau$ containing $4 - m$ $q \bar{q}$ pairs with  $m \geq n$.

In the following we will need to know how complex conjugation, hermitian conjugation (in s-channel), and permutation operators $\sigma_{ij}$ (exchanging the `generalized partons $\Irrep{27}{}$' $i$ and $j$), act on the 42 tensors $\tau \in {\cal T}$. This can be obtained {\it visually} by applying these operators to the birdtracks listed in Appendix~\ref{app:B1}. The results are collected in Appendix~\ref{app:B2}. We will also need the traces of the tensors $\tau$, which are given in Appendix~\ref{app:B3}.

\subsection{Action of $\hat{C}_2$ in basis ${\cal T}$} 
\label{sec:Casimir-action}

From Table~\ref{table:irreps}, we can see that a big step towards determining Hermitian projectors and transition operators would be taken if we knew the eigenspaces of the quadratic Casimir operator $\hat{C}_2$. The goal of the present section is to determine the action of $\hat{C}_2$ in the basis ${\cal T}$ of the full color space, from which the eigensystem of $\hat{C}_2$ can be directly obtained using an appropriate symbolic calculation code. (The eigenspaces of $\hat{C}_2$ will then be used in section~\ref{sec:filtering} as an efficient filter of the full set of operators $\{ \Proj{\al}, \Trans{\al \to \beta} \}$, the final filtering being done with the help of complex conjugation, hermitian conjugation, and permutation operators...) 

By taking the example of a specific tensor, we show in section~\ref{sec:C2action} how the action of $\hat{C}_2$ on a tensor $\tau \in {\cal T}$ can be simply obtained pictorially. The results for $\hat{C}_2 \cdot \tau$ found using this procedure for all tensors of the basis are presented in section~\ref{sec:C2matrix}.

\subsubsection{Pictorial derivation}
\label{sec:C2action}

The action of $\hat{C}_2$ on a tensor $\tau$ can be defined pictorially as: 
\be
\label{C2hat-def}
\hat{C}_2 \cdot \tau \ \equiv \ \Casimirdef{\tau} \ \ , 
\ee
where a grey ellipse crossed by a set of parton lines denotes the sum of all attachments of the gluon to those lines. 

Let's illustrate how this can be simply evaluated on the example of the tensor $Z_3$ (defined in Table~\ref{table:TensorBasis}): 
\begin{align}
\label{Z3-example}
\hat{C}_2 \cdot Z_3 \ = \ \TFPscaled{-1.1}{1.1}{0} \ \ = \ \ \TFPscaled{-0.4}{0.}{-0.2} \ \ . 
\end{align}
Here we used color conservation (see footnote~\ref{foot:birdtrack-conv}), which in diagrammatic form simply means that a grey ellipse can be moved horizontally to be in any intermediate state of the tensor $Z_3$. It is convenient to place the two ellipses in the middle intermediate state of $Z_3$, as in the r.h.s.~of~\eq{Z3-example}.
In this way the operator $\hat{C}_2$ acts on a state containing four partons instead of the eight incoming partons. Let's number these four partons from top to bottom by $i = 1 \dots 4$. Thus, the operator $\hat{C}_2$ can be formally written as:
\be
\hat{C}_2  \ = \ \sum_{i,j} T_i^a T_j^a \ = \ \sum_{i} T_i^a T_i^a + 2 \sum_{i<j} T_i^a T_j^a \ ,
\ee
where $T_i^a$ denotes the $\sun$ generator in the representation of parton $i$. Since parton $i$ is either a quark or an antiquark, we have $T_i^a T_i^a = C_F$ and we obtain 
\be
\label{C2on4quarks}
\hat{C}_2 \cdot \left( \, \IDqqqq \, \right) = 4 C_F  \left( \, \IDqqqq \, \right) 
- 2 \left( \, \potential{2}{1} \, + \, \potential{2}{-1} \, + \, \potential{2}{-2} \, + \, \potential{1}{-1} \, + \, \potential{1}{-2}\, + \, \potential{-1}{-2} \right) \, .
\ee
Note that in the r.h.s.~of~\eq{C2on4quarks}, the minus sign in front of the second term simply follows from our sign convention to define the birdtrack for the antiquark generator (see footnote~\ref{foot:birdtrack-conv}). 

The color graph representing a gluon exchange between partons $i$ and $j$ is sometimes referred to as the `color potential' between the two partons~\cite{Dokshitzer:1995fv}. We will denote this operator, including the factor $-2$, as $\mathcal{V}_{ij}$. 
Inserting~\eq{C2on4quarks} into~\eq{Z3-example} yields several contributions. The first term of \eq{C2on4quarks} is simply proportional to the identity operator, and its contribution to~\eq{Z3-example} is therefore $4C_F Z_3$. 
The contributions of the `color potentials' $\mathcal{V}_{ij}$ can be evaluated using a standard algorithm allowing to systematically simplify color graphs (see \eg~\cite{Keppeler:2017kwt}), which in the present case simply amounts to get rid of the exchanged gluon by using the Fierz identity. 

Let us first consider the potential acting on a quark pair or an antiquark pair, namely, the terms $\mathcal{V}_{13}$ and $\mathcal{V}_{24}$ in~\eq{C2on4quarks}. One uses the Fierz identity in the form: 
\be
\label{eq:V13}
-2\ \Vonethree{1} \ = \ \X{1} \ - \frac{1}{\Nc}\ \Iqq{1} \ , 
\ee
or its complex conjugate (obtained by reversing quark arrows).
When inserting \eq{eq:V13} in \eq{Z3-example}, we see that the final parton lines of \eq{eq:V13} enter the same (quark or antiquark) symmetrizer on the right side of the birdtrack in~\eq{Z3-example}. Therefore, for this contribution the permutation operator $\,\X{0.7}\,$ can be replaced by the identity $\Iqq{0.6}\,$, and the contributions to \eq{Z3-example} from the terms $\mathcal{V}_{13}$ and $\mathcal{V}_{24}$ in \eq{C2on4quarks} sum up to $2(1-\frac{1}{\Nc}) Z_3\,$. 

Let us now consider the potential acting on a $q\bar{q}$ pair, such as the terms $\mathcal{V}_{12}$, $\mathcal{V}_{14}$, $\mathcal{V}_{23}$ and $\mathcal{V}_{34}$.
One uses the Fierz identity in the form: 
\be
\label{eq:V12}
-2\ \Vonetwo{1} \ = \ - \ \Sqq{1} \ + \frac{1}{\Nc}\ \Iqqbar{1} \ .
\ee
Using \eq{eq:V12} to evaluate the contribution of $\mathcal{V}_{12}$ and $\mathcal{V}_{34}$ to \eq{Z3-example}, we see that the first (singlet) term in the r.h.s.~of \eq{eq:V12} connects to a $q\bar{q}$ pair taken in the same projector $\Proj{\Irrep{27}{}}$, and can be dropped since $\Proj{\Irrep{27}{}}$ is traceless. The sum of these two contributions is thus simply $\frac{2}{\Nc} Z_3$. 
As for the contributions from $\mathcal{V}_{14}$ and $\mathcal{V}_{23}$, the $q\bar{q}$ pair connects to distinct projectors $\Proj{\Irrep{27}{}}$ (on both left and right of the birdtrack in~\eq{Z3-example}), and the first term in the r.h.s.~of \eq{eq:V12} must be kept in this case. In \eq{Z3-example}, that term contributes a tensor corresponding to an additional trace when compared to $Z_3$, thus belonging to the set $\{ B_1, B_2, B_3, B_4 \}$ (see Table~\ref{table:TensorBasis}). We readily identify that the contributions to \eq{Z3-example} of the terms $\mathcal{V}_{14}$ and $\mathcal{V}_{23}$ are $\frac{1}{\Nc} Z_3 - B_1$ and $\frac{1}{\Nc}Z_3 - B_4$, respectively. 

Adding all the above contributions to \eq{Z3-example}, we obtain
\begin{align}
\label{eq:C2Z3}
\hat{C}_2 \cdot Z_3 \ = \ 2(\Nc+1) Z_3 - B_1 - B_4 \ .  
\end{align}
The example of the tensor $Z_3$ illustrates how the action of $\hat{C}_2$ on any element $\tau \in {\cal T}$ can be evaluated pictorially, using color conservation  and the Fierz identity. 

\subsubsection{Full expression of $\hat{C}_2$ in basis ${\cal T}$}
\label{sec:C2matrix}

The procedure used to calculate $\hat{C}_2 \cdot Z_3$ can be applied to any tensor $\tau$, including tensors in a class with a smaller $m$ than $Z_3$, \ie, having a larger number of $q\bar{q}$ pairs in the middle intermediate state. For these tensors, the procedure generates more terms in $\hat{C}_2 \cdot \tau$, due to a larger number of color potentials, but is otherwise exactly similar.

Removing the exchanged gluon with the help of the Fierz identity (as illustrated in the previous section) produces only one of the following three cases: the potential $\mathcal{V}_{k\ell}$ is replaced by (i) the identity; (ii) a permutation; (iii) a trace. Thus, if $\tau$ is a tensor in class $m$, $\hat{C}_2 \cdot \tau$ must be a linear combination of tensors belonging to classes $m$ and $m+1$. Representing the operator $\hat{C}_2$ as a $42 \times 42$ matrix acting on the basis ${\cal T}$, it follows that this matrix is not only sparse,\footnote{This is simply due to the limited number of potential terms, and to the fact that the Fierz identity generates few terms.} but also {\it block upper triangular}, with blocks on the diagonal of sizes $1, 4, 12, 16, 9$ -- the numbers of tensors of class $m$ with decreasing $m$, see Table~\ref{table:TensorBasis}. 

Let us note that in order to obtain $\hat{C}_2 \cdot \tau$ for all $\tau$, we do not need to repeat 41 times the procedure of the previous section. Indeed, within a class $m$ of tensors, the expressions $\hat{C}_2 \cdot \tau$ for some subset of tensors are simply related by permutations $\sigma_{12}$, $\sigma_{34}$, Hermitian conjugation $\sigma_{\rm hc}$, or complex conjugation $\sigma_{\rm cc}$, whose actions on the basis ${\cal T}$ are known (see Table~\ref{tab:CrossingR} in Appendix~\ref{app:B2}). 

As an example, consider the relation \eq{eq:C2Z3} giving the action of $\hat{C}_2$ on $Z_3$. By applying successively $\sigma_{34}$, $\sigma_{\rm hc}$, and $\sigma_{12}$ to this equation and using Table~\ref{tab:CrossingR}, we directly obtain the action of $\hat{C}_2$ on $Z_4$, $Z_2$ and $Z_1$, given in~\eq{EQ:C2Z} below. (The Casimir operator obviously commutes with $\sigma_{\rm hc}$ and $\sigma_{\rm cc}$, but also with $\sigma_{12}$ and $\sigma_{34}$, which follows from the definition~\eq{C2hat-def} and color conservation.) As another example, the tensors $J_1$, $J_2$, $K_1$, $K_2$ are simply related by the action of $\sigma_{12}$, $\sigma_{34}$, $\sigma_{\rm hc}$, $\sigma_{\rm cc}$. Thus, only one of the relations \eq{C2J1}, \eq{C2J2}, \eq{C2K1}, \eq{C2K2} given below needs to be derived as in section~\ref{sec:C2action} to find all four. 

The results for $\hat{C}_2 \cdot \tau$ for any $\tau \in {\cal T}$ are collected below. 
\begin{align}
\label{EQ:C2S} \hat{C}_2 S & = 0  
\end{align}
\vspace{-7mm}
\begin{align}
\label{EQ:C2B} \hat{C}_2 B_i  &= \Nc B_i - S  \hskip 17mm (i=1,\ldots,4)   
\end{align}
\vspace{-7mm}
\begin{align}
\label{EQ:C2C} \hat{C}_2 C_i &= 2(\Nc+1) C_i - 4 B_i  \hskip 1cm (i=1,\ldots,4) \ \ \ 
\end{align}
\vspace{-7mm}
\begin{subequations} \label{EQ:C2Z}
\begin{align}
\hat{C}_2 Z_1 = 2(\Nc+1)Z_1 - (B_3 + B_1), \qquad  \hat{C}_2 Z_3 = 2(\Nc+1)Z_3 - (B_4 + B_1) \\
\hat{C}_2 Z_2 = 2(\Nc+1)Z_2 - (B_4 + B_2), \qquad \hat{C}_2 Z_4 = 2(\Nc+1)Z_4 - (B_3 + B_2)
\end{align}
\end{subequations}
\vspace{-7mm}
\begin{subequations}\label{EQ:C2H}
\begin{align}
\hat{C}_2 H_1 &= 2\Nc H_1 + H_3 + H_4 - B_1 - B_2, \qquad \ \ \hat{C}_2 H_3 = 2\Nc H_3 + H_1 + H_2 \\
\hat{C}_2 H_2 &= 2\Nc H_2 + H_3 + H_4 - B_3 - B_4, \qquad \ \ \hat{C}_2 H_4 = 2\Nc H_4 + H_1 + H_2
\end{align}
\end{subequations}
\vspace{-7mm}
\begin{subequations}\label{EQ:C2J}
\begin{align}
\hat{C}_2 J_1 &= (3\Nc+2)J_1 +2J_3 +2J_5 -C_1 -4H_1 \hskip 20mm  \label{C2J1} \\
\hat{C}_2 J_3 &= (3\Nc+3)J_3 +J_1 +2J_7 -2H_4 \\
\hat{C}_2 J_5 &= (3\Nc+3)J_5 +J_1 +2J_7 -2H_3 \\
\hat{C}_2 J_7 &= (3\Nc+4)J_7 +J_3 +J_5 -Z_1 -Z_3 -H_2
\end{align}
\vspace{-7mm}
\begin{align}
\hat{C}_2 J_2 &= (3\Nc+2)J_2 +2J_6 +2J_4 -C_2 -4H_1  \hskip 20mm \label{C2J2} \\
\hat{C}_2 J_4 &= (3\Nc+3)J_4 +J_2 +2J_8 -2H_4 \\ 
\hat{C}_2 J_6 &= (3\Nc+3)J_6 +J_2 +2J_8 -2H_3 \\
\hat{C}_2 J_8 &= (3\Nc+4)J_8 +J_6 +J_4 -Z_2 -Z_4 -H_2
\end{align}
\end{subequations}
\vspace{-7mm}
\begin{subequations}\label{EQ:C2K}
\begin{align}
\hat{C}_2 K_1 &= (3\Nc+2)K_1 +2K_3 +2K_5 -C_3 -4H_2  \hskip 18mm \label{C2K1} \\ 
\hat{C}_2 K_3 &= (3\Nc+3)K_3 +2K_7 +K_1 -2H_3\\
\hat{C}_2 K_5 &= (3\Nc+3)K_5 +2K_7 +K_1 -2H_4\\
\hat{C}_2 K_7 &= (3\Nc+4)K_7 +K_3 +K_5 -Z_1 -Z_4 -H_1
\end{align}
\vspace{-7mm}
\begin{align}
\hat{C}_2 K_2 &= (3\Nc+2)K_2 +2K_4 +2K_6 -C_4 -4H_2  \hskip 18mm  \label{C2K2} \\ 
\hat{C}_2 K_4 &= (3\Nc+3)K_4 +2K_8 +K_2 -2H_3\\
\hat{C}_2 K_6 &= (3\Nc+3)K_6 +2K_8 +K_2 -2H_4\\
\hat{C}_2 K_8 &= (3\Nc+4)K_8 +K_4 +K_6 -Z_2 -Z_3 -H_1
\end{align}
\end{subequations}
\vspace{-7mm}
\begin{subequations}\label{EQ:C2L}
\begin{align}
\hat{C}_2L_1 &= (4\Nc+4)L_1 +4(L_4+L_5-J_1-J_2) \\
\hat{C}_2L_2 &= (4\Nc+4)L_2 +4(L_6+L_7-K_1-K_2) \\
\hat{C}_2L_3 &= (4\Nc+8)L_3 +(L_4+L_5+L_6+L_7) -(J_7+J_8+K_7+K_8) \\
\hat{C}_2L_4 &= (4\Nc+6)L_4 +4L_3 +L_1 +L_8 -2(J_6+J_5)\\
\hat{C}_2L_5 &= (4\Nc+6)L_5 +4L_3 +L_1 +L_9 -2(J_3+J_4)\\
\hat{C}_2L_6 &= (4\Nc+6)L_6 +4L_3 +L_2 +L_9 -2(K_5+K_6)\\
\hat{C}_2L_7 &= (4\Nc+6)L_7 +4L_3 +L_2 +L_8 -2(K_3+K_4) \\ 
\hat{C}_2L_8 &= (4\Nc+4)L_8 +4(L_4+L_7)\\
\hat{C}_2L_9 &= (4\Nc+4)L_9 +4(L_5+L_6)
\end{align}
\end{subequations}

Note that we have presented the 42 expressions above to emphasize that in the basis ${\cal T}$, the expected large blocks on the diagonal of the (block upper triangular) matrix $\hat{C}_2$ are actually made of smaller blocks. The block of size 12 reduces to blocks of size 4 (see Eqs.~\eq{EQ:C2C}, \eq{EQ:C2Z} and \eq{EQ:C2H}) as well as the block of size 16 (see Eqs.~\eq{EQ:C2J} and \eq{EQ:C2K}). As for the block of size 9 (see \eq{EQ:C2L}), it can be easily reduced to two blocks of size 6 and 3 by a simple change of basis, trading the tensors $L_i$ ($i=1\ldots 9$) for the sets $\{ L_1, L_2, L_3, L_4+L_5, L_6+L_7, L_8+L_9 \}$ and $\{ L_4-L_5, L_6-L_7, L_8-L_9 \}$.

The above relations allow one to express $\hat{C}_2$ as a $42 \times 42$ matrix, which can be implemented using a symbolic calculation software (\eg\ Mathematica \cite{Mathematica}). 
Although $\hat{C}_2$ is a rather large matrix, the determination of its eigenvalues and eigenspaces is very fast due to its sparse and block upper triangular structure. 
One can readily check that the obtained $\hat{C}_2$ eigenvalues and associated multiplicities are correct, as specified by Table~\ref{table:irreps}. 
 
\section{Full color space of $\Irrep{27}{} \otimes \Irrep{27}{}\,$: projectors and transition operators}  
\label{sec:filtering}

The aim of this section is to explicitly derive the 26 Hermitian projectors $\Proj{\al}$ associated with the irreps $\al$ of the decomposition \eq{27times27-deco}, as well as the 16 transition operators between equivalent irreps. 

As we can see from Table~\ref{table:irreps}, the knowledge of the eigenspaces of $\hat{C}_2$ already provides some of the Hermitian projectors (up to their overall normalization which can be trivially fixed),  namely, the projectors on the irreps of Casimir $C_R=0, \, 2\Nc -2,  \, 4 \Nc, \, 4 \Nc+4, \, 4 \Nc+12$, and a good `filtering' of other projectors and transition operators, in subsets corresponding to the $\hat{C}_2$ eigenspaces of dimension greater than unity.

The separation of the latter eigenspaces into 1-dimensional subspaces, each corresponding to one projector or transition operator, will be done as follows. In the case of an eigenspace of $\hat{C}_2$ of dimension 2, corresponding to a pair of complex conjugated irreps $R$ and $\overline{R}$, the precise linear combination of eigenvectors proportional to $\Proj{R}{}$ is identified by requiring its symmetry properties w.r.t.~quark and antiquark indices to coincide with those characterizing the irrep $R$. For an eigenspace of $\hat{C}_2$ of dimension 4 spanned by the two projectors and two transition operators associated to a pair of equivalent irreps, the two irreps are distinguished by their different `symmetry signum' under $\psi \leftrightarrow \chi$. 
The projector on each irrep is thus uniquely determined by the (properly normalized) eigenvector of $\hat{C}_2$ having a specific $\psi \leftrightarrow \chi$ symmetry, and a given transition operator by the eigenvector with specific {\it distinct} symmetries on the left and right under $\psi \leftrightarrow \chi$. 

The sector containing {\it three} equivalent irreps $\Irrep{27}{}$ (see the decomposition~\eq{27times27-deco}) requires a little more work. In that case, the $\psi \leftrightarrow \chi$ symmetry does not allow to distinguish all equivalent irreps ($\Irrep{27}{+}$ and $\Irrep{27'}{\!\!+}$ being both symmetric under $\psi \leftrightarrow \chi$). As a consequence, the projectors $\Proj{\Irrep{27}{+}}$ and $\Proj{\Irrep{27'}{\!\!+}}$ are not uniquely determined separately -- only their sum is. In order to address this case, and in particular to derive all solutions for the pair $(\Proj{\Irrep{27}{+}},\Proj{\Irrep{27'}{\!\!+}})$, it will be convenient to express projectors and transition operators in terms of {\it clebsches}~\cite{Cvitanovic:2008zz}. So we start with a reminder of clebsches in section~\ref{sec:clebsches}, before deriving all projectors and transition operators in sections~\ref{sec:eigenspaces-dim1} to \ref{sec:eigenspaces-dim9}.\footnote{Note that in the sectors different from the $\Irrep{27}{}$'s, the $\hat{C}_2$ eigenvectors with specific symmetries (with respect to quark/antiquark indices or under $\psi \leftrightarrow \chi$) give all projectors and transition operators directly, without the aid of clebsches. However, clebsches can still be useful if we want to satisfy a specific normalization convention for transition operators, \eg\ \eq{eq:trans-norm}. Although a little anecdotal in our study, we will show how to adjust this normalization for the transition operators encountered in section~\ref{sec:eigenspaces-dim4} and following.} 

\subsection{A little help from clebsches} 
\label{sec:clebsches}

Suppose that $\al$ is an irrep corresponding to $n$ traces in the decomposition of $\psi \otimes \chi \sim \Irrep{27}{} \otimes \Irrep{27}{}$. It is straightforward to show that the projector~\eq{eq:proj-alpha} can be put in the form 
\be
\label{eq:proj-alpha-G}
\Proj{\al} \ = \ \projdashG{G}{\al}{0.26} \ , 
\ee
where $\,\irrepdash{\al}{0.26} \,$ denotes the Hermitian projector on the irrep $\al$ of a system containing $4-n$ $q \bar{q}$ pairs. (Recall that a dashed line without an irrep label denotes the ubiquitous irrep $\Irrep{27}{}$, see \eq{dashproj-def}.) A more compact notation for~\eq{eq:proj-alpha-G} is~\cite{Cvitanovic:2008zz}:
\be
\label{eq:proj-alpha-CG}
\Proj{\al} \ = \ \projdashCG{}{}{\al}{0.26}{1} \  = \ c_\al^\dagger c_\al  \ , 
\ee
where the operator
\be
\label{eq:clebsch-alpha}
c_\al \ = \ \CGdashG{G}{\al}{0.26} \ \equiv \ \CGdash{}{\al}{0.26}{1}
\ee
is a so-called Clebsch-Gordan operator or simply {\it clebsch}~\cite{Cvitanovic:2008zz}, mapping $\Irrep{27}{} \otimes \Irrep{27}{}$ to the irrep $\al$ of the $(q \bar{q})^{\otimes (4-n)}$ system.\footnote{\label{foot:rightop-applies-first}In our convention, in the product $c_\al^\dagger c_\al$ the operator $c_\al$ applies first and is therefore drawn on the left in the pictorial representation of the projector \eq{eq:proj-alpha-CG}.} Note that the correct normalization of $\Proj{\al}$ follows automatically from \eq{eq:proj-alpha-CG}. Indeed, using Schur's lemma we can write:
\be
\Proj{\al}^2 \ = \ \projdashCG{}{}{\al}{0.26}{0.6} \projdashCG{}{}{\al}{0.26}{0.6} = \frac{\tr{\left( \CGdashdag{}{\al}{0.26}{0.8} \CGdash{}{\al}{0.26}{0.8} \right) }}{K_\al} \, c_\al^\dagger c_\al = \frac{\tr{ c_\al c_\al^\dagger}}{K_\al} \, c_\al^\dagger c_\al = \frac{\tr{\Proj{\al}}}{K_\al} \, \Proj{\al} = \Proj{\al}  \ ,
\ee
where $K_\al$ is the dimension of the irrep $\al$. 

Let's now suppose that the decomposition of $\Irrep{27}{} \otimes \Irrep{27}{}$ contains two irreps equivalent to the irrep $\al$, of associated projectors $\Proj{\al 1}$ and $\Proj{\al 2}$ (satisfying $\Proj{\al 1} \Proj{\al 2} = \Proj{\al 2} \Proj{\al 1}=0$). 
The projectors $\Proj{\al 1}$ and $\Proj{\al 2}$ can both be put into the form \eq{eq:proj-alpha-CG},\footnote{In~\eq{eq:P1-CG} and~\eq{eq:P2-CG}, the intermediate state $\irrepdash{\al}{0.26}$ is the same, since for the system $(q \bar{q})^{\otimes (4-n)}$, the irrep $\al$ must appear only once.}
\begin{subequations}
\label{eq:P12-CG} 
\begin{align}
\Proj{\al 1} \ &= \ \projdashCG{1}{1}{\al}{0.26}{1} \  = \ c_{\al 1}^\dagger c_{\al 1}  \ ,  \label{eq:P1-CG}  \\[2mm] 
\Proj{\al 2} \ &= \ \projdashCG{2}{2}{\al}{0.26}{1} \ = \  c_{\al 2}^\dagger c_{\al 2} \ , \label{eq:P2-CG} 
\end{align}
\end{subequations}
and we thus have two different clebsches: 
\be
\label{eq:CG-def}
c_{\al i} \ = \ \CGdashG{G_i}{\al}{0.26} \ \equiv \ \CGdash{i}{\al}{0.26}{1} \ . 
\ee
Expressing the product $\Proj{\al 2} \Proj{\al 1}$ as (using again Schur's lemma): 
\begin{align}
\Proj{\al 2} \Proj{\al 1} \ &= \ \projdashCG{1}{1}{\al}{0.26}{1} \projdashCG{2}{2}{\al}{0.26}{1} \ = \ \frac{ \tr{(c_{\al 2} c_{\al 1}^\dagger)} }{K_\al} \, c_{\al 2}^\dagger c_{\al 1} \ , 
\end{align}
we infer that the orthogonality condition $\Proj{\al 2} \Proj{\al 1} = 0$ is equivalent to $\tr{(c_{\al 1}^\dagger c_{\al 2})}=0$. 

As for the transition operator $\Trans{\al 1 \to \al 2}$ defined by \eq{trans-def}, it directly follows from \eq{eq:P12-CG} and Schur's lemma that $\Trans{\al 1 \to \al 2} \propto c_{\al 2}^\dagger c_{\al 1}$. Although the overall normalization of a transition operator is arbitrary, a common choice is to define $\Trans{\al 1 \to \al 2} \equiv c_{\al 2}^\dagger c_{\al 1}$. Within this convention, the transition operators $\Trans{\al 1 \to \al 2}$ and $\Trans{\al 2 \to \al 1} = \Trans{\al 1 \to \al 2}^\dagger$ are normalized as
\begin{subequations}
\label{eq:trans-norm} 
\begin{align}
\Trans{\al 1 \to \al 2}^\dagger \Trans{\al 1 \to \al 2}  \ &= 
\ \projdashCG{1}{2}{\al}{0.26}{1}  \projdashCG{2}{1}{\al}{0.26}{1} \ = \  \Proj{\al 1} \ , \\[2mm]
\Trans{\al 2 \to \al 1}^\dagger \Trans{\al 2 \to \al 1}  \ &= 
\ \projdashCG{2}{1}{\al}{0.26}{1}  \projdashCG{1}{2}{\al}{0.26}{1} \ = \  \Proj{\al 2} \ .
\end{align}
\end{subequations}

In summary, in the space of tensors mapping $\Irrep{27}{} \otimes \Irrep{27}{} \to \Irrep{27}{} \otimes \Irrep{27}{}$, the `sector' of the equivalent irreps $\al 1$ and $\al 2$, spanned by the tensors $\{ \Proj{\al 1} , \Proj{\al 2}, \Trans{\al 1 \to \al 2}, \Trans{\al 2 \to \al 1}\}$, is fully determined by the clebsches $c_{\al 1}$ and $c_{\al 2}$. For convenience we may collect the operators $\{ \Proj{\al 1} , \Proj{\al 2}, \Trans{\al 1 \to \al 2}, \Trans{\al 2 \to \al 1}\}$ in the $2\times 2$ {\it matrix of operators}~\cite{Alcock-Zeilinger:2016cva}
\begin{align}
\label{eq:P12T12-CG} 
& {\rm M}_\al \ \equiv \ 
\left( \begin{matrix}  \Proj{\al 1}  \ & \ \Trans{\al 2 \to \al 1}\  \\[2mm] 
\Trans{\al 1 \to \al 2} \ & \  \Proj{\al 2}\end{matrix}  \right)  \ = \ 
\left( \begin{matrix} c_{\al 1}^\dagger c_{\al 1} \  & \ c_{\al 1}^\dagger c_{\al 2} \  \\[2mm]
c_{\al 2}^\dagger c_{\al 1}  \ & \ c_{\al 2}^\dagger c_{\al 2} \end{matrix}  \right)  
\ = \ \left( \begin{matrix} c_{\al 1}^\dagger \\[2mm] c_{\al 2}^\dagger \end{matrix}  \right) 
\left( \begin{matrix} c_{\al 1} \\[2mm] c_{\al 2}  \end{matrix}  \right)^{\!\!\intercal} \ .
\end{align} 
The fact that the entries of this matrix are projectors and transition operators satisfying the required normalization and orthogonality conditions directly follows from Schur's lemma and the relation
\begin{align}
\label{eq:ccdag-traces} 
& \tr \left[ ({\rm M}_\al)_{ij} \right] = \tr \left[ c_{\al i}^\dagger c_{\al j} \right]  =  K_\al \, \delta_{ij}  \  ,
\end{align} 
where the trace symbol $\tr$ here designates the trace of birdtrack operators, and thus applies to each matrix element of ${\rm M}_\al$. 

A standard method for obtaining the projectors $\Proj{\al 1}$ and $\Proj{\al 2}$ is to start from two independent `vertices' $v_{\al 1}$ and $v_{\al 2}$, corresponding to different birdtrack operators mapping $\Irrep{27}{} \otimes \Irrep{27}{}$ to the irrep $\al$ of a $(q \bar{q})^{\otimes (4-n)}$ system. These vertices are usually easy to find (see some examples in sections~\ref{sec:eigenspaces-dim4}, \ref{sec:eigenspaces-dim8} and \ref{sec:eigenspaces-dim9}), and must be linear combinations of the clebsches $c_{\al 1}$ and $c_{\al 2}$ constructing the projectors and transition operators collected in ~\eq{eq:P12T12-CG}. So, given a certain choice for $v_{\al 1}$ and $v_{\al 2}$, we need to find some linear combinations 
\begin{subequations}
\label{CG-LC}
\begin{align}
c_{\al 1} \ &= \ a \, v_{\al 1} + b \, v_{\al 2}  \ , \\ 
c_{\al 2} \ &= \ c \, v_{\al 1} + d \, v_{\al 2} \  , 
\end{align}
\end{subequations}
such that \eq{eq:ccdag-traces} holds, which ensures that the operators $c_{\al i}^\dagger c_{\al j}$ are the desired projectors and transition operators. 

It should be noted that clebsches are not unique~\cite{Cvitanovic:2008zz}. Indeed, given a solution $(c_{\al 1},c_{\al 2})$ (thus satisfying~\eq{eq:ccdag-traces}), we directly check that the rotated version 
\begin{align}
\label{CG-rot}
\left( \begin{matrix} c_{\al 1}' \\[2mm] c_{\al 2}'  \end{matrix}  \right) 
= A \cdot \left( \begin{matrix} c_{\al 1} \\[2mm] c_{\al 2}  \end{matrix}  \right) \ , 
\end{align}
where $A$ is an orthogonal matrix ($A^{-1} = A^\intercal$), also satisfies~\eq{eq:ccdag-traces}, namely: 
\begin{subequations}
\begin{align}
\label{eq:ccdag-traces-prime} 
 \tr \left[ ({\rm M}_\al')_{ij} \right] &= \tr \left[ c_{\al i}'^\dagger c_{\al j}' \right]  = K_\al \, \delta_{ij} \ ,
 \\[1mm]
\label{Malphaprime} 
{\rm M}_\al' &\equiv A \cdot {\rm M}_\al \cdot A^\intercal \ . 
\end{align} 
\end{subequations}
This implies that the set of operators $\{ \Proj{\al 1}' , \Proj{\al 2}', \Trans{\al 1 \to \al 2}', \Trans{\al 2 \to \al 1}'\}$  collected by the matrix ${\rm M}_\al'$ is another fully legitimate solution for the two projectors and associated transition operators in the sector of the irrep $\al$. We stress that the rotation~\eq{CG-rot} acts as a similarity transformation for the matrix ${\rm M}_\al$. The trace of ${\rm M}_\al$ (which should not be confused with the trace of birdtrack operators denoted by $\tr$ and acting on the separate {\it matrix elements} of ${\rm M}_\al$) is thus invariant, 
\begin{align}
\label{eq:Malpha-trace}
\Proj{\al 1}' + \Proj{\al 2}' \ = \ 
c_{\al 1}'^\dagger c_{\al 1}' + c_{\al 2}'^\dagger c_{\al 2}' \ = \ c_{\al 1}^\dagger c_{\al 1} + c_{\al 2}^\dagger c_{\al 2} \  =  \ \Proj{\al 1} + \Proj{\al 2} \ .  
\end{align}
The projectors on the two equivalent irreps are not uniquely defined, but {\it their sum is}. 

The above discussion directly generalizes to any sector associated with $n>2$ irreps equivalent to some irrep $\al$. Such a sector is determined by $n$ independent vertices $v_{\al i}$, allowing to build $n$ clebsches $c_{\al i}$, and thus $n$ orthogonal projectors $\Proj{\al i} = c_{\al i}^\dagger c_{\al i}$ and $n(n-1)$ transition operators $\Trans{\al i \to \al j} = c_{\al j}^\dagger c_{\al i}$ (for $i \neq j$), normalized as $\Trans{\al i \to \al j}^\dagger \Trans{\al i \to \al j} = \Proj{\al i}\,$.

\subsection{Eigenspaces of $\hat{C}_2$ of dimension 1}
\label{sec:eigenspaces-dim1}

The Casimir operator $\hat{C}_2$ has five eigenspaces of dimension 1, associated to the eigenvalues 
$C_R=0, \, 2\Nc -2,  \, 4 \Nc, \, 4 \Nc+4, \, 4 \Nc+12$ (see Table~\ref{table:irreps}), corresponding respectively to the irreps $\Irrep{1}{}$, $\Irrep{0}{a}$, $\Irrep{0}{c}$, $\Irrep{0}{e}$ and $\Irrep{125}{}$ appearing only once in the decomposition~\eq{27times27-deco}.  

As already mentioned, in those cases the eigenvector of $\hat{C}_2$ of eigenvalue $C_R$, which is directly obtained from the matrix $\hat{C}_2$ determined in section~\ref{sec:C2matrix}, is proportional to the projector on the corresponding irrep. Using the traces of all tensors of the basis ${\cal T}$ given in Appendix~\ref{app:B3}, the overall normalization of each projector is then fixed to satisfy $\tr \Proj{\al} = K_\al$, where the dimension $K_\al$ of the irrep $\al$ is given in Table~\ref{table:irreps}. 

Using the abbreviated notations (for any tensor $\tau \in \mathcal{T}$)  
\begin{align}
\label{tau-ij-notation} 
& \tau_{ij} \ \equiv \ \tau_i + \tau_j \ \ ; \ \ \ \ \overline{\tau}_{ij} \ \equiv \ \tau_i - \tau_j  \ ,
\end{align}
the obtained projectors can be written as:
\begin{align}
\label{eq:projP1}
\Proj{\Irrep{1}{}} \ &= \ \frac{4\, S}{(\Nc-1)\Nc^2(\Nc+3)}  \\[4mm]
\label{eq:projP0a}
\Proj{\Irrep{0}{a}} \ &= \ \frac{4}{(\Nc+1) (\Nc+3)} \left\{ H_{12}-H_{34} - \frac{B_{12}+B_{34}}{\Nc-2} + \frac{2 \, S}{(\Nc-2) (\Nc-1)} \right\} \\[6mm]
\label{eq:projP0c}
\Proj{\Irrep{0}{c}} \ &= \ \frac{L_{12} + 4 L_3 -2 (L_{45}+L_{67})+L_{89}}{9}   \notag \\[2mm]
&- \frac{4 (J_{12}+K_{12} +J_{78}+K_{78} - J_{34}-J_{56}-K_{34}-K_{56})}{9\Nc} \notag \\[2mm]
&+ \frac{2 \left(C_{12}+C_{34}\right)+4\left(Z_{12}+Z_{34}\right)+2 \left(H_{12}+H_{34}\right)}{9(\Nc-1) \Nc}+\frac{18 \left(H_{12}-H_{34}\right)}{9\Nc (\Nc+1)} \notag \\[2mm]
&- \frac{12 \left(B_{12}+B_{34}\right)}{9(\Nc-1) \Nc (\Nc+1)} +\frac{12 S}{9(\Nc-1) \Nc^2 (\Nc+1)} 
\end{align}
\begin{align} 
\label{eq:projP0e}
\Proj{\Irrep{0}{e}} \ &= \ \frac{L_{12}-L_{89}}{4} \nn \\[2mm]
& - \frac{\left(\Nc^2+\Nc-4\right) \left(J_{12}+K_{12}\right) + 8 \left(J_{78}+K_{78}\right) + 2 \Nc  \left(J_{34}+J_{56}+K_{34}+K_{56}\right)}{(\Nc-2)(\Nc+1) (\Nc+4)} \nn \\[2mm]
&+ \frac{\left(C_{12}+C_{34}\right) \left(\Nc^2+\Nc-4\right)+16 \left(Z_{12}+Z_{34}\right)+4 \left(H_{12}+H_{34}\right) \left(\Nc^2+3 \Nc-2\right)}{2 (\Nc-2) (\Nc+1)^2 (\Nc+4)} \nn \\[2mm]
& +\frac{2 \left(H_{12}-H_{34}\right)}{(\Nc+3) (\Nc+4)} - \frac{2 \left(B_{12}+B_{34}\right)}{(\Nc-2) (\Nc+3) (\Nc+4)}+\frac{2 S}{(\Nc-2)(\Nc+1) (\Nc+3) (\Nc+4)} \nn \\[2mm] 
\\ \nn 
\label{eq:projP125}
\Proj{\Irrep{125}{}} \ &= \ \frac{L_{12} + 16  L_3 + 4 (L_{45}+L_{67}) +L_{89}}{36}  \nn \\[2mm]     
&- \frac{J_{12}+K_{12}+ 4 \left(J_{78}+K_{78}\right) + 2 \left(J_{34}+J_{56} + K_{34}+K_{56} \right)}{9(\Nc+6)} \nn \\[2mm]    
&+ \frac{C_{12}+C_{34} + 8 \left(Z_{12}+Z_{34}\right) + 16   \left(H_{12}+ H_{34}\right)}{18 (\Nc+5) (\Nc+6)}  \nn \\[2mm]    
& - \frac{2\left(B_{12}+B_{34}\right)}{3 (\Nc+4) (\Nc+5) (\Nc+6)} + \frac{2 \, S}{3 (\Nc+3)(\Nc+4) (\Nc+5) (\Nc+6)} \\ \nn 
\end{align}

In each of the above expressions, the tensors are ordered with an increasing number of `traces'. The tensors appearing first are those associated with the same number of traces as that characterizing the irrep, and the relative weights of these tensors are reminiscent from the symmetry properties of the irrep. For a given irrep, we will call these tensors the `leading tensors'. For instance, the leading tensors contributing to $\Proj{\Irrep{0}{a}}$ given by \eq{eq:projP0a} are the tensors $H_i$ defined in Table~\ref{table:TensorBasis}. The irrep $\Irrep{0}{a} \sim T^{ \left[ jn \right] }_{ \left[ lq \right]}$ (see~\eq{Tjnlq}) is characterized by two antisymmetrized upper and two antisymmetrized lower indices, and we readily check that this is precisely the symmetry of the linear combination $H_{12}-H_{34} = H_1+H_2-H_3-H_4$ appearing in \eq{eq:projP0a}. 

The terms following the `leading tensors' correspond to the subtraction of unwanted irreps in the intermediate state (corresponding to a larger number of traces than the leading tensors), required to make the complete linear combination an eigenvector of $\hat{C}_2$ (and also ensuring orthogonality between all projectors). The subtraction terms, which can lead to rather long expressions of a given projector (or transition operator), are provided directly by the eigenvectors of the Casimir operator. For the sake of clarity, in most of the long expressions that follow, the leading tensors will be written on the first line (as in \eq{eq:projP0c}--\eq{eq:projP125}), and the subtraction terms on the following lines.

\subsection{Eigenspaces of $\hat{C}_2$ of dimension 2}
\label{sec:eigenspaces-dim2}

There are four eigenspaces of $\hat{C}_2$ of dimension 2, associated respectively to the eigenvalues $C_R=2\Nc$, $4\Nc+2$, $4\Nc+6$, and $4\Nc+8$ (see Table~\ref{table:irreps}), and corresponding to the sets of irreps $\{ \Irrep{10}{}, \Irrep{\overline{10}}{} \}$, 
$\{ \Irrep{0}{d}, \Irrep{\overline{0}}{d} \}$, $\{ \Irrep{28}{}, \Irrep{\overline{28}}{} \}$ and $\{ \Irrep{81}{}, \Irrep{\overline{81}}{} \}$. 

In each case $\{ R, \overline{R} \}$ considered below, we first give the $\hat{C}_2$ eigenvectors $e_1$ and $e_2$ of Casimir $C_R$ being respectively {\it even} and {\it odd} under complex conjugation $\sigma_{\rm cc}$ (namely, $e_1^* =  e_1$ and $e_2^* = - e_2$). These eigenvectors are given by the (one-dimensional) kernel of the operator\footnote{\label{foot:C2sigcc-com}This is easily shown by choosing a basis of $\Irrep{27}{} \otimes \Irrep{27}{}$ where the operators $\hat{C}_2$ and $\sigma_{\rm cc}$, which are diagonalizable and commute, are both diagonal.}
\begin{align}
\label{eq:null-space-2dim}
& \frac{\unit + \lambda_{\rm cc} \sigma_{\rm cc}}{2}  \cdot \hat{C}_2 - C_R \, \unit \ , 
\end{align}
where $\lambda_{\rm cc} = \pm 1$ for $e_1$ and $e_2$, respectively, and $\unit$ and $\sigma_{\rm cc}$ denote the identity and complex conjugation operators acting on the basis ${\cal T}$.
The eigenvectors $e_1$ and $e_2$ thus result directly from the expressions of the matrix $\hat{C}_2$ obtained in section~\ref{sec:C2matrix} and of the matrix $\sigma_{\rm cc}$ which follows from Table~\ref{tab:CrossingR}. 

We then write the projectors on the irreps $R$ and $\overline{R}$ in the form
\begin{align}
& \Proj{{R}{}/{\overline{R}}{}} \ \propto \ e_1 \pm \lambda e_2 \ ,
\end{align}
determine the coefficient $\lambda$ from the specific symmetry properties of the irreps, and the overall normalization by requiring $\tr \Proj{\al} = K_\al$. 

Note that in each case $\{ R, \overline{R} \}$ listed below, the eigenspace of $\hat{C}_2$ is spanned by two Hermitian projectors $\Proj{{R}{}}$ and $\Proj{{\overline{R}}{}}$, and $e_1$ and $e_2$ are a fortiori Hermitian, $e_1^\dagger = e_1$, $e_2^\dagger = e_2$.

\subsubsection{$\{\Irrep{10}{}, \Irrep{\overline{10}}{}\}$}

Let's start with the eigenspace of $\hat{C}_2$ corresponding to $C_R=2\Nc$. The eigenvectors $e_1$ and $e_2$ defined above are:
\begin{align}
& e_1 = \Hbar_{12} -\frac{1}{\Nc}  \left(B_{12}-B_{34} \right) \  \ ; \ \ \ \  e_2 = \Hbar_{34} \ , 
\end{align}
implying $\Proj{\Irrep{10}{}/\Irrep{\overline{10}}{}} \propto e_1 \pm \lambda e_2$. The irrep $\Irrep{10}{}$ is characterized by a tensor $\sim T^{ \left\{ jn \right\} }_{ \left[ lq \right] }$ (see  \eq{Tjnlq}), \ie\ it is symmetric in the two quark indices. Thus, in $\Proj{\Irrep{10}{}}$ the tensors $H_1$ and $H_4$ (see Table~\ref{table:TensorBasis}) can only appear in the combination $H_1 + H_4$. Hence, $\lambda =-1$ and we obtain: 
\begin{align}
\label{eq:projP10}
& \Proj{\Irrep{10}{}/\Irrep{\overline{10}}{}} = \frac{4(\Nc+2)^2}{\Nc(\Nc+1)(\Nc+3)(\Nc+4)}
\left\{ \Hbar_{12}  - \frac{B_{12}-B_{34}}{\Nc}  \mp \   \Hbar_{34}  \right\} \ , 
\end{align}
where the overall factor follows from $\tr \Proj{\Irrep{10}{}} = \K{\Irrep{10}{}}$, using the dimensions $K_\al$ of Table~\ref{table:irreps} and the traces of tensors of Appendix~\ref{app:B3}. 
Note that in \eq{eq:projP10} and in what follows, our convention of ordering tensors with respect to an increasing number of `traces' is applied separately to the terms appearing before and after some $\pm$ or $\mp$ symbol.  

\subsubsection{$\{ \Irrep{0}{d}, \Irrep{\overline{0}}{d} \}$ }

In this case, $C_R= 4\Nc+2$, and the kernel of the operator \eq{eq:null-space-2dim} gives (for $\lambda_{\rm cc} = \pm 1$):
\begin{align}
& e_1 = \Lbar_{12} -L_{45}+L_{67} + \ldots \ \ ; \ \ \ \ 
e_2 = \Lbar_{45}-\Lbar_{67}-\Lbar_{89} + \ldots  \ , 
\end{align}
where the dots represent `subtraction terms', which are not relevant to the discussion below (but will be made explicit in \eq{eq:projP0d}). 

Similarly to the previous case, in order to fix $\lambda$ in $\Proj{\Irrep{0}{d}/\Irrep{\overline{0}}{d}} \propto e_1 \pm \lambda e_2$, we observe from Eqs.~\eq{eq:0trace-indices} and \eq{irreps-0trace} that the irrep $\Irrep{0}{d}$ has a subset of quark indices in the irrep $\Irrep{15}{}$ (see \eq{6times6}), \ie, with two quark indices being symmetrized and the other two being antisymmetrized. This means that in order to obtain the expression of $\Proj{\Irrep{0}{d}}$, we are free to replace the identity of $[qq]_s\otimes[qq]_s \to [qq]_s\otimes[qq]_s$ appearing in the intermediate state of any tensor $L_1 \ldots L_9$ by the structure: 
\begin{align}
\begin{tikzpicture}[scale=0.7]
\draw[thick] (-3,.75) -- ++(2,0); \draw[thick] (3,.75) -- ++(-2,0); 
\draw[thick] (-2.5,.75) -- ++(-.1,.1); \draw[thick] (-2.5,.75) -- ++(-.1,-.1);
\draw[thick] (-3,.25) -- ++(2,0); \draw[thick] (3,.25) -- ++(-2,0); 
\draw[thick] (-2.5,.25) -- ++(-.1,.1); \draw[thick] (-2.5,.25) -- ++(-.1,-.1);
\draw[thick] (-3,-.25) -- ++(2,0); \draw[thick] (3,-.25) -- ++(-2,0); 
\draw[thick] (-2.5,-.25) -- ++(-.1,.1); \draw[thick] (-2.5,-.25) -- ++(-.1,-.1);
\draw[thick] (-3,-.75) -- ++(2,0); \draw[thick] (3,-.75) -- ++(-2,0); 
\draw[thick] (-2.5,-.75) -- ++(-.1,.1); \draw[thick] (-2.5,-.75) -- ++(-.1,-.1);
\draw[fill=white] (-2.1,.85) rectangle (-1.9,.15);
\draw[fill=white] (-2.1,-.85) rectangle (-1.9,-.15);
\draw[fill=white] (2.1,.85) rectangle (1.9,.15);
\draw[fill=white] (2.1,-.85) rectangle (1.9,-.15);
\draw[thick] (-1,.75) to[out=0,in=180] (1,.75);
\draw[thick] (-1,.25) to[out=0,in=180] (0,-.25) to[out=0,in=180] (1,.25);
\draw[thick] (-1,-.25) to[out=0,in=180] (0,.25) to[out=0,in=180] (1,-.25);
\draw[thick] (-1,-.75) to[out=0,in=180] (1,-.75);
\draw[fill=white] (-.1,.85) rectangle (.1,.15);
\draw[fill=white!50!black] (-.1,-.85) rectangle (.1,-.15);
\end{tikzpicture} \ \ . 
\end{align}
Doing so with the tensor $L_1$ (see Table~\ref{table:TensorBasis}), we see that it is changed as $L_1 \rightarrow \frac{1}{4} (L_1 - L_9)$. We infer that in $\Proj{\Irrep{0}{d}}$, the tensors $L_1$ and $L_9$ must  appear in the combination $L_1 - L_9$, hence, $\lambda =-1$. 
One finds: 
\begin{align}
\label{eq:projP0d}
\Proj{\Irrep{0}{d}/\Irrep{\overline{0}}{d}} 
\ &=\ \frac{\Lbar_{12}-L_{45}+L_{67}}{6} \nn \\[2mm] 
&- \frac{2 \Nc \left(J_{12}-K_{12}\right)   - 4 \left(J_{78}-K_{78}\right) - (\Nc-2) \left(J_{34}+J_{56} - K_{34}-K_{56} \right) }{3 (\Nc-1) (\Nc+2)} \nn \\[2mm] 
& + \frac{C_{12}-C_{34}}{3 (\Nc-1) (\Nc+2)}+\frac{4 \Hbar_{12} (2 \Nc+1)}{3 (\Nc-1) (\Nc+1)(\Nc+2)} -\frac{4 \left(B_{12}-B_{34}\right)}{3 (\Nc-1) (\Nc+1) (\Nc+2)} \nn \\[3mm]
& \mp \left[ \, \frac{\Lbar_{45}-\Lbar_{67}-\Lbar_{89}}{6} + \frac{J_{34}-J_{56}-K_{34}+K_{56} }{3(\Nc-1)}+ \frac{4 \Hbar_{34}}{3 (\Nc-1) (\Nc+1)} \, \right]  \ . 
\end{align}

\subsubsection{$\{ \Irrep{28}{}, \Irrep{\overline{28}}{} \}$ }

Here we have $C_R = 4\Nc+6$, and one finds: 
\begin{align}
& e_1 = L_{12}-8 L_3+L_{45}+L_{67}+L_{89} + \ldots \ \ ; \ \ \ \ 
e_2 = \Lbar_{45}+\Lbar_{67} + \ldots \ .
\end{align}

To determine $\lambda$ in $\Proj{\Irrep{28}{}/\Irrep{\overline{28}}{}} \propto e_1 \pm \lambda e_2$, note that the irrep $\Irrep{28}{}$ has four quark indices in the irrep $\Irrep{15'}{}$, \ie, fully symmetrized (see~Eqs.~\eq{eq:0trace-indices}--\eq{irreps-0trace} and \eq{6times6}). The expression of $\Proj{\Irrep{28}{}}$ is thus unchanged when replacing the identity of $[qq]_s\otimes[qq]_s \to [qq]_s\otimes[qq]_s$ in any tensor $L_1 \ldots L_9$ by the four-quark symmetrizer. For instance, when inserting in $L_1$ the four-quark symmetrizer which is a sum of $4!$ permutation operators, 4 permutations contribute to $L_1$, 16 to $L_5$, and 4 to $L_9$, and we have $L_1 \rightarrow \frac{1}{6} (L_1 + 4 L_5 + L_9)$. Thus, only the latter combination can enter $\Proj{\Irrep{28}{}}$, and we get $\lambda =-3$. As a result: 
\begin{align}
\label{eq:projP28}
\Proj{\Irrep{28}{}/\Irrep{\overline{28}}{}} 
\ &=\ 
\frac{L_{12} + L_{45} - 8 L_3 + L_{67}+ L_{89}}{18} \nn \\[2mm] 
&-\frac{2 \left(J_{12}+K_{12}\right) - 4 \left(J_{78}+K_{78}\right) +J_{34}+J_{56} +K_{34}+K_{56} }{9 (\Nc+3)} \nn \\[2mm] 
&+ \frac{C_{12}+C_{34} - 4 \left(Z_{12}+Z_{34}\right)+4 \left(H_{12}+H_{34}\right) }{9 (\Nc+2) (\Nc+3)}
\nn \\[2mm] 
&\mp \left[ \frac{\Lbar_{45}+ \Lbar_{67}}{6} + \frac{J_{34}-J_{56}+K_{34}-K_{56}}{3(\Nc+3)} \right] \ . 
\end{align}

\subsubsection{$\{ \Irrep{81}{}, \Irrep{\overline{81}}{} \}$ }

For $C_R = 4\Nc+8$, we obtain 
\begin{align}
& e_1 = -\frac{\bar{L}_{12}}{2}  -L_{45}+L_{67} + \ldots \ \ ; \ \ \ \ 
e_2 =  -2 \bar{L}_{45}+2 \bar{L}_{67}-\bar{L}_{89} + \ldots \ .
\end{align}

Similarly to the irrep $\Irrep{28}{}$ considered in the last paragraph, the irrep $\Irrep{81}{}$ is fully symmetric in the four quark indices (see~\eq{eq:0trace-indices}, \eq{irreps-0trace}, and \eq{6times6}). Following exactly the same discussion as above, only the combination $(L_1 + 4 L_5 + L_9)$ can enter $\Proj{\Irrep{81}{}}$, implying $\lambda =-\frac{1}{2}$ in $\Proj{\Irrep{81}{}/\Irrep{\overline{81}}{}} \propto e_1 \pm \lambda e_2$. After adjusting the normalization of projectors we find: 
\begin{align}
\label{eq:projP81}
\Proj{\Irrep{81}{}/\Irrep{\overline{81}}{}} 
\ &=\ \frac{L_{45}}{6}-\frac{L_{67}}{6}+\frac{\bar{L}_{12}}{12}
\nn \\[2mm] 
&-\frac{(\Nc+3) \left(J_{12}-K_{12}\right) + 4 \left(J_{78}-K_{78}\right)}{3
   (\Nc+2) (\Nc+5)}-\frac{(\Nc+4)\left(J_{34}+J_{56}-K_{34}-K_{56}\right)}{3 (\Nc+2)
   (\Nc+5)}
\nn \\[2mm] 
&+\frac{C_{12}-C_{34}}{6 (\Nc+2) (\Nc+5)}+\frac{4 \Hbar_{12}}{3 (\Nc+4)
   (\Nc+5)}-\frac{2 \left(B_{12}-B_{34}\right)}{3 (\Nc+2) (\Nc+4) (\Nc+5)}
\nn \\[2mm] 
&\pm \left[
-\frac{\bar{L}_{45}}{6}+\frac{\bar{L}_{67}}{6}-\frac{\bar{L}_{89}}{12}-\frac{J_{34}-J_{56}-K_{34}+K_{56}}{3(\Nc+5)}-\frac{4 \Hbar_{34}}{3 (\Nc+4) (\Nc+5)}\right] \ .
   \\ \nn
\end{align}

\subsection{Eigenspaces of $\hat{C}_2$ of dimension 4}
\label{sec:eigenspaces-dim4}

In this section we consider the three pairs of irreps $\{ \Irrep{8}{+}, \Irrep{8}{-} \}$, $\{ \Irrep{0}{b+}, \Irrep{0}{b-} \}$, and $\{ \Irrep{64}{+}, \Irrep{64}{-} \}$, of Casimir $C_R = \Nc$, $3\Nc$, $3\Nc +6$, respectively (see Table~\ref{table:irreps}). In each case, $\{ {R}_{+}, {R}_{-} \}$ is a pair of {\it equivalent} irreps, and the eigenspace of $\hat{C}_2$ associated to the eigenvalue $C_R$ is of dimension 4. Indeed, the transition operators $\Trans{{R}_{+} \to {R}_{-}}$ and $\Trans{{R}_{-} \to {R}_{+}}$ are eigenvectors of $\hat{C}_2$ sharing the same eigenvalue as the projectors $\Proj{{R}_{+}}$ and $\Proj{{R}_{-}}$, as a direct consequence of color conservation and the general form~\eq{trans-def} of transition operators. It is also obvious from~\eq{trans-def} that transition operators and projectors are distinguished by different symmetries w.r.t.~permutations $\sigma_{12}$ and $\sigma_{34}$, so that their determination is straightforward. 

For a given pair $\{ {R}_{+}, {R}_{-} \}$, the operators $\Proj{{R}_{+}}$, $\Proj{{R}_{-}}$, $\Trans{{R}_{+} \to {R}_{-}}$, $\Trans{{R}_{-} \to {R}_{+}}$ are eigenvectors of both $\sigma_{12}$ and $\sigma_{34}$, with eigenvalues $(s_{12}, s_{34})=(+,+)$, $(-,-)$, $(+,-)$, $(-,+)$, respectively, and the operator corresponding to a given `symmetry signum' $(s_{12}, s_{34})$ is obtained from the kernel of the operator\footnote{This directly follows from the fact that $\hat{C}_2$, $\sigma_{12}$ and $\sigma_{34}$ are diagonalizable and commute.} 
\begin{align}
\label{eq:null-space-dim4}
& \frac{\unit + s_{12} \sigma_{12}}{2} \cdot \frac{\unit + s_{34}  \sigma_{34}}{2} \cdot \hat{C}_2 - C_R \, \unit \ .
\end{align}

For each pair $\{ {R}_{+}, {R}_{-} \}$, we give below the results for $\{ \Proj{{R}_{+}}, \Proj{{R}_{-}}, \Trans{{R}_{+} \to {R}_{-}}, \Trans{{R}_{-} \to {R}_{+}} \}$. As usual, the normalization of a projector is fixed by $\tr \Proj{{R}{}} = K_R$. As for transition operators, we will present them first without requiring any specific normalization (which is totally arbitrary, as recalled in section~\ref{sec:generalities}), with the exception of the trivial relation $\Trans{{R}_{-} \to {R}_{+}} = \Trans{{R}_{+} \to {R}_{-}}^\dagger$. 
Although not essential, we will then adjust their normalization according to the convention~\eq{eq:trans-norm}, which can easily be done using clebsches (see section~\ref{sec:clebsches}).  
Transition operators will be presented by distinguishing their Hermitian and anti-Hermitian parts, so that $\Trans{{R}_{+} \to {R}_{-}} = \Trans{}$ and $\Trans{{R}_{-} \to {R}_{+}} = \Trans{}^\dagger$ can be given at once in the form
\begin{align}
 \Trans{{R}_{+} \to {R}_{-} / {R}_{-} \to {R}_{+}} = \left( \frac{\Trans{}+\Trans{}^\dagger}{2} \right) \pm  \left( \frac{\Trans{}-\Trans{}^\dagger}{2} \right) \ . 
\end{align}

\subsubsection{Sector of the $\Irrep{8}{}$}
\label{sec:sector-8}

The eigenvectors of $\hat{C}_2$ with eigenvalue $C_R = \Nc$ and specific symmetries under $\sigma_{12}$ and $\sigma_{34}$ give the following projectors and transition operators: 
\begin{subequations}
\label{proj-8p8m}
\begin{align}
& \Proj{\Irrep{8}{+}} = \ \frac{4(\Nc+2)}{(\Nc-2)\Nc(\Nc+3)(\Nc+4)} \left( B_{12}+B_{34} -\frac{4 \, S}{\Nc}\right)  \ , \label{proj-8p} \\[2mm]
& \Proj{\Irrep{8}{-}} = \ \frac{4}{\Nc^2(\Nc+3)}\left(B_{12}-B_{34} \right) \label{proj-8m} \ , 
\end{align}
\end{subequations}
\vspace{-5mm}
\begin{align}
\Trans{\Irrep{8}{+} \to \Irrep{8}{-} / \Irrep{8}{-} \to \Irrep{8}{+}} & = \ \Bbar_{12} \mp \Bbar_{34}  \ . \hskip 40mm
\label{trans-8pm}
\end{align}

We now adjust the normalization of the latter transition operators according to~\eq{eq:trans-norm}. 
Since there are two equivalent irreps $\Irrep{8}{+}$ and $\Irrep{8}{-}$, there are two clebsches $c_{\al 1}$ and $c_{\al 2}$, and thus two vertices $v_{\al 1}$ and $v_{\al 2}$ mapping $\Irrep{27}{} \otimes \Irrep{27}{}$ to the irrep $\al = \Irrep{8}{}$ of a $q \bar{q}$ pair. An obvious choice is (for simplicity we  omit the subscript $\al$ in clebsches and vertices) 
\begin{align}
\label{eq:vvdag-8}
v_1 = \VextexOctetONE \ \ , \ \ v_2 = \VextexOctetTWO \ \ , 
\end{align}
where the octet $q\bar{q}$ projector reads:
\begin{align}
& \irrepdash{\Irrep{8}{}}{0.3} \ \equiv \ \Proj{\Irrep{8}{}}^{q\bar{q}}  \ = \  \Iqqbar{0.8} - \frac{1}{\Nc} \ \Sqq{1}  \ . \label{eq:projdot-8}
\end{align}

We need to find the clebsches \eq{CG-LC} constructing the matrix of operators \eq{eq:P12T12-CG}. Under $\psi \leftrightarrow \chi$ exchange, we have $v_1 \leftrightarrow v_2$, and $\Proj{\Irrep{8}{+}}$ and $\Proj{\Irrep{8}{-}}$ are respectively symmetric and antisymmetric. Thus, $c_1 = a (v_1+v_2)$ and $c_2 = b (v_1-v_2)$. This implies: 
\begin{subequations}
\label{eq:P12-CG-8}
\begin{align}
\Proj{\Irrep{8}{+}} \ &= \ c_1^\dagger c_1 \ = \ a^2 (v_1+v_2)^\dagger (v_1+v_2) \ = \ a^2 (v_{1}^\dagger v_{1} + v_{2}^\dagger v_{2} +v_{1}^\dagger v_{2} +v_{2}^\dagger v_{1}) \ , \\[2mm]
\Proj{\Irrep{8}{-}} \ &= \ c_2^\dagger c_2 \ = \ b^2 (v_1-v_2)^\dagger (v_1-v_2)  \ = \ b^2 (v_{1}^\dagger v_{1} + v_{2}^\dagger v_{2} -v_{1}^\dagger v_{2} -v_{2}^\dagger v_{1}) \ , 
\end{align}
\end{subequations}
and
\begin{align}
\label{eq:T12-CG-8}
\Trans{\Irrep{8}{+} \to \Irrep{8}{-}} \ &= \ c_2^\dagger c_1\ = \ ab (v_1-v_2)^\dagger (v_1+v_2) = \ ab (v_1^\dagger v_1 - v_2^\dagger v_2 +v_1^\dagger v_2 -v_2^\dagger v_1)  \ .
\end{align}
Using the octet $q\bar{q}$ projector~\eq{eq:projdot-8} in \eq{eq:vvdag-8} we find:
\be
v_1^\dagger v_1 = B_1 -\frac{S}{\Nc}  \ \ ; \ \ 
v_2^\dagger v_2 = B_2 -\frac{S}{\Nc}  \ \ ; \ \ 
v_1^\dagger v_2 = B_4 -\frac{S}{\Nc}  \ \ ; \ \ 
v_2^\dagger v_1 = B_3 -\frac{S}{\Nc}  \ .
\ee
Inserting the latter identities in \eq{eq:P12-CG-8} and comparing to \eq{proj-8p8m}, we obtain:
\be
a = \sqrt{\frac{4(\Nc+2)}{(\Nc-2)\Nc(\Nc+3)(\Nc+4)}} \ \ ; \ \ b = \sqrt{\frac{4}{\Nc^2(\Nc+3)}} \ \ . 
\ee
Comparing now \eq{eq:T12-CG-8} and \eq{trans-8pm}, we see that  \eq{trans-8pm} should be multiplied by $ab$ to conform with the normalization convention~\eq{eq:trans-norm}.

\subsubsection{Sector of the $\Irrep{0}{b}$}
\label{sec:sector-0b}

In this sector, $C_R = 3 \Nc$ and the kernel of \eq{eq:null-space-dim4} for various values of $(s_{12},s_{34})$ gives 
\begin{subequations}
\label{eq:projP0b+0b-}
\begin{align}
\label{eq:projP0b+}
&\Proj{\Irrep{0}{b+}} \ = \  \frac{8 (\Nc+2)}{9 \Nc (\Nc+4)} 
\left\{ \phantom{\hskip -2mm \frac{1}{1}} 
J_{12}+K_{12}+J_{78}+K_{78}-(J_{34}+J_{56}+K_{34}+K_{56})  \right. \nn \\[2mm]
&-\frac{C_{12}+C_{34}+2 \left(Z_{12}+Z_{34}\right) +H_{12}+H_{34}}{\Nc-2}-\frac{9 \left(H_{12}-H_{34}\right)}{\Nc+2} \nn \\[1mm]
&+ \left. \frac{9 \left(B_{12}+B_{34}\right)}{(\Nc-2) (\Nc+2)}-\frac{12 \, S}{(\Nc-2) \Nc (\Nc+2)} \right\} \ , \\[2mm]
\label{eq:projP0b-}
&\Proj{\Irrep{0}{b-}} \ =\ \frac{8}{9(\Nc+2)} \left\{ \phantom{\hskip -3mm \frac{1}{1}}
\frac{}{} J_{12}-K_{12} +J_{78}-K_{78} -(J_{34}+J_{56}-K_{34}-K_{56}) \right. \notag \\[1mm]
&\left.-\frac{1}{\Nc-2} (C_{12}-C_{34}) -\frac{6}{\Nc} \Hbar_{12} + \frac{5\Nc-6}{\Nc^2(\Nc-2)}(B_{12}-B_{34}) \right\} \ ,
\end{align}
\end{subequations}
\begin{align} 
\label{eq:trans0bpm}
&\Trans{\Irrep{0}{b+} \to \Irrep{0}{b-} / \Irrep{0}{b-} \to \Irrep{0}{b+}} \ = \ 
\Jbar_{12}+\Jbar_{78}-J_{35}+J_{46}-\frac{\Cbar_{12}+Z_{13}-Z_{24}}{\Nc-2}+\frac{3 \Bbar_{12}}{(\Nc-2) \Nc} \notag\\
& \mp \left( 
\Kbar_{12}+\Kbar_{78}-K_{35}+K_{46}-\frac{\Cbar_{34}+Z_{14}-Z_{23}}{\Nc-2}+\frac{3 \Bbar_{34}}{(\Nc-2) \Nc} \right)  \ . 
\end{align}

If one wants to normalize the transition operators according to \eq{eq:trans-norm}, one proceeds similarly to the case of the octet sector in the previous section. In the sector of the $\Irrep{0}{b}$, one may choose the two vertices (related by $\psi \leftrightarrow \chi$ exchange) 
\begin{align}
\label{eq:CG-0b-12}
v_1=\VextexZerobONE \ \ , \ \ v_2=\VextexZerobTWO \ \ , 
\end{align}
depending on the projector $\Proj{\Irrep{0}{b}}$ on the irrep $\Irrep{0}{b}$ of the $qqq\bar{q}\bar{q}\bar{q}$ system. 

We need to calculate the four quantities $v_{i}^\dagger  v_{j}$. Then, using the analog of \eq{eq:P12-CG-8} and comparing to $\Proj{\Irrep{0}{b+}}$ and $\Proj{\Irrep{0}{b-}}$ given by \eq{eq:projP0b+0b-}, we can determine $a$ and $b$. In fact, since $\Proj{\Irrep{0}{b+}}$ and $\Proj{\Irrep{0}{b-}}$ are already fully determined (thanks to the `filtering power' of the Casimir operator), it is sufficient to calculate only the {\it leading terms} of $v_{i}^\dagger  v_{j}$ to extract $a$ and $b$. 

Let's start with $v_{1}^\dagger  v_{1}$: 
\begin{align}
\label{eq:0b-c1dagc1}
v_1^\dagger v_1 = \VoneDagVone \ \ . 
\end{align}

From section~\ref{sec:one-trace}, the irrep ${\Irrep{0}{b}}$ of the system $qqq\bar{q}\bar{q}\bar{q}$ arises from $\Syst{qqq}{\Irrep{8}{}}{} \otimes \Syst{\bar{q}\bar{q}\bar{q}}{\Irrep{8}{}}{}$, with $\Syst{qqq}{\Irrep{8}{}}{}$ (resp.~$\Syst{\bar{q}\bar{q}\bar{q}}{\Irrep{8}{}}{}$) built from $\Syst{qq}{\Irrep{6}{}}{q}$ (resp.~$\Syst{\bar{q}\bar{q}}{\Irrep{\bar{6}}{}}{\bar{q}}$). 
The decomposition of $\Syst{qq}{\Irrep{6}{}}{q}$ into irreps reads
\be
\label{6times3-tensor}
\bm 6 \otimes \bm 3 \ = \ \bm{10} \oplus \bm{8}_{\mathsize{7}{}} \ ,
\ee
where the associated projectors are given by~\cite{Peigne:2024srm}: 
\begin{align}
\label{eq:qqq-10}
\Proj{\bm{10}} \ = \  \PqqqSym{1} \ &= \  \frac{1}{3} \ \Idqqalq{white}{qua}{1} + \frac{2}{3} \ \Aqqq{1}
\ \ ,  \\[3mm]
\label{eq:qqq-8}
\Proj{\Irrep{8}{}} \ = \  \Pqqqplus{1} \ &= \  \frac{2}{3} \ \Idqqalq{white}{qua}{1} - \frac{2}{3} \  \Aqqq{1} \ \ . 
\end{align}
The projector $\Proj{\Irrep{0}{b}}$ to use in \eq{eq:0b-c1dagc1} is thus of the form 
\begin{align}
\label{eq:qqqqqqP0b}
\Proj{\Irrep{0}{b}} \ =\  \PZerob \ + \  \ldots  \ , 
\end{align}
where the subleading terms indicated by dots contribute only to subleading terms in $v_1^\dagger v_1$ and are thus irrelevant for our present purpose, and a blob $\Irrep{8}{}$ denotes either the projector \eq{eq:qqq-8} for the $\Syst{qq}{\Irrep{6}{}}{q}$ system, or its complex conjugate for the $\Syst{\bar{q}\bar{q}}{\Irrep{\bar{6}}{}}{\bar{q}}$ system.\footnote{For visual convenience when inserting \eq{eq:qqqqqqP0b} into \eq{eq:0b-c1dagc1}, the birdtrack for the projector $\Proj{\Irrep{8}{}}$ of the system $\Syst{\bar{q}\bar{q}}{\Irrep{\bar{6}}{}}{\bar{q}}$ is also reflected w.r.t.~a horizontal axis.} 

One finds for the leading terms of $v_{1}^\dagger  v_{1}$, and similarly $v_{2}^\dagger  v_{2}$, $v_{2}^\dagger  v_{1}$, $v_{1}^\dagger  v_{2}$:
\begin{subequations}
\begin{align}
v_{1}^\dagger  v_{1} \ &= \ \frac{4}{9} \left\{ J_1+ J_7 - J_3 - J_5 \right\} \ + \ \ldots \\ 
v_{2}^\dagger  v_{2} \ &= \ \frac{4}{9} \left\{ J_2+ J_8 - J_4 - J_6 \right\} \ \ + \ \ldots \\ 
v_{2}^\dagger  v_{1} \ &= \ \frac{4}{9} \left\{ K_1+ K_7 - K_3 - K_5 \right\} \ + \ \ldots \\ 
v_{1}^\dagger  v_{2} \ &= \ \frac{4}{9} \left\{ K_2+ K_8 - K_4 - K_6 \right\}  \ + \ \ldots 
\end{align}
\end{subequations}
Inserting the latter expressions into \eq{eq:P12-CG-8} and \eq{eq:T12-CG-8} (strictly speaking, their analogs for the sector of the $\Irrep{0}{b}\,$), we obtain 
\begin{subequations}
\begin{align}
&\Proj{\Irrep{0}{b+}} = a^2 \frac{4}{9} 
\left\{ J_{12}+K_{12}+J_{78}+K_{78}-(J_{34}+J_{56}+K_{34}+K_{56}) + \ldots  \right\} \ , \hskip 10mm \\[2mm]
&\Proj{\Irrep{0}{b-}} = b^2 \frac{4}{9}  \left\{ J_{12}-K_{12} +J_{78}-K_{78} -(J_{34}+J_{56}-K_{34}-K_{56}) + \ldots  \right\} \ , \hskip 10mm
\end{align}
\end{subequations}
\vspace{-6mm}
\begin{align}
&\Trans{\Irrep{0}{b+} \to \Irrep{0}{b-} / \Irrep{0}{b-} \to \Irrep{0}{b+}} = ab \frac{4}{9} \left\{ \Jbar_{12}+\Jbar_{78}-J_{35}+J_{46} 
\mp \left( \Kbar_{12}+\Kbar_{78}-K_{35}+K_{46} \right) + \ldots \right\} \ . 
\end{align}
 
Comparing with \eq{eq:projP0b+0b-} gives the values of $a$ and $b$ (in addition to providing a non-trivial cross-check of the leading terms of the projectors):
\begin{align}
a = \sqrt{\frac{2 (\Nc+2)}{\Nc (\Nc+4)}} \ \ ; \ \ b = \sqrt{\frac{2}{\Nc+2}}  \  . 
\end{align}
The transition operator \eq{eq:trans0bpm} should be multiplied by $\frac{4}{9} ab$ to be normalized as in~\eq{eq:trans-norm}. 

\subsubsection{Sector of the $\Irrep{64}{}$}
\label{sec:sector-64}

For $C_R = 3 \Nc +6$, the kernel of \eq{eq:null-space-dim4} for different $(s_{12},s_{34})$ gives: 
\begin{subequations}
\label{eq:projP64+64-}
\begin{align}
\label{eq:projP64+}
\Proj{\Irrep{64}{+}} \ &=\
\left.\frac{2(\Nc+2)}{9(\Nc-2)(\Nc+6)}\right\{ (J_{12}+K_{12})  +4 (J_{78}+K_{78}) + 2(J_{34}+J_{56} +K_{34}+K_{56}) \notag \\[2mm]
&- \frac{C_{12}+C_{34} +8(Z_{12} + Z_{34})+16(H_{12}+H_{34}) }{\Nc+4} \notag \\[0mm]
&+ \left.  \frac{18(B_{12}+B_{34})}{(\Nc+3)(\Nc+4)} - \frac{24 S}{(\Nc+2)(\Nc+3)(\Nc+4)} \right\} \ , 
\\[3mm]
\label{eq:projP64-}
\Proj{\Irrep{64}{-}} &=\
\left.\frac{2}{9(\Nc+2)}\right\{ J_{12}-K_{12} +4 (J_{78}-K_{78}) + 2(J_{34}+J_{56}-K_{34}-K_{56}) \notag \\[1mm]
&- \left.\frac{C_{12}-C_{34}}{\Nc+4} + \frac{2(B_{12}-B_{34})}{(\Nc+3)(\Nc+4)} \right\} \ , 
\end{align}
\end{subequations}
\vspace{-5mm}
\begin{align}
\label{eq:trans64pm}
& \Trans{\Irrep{64}{+} \to \Irrep{64}{-} / \Irrep{64}{-} \to \Irrep{64}{+}} =
\Jbar_{12}+4 \Jbar_{78}+2 \left(J_{35}-J_{46}\right)-\frac{\Cbar_{12}+4 \left(Z_{13}-Z_{24}\right)}{\Nc+4}+\frac{6 \Bbar_{12}}{(\Nc+3) (\Nc+4)}
 \notag\\[2mm]
&\mp \left( \Kbar_{12}+4 \Kbar_{78}+2 \left(K_{35}-K_{46}\right)-\frac{\Cbar_{34}+4 \left(Z_{14}-Z_{23}\right)}{\Nc+4}+\frac{6 \Bbar_{34}}{(\Nc+3) (\Nc+4)} \right)  \ . 
\end{align}

To normalize the transition operators as in~\eq{eq:trans-norm}, one can proceed exactly as in the sector of the $\Irrep{0}{b}$ (see section~\ref{sec:sector-0b}). Up to the replacement $\Irrep{0}{b} \to \Irrep{64}{}$, the two vertices are given by \eq{eq:CG-0b-12}, and $v_{1}^\dagger v_{1}$ by \eq{eq:0b-c1dagc1}. 
From the symmetries of the irrep ${\Irrep{64}{}}$ in quark and antiquark indices (see section~\ref{sec:one-trace}), the projector on the irrep ${\Irrep{64}{}}$ of $qqq\bar{q}\bar{q}\bar{q}$ is of the form \eq{eq:qqqqqqP0b}, but with the quark (resp.~antiquark) sub-system in the irrep $\Irrep{10}{}$ (resp.~$\Irrep{\overline{10}}{}$) of projector given by \eq{eq:qqq-10} (resp.~its complex conjugate). One finds: 
\begin{align}
v_{1}^\dagger  v_{1} \ &= \ \frac{1}{9} \left\{ J_1+ 2 J_5 +2 J_3 +4 J_7 \right\} \ + \ \ldots \ . 
\end{align}
This information is sufficient to obtain $a$ and $b$ by comparing \eq{eq:projP64+64-} with \eq{eq:P12-CG-8}: 
\begin{align}
a = \sqrt{\frac{2(\Nc+2)}{(\Nc-2)(\Nc+6)}} \ \ ; \ \ b = \sqrt{\frac{2}{(\Nc+2)}}  \ . 
\end{align}
The transition operator \eq{eq:trans64pm} should be multiplied by $\frac{1}{9} ab$ to satisfy the normalization convention \eq{eq:trans-norm}.

\subsection{Eigenspace of $\hat{C}_2$ of dimension 8: sector of the $\Irrep{35}{}$}
\label{sec:eigenspaces-dim8}

Four irreps have $C_R= 3\Nc + 3$, namely, $\Irrep{35}{+}$, $\Irrep{35}{-}$, $\Irrep{\overline{35}}{+}$, $\Irrep{\overline{35}}{-}$ (see Table~\ref{table:irreps}), which can be grouped into two pairs of irreps linked by complex conjugation, $\{ \Irrep{35}{+}, \Irrep{\overline{35}}{+}  \}$ and $\{ \Irrep{35}{-}, \Irrep{\overline{35}}{-} \}$, or into two pairs of equivalent irreps, $\{ \Irrep{35}{+}, \Irrep{35}{-} \}$ and $\{ \Irrep{\overline{35}}{+}, \Irrep{\overline{35}}{-} \}$. In the full color space of $\Irrep{27}{} \otimes \Irrep{27}{}$, the $\hat{C}_2$ eigenspace of eigenvalue $3\Nc + 3$ is thus of dimension 8, spanned by four projectors ($\Proj{\Irrep{35}{+}}, \Proj{\Irrep{35}{-}}, \Proj{\Irrep{\overline{35}}{+}}, \Proj{\Irrep{\overline{35}}{-}}$) and four transition operators ($\Trans{\Irrep{35}{+} \to \Irrep{35}{-}}, \Trans{\Irrep{35}{-} \to \Irrep{35}{+}}, \Trans{\Irrep{\overline{35}}{+} \to \Irrep{\overline{35}}{-}}, \Trans{\Irrep{\overline{35}}{-} \to \Irrep{\overline{35}}{+}}$). These operators can be easily found using a combination of the procedures seen previously in sections~\ref{sec:eigenspaces-dim2} and~\ref{sec:eigenspaces-dim4}. 

For instance, the kernel of the operator \eq{eq:null-space-dim4} for $C_R =3\Nc + 3$ and $(s_{12}, s_{34})=(+,+)$ directly gives two $\hat{C}_2$ eigenvectors which must be linear combinations of $\Proj{\Irrep{35}{+}}$ and $\Proj{\Irrep{\overline{35}}{+}}$. 
Projecting these eigenvectors using $\frac{\unit \pm \sigma_{\rm cc}}{2}$ to make them either even or odd under complex conjugation, we get two eigenvectors $e_1 \propto \Proj{\Irrep{35}{+}} + \Proj{\Irrep{\overline{35}}{+}}$ and $e_2 \propto \Proj{\Irrep{35}{+}}-\Proj{\Irrep{\overline{35}}{+}}$. Of course, $e_1$ and $e_2$ can be obtained in a single operation by searching for the kernel of the operator
\begin{align}
\label{eq:null-space-dim8}
&\frac{\unit + s_{12} \sigma_{12}}{2} \cdot \frac{\unit + s_{34}  \sigma_{34}}{2} \cdot \frac{\unit + \lambda_{\rm cc} \sigma_{\rm cc}}{2} \cdot \hat{C}_2 - C_R \, \unit \ ,
\end{align}
for $C_R =3\Nc + 3$ and setting $(s_{12}, s_{34},\lambda_{\rm cc})=(+,+,+)$ and $(+,+,-)$, respectively. 

Then, the specific linear combinations $\Proj{\Irrep{35}{+}/\Irrep{\overline{35}}{+}} \propto e_1 \pm \lambda e_2$ are obtained as in section~\ref{sec:eigenspaces-dim2}, by observing that the irrep $\Irrep{35}{}$ is characterized by three symmetrized quark indices, see~\eq{irreps-1trace}. 
Symmetrizing the three-quark intermediate state of any of the tensors $J_1 \ldots J_8$ and $K_1 \ldots K_8$ (corresponding to the class of tensors with `one trace', see Table~\ref{table:TensorBasis}, containing the leading tensors of the irrep $\Irrep{35}{}$) should thus leave the expression of $\Proj{\Irrep{35}{+}}$ unchanged. 
When inserting in $J_1$ the three-quark symmetrizer~\eq{eq:qqq-10}, we have $J_1 \rightarrow \frac{1}{3} ( J_1 + 2 J_3)$.
Thus, $J_1$ can enter the expression of $\Proj{\Irrep{35}{+}}$ only via the latter linear combination, and this information suffices to determine the parameter $\lambda$ in $\Proj{\Irrep{35}{+}/\Irrep{\overline{35}}{+}} \propto e_1 \pm \lambda e_2$. Finally, the normalization of projectors is fixed as usual.  

Using the same procedure for $(s_{12}, s_{34},\lambda_{\rm cc})=(-,-,\pm)$, $(+,-,\pm)$ and $(-,+,\pm)$, we obtain the expressions of $\{ \Proj{\Irrep{35}{-}} , \Proj{\Irrep{\overline{35}}{-}} \}$, $\{ \Trans{\Irrep{35}{+} \to \Irrep{35}{-}}, \Trans{\Irrep{\overline{35}}{+} \to \Irrep{\overline{35}}{-}} \}$ and $\{ \Trans{\Irrep{35}{-} \to \Irrep{35}{+}}, \Trans{\Irrep{\overline{35}}{-} \to \Irrep{\overline{35}}{+}} \}$, respectively. Note that since the transition operators involving the $\Irrep{35}{}$ (\ie\ $\Trans{\Irrep{35}{+} \to \Irrep{35}{-}}$ and $\Trans{\Irrep{35}{-} \to \Irrep{35}{+}}$) are built from $\Proj{\Irrep{35}{+}}$ and $\Proj{\Irrep{35}{-}}$ (see~\eq{trans-def}), they have the same symmetries in quark/antiquark indices, and must also depend on the combination $J_1 + 2 J_3$. 

The eight operators sought are as follows: 
\begin{align}
\label{eq:projP35+}
\Proj{\Irrep{35}{+}/\Irrep{\overline{35}}{+}} 
&=\ \frac{2(\Nc+2)}{9(\Nc+1)(\Nc+3)} \left\{ \phantom{\frac{1}{1}} \hskip -2mm 2(J_{12}+K_{12}) -4(J_{78}+K_{78}) +J_{34} + J_{56} + K_{34} + K_{56} \right. \notag \\[2mm]
& \hskip -12mm \left. - \frac{2( C_{12}+C_{34}) -8(Z_{12}+Z_{34})+8(H_{12}+H_{34})}{\Nc+1} 
\ \pm 3 (J_{34}-J_{56}+K_{34}-K_{56}) \phantom{\frac{1}{1}} \hskip -2mm \right\} \\[4mm]
\label{eq:projP35-}
\Proj{\Irrep{35}{-}/\Irrep{\overline{35}}{-}} 
&=\ \frac{2(\Nc+2)}{9(\Nc-1)(\Nc+5)} \left\{ \phantom{\frac{1}{1}} \hskip -2mm 2(J_{12} - K_{12}) -4(J_{78}-K_{78}) +J_{34}+J_{56} -K_{34}-K_{56} \right. \notag \\[2mm]
& \hskip -12mm \left. - \frac{2(C_{12}-C_{34})}{\Nc+1}-\frac{24 \Hbar_{12}}{\Nc+3} +\frac{16(B_{12}-B_{34})}{(\Nc+1)(\Nc+3)} \ \pm 3 \left(J_{34}-J_{56}-K_{34}+K_{56} +\frac{8\Hbar_{34}}{\Nc+3} \right) \right\} \notag \\[1mm]
\end{align}
\begin{subequations}
\label{eq:trans35}
\begin{align}
\label{eq:transP35pmmp}
\Trans{\Irrep{35}{+} \to \Irrep{35}{-} / \Irrep{35}{-} \to \Irrep{35}{+}} =&\  2 \Jbar_{12} - 4 \Jbar_{78} +J_{35}-J_{46} + 3(J_{36}-J_{45}) - \frac{2 \Cbar_{12} -4 Z_{13} +4Z_{24}}{\Nc+1}  \notag \\
& \hskip -30mm  \pm \left( - 2 \Kbar_{12} + 4 \Kbar_{78} -K_{35}+K_{46} - 3(K_{36}-K_{45}) + \frac{2 \Cbar_{34} -4 Z_{14} +4Z_{23}}{\Nc+1} \right) \\[2mm]
\label{eq:transP35barpmmp}
\Trans{\Irrep{\overline{35}}{+} \to \Irrep{\overline{35}}{-} / \Irrep{\overline{35}}{-} \to \Irrep{\overline{35}}{+}} =&\  -2 \Jbar_{12} +4 \Jbar_{78} -J_{35}+J_{46} + 3(J_{36}-J_{45}) +\frac{2 \Cbar_{12} -4 Z_{13} +4Z_{24}}{\Nc+1}  \notag \\
& \hskip -30mm  \pm \left(  2 \Kbar_{12} - 4 \Kbar_{78} +K_{35}-K_{46} - 3(K_{36}-K_{45}) - \frac{2 \Cbar_{34} -4 Z_{14} +4Z_{23}}{\Nc+1} \right) 
\end{align}
\end{subequations}

Normalizing transition operators as in \eq{eq:trans-norm} can be done as in sections~\ref{sec:sector-0b} and \ref{sec:sector-64} for the sectors of the $\Irrep{0}{b}$ and of the $\Irrep{64}{}$. Replacing $\Irrep{0}{b} \to \Irrep{35}{}$ in \eq{eq:CG-0b-12} and in \eq{eq:0b-c1dagc1}, one gets the two vertices $v_1$ and $v_2$ and the expression of $v_{1}^\dagger v_{1}$.   
The projector on the irrep ${\Irrep{35}{}}$ of $qqq\bar{q}\bar{q}\bar{q}$ is of the form \eq{eq:qqqqqqP0b}, with the quark (resp.~antiquark) sub-system in the irrep $\Irrep{10}{}$ (resp.~$\Irrep{8}{}$) of projector given by \eq{eq:qqq-10} (resp.~by the conjugate of \eq{eq:qqq-8}). We obtain: 
\begin{align}
v_{1}^\dagger  v_{1} \ &= \ \frac{2}{9} \left\{ J_1 - J_5 +2 J_3 -2 J_7 \right\} \ + \  \ldots
\end{align}
Comparing \eq{eq:projP35+}-\eq{eq:projP35-} with \eq{eq:P12-CG-8} we get: 
\begin{align}
a = \sqrt{\frac{2(\Nc+2)}{(\Nc+1)(\Nc+3)}} \ \ ; \ \ b = \sqrt{\frac{2(\Nc+2)}{(\Nc-1)(\Nc+5)}} \ ,
\end{align}
and the expressions \eq{eq:trans35} should be multiplied by $\frac{1}{9} ab$ to satisfy the normalization \eq{eq:trans-norm}.

\subsection{Eigenspace of $\hat{C}_2$ of dimension 9: sector of the $\Irrep{27}{}$}
\label{sec:eigenspaces-dim9}

We will now consider the eigenspace of $\hat{C}_2$ associated to the eigenvalue $C_R = 2\Nc+2$ (the Casimir of the irrep $\Irrep{27}{}$, see Table~\ref{table:irreps}). Since the $\Irrep{27}{}$ appears three times in the decomposition of $\Irrep{27}{} \otimes \Irrep{27}{}$, this eigenspace, denoted by $\E{\Irrep{27}{}}$ and dubbed as the `sector of the $\Irrep{27}{}$', is of dimension 9: it is spanned by three projectors ($\Proj{\Irrep{27}{-}}$, $\Proj{\Irrep{27}{+}}$, $\Proj{\Irrep{27'}{\!\!+}}$) and six transition operators ($\Trans{\Irrep{27}{-} \to \Irrep{27}{+}}$, $\Trans{\Irrep{27}{-} \to \Irrep{27'}{\!\!+}}$, $\Trans{\Irrep{27}{+} \to \Irrep{27'}{\!\!+}}$ and their Hermitian conjugates). 

The projector $\Proj{\Irrep{27}{-}}$ can be obtained directly from the kernel of the operator \eq{eq:null-space-dim4} for $C_R = 2\Nc+2$ and $(s_{12}, s_{34})=(-,-)$, and using Appendix~\ref{app:B3} to fix the normalization $\tr \Proj{\Irrep{27}{-}} = \K{\Irrep{27}{}}$. One finds
\begin{align}
\label{eq:P27minus}
\Proj{\Irrep{27}{-}} = \frac{(\Nc+2)}{(\Nc-2)(\Nc+1) (\Nc+4)} \left\{ C_{12}-C_{34}-\frac{4 \left(B_{12}-B_{34}\right)}{\Nc+2} \right\} \ . 
\end{align}

However, the method used in the previous sections proves insufficient to find the two other projectors and the six transition operators in $\E{\Irrep{27}{}}$. For instance, in order to look for the Hermitian projectors $\Proj{\Irrep{27}{+}}$ and $\Proj{\Irrep{27'}{\!\!+}}$, one may consider the kernel of\footnote{Recall that $\sigma_{\rm hc}$ denotes the Hermitian conjugation operator.}
\be
\label{eq:HphiZ-kernel}
\frac{\unit + s_{12} \sigma_{12}}{2} \cdot \frac{\unit + s_{34}  \sigma_{34}}{2} \cdot \frac{\unit + \lambda_{\rm hc} \sigma_{\rm hc}}{2} \cdot \hat{C}_2 - C_R \, \unit \ \ 
\ee
for $(s_{12}, s_{34}, \lambda_{\rm hc})=(+,+,+)$. But this subspace of $\E{\Irrep{27}{}}$ is of dimension 3. It is spanned by $\Proj{\Irrep{27}{+}}$, $\Proj{\Irrep{27'}{\!\!+}}$, and the sum of transition operators $\Trans{\Irrep{27}{+} \to \Irrep{27'}{\!\!+}}+\Trans{\Irrep{27'}{\!\!+} \to \Irrep{27}{+}}$, and it is not possible to further separate this subspace using symmetry arguments. 
Due to the presence of three equivalent irreps $\Irrep{27}{}$, two of which have a fortiori the same symmetry with respect to the permutation $\sigma_{12}$ (namely $\Irrep{27}{+}$ and $\Irrep{27'}{\!\!+}$), the sector of the $\Irrep{27}{}$ is a little more complex than the other eigenspaces of $\hat{C}_2$ considered previously. 

To determine the projectors $\Proj{\Irrep{27}{+}}$ and $\Proj{\Irrep{27'}{\!\!+}}$, as well as the six transition operators, we first need to derive the multiplication table between the tensors of $\E{\Irrep{27}{}}$.

\subsubsection{Vertices for $\Irrep{27}{} \otimes \Irrep{27}{} \to \Irrep{27}{}$}

From the knowledge of $\hat{C}_2$, we can directly write nine eigenvectors of $\hat{C}_2$ associated to the eigenvalue $C_R = 2\Nc+2$, \ie, a basis of $\E{\Irrep{27}{}}$: 
\begin{subequations}
\label{eq:tensors27}
\begin{align}
\tilde{H} \equiv&\ \frac{H_{12}+H_{34}}{4} - \frac{B_{12}+B_{34}}{4(\Nc+2)} + \frac{S}{2(\Nc+1)(\Nc+2)} \ , \\[1mm]
\tilde{Z}_1 \equiv&\  Z_1 - \frac{B_{13}}{\Nc+2} + \frac{S}{(\Nc+1)(\Nc+2)}  \ ,  \\[1mm]
\tilde{Z}_2 \equiv&\  Z_2 - \frac{B_{24}}{\Nc+2} + \frac{S}{(\Nc+1)(\Nc+2)} \ ,   \\[1mm]
\tilde{Z}_3 \equiv&\  Z_3 - \frac{B_{14}}{\Nc+2} + \frac{S}{(\Nc+1)(\Nc+2)} \ ,   \\[1mm]
\tilde{Z}_4 \equiv&\  Z_4 - \frac{B_{23}}{\Nc+2} + \frac{S}{(\Nc+1)(\Nc+2)} \ ,  \\[1mm]
\tilde{C}_i \equiv&\  C_i - \frac{ 4 B_{i}}{\Nc+2} + \frac{2\,S}{(\Nc+1)(\Nc+2)} \hskip 5mm ({\rm for}\ i=1,\ldots,4) \ . 
\end{align}
\end{subequations}

From the birdtrack expressions of the tensors given in Table~\ref{table:TensorBasis}, we see that in each of the above expressions, the `leading term' (involving the tensors associated to the lowest number of traces) corresponds to an intermediate state made of a symmetric quark pair and a symmetric antiquark pair, whose decomposition into irreps is given by \eq{6times6bar-deco}--\eq{6times6bar-comp}. Subtracting the irreps $\Irrep{1}{}$ and $\Irrep{8}{}$ from this leading term (\ie, replacing the intermediate state by the projector~\eq{eq:proj27-def}) directly yields each of the expressions \eq{eq:tensors27}. This shows that the nine eigenvectors \eq{eq:tensors27} can be represented pictorially as the birdtrack for the leading term, up to the replacement of the intermediate state by the blob denoting the projector~\eq{eq:proj27-def} on the irrep $\Irrep{27}{}$. For instance:
\begin{align}
\label{C2C3Z4}
&\tilde{C}_2 \  =  \ \TCMblob  \ \ ; \ \ \tilde{C}_3 \  =  \ \TDPblob  \ \ ; 
\ \ \tilde{Z}_4 \  = \ \TFMblob \ \ , \ {\rm etc\ldots} 
\end{align}

Since there are three equivalent irreps $\Irrep{27}{}$, all vectors of $\E{\Irrep{27}{}}$ can be expressed in terms of three vertices $v_i$ mapping $\Irrep{27}{} \otimes \Irrep{27}{} \to \Irrep{27}{}$ (see section~\ref{sec:clebsches}). Choosing 
\begin{align}
\label{Clebsches-27}
v_1 \ \equiv \ \ClebschONE \ \ ; \ \ v_2 \ \equiv \ \ClebschTWO \ \ ; \ \ v_3 \ \equiv \ \ClebschTHREE \ \ ,
\end{align}
and renaming the eigenvectors \eq{eq:tensors27} as
\begin{align}
\label{tensors27-renaming}
\{ \tilde{H},\tilde{Z}_1,\tilde{Z}_2,\tilde{Z}_3,\tilde{Z}_4,\tilde{C}_1,\tilde{C}_2,\tilde{C}_3,\tilde{C}_4 \} \equiv \{ E_{11},E_{12},E_{13},E_{21},E_{31},E_{22},E_{33},E_{32},E_{23} \} \ , 
\end{align}
we quickly verify {\it visually} that the nine eigenvectors are related to the $v_i$'s as follows:\footnote{Recall that in the product $v_i^\dagger v_j$, the operator $v_j$ applies first and is drawn to the left in the corresponding birdtrack expression, see footnote~\ref{foot:rightop-applies-first}.} 
\begin{align}
\label{Eij-def}
E_{ij} \ = \ v_i^\dagger v_j  \ . 
\end{align}

In the following we will need the traces of the $E_{ij}$'s, which using \eq{tensors27-renaming} and \eq{eq:tensors27} are directly obtained from the traces of the tensors $\tau$ listed in Appendix~\ref{app:B3}. 
Since those traces are real, we can write $ \tr ( E_{ij} ) =  \tr ( v_i^\dagger v_j ) = \tr ( v_j^\dagger v_i ) = \tr ( E_{ji} )$, so that 
\begin{align}
\label{nij-def}
n_{ij} \equiv \frac{\tr ( E_{ij} )}{\K{\Irrep{27}{}}} 
\end{align}
is a real, symmetric $3 \times 3$ matrix. We find: 
\begin{subequations}
\label{nij-mat}
\begin{align}
n_{11} =& \ \frac{\Nc^5+9 \Nc^4+23 \Nc^3+11 \Nc^2+8 \Nc+28}{16 (\Nc+1) (\Nc+2)^2} \ , \\[2mm]
n_{22} =& \ n_{33} = \ \frac{\Nc^5+6 \Nc^4+5 \Nc^3-20 \Nc^2-20 \Nc+8}{2 (\Nc+1) (\Nc+2)^2} \ , \\[2mm]
n_{12} =&\  n_{13} = \ -\frac{(2 \Nc+3) \left(\Nc^2+3 \Nc-2\right)}{2 (\Nc+1) (\Nc+2)^2} \ , \\[2mm]
n_{23} =& \ \frac{4 (2 \Nc+3)}{(\Nc+1) (\Nc+2)^2} \ . 
\end{align}
\end{subequations}

The multiplication table of the eigenvectors \eq{tensors27-renaming} then reads 
\begin{align}
\label{mult-table}
E_{ij} E_{kl}  \ =&\ v_i^\dagger v_j  v_k^\dagger v_l = n_{jk} v_i^\dagger v_l =   n_{jk} E_{il} \ , 
\end{align}
where we used $v_j  v_k^\dagger = \frac{1}{\K{\Irrep{27}{}}} \, \tr ( v_j  v_k^\dagger ) \, \Proj{\Irrep{27}{}}$ (a direct consequence of Schur's lemma, with $\Proj{\Irrep{27}{}}$ defined by \eq{eq:proj27-def}), and $\tr ( v_j  v_k^\dagger ) = \tr ( v_k^\dagger v_j ) = \K{\Irrep{27}{}} \, n_{jk}\,$.

\subsubsection{Projectors and transition operators of $\E{\Irrep{27}{}}$}
\label{sec:proj27}

In this section we derive the three projectors and six transition operators of $\E{\Irrep{27}{}}$, starting from their general expression in terms of the vertices \eq{Clebsches-27}. 

Under $\psi \leftrightarrow \chi$, $v_1$ is invariant and $v_2 \leftrightarrow v_3$. From the symmetries of $\Irrep{27}{-}$, $\Irrep{27}{+}$ and $\Irrep{27'}{\!\!+}$ under $\psi \leftrightarrow \chi$, the corresponding projectors must be of the form (see section~\ref{sec:clebsches}):
\begin{subequations}
\label{eq:P27s-CG} 
\begin{align}
\label{eq:P27p-CG}
\Proj{\Irrep{27}{+}} &= \ \left[ a v_1 + b (v_2 + v_3) \right]^\dagger \left[ a v_1 + b (v_2 + v_3) \right] \ ,  \\[2mm]
\label{eq:P27pp-CG}
\Proj{\Irrep{27'}{\!\!+}} &= \ \left[ a' v_1 + b' (v_2 + v_3) \right]^\dagger \left[ a' v_1 + b' (v_2 + v_3) \right] \ , \\[2mm] 
\label{eq:P27m-CG}
\Proj{\Irrep{27}{-}} &= \ \left[ r (v_2 - v_3) \right]^\dagger \left[ r (v_2 - v_3) \right] \ .
\end{align}
\end{subequations}
We need to find the coefficients $a, \, b, \, a', \, b', \, r$ ensuring that each projector indeed projects on an irrep $\Irrep{27}{}$, and that the three projectors are orthogonal to each other. 

First, the coefficient $r$ in the projector $\Proj{\Irrep{27}{-}}$ is directly obtained by expanding \eq{eq:P27m-CG}, 
\begin{align}
\Proj{\Irrep{27}{-}} &= \ r^2 \left( v_2^\dagger v_2 + v_3^\dagger v_3 - v_2^\dagger v_3  - v_3^\dagger v_2  \right) \ , 
\end{align} 
and imposing $\tr \Proj{\Irrep{27}{-}} = \K{\Irrep{27}{}}$. Recalling \eq{Eij-def} and \eq{nij-def}, and the fact that $n_{22} = n_{33}$ (see \eq{nij-mat}), we obtain $r^2 = \frac{1}{2(n_{22}-n_{23})}$. Thus,
\begin{align}
\label{eq:P27m-CG-2}
\Proj{\Irrep{27}{-}} &= \  c_a^\dagger c_a \ = \ \frac{\tilde{C}_1 + \tilde{C}_2 - \tilde{C}_3 - \tilde{C}_4}{2(n_{22}-n_{23})} \ , 
\end{align} 
where $c_a$ is the `antisymmetric clebsch' defined by
\begin{align}
\label{eq:ca} 
c_a & \equiv\ \frac{v_2 - v_3}{\sqrt{2(n_{22}-n_{23})}} \ . 
\end{align}

Using \eq{nij-mat}, we readily check that
\be
2(n_{22}-n_{23}) = \frac{(\Nc-2)(\Nc+1) (\Nc+4)}{\Nc+2} \ , 
\ee
and \eq{eq:P27m-CG-2} thus duly coincides with \eq{eq:P27minus} (obtained using the Casimir operator). 
As a non-trivial cross-check one can verify using the multiplication table \eq{mult-table} that the r.h.s.~of~\eq{eq:P27m-CG-2} is indeed a projector, $\Proj{\Irrep{27}{-}}^2 = \Proj{\Irrep{27}{-}}$. 
The fact that $\Proj{\Irrep{27}{-}}$ is orthogonal to both $\Proj{\Irrep{27}{+}}$ and $\Proj{\Irrep{27'}{\!\!+}}$ follows from their different symmetries under $\psi \leftrightarrow \chi$, or alternatively from \eq{eq:P27s-CG} by noting that $\tr [ v_1^\dagger (v_2-v_3) ] = \tr [ (v_2+v_3)^\dagger (v_2-v_3) ] = 0$ (see~\eq{eq:ccdag-traces}). 

We now look for the projectors $\Proj{\Irrep{27}{+}}$ and $\Proj{\Irrep{27'}{\!\!+}}$, which after some rescaling of the parameters $a, b, a', b'$ in \eq{eq:P27p-CG}-\eq{eq:P27pp-CG} can be written as 
\begin{subequations}
\label{eq:P27ppp-ab}
\begin{align}
\Proj{\Irrep{27}{+}} &= \ c_1^\dagger c_1 \ , \\[1mm]
\Proj{\Irrep{27'}{\!\!+}} &= \ c_2^\dagger c_2 \ , 
\end{align}
\end{subequations}
in terms of the `symmetric clebsches' 
\begin{subequations}
\label{eq:c1c2}
\begin{align}
c_1 &\equiv\ a\, \frac{v_1}{\sqrt{n_{11}}} + b\, \frac{v_2 + v_3}{\sqrt{2(n_{22}+n_{23})}} \ , \\ 
c_2 &\equiv\ a' \frac{v_1}{\sqrt{n_{11}}} + b'\, \frac{v_2 + v_3}{\sqrt{2(n_{22}+n_{23})}} \ . 
\end{align}
\end{subequations}
Expanding the expressions \eq{eq:P27ppp-ab} and using \eq{Eij-def} and \eq{tensors27-renaming} we get:\footnote{The minus sign in front of the third term of~\eq{eq:P27p-basis-tau} or \eq{eq:P27pp-basis-tau} arises from $n_{12} <0$, see~\eq{nij-mat}.}
\begin{subequations}
\label{eq:P27ppp-basis-tau}
\begin{align}
\label{eq:P27p-basis-tau}
\Proj{\Irrep{27}{+}} \ &= \  a^2 \frac{\tilde{H}}{n_{11}}\, + b^2 \frac{\tilde{C}}{2(n_{22}+n_{23})}\, - \frac{2ab}{\sqrt{\rho}} \frac{\tilde{Z}}{4 n_{12}} \ ,  \\[2mm]
\label{eq:P27pp-basis-tau}
\Proj{\Irrep{27'}{\!\!+}} \ &= \  a'^{\,2} \frac{\tilde{H}}{n_{11}}\, + b'^{\,2} \frac{\tilde{C}}{2(n_{22}+n_{23})}\, - \frac{2a'b'}{\sqrt{\rho}} \frac{\tilde{Z}}{4 n_{12}} \ ,
\end{align}
\end{subequations}
where the tensors $\tilde{C}$ and $\tilde{Z}$ are given by (see~\eq{eq:tensors27})
\begin{subequations}
\label{CZtilde-def}
\begin{align}
\tilde{C} \equiv  \sum_{i=1}^{4}  \tilde{C}_i  \ =& \ C_{12}+C_{34}  - \frac{4(B_{12}+B_{34})}{\Nc+2} + \frac{8 \, S}{(\Nc+1)(\Nc+2)} \ , \\
\tilde{Z} \equiv  \sum_{i=1}^{4}  \tilde{Z}_i  \ =& \ Z_{12}+Z_{34}  - \frac{2(B_{12}+B_{34})}{\Nc+2} + \frac{4 S}{(\Nc+1)(\Nc+2)}\ ,
\end{align}
\end{subequations}
and the parameter $\rho$ by\footnote{$\rho > 1$ for any $\Nc \geq 3$ follows from the identity $4n_{12}^2(\rho-1) = \frac{(\Nc-2)(\Nc^2-1)(\Nc+4)(\Nc+5)}{16(\Nc+2)}\,$.} 
\begin{align}
\label{rho-def}
\rho \ \equiv \ \frac{n_{11}(n_{22}+n_{23})}{2n_{12}^2}  \ > \, 1  \ .
\end{align}
The fact that $\Proj{\Irrep{27}{+}}$ and $\Proj{\Irrep{27'}{\!\!+}}$ are linear combinations of $\tilde{H} $, $\tilde{C}$ and $\tilde{Z}$ is not a surprise: as one can verify, the set $\{ \tilde{H},\tilde{C},\tilde{Z} \}$ is a basis of the kernel of \eq{eq:HphiZ-kernel} for $(s_{12}, s_{34}, \lambda_{\rm hc})=(+,+,+)$, \ie, of the subspace of $\E{\Irrep{27}{}}$ spanned by the tensors that are Hermitian and  symmetric under permutations $\sigma_{12}$ and $\sigma_{34}$. 

We now determine the coefficients $a, \, b, \, a', \, b'$ appearing in \eq{eq:c1c2} and \eq{eq:P27ppp-basis-tau}. Requiring $\tr \Proj{\Irrep{27}{+}} = \tr \Proj{\Irrep{27'}{\!\!+}} = \K{\Irrep{27}{}}$, we obtain:\footnote{In~\eq{eq:P27ppp-basis-tau}, use $\tr \frac{\tilde{H}}{n_{11}} = \tr \frac{\tilde{C}}{2(n_{22}+n_{23})} = \tr \frac{\tilde{Z}}{4 n_{12}} = \K{\Irrep{27}{}}$, which simply follows from~\eq{tensors27-renaming} and \eq{nij-def}.}
\begin{align}
\label{syst-ab}
& a^2 + b^2 -  \frac{2ab}{\sqrt{\rho}} \ = \  a'^{\,2} + b'^{\,2} -  \frac{2a'b'}{\sqrt{\rho}}  \ = \ 1 \ .
\end{align}
The orthogonality of $\Proj{\Irrep{27}{+}}$ and $\Proj{\Irrep{27'}{\!\!+}}$ is equivalent to
$\tr [ c_2^\dagger c_1 ] =0$ (see section~\ref{sec:clebsches} and Eq.~\eq{eq:ccdag-traces}), which leads to: 
\begin{align}
\label{syst-ab-ortho}
a a' + b b' -  \frac{a' b + a b'}{\sqrt{\rho}} \ &= \ 0 \  . 
\end{align}
Equations~\eq{syst-ab}-\eq{syst-ab-ortho} form a system of three equations with four unknowns, whose set of solutions can be found in Appendix~\ref{app:C1}, see Eq.~\eq{eq:abab}. 

All solutions for the projectors \eq{eq:P27ppp-basis-tau} are thus of the form 
\begin{subequations}
\label{eq:P27ppp-param}
\vspace{-2mm}
\begin{align}
\label{eq:P27p-param}
\Proj{\Irrep{27}{+}} &\equiv P(t) =  \frac{\rho}{\rho-1} \left[ \frac{\cos^2{(\theta +t)}}{n_{11}} \, \tilde{H} + \frac{\cos^2{(\theta -t)}}{2(n_{22}+n_{23})}\, \tilde{C} \right] - \frac{1+\sqrt{\rho} \cos{2t}}{4 n_{12}(\rho-1)} \, \tilde{Z} \ ,  \\[2mm]
\label{eq:P27pp-param}
\Proj{\Irrep{27'}{\!\!+}} &\equiv P'(t) =  \frac{\rho}{\rho-1} \left[ \frac{\sin^2{(\theta +t)}}{n_{11}} \, \tilde{H} + \frac{\sin^2{(\theta -t)}}{2(n_{22}+n_{23})}\, \tilde{C} \right] - \frac{1-\sqrt{\rho} \cos{2t}}{4 n_{12}(\rho-1)} \, \tilde{Z} \ , \\[-3mm] \nn 
\end{align}
\end{subequations}
where the fixed parameter $\theta \in \ ] 0,  \frac{\pi}{4} [$ is defined by~\eq{theta-def}, and the parameter $t$ can a priori vary within $[0,  2 \pi [$. Noting that $P(t + \pi) = P(t)$ and $P'(t + \pi) = P'(t)$, the latter interval can however be reduced to $[ 0, \pi [$.

The non-unicity of $\Proj{\Irrep{27}{+}}$ and $\Proj{\Irrep{27'}{\!\!+}}$ arises from the non-unicity of clebsches~\cite{Cvitanovic:2008zz}, and the fact that the equivalent irreps $\Irrep{27}{+}$ and $\Irrep{27'}{\!\!+}$ cannot be distinguished from their `external' symmetry under $\psi \leftrightarrow \chi$.\footnote{A similar situation illustrating the non-unicity of projectors associated with equivalent irreps is encountered when decomposing simple systems of three partons, such as $qqq$, $qq\bar{q}$, $\Syst{qg}{\Irrep{15}{}}{g}$, see Ref.~\cite{Peigne:2024srm}.}
As expected from section~\ref{sec:clebsches} (see~\eq{CG-rot} and the following discussion) and checked in Appendix~\ref{app:C1} (see~\eq{c1c2-rot}), the set of solutions for $(\Proj{\Irrep{27}{+}},  \Proj{\Irrep{27'}{\!\!+}})$ is obtained by rotating a particular solution for the clebsches~\eq{eq:c1c2}. As a result (see~\eq{eq:Malpha-trace}), the sum of projectors,
\begin{align}
\label{eq:Psum}
\Proj{{\rm sum}}{} \equiv P(t)+P'(t)   \ =& \ 
\frac{\rho}{\rho-1}  \left[ \frac{\tilde{H}}{n_{11}} + \frac{\tilde{C}}{2(n_{22}+n_{23})}  \right] - \frac{1}{\rho-1} \, \frac{\tilde{Z}}{2 n_{12}}  \ , 
\end{align}
is independent of $t$, \ie, uniquely defined.\footnote{In the present study of $\Irrep{27}{} \otimes \Irrep{27}{}$, this also follows trivially from the completeness relation~\eq{comp-rel}, since to the exception of $\Proj{\Irrep{27}{+}}$ and $\Proj{\Irrep{27'}{\!\!+}}$, all projectors on the irreps of $\Irrep{27}{} \otimes \Irrep{27}{}$ are uniquely determined.} 
The pair $(P(t),P'(t))$ is thus determined by giving either $P(t)$ or $P'(t)$, and since $P(t + \frac{\pi}{2}) = P'(t)$, it is sufficient to vary $t$ in $[ 0,  \frac{\pi}{2} [$ to find all solutions. A few convenient choices for $(\Proj{\Irrep{27}{+}},\Proj{\Irrep{27'}{\!\!+}})$, corresponding to some specific values of $t$, are listed in Appendix~\ref{app:C2}. 

We end this section by giving the expressions of the six transition operators of $\E{\Irrep{27}{}}$. These operators all depend on $\Proj{\Irrep{27}{+}}$ or $\Proj{\Irrep{27'}{\!\!+}}$, and thus on the continuous parameter $t$. 
From section~\ref{sec:clebsches} and the expressions of $\Proj{\Irrep{27}{-}}$ (resp.~$\Proj{\Irrep{27}{+}}$ and $\Proj{\Irrep{27'}{\!\!+}}$) given in \eq{eq:P27m-CG-2} (resp.~\eq{eq:P27ppp-ab}) in terms of the antisymmetric clebsch~\eq{eq:ca} (resp.~symmetric clebsches~\eq{eq:c1c2}), the transition operators normalized according to \eq{eq:trans-norm} are:
\begin{subequations}
\label{eq:trans27-CG}
\begin{align}
\label{eq:trans27-ppp-CG}
\Trans{\Irrep{27}{+} \to \Irrep{27'}{\!\!+}}  &\equiv \ T(t) \ = \ c_2^\dagger c_1 \ , \\[1mm]
\label{eq:trans27-pm-CG}
\Trans{\Irrep{27}{+} \to \Irrep{27}{-}} &\equiv \ T_{a}(t) \ = \ c_a^\dagger c_1 \ , \\[1mm]
\label{eq:trans27-ppm-CG}
\Trans{\Irrep{27'}{\!\!+} \to \Irrep{27}{-}} &\equiv \ T'_{a}(t) \ = \ c_a^\dagger c_2 \ ,
\end{align}
\end{subequations}
the other three transition operators being obtained by Hermitian conjugation.
Expanding \eq{eq:trans27-CG} and using \eq{tensors27-renaming}, \eq{Eij-def} and \eq{rho-def} we obtain: 
\begin{subequations}
\label{eq:trans27-ab}
\begin{align}
\label{eq:trans27-ppp-ab}
T(t) &= \  a a' \frac{\widetilde{H}}{n_{11}} + b b'  \frac{\widetilde{C}}{2(n_{22}+n_{23})} -\frac{a' b(\widetilde{Z}_{1}+\widetilde{Z}_{2}) + a b' (\widetilde{Z}_{3}+\widetilde{Z}_{4}) }{2n_{12} \sqrt{\rho}} \ , \\[2mm]
\label{eq:trans27-pm-ab}
T_{a}(t)  &=\ \frac{1}{\sqrt{2(n_{22}-n_{23})}} \left[ a \, \frac{\widetilde{Z}_3-\widetilde{Z}_4}{\sqrt{n_{11}}} + b \, \frac{\widetilde{C}_{1}+\widetilde{C}_{4} - \widetilde{C}_{2}- \widetilde{C}_{3}}{\sqrt{2(n_{22}+n_{23})}} \right] \ , \\[2mm]
\label{eq:trans27-ppm-ab}
T'_{a}(t) &= \ \frac{1}{\sqrt{2(n_{22}-n_{23})}} \left[ a' \, \frac{\widetilde{Z}_3-\widetilde{Z}_4}{\sqrt{n_{11}}} + b' \, \frac{\widetilde{C}_{1}+\widetilde{C}_{4} - \widetilde{C}_{2}- \widetilde{C}_{3}}{\sqrt{2(n_{22}+n_{23})}} \right] \ .
\end{align}
\end{subequations}
With $a, b, a', b'$ given by \eq{eq:abab}, the `tilde tensors' defined by \eq{eq:tensors27}, and the Hermitian conjugates being obtained using Table~\ref{tab:CrossingR}, the expressions \eq{eq:trans27-ab} provide the six transition operators of $\E{\Irrep{27}{}}$ in explicit form, for any choice of the parameter $t$. 

In the next section we will choose a value of $t$ to define $\Proj{\Irrep{27}{+}}=P(t)$, $\Proj{\Irrep{27'}{\!\!+}}=P'(t)$ and the six transition operators ($T(t),\, T_{a}(t),\, T'_{a}(t)$ and their Hermitian conjugates), this set of eight operators being defined in terms of clebsches by \eq{eq:P27ppp-ab} and \eq{eq:trans27-CG}. In case we want to make another choice, the relation between sets corresponding to different values of $t$ directly follows from the symmetric clebsches transforming under a simple rotation $R(t) \in {\rm SO}(2)$ when $t$ changes, see \eq{Rt-def}-\eq{c1c2-rot}. 

Consider first the subset $\{ P(t),\, P'(t),\, T(t),\, T(t)^\dagger \}$. According to \eq{Malphaprime}, the 
matrix of operators \eq{eq:P12T12-CG} for a general parameter $t$ is related to that for $t=0$ by 
\begin{align}
\label{eq:PTPT}
\left( \begin{matrix}  P(t)  \ & \ T(t)^\dagger \  \\[2mm] T(t)  \ & \  P'(t) \end{matrix}  \right)  \ = \ 
R(t) \cdot 
\left( \begin{matrix}  P(0)  \ & \ T(0)^\dagger \  \\[2mm] T(0)  \ & \ P'(0) \end{matrix}  \right)
\cdot R(t)^{\intercal} \ .
\end{align} 
Using \eq{Rt-def} and performing the matrix product in \eq{eq:PTPT}, we find  
\begin{subequations}
\begin{align}
\label{eq:PpP} P(t) + P'(t) &= \ P(0) + P'(0) \ , \\
\label{eq:TmT} T(t) - T(t)^\dagger &= \ T(0) - T(0)^\dagger \ , \\[2mm]
\label{eq:PmP-TpT}
\left( \begin{matrix} P(t)-P'(t) \\[2mm] T(t) + T^\dagger(t) \end{matrix}  \right) 
= \ & R(2t) \cdot \left( \begin{matrix} P(0)-P'(0) \\[2mm] T(0) + T(0)^\dagger \end{matrix}  \right) \ .
\end{align}
\end{subequations}
We see that not only $P+P'$, but also the linear combination $T - T^\dagger$ is independent of $t$,\footnote{The difference of the off-diagonal elements of a $2\times 2$ matrix is indeed invariant under a similarity transformation when the latter is a rotation, as in \eq{eq:PTPT}. But the unicity of $T - T^\dagger$ can also be understood as follows. In the subspace of $\E{\Irrep{27}{}}$ being symmetric under $\sigma_{12}$ and $\sigma_{34}$, of basis $\{ P(t), P'(t), T(t), T(t)^\dagger \}$, the operator $T - T^\dagger$ is the only possible {\it anti-Hermitian} one, corresponding to the kernel of \eq{eq:HphiZ-kernel} being one-dimensional for $(s_{12}, s_{34}, \lambda_{\rm hc})=(+,+,-)$. Non-unicity only appears in the sector of Hermitian tensors -- the kernel of \eq{eq:HphiZ-kernel} for $(s_{12}, s_{34}, \lambda_{\rm hc})=(+,+,+)$, of basis $\{ P(t), P'(t), T(t)+T(t)^\dagger \}$. 
Note also that $i(T - T^\dagger)$ is the `square root' of $P+P'$, which simply follows from \eq{eq:trans-norm}: 
\begin{align}
(T - T^\dagger)^2 &= - (P+P') \ . \nn 
\end{align}
} 
and that the doublet $(P - P', T + T^\dagger)$ transforms under a rotation of parameter $2t$ when $t$ varies. 
As a cross-check, we easily verify that the general expressions \eq{eq:P27ppp-param} and \eq{eq:trans27-ppp-ab} found for $P(t)$, $P'(t)$ and $T(t)$ indeed satisfy \eq{eq:TmT} and \eq{eq:PmP-TpT}.  

As for the subset $\{T_{a}(t),\, T'_{a}(t), T_{a}(t)^\dagger,\, T'_{a}(t)^\dagger \}$,  we directly find using \eq{eq:trans27-pm-CG}-\eq{eq:trans27-ppm-CG} and \eq{c1c2-rot} that the two doublets $(T_{a}(t),\, T'_{a}(t) )$ and $( T_{a}(t)^\dagger,\, T'_{a}(t)^\dagger )$ transform like the doublet of symmetric clebsches $(c_1(t),c_2(t))$ when $t$ varies: 
\begin{align}
\label{eq:trans-pmppm}
&\left( \begin{matrix} T_{a}(t) \\[2mm] T'_{a}(t)  \end{matrix}  \right) = c_a^\dagger \left( \begin{matrix} c_1(t) \\[2mm] c_2(t)  \end{matrix}  \right) 
= R(t) \cdot \left( \begin{matrix} T_{a}(0) \\[2mm] T'_{a}(0)  \end{matrix}  \right) ; \ 
\left( \begin{matrix} T_{a}(t)^\dagger \\[2mm] T'_{a}(t)^\dagger \end{matrix}  \right) =
\left( \begin{matrix} c_1(t)^\dagger \\[2mm] c_2(t)^\dagger  \end{matrix}  \right) c_a = R(t) \cdot 
\left( \begin{matrix} T_{a}(0)^\dagger \\[2mm] T'_{a}(0)^\dagger \end{matrix}  \right) \ . \nn \\
\end{align}
It follows that any linear combination of the doublets $( T_{a}(t),\, T'_{a}(t) )$ and $( T_{a}(t)^\dagger,\, T'_{a}(t)^\dagger )$ will transform in the same way under a change of parameter $t$.

\subsection{Sum up}
\label{subsec:Sum-up}

We have found explicit expressions of all projectors and transition operators spanning the full color space of $\Irrep{27}{} \otimes \Irrep{27}{} \to \Irrep{27}{} \otimes \Irrep{27}{}$, as linear combinations of the 42 tensors defined in Table~\ref{table:TensorBasis} of Appendix~\ref{app:B}. 

The 26 Hermitian projectors are given by Eqs.~\eq{eq:projP1}--\eq{eq:projP125}, \eq{eq:projP10}, \eq{eq:projP0d}, \eq{eq:projP28}, \eq{eq:projP81}, \eq{proj-8p8m}, \eq{eq:projP0b+0b-}, \eq{eq:projP64+64-}, \eq{eq:projP35+}, \eq{eq:projP35-}, \eq{eq:P27minus}, \eq{eq:P27ppp-param}, and the 16 transition operators by Eqs.~\eq{trans-8pm}, \eq{eq:trans0bpm}, \eq{eq:trans64pm}, \eq{eq:trans35}, \eq{eq:trans27-ab}. It is gratifying to check that all 26 Hermitian projectors explicitly satisfy the completeness relation \eq{comp-rel}.

\section{Soft anomalous dimension matrix}
\label{sec:sADM}

A rigorous treatment of hard QCD processes requires including corrections from induced gluon radiation. Soft and quasi-collinear gluon radiation gives rise to double logarithmic (DL) contributions, resummed in (exponential) Sudakov form factors, one for each of the external partons (the Sudakov factor associated to a given parton depending on $\Nc$ only through the color charge of that parton). For hard processes where the external particles comprise only two or three partons (with possibly other electroweak particles), it is possible, thanks to the {\it color triviality}~\cite{Dokshitzer:2005ig} of such processes, to  account for both DL and SL corrections by a product of Sudakov form factors. However, already for processes involving four partons, certain SL contributions from soft and large-angle gluon radiation cannot be taken into account in this way~\cite{Botts:1989kf}, and require the introduction of a ``fifth form factor''~\cite{Dokshitzer:2005ig}, obtained by exponentiating the so-called soft anomalous dimension matrix (sADM)~\cite{Botts:1989kf,Sotiropoulos:1993rd,Contopanagos:1996nh,Kidonakis:1998nf,Oderda:1999kr,Kyrieleis:2005dt,Dokshitzer:2005ig}. 

The sADM associated with a $2 \to 2$ partonic process depends on the different color states available in the $s, t, u$ channels of the process, and therefore presents a {\it non-trivial} color structure, requiring knowledge of the Hermitian projectors on the two-parton states (and in general of transition operators between them) in the different channels. In this section we consider the sADM of $\Irrep{27}{} \otimes \Irrep{27}{} \to \Irrep{27}{} \otimes \Irrep{27}{}$, defined in the $s$-channel as 
\begin{align}
\label{s-chan-def}
\schanprocess{\cal M}{0.3} \ \ ,
\end{align}
where the `generalized partons' $\Irrep{27}{}$ (drawn using the notational convention \eq{eq:color-map}) are numbered counter-clockwise. The sADM will directly follow from the projectors and transition operators derived previously. 

We first discuss the crossing symmetry of this process (section~\ref{sec:quantum-numbers}), then the block-diagonal structure of the sADM (section~\ref{sec:sADM-structure}) and explicitly give its characteristic polynomial (section~\ref{sec:sADM-char-pol}).

\subsection{Quantum numbers and crossing symmetry}
\label{sec:quantum-numbers}

The ($s$-channel) quadratic Casimir operator $\hat{C}_2$ and the operators $\sigma_{\rm cc}$, $\sigma_y \equiv \sigma_{14} \sigma_{23}$, $\sigma_x \equiv \sigma_{12} \sigma_{34}$ are diagonalizable and commute.\footnote{Note that $\sigma_x$ and $\sigma_y$ belong to the group $\{ \unit, \sigma_x, \sigma_y, \sigma_x \sigma_y \}$, the symmetry group of a (non-square) rectangle (which is an abelian subgroup of the symmetric group ${\rm S}_4$). This motivates the notations $\sigma_x$ and $\sigma_y$ for the reflections in the horizontal and vertical axes.} It is therefore convenient to organize the eigenvectors of $\hat{C}_2$ (namely, the 42 Hermitian projectors and transition operators in the {\it $s$-channel}) as being simultaneously eigenvectors of $\sigma_{\rm cc}$, $\sigma_y$, $\sigma_x$, \ie\ having specific $\pm 1$ ``quantum numbers'' with respect to these operators. 

\begin{table}[t]
\renewcommand{\arraystretch}{1.2} 
\setlength\tabcolsep{4pt}
\hskip 0mm
\begin{center}
\begin{tabular}{|c|c|c|}
\hline
$\ba{c}  {\rm quantum\  numbers}  \\[-2mm]  
\lambda=(\sigma_{\rm cc}, \sigma_y,\sigma_x) \ea$ &   $\mathcal{V}_\lambda^{(s)}$  &  
$\ba{c} \mathcal{T}'_{\lambda}  \ea$ \\
\hline
$(+,+,+)$  &  $\ba{c} {\rm 21 \ operators:} \\[-1mm]  
{\rm 20 \ projectors \ and \ } 
\TpTdag{\Irrep{27}{+} \to \Irrep{27'}{\!\!+}}  \ea$ &
$\ba{c} {\rm 21 \ operators:} \\[-1mm]  
S,B_{12},B_{34},C_{12},C_{34}, \\[-2mm]  Z_{12}+Z_{34}, H_{1}, H_{2}, H_{34}, \\[-2mm] J_{12},J_{34}+J_{56},J_{78}, \\[-2mm] K_{12},K_{34}+K_{56},K_{78}, \\[-2mm] 
L_{1}, L_{2}, L_3, L_{45}, L_{67}, L_{89} \ea$  \\ 
\hline
$(+,+,-)$  &   $\ba{c} {\rm 1 \ operator:} \\[-1mm]  
\TpTdag{\Irrep{35}{+} \to \Irrep{35}{-}} \!\!+ 
\TpTdag{\Irrep{\overline{35}}{+} \to \Irrep{\overline{35}}{-}}  \ea$ & 
$\ba{c} {\rm 1 \ operator:} \\[-1mm]  \Jbar_{34} - \Jbar_{56} \ea$ \\ 
\hline
$(+,-,+)$  &   $\ba{c} {\rm 1 \ operator:} \\[-1mm]  
\TmTdag{\Irrep{27}{+} \to \Irrep{27'}{\!\!+}} \ea$ & 
$\ba{c} {\rm 1 \ operator:} \\[-1mm]  Z_{12} - Z_{34}  \ea$  \\ 
\hline
$(+,-,-)$  &  $\ba{c} {\rm 1 \ operator:} \\[-1mm]  
\TmTdag{\Irrep{35}{+} \to \Irrep{35}{-}} \!\!
+ \TmTdag{\Irrep{\overline{35}}{+} \to \Irrep{\overline{35}}{-}}  \ea$ & 
$\ba{c} {\rm 1 \ operator:} \\[-1mm]  \Kbar_{34} - \Kbar_{56}  \ea$  \\ 
\hline
$(-,+,+)$  & $\ba{c} \ \\[-3mm] {\rm 0 \ operator} \\[-3mm] \ \ea$ &  $\ba{c} \ \\[-3mm] {\rm 0 \ operator} \\[-3mm] \  \ea$ \\ 
\hline
$(-,+,-)$  &  $\ba{c} {\rm 6 \ operators:} \\[-1mm]  
\TmTdag{\Irrep{\alpha}{+} \to \Irrep{\alpha}{-}}  \ (\alpha = 1 \ldots 5) \\[-2mm]
\TmTdag{\Irrep{35}{+} \to \Irrep{35}{-}} \!\! - 
\TmTdag{\Irrep{\overline{35}}{+} \to \Irrep{\overline{35}}{-}} \ea$ &  
$\ba{c} {\rm 6 \ operators:} \\[-1mm] 
\Bbar_{34},\Cbar_{34}, \Zbar_{12}-\Zbar_{34},\\[-2mm]  \Kbar_{12},\Kbar_{34}+\Kbar_{56},\Kbar_{78}
\ea$ \\ 
\hline
$(-,-,+)$  &  $\ba{c} {\rm 6 \ operators:} \\[-1mm]  \Proj{R}-\Proj{\overline{R} } \ea$ & 
$\ba{c} {\rm 6 \ operators:} \\[-1mm] 
\Hbar_{34}, J_{34} - J_{56} ,K_{34}-K_{56}, \\[-2mm]  \Lbar_{45},\Lbar_{67},\Lbar_{89} \ea$ \\ 
\hline
$(-,-,-)$  &  $\ba{c} {\rm 6 \ operators:} \\[-1mm]  
\TpTdag{\Irrep{\alpha}{+} \to \Irrep{\alpha}{-}}  \ (\alpha = 1 \ldots 5) \\[-2mm]
\TpTdag{\Irrep{35}{+} \to \Irrep{35}{-}} \!\! - 
\TpTdag{\Irrep{\overline{35}}{+} \to \Irrep{\overline{35}}{-}} \ea$ &  
$\ba{c} {\rm 6 \ operators:} \\[-1mm] 
\Bbar_{12},\Cbar_{12}, \Zbar_{12} + \Zbar_{34}, \\[-2mm]  \Jbar_{12},\Jbar_{34}+\Jbar_{56},\Jbar_{78}
\ea$ \\ 
\hline
\end{tabular}
\end{center}
\caption{\label{table:q-numbers} 
Two bases of the $\Irrep{27}{} \otimes \Irrep{27}{}$ color space, organized in sets corresponding to different quantum numbers $\lambda$ (first column) under the action of $\{\sigma_{\rm cc},\sigma_y, \sigma_x \}$. Second column: basis $\mathcal{V}_\lambda^{(s)}$, whose elements are eigenvectors of $\hat{C}_2$. For the entries $\lambda = (-,+,-)$ and $\lambda = (-,-,-)$, $\alpha = 1 \ldots 5$ denotes the five pairs of equivalent irreps in \eq{27times27-deco} having different symmetries under $\sigma_{12}$ and being distinct from the $\Irrep{35}{}$ and $\Irrep{\overline{35}}{}$, namely, $(\Irrep{8}{+}, \Irrep{8}{-})$, $(\Irrep{27}{+},\Irrep{27}{-})$, $(\Irrep{27'}{\!\!+},\Irrep{27}{-})$, $(\Irrep{0}{b+}, \Irrep{0}{b-})$ and $(\Irrep{64}{+}, \Irrep{64}{-})$. Third column: basis $\mathcal{T}'_{\lambda}$ in terms of the tensors of Table~\ref{table:TensorBasis} (using the shorthand notations defined in \eq{tau-ij-notation}).} 
\end{table}

In Table~\ref{table:q-numbers} (second column), a basis of eigenvectors of $\hat{C}_2$ is arranged in bases $\mathcal{V}_\lambda^{(s)}$ of the subspaces corresponding to specific values of $\lambda \equiv  (\sigma_{\rm cc},\sigma_y,\sigma_x)$. For each subspace with quantum numbers $\lambda$, we give in the third column of Table~\ref{table:q-numbers} an alternative basis $\mathcal{T}'_{\lambda}$, whose elements are not eigenvectors of $\hat{C}_2$ but expressed in terms of simple linear combinations of the tensors of the basis ${\cal T}$ (see Table~\ref{table:TensorBasis}).\footnote{For $\lambda = (\lambda_1,\lambda_2,\lambda_3)$, $\mathcal{T}'_{\lambda}$ is obtained from the kernel of the operator 
$\frac{\unit + \lambda_1 \sigma_{\rm cc}}{2}  \cdot \frac{\unit + \lambda_2 \sigma_y}{2} \cdot \frac{\unit + \lambda_3 \sigma_x}{2} - \unit$.}
For instance, for $\lambda = (+,+,+)$ we have\footnote{The elements of $\mathcal{V}_\lambda^{(s)}$ are ordered similarly to the irreps in Table~\ref{table:irreps}, and we denote $\Proj{R + \overline{R}}{} \equiv \Proj{R}{}+ \Proj{\overline{R}}{}\,$.}
\begin{align}
\mathcal{V}_{(+,+,+)}^{(s)} \ &=\  \left\{ \Proj{\Irrep{1}{}}, \Proj{\Irrep{8}{+}}, \Proj{\Irrep{8}{-}}, \Proj{\Irrep{0}{a}}, \Proj{\Irrep{10}{}+\Irrep{\overline{10}}{}},  \Proj{\Irrep{27}{+}}, \Proj{\Irrep{27'}{\!\!+}}, \Proj{\Irrep{27}{-}}, \Proj{\Irrep{0}{b+}}, \Proj{\Irrep{0}{b-}}, \Proj{\Irrep{35}{+}+\Irrep{\overline{35}}{+}}, \Proj{\Irrep{35}{-}+\Irrep{\overline{35}}{-}},   \right.  \nn \\  
& \hskip 7mm \left. \Proj{\Irrep{64}{+}}, \Proj{\Irrep{64}{-}}, \Proj{\Irrep{0}{c}}, \Proj{\Irrep{0}{d}+\Irrep{\overline{0}}{d}}, \Proj{\Irrep{0}{e}}, \Proj{\Irrep{28}{}+\Irrep{\overline{28}}{}}, \Proj{\Irrep{81}{}+\Irrep{\overline{81}}{}},  \Proj{\Irrep{125}{}},  \TpTdag{\Irrep{27}{+} \to \Irrep{27'}{\!\!+}}  \right\} \, , \label{Vsppp}  \\[2mm]
\mathcal{T}'_{(+,+,+)} \ &=\ \{ S, B_{12}, B_{34}, \ldots, L_{45}, L_{67}, L_{89} \} \ ,
\end{align}
and for $\lambda = (-,-,+)$ : 
\begin{align}
\mathcal{V}_{(-,-,+)}^{(s)} \ &=\ \{ \Proj{\Irrep{10}{}}\!-\Proj{\Irrep{\overline{10}}{}}, \,\Proj{\Irrep{35}{+}}\!\! -\Proj{\Irrep{\overline{35}}{+}}, \, \Proj{\Irrep{35}{-}}\!\!-\Proj{\Irrep{\overline{35}}{-}}, \, \Proj{\Irrep{0}{d}}\!\!-\Proj{\Irrep{\overline{0}}{d}}, \, \Proj{\Irrep{28}{}}\!-\Proj{\Irrep{\overline{28}}{}}, \, \Proj{\Irrep{81}{}}\!-\Proj{\Irrep{\overline{81}}{}} \} \ , \label{Vsmmp} \\[2mm]
\mathcal{T}'_{(-,-,+)} \ &=\ \{ \Hbar_{34}, \, J_{34} - J_{56}, \, K_{34}-K_{56}, \, \Lbar_{45},\, \Lbar_{67}, \, \Lbar_{89}  \} \ . 
\end{align}
We observe that among the eight possible values of $\lambda$, the case $\lambda=(-,+,+)$ is not realized for the process under consideration.  

The bases $\mathcal{V}^{(s)}$ and $\mathcal{T}'$ of the full color space obtained by joining either the bases $\mathcal{V}_\lambda^{(s)}$ or the bases $\mathcal{T}'_{\lambda}$ in a (column) vector, 
\begin{align}
\label{two-bases}
& \mathcal{V}^{(s)} \equiv 
\left( \begin{matrix} \mathcal{V}_\lambda^{(s)} \end{matrix}  \right)  \ \ ; \ \ \ \ {\mathcal T}' \equiv \left( \begin{matrix} \mathcal{T}'_\lambda \end{matrix}  \right) \ , 
\end{align}
are related by a transition matrix $G'$ 
\begin{align}
\label{s-chan-tau-prime}
&  \mathcal{V}^{(s)} = G' \cdot {\cal T}'  \ , 
\end{align}
which from Table~\ref{table:q-numbers} is a block-diagonal matrix, with blocks of size $(21,1,1,1,6,6,6)$. This matrix is known explicitly from the expressions of all projectors and transition operators found previously. 

Let's emphasize that~\eq{s-chan-tau-prime} relates the {\it $s$-channel} (projection and transition) operators of the basis $\mathcal{V}^{(s)}$ to the basis ${\cal T}'$, where the $s$-channel process $12 \to 43$ is defined by \eq{s-chan-def}. We will define the $t$ and $u$-channel processes as $32 \to 41$ and $42 \to 13$, obtained from the $s$-channel by the permutation operators $\sigma_{13}$ and $\sigma_{14}$, respectively. In the $t$-channel (resp.~$u$-channel), the Hermitian projectors and transition operators collected in $\mathcal{V}^{(t)}$ (resp.~$\mathcal{V}^{(u)}$) are thus given by:
\begin{align}
\label{tu-chan-tau-prime}
& \mathcal{V}^{(t)} = G' \cdot \sigma_{13}' \cdot {\cal T}'  \ \ ; \ \ \ \ 
\mathcal{V}^{(u)} = G' \cdot \sigma_{14}' \cdot {\cal T}'  \ , 
\end{align}
where $\sigma_{13}'$ and $\sigma_{14}'$ are the matrices for the permutation operators in the basis ${\cal T}'$.\footnote{They are given by $\sigma_{13}' = M \cdot \sigma_{13} \cdot M^{-1}$ and $\sigma_{14}' = M \cdot \sigma_{14} \cdot M^{-1}$, where the matrices $\sigma_{13}$ and $\sigma_{14}$ in the basis ${\cal T}$ are given (see Table~\ref{tab:CrossingR}), and $M$ is the transition matrix between the bases ${\cal T}$ and ${\cal T}' = M \cdot {\cal T}$, which can be directly obtained from the third column of Table~\ref{table:q-numbers}.} Using \eq{s-chan-tau-prime} and \eq{tu-chan-tau-prime} we obtain the transition matrix $K_{ts}$ (resp.~$K_{us}$) between the $s$ and $t$-channel (resp.~$u$-channel) bases, 
\begin{subequations}
\label{KtsKus}
\begin{align}
\label{Kts}
& \mathcal{V}^{(t)} = K_{ts} \cdot \mathcal{V}^{(s)}  \ \ \ ; \ \ \ \ K_{ts} = G' \cdot \sigma_{13}' \cdot G'^{-1} =  K_{ts}^{-1} \ ,  \\ 
\label{Kus}
& \mathcal{V}^{(u)} = K_{us} \cdot \mathcal{V}^{(s)}  \ \ ; \ \ \ \ K_{us} = G' \cdot \sigma_{14}' \cdot G'^{-1} =  K_{us}^{-1} \ . 
\end{align}
\end{subequations}

Although the matrices $K_{ts}$ and $K_{us}$ follow directly from knowledge of $G'$, $\sigma_{13}'$ and $\sigma_{14}'$, it is interesting to observe that their general structure can be deduced from simple reasoning about quantum numbers. Indeed, as is the case for the eigenvectors $\mathcal{V}_\lambda^{(s)}$ of the $s$-channel Casimir operator $\hat{C}_2$ (second column of Table~\ref{table:q-numbers}), the $t$-channel operators can be organized in subsets $\mathcal{V}_\lambda^{(t)}$ of given {\it $t$-channel quantum numbers} $\lambda$ defined w.r.t.~the operators $\sigma_{\rm cc}$, $\sigma_{12}  \sigma_{34} = \sigma_x$, $\sigma_{14}  \sigma_{23} =  \sigma_y$, which are the $t$-channel analogs of the $s$-channel operators $\sigma_{\rm cc}$, $\sigma_y$, $\sigma_x$. Thus, an eigenspace of the $t$-channel Casimir operator $\hat{C}_2^{(t)}$ 
with $\lambda=(\lambda_1,\lambda_2,\lambda_3)$ is matched to an eigenspace of $\hat{C}_2$ with $\lambda = (\lambda_1,\lambda_3,\lambda_2)$. Similarly, one finds that an eigenspace of $\hat{C}_2^{(u)}$ with
$\lambda=(\lambda_1,\lambda_2,\lambda_3)$ is matched to an eigenspace of $\hat{C}_2$ with $\lambda=(\lambda_1,\lambda_2,\lambda_2 \lambda_3)$. 

These simple rules have interesting consequences. For instance, by a simple inspection of the first two columns of Table~\ref{table:q-numbers}, one infers that: 
\bi
\item{} The $t$-channel analog of the basis~\eq{Vsppp}, \ie, $\{ \Proj{\Irrep{1}{}}^{(t)}, \ldots, \Proj{\Irrep{125}{}}^{(t)},  \TpTdag{\Irrep{27}{+} \to \Irrep{27'}{\!\!+}}^{(t)} \}$, must correspond to the $t$-channel quantum numbers $(+,+,+)$, and thus to the same $s$-channel quantum numbers. The 21 operators $\mathcal{V}_{^{(+,+,+)}}^{(t)}$ (and similarly $\mathcal{V}_{^{(+,+,+)}}^{(u)}$) are thus linear combinations of the operators $\mathcal{V}_{^{(+,+,+)}}^{(s)}$ in the $s$-channel.\footnote{In the $\Irrep{8}{} \otimes \Irrep{8}{} \to \Irrep{8}{} \otimes \Irrep{8}{}$ case of Ref.~\cite{Dokshitzer:2005ig}, after grouping the $\Irrep{10}{}$ and $\Irrep{\overline{10}}{}$, it was possible to express (self-conjugate) $t$-channel projectors in terms of $s$-channel ones only. In the present case, it is not possible, due to the presence of two equivalent irreps of same symmetry under permutation ($\Irrep{27}{+} $ and $\Irrep{27'}{\!\!+}$), leading to a mixing of the Hermitian (and self-conjugate) projectors with the Hermitian sum $\TpTdag{\Irrep{27}{+} \to \Irrep{27'}{\!\!+}}$ of transition operators.} This could also be inferred from \eq{tu-chan-tau-prime} by noting that the set $\mathcal{T}'_{^{(+,+,+)}}$ (third column of Table~\ref{table:q-numbers}) is a representation of the symmetric group ${\rm S}_4$ (as can be readily verified using Table~\ref{tab:CrossingR}) and thus globally invariant under $\sigma_{13}$ and $\sigma_{14}$. 
\item{} The $t$-channel analog of \eq{Vsmmp}, namely $\mathcal{V}_{^{(-,-,+)}}^{(t)} = \{ \Proj{\Irrep{10}{}}^{(t)}\!-\Proj{\Irrep{\overline{10}}{}}^{(t)}, \ldots, \,\Proj{\Irrep{81}{}}^{(t)}\!-\Proj{\Irrep{\overline{81}}{}}^{(t)} \}$, has $t$-channel quantum numbers $(-,-,+)$, \ie\ $(-,+,-)$ in the $s$-channel. Thus, we see from Table~\ref{table:q-numbers} that the six operators in $\mathcal{V}_{^{(-,-,+)}}^{(t)}$ are linear combinations of the $s$-channel operators 
$\TmTdag{\Irrep{\alpha}{+} \to \Irrep{\alpha}{-}} $ ($\alpha = 1 \ldots 5$) and $\TmTdag{\Irrep{35}{+} \to \Irrep{35}{-}} \!\! - \TmTdag{\Irrep{\overline{35}}{+} \to \Irrep{\overline{35}}{-}}$. 
\ei

With similar reasoning for all quantum numbers $\lambda$ in the $t$-channel, as well as in the $u$-channel, we find that the matrices $K_{ts}$ and $K_{us}$ defined by \eq{KtsKus} have the block structure:
\begin{align}
\label{KtsKus-block}
& K_{ts} = 
\mbox{\fontsize{12}{2}\selectfont $
\begin{psmallmatrix} \ \tikeqbis{\begin{scope}[scale=2] \draw (-0.2,-0.2) rectangle (0.2,0.2); 
\node at (0,0) {\scriptsize $B$}; \end{scope}} &  &  &  &  \\[0mm]
  & \tikeqbis{\begin{scope}[scale=1.5] 
 \node at (0,0) {$\mathsize{6}{\begin{matrix} \ 0  \ &  \ a\  & \ 0 \   \\[2mm]  \  \frac{1}{a} \ & \ 0 \ & \ 0 \   \\[2mm]  \ 0 \ & \ 0\  & \ 1\ \end{matrix}}$};  \end{scope}} & & &  \\[0mm]
  &  &  0 & \tikeqbis{\begin{scope}[scale=1.8] \draw (-0.2,-0.2) rectangle (0.2,0.2); 
\node at (0,0) {\scriptsize $C$}; \end{scope}}  & 0  \\[0mm]
  &  &  \tikeqbis{\begin{scope}[scale=1.8] \draw (-0.2,-0.2) rectangle (0.2,0.2); 
\node at (0,0) {\scriptsize $C^{-1}$}; \end{scope}} & 0  & 0  \\[0mm]
  &  &  0 & 0  & \ \tikeqbis{\begin{scope}[scale=1.8] \draw (-0.2,-0.2) rectangle (0.2,0.2); 
\node at (0,0) {\scriptsize $D$}; \end{scope}}  
\ \end{psmallmatrix}$}, \ K_{us} = 
\mbox{\fontsize{12}{2}\selectfont $
\begin{psmallmatrix} \ \tikeqbis{\begin{scope}[scale=2] \draw (-0.2,-0.2) rectangle (0.2,0.2); 
\node at (0,0) {\scriptsize $B'$}; \end{scope}} &  &  &  &  \\[0mm]
  & \tikeqbis{\begin{scope}[scale=1.5] 
 \node at (0,0) {$\mathsize{6}{\begin{matrix} 1  &  0 & 0  \\[2mm]   0 & 0 & -\frac{1}{a}  \\[2mm]  0 & -a & 0 \end{matrix}}$};  \end{scope}}  & & &  \\[0mm]
 & &  \tikeqbis{\begin{scope}[scale=1.8] \draw (-0.2,-0.2) rectangle (0.2,0.2); 
\node at (0,0) {\scriptsize $D$}; \end{scope}}  & 0 & 0   \\[0mm]
  & & 0 & 0 &  \tikeqbis{\begin{scope}[scale=1.8] \draw (-0.2,-0.2) rectangle (0.2,0.2); 
\node at (0,0) {\scriptsize $C'^{-1}$}; \end{scope}} \\[0mm]
  & & 0  &  \tikeqbis{\begin{scope}[scale=1.8] \draw (-0.2,-0.2) rectangle (0.2,0.2); 
\node at (0,0) {\scriptsize $C'$}; \end{scope}}  & 0 
\ \end{psmallmatrix} $} , \nn \\[2mm]
\end{align}
where $B, B'$ are $21\times 21$ matrices, $C, C', D$ are $6\times 6$ matrices, and $a$ is some number. The latter quantities depend only on $\Nc$ and are explicitly known from the expressions of all ($s$-channel) projectors and transition operators.

The forms~\eq{KtsKus-block} of the matrices $K_{ts}$ and $K_{us}$ are useful for visualizing how sets with different quantum numbers are matched when moving from the $\mathcal{V}^{(s)}$ basis to the $\mathcal{V}^{(t)}$ or $\mathcal{V}^{(u)}$ basis, and also for deducing the general structure of the sADM (see next section).

\subsection{Structure of the sADM}
\label{sec:sADM-structure}

The sADM associated to the $\Irrep{27}{} \otimes \Irrep{27}{} \to \Irrep{27}{} \otimes \Irrep{27}{}$ process is obtained from the operator $\mathcal{Q}$ defined as~\cite{Dokshitzer:2005ig}: 
\begin{align}
\label{eq:Qop}
\mathcal{Q}(b) \ = \ \frac{1+b}{2\Nc} \, T_t^2 + \frac{1-b}{2\Nc} \, T_u^2 \ ,
\end{align}
where $b$ is a kinematic variable depending on the Mandelstam variables $s,t,u$ of the $2 \to 2$ process,\footnote{Namely, $b = \frac{T-U}{T+U}$ where $T= \ln \frac{s}{-t} -i\pi$ and $U= \ln \frac{s}{-u} -i\pi$~\cite{Dokshitzer:2005ig}.} and $T_t^2 \equiv \hat{C}_2^{(t)}$ (resp.~$T_u^2 \equiv \hat{C}_2^{(u)}$) is the $t$-channel (resp.~$u$-channel), $\sun$ invariant Casimir operator (defined below in terms of birdtracks).

The elements of the $t$-channel basis $\mathcal{V}^{(t)}$ are eigenvectors of $T_t^2$,  ordered as for the $s$-channel (second column of Table~\ref{table:q-numbers}). In the $t$-channel basis, the operator $T_t^2$ is thus simply represented by the diagonal matrix $\mathcal{C}$ of the corresponding Casimirs. In the $u$-channel basis $\mathcal{V}^{(u)}$, the operator $T_u^2$ is obviously represented by the same matrix $\mathcal{C}$. Using the $K_{ts}$ and $K_{us}$ transition matrices defined by \eq{KtsKus}, the matrix $\mathcal{Q}^{(s)}$ of the operator $\mathcal{Q}$ in the $s$-channel basis $\mathcal{V}^{(s)}$ thus reads: 
\begin{align}
\label{eq:Qmat}
\mathcal{Q}^{(s)}(b) \ &= \ \frac{1+b}{2\Nc} \, K_{ts} \cdot {\cal C} \cdot K_{ts} + \frac{1-b}{2\Nc} \, K_{us} \cdot {\cal C} \cdot K_{us} \ . 
\end{align}
From the general form~\eq{KtsKus-block} of $K_{ts}$ and $K_{us}$, it follows directly that $\mathcal{Q}^{(s)}$ is block-diagonal, with blocks of size $(21,1,1,1,6,6,6)$ (similarly to the matrix $G'$, see \eq{s-chan-tau-prime}). Thus, each subspace with given quantum numbers $\lambda$ is invariant under the action of the sADM. 

It is interesting to prove this last result in a basis-independent way. To this end, it is convenient to represent operators acting on the color space of $\Irrep{27}{} \otimes \Irrep{27}{} \to \Irrep{27}{} \otimes \Irrep{27}{}$ (such as the $s$-channel Casimir operator $\hat{C}_2$ acting on the basis ${\cal T}$ of tensors, see \eq{C2hat-def}) as operators acting on the space $\Irrep{27}{}^{\otimes 4}$ of all color singlet states which can be built from four $\Irrep{27}{}$'s. In birdtrack notation, a basis of this space is simply obtained from the basis $\mathcal{T}$ (given in Table~\ref{table:TensorBasis}) by stretching the external ``parton legs'' of a given tensor $\tau$ to the same side, and thus build vectors in a ``bra/ket'' notation. For example, the tensor $\tbp$ (a linear map $\Irrep{27}{} \otimes \Irrep{27}{} \rightarrow \Irrep{27}{} \otimes \Irrep{27}{}$) is traded for the ket $|\tbp \rangle$ (a singlet state of $\Irrep{27}{}^{\otimes 4}$) according to: 
\begin{align}
& \tbp \ = \ 
\begin{tikzpicture}
\node at (0,0) {\TBP};
\node at (-1.4,.4) {$\psi_1 $};
\node at (-1.4,-.4) {$\chi_2 $};
\node at (+1.4,-.4) {$\chi_3 $};
\node at (+1.4,.4) {$\psi_4 $};
\end{tikzpicture}
\  \  \longrightarrow \  \ 
|\tbp \rangle \ = \  
\begin{tikzpicture}[scale=\setscaletikz]
\draw[thick] (-3,5.5) -- ++(-1,0);
\draw[thick] (-3,5) -- ++(-1,0);
\draw[fill=white] (-4,5.7) rectangle ++(-.4,-.9);
\draw[thick] (-3,4) -- ++(-1,0);
\draw[thick] (-3,3.5) -- ++(-1,0);
\draw[fill=white] (-4,4.2) rectangle ++(-.4,-.9);
\draw[thick] (-3,2.5) -- ++(-1,0);
\draw[thick] (-3,2) -- ++(-1,0);
\draw[fill=white] (-4,2.7) rectangle ++(-.4,-.9);
\draw[thick] (-3,1) -- ++(-1,0);
\draw[thick] (-3,.5) -- ++(-1,0);
\draw[fill=white] (-4,1.2) rectangle ++(-.4,-.9);
\draw[thick] (-3,-5.5) -- ++(-1,0);
\draw[thick] (-3,-5) -- ++(-1,0);
\draw[fill=white] (-4,-5.7) rectangle ++(-.4,.9);
\draw[thick] (-3,-4) -- ++(-1,0);
\draw[thick] (-3,-3.5) -- ++(-1,0);
\draw[fill=white] (-4,-4.2) rectangle ++(-.4,.9);
\draw[thick] (-3,-2.5) -- ++(-1,0);
\draw[thick] (-3,-2) -- ++(-1,0);
\draw[fill=white] (-4,-2.7) rectangle ++(-.4,.9);
\draw[thick] (-3,-1) -- ++(-1,0);
\draw[thick] (-3,-.5) -- ++(-1,0);
\draw[fill=white] (-4,-1.2) rectangle ++(-.4,.9);
\draw[thick] (-3,5.5) -- ++(1.5,0)-- ++(0,-5) -- ++(-1.5,0);
\draw[thick] (-3,5) -- ++(1,0)-- ++(0,-4) -- ++(-1,0);
\draw[thick] (-3,4) -- ++(.5,0)-- ++(0,-2) -- ++(-.5,0);
\draw[thick] (-3,-5.5) -- ++(1.5,0)-- ++(0,5) -- ++(-1.5,0);
\draw[thick] (-3,-5) -- ++(1,0)-- ++(0,4) -- ++(-1,0);
\draw[thick] (-3,-4) -- ++(.5,0)-- ++(0,2) -- ++(-.5,0);
\draw[thick,\qbcolor] (-3,3.5) -- ++(5,0) --++(0,-7) --++(-5,0);
\draw[thick,\qcolor]  (-3,2.5) -- ++(4,0) --++(0,-5) --++(-4,0);
\node at (-6,4.5) {$\psi_1$};
\node at (-6,1.5) {$\chi_2$};
\node at (-6, -1.5) {$\bar{\chi}_3$};
\node at (-6,-4.5) {$\bar{\psi}_4$};
\end{tikzpicture} \ \ .
\end{align}
We recall that the generalized partons $\psi_1$, $\chi_2$, $\chi_3 $, $\psi_4$ belong to the irrep $\Irrep{27}{}$, as specified by the abbreviated notation~\eq{eq:tensor-abbrev}. Note that since the irrep $\Irrep{27}{}$ is self-conjugate, in the birdtrack for the ket $|\tbp \rangle $ we could write $\bar{\chi}_3 = \chi_3$ and $\bar{\psi}_4 = \psi_4$.

With the above pictorial notations, an operator acting on kets can be viewed as a map 
$\psi_1 \chi_2  \bar{\chi}_3 \bar{\psi}_4 \to \psi_1' \chi_2'  \bar{\chi}_3' \bar{\psi}_4'$. For instance, the inner product in the space of kets, $\langle \tau_i| \tau_j \rangle = \langle \tau_i|\unit| \tau_j \rangle$, involves the identity operator 
\begin{align}
\label{inner-prod}
\unit \ &= \ \tikeq{
\draw[thick,dashed] (-1,.5) -- ++(2,0);
\draw[thick,dashed] (-1,1) -- ++(2,0);
\draw[thick,dashed] (-1,0) -- ++(2,0);
\draw[thick,dashed] (-1,-.5) -- ++(2,0);
\node at (-1.5,1) {\scriptsize $\psi_1$};
\node at (1.5,1) {\scriptsize$\psi_1'$};
\node at (-1.5,.5) {\scriptsize $\chi_2$};
\node at (1.5,.5) {\scriptsize$\chi_2'$};
\node at (-1.5,0) {\scriptsize $\bar{\chi}_3$};
\node at (1.5,0) {\scriptsize$\bar{\chi}_3'$};
\node at (-1.5,-.5) {\scriptsize $\bar{\psi}_4$};
\node at (1.5,-.5) {\scriptsize$\bar{\psi}_4'$};
} \ \ , 
\end{align}
where parton numbering goes from top to bottom, and a dashed line represents the irrep $\Irrep{27}{}$ according to the conventions of section~\ref{sec:generalities}. This allows one to write the Casimir operator in different channels in the following form: 
\begin{subequations}
\label{eq:TtTuTs}
\begin{align}
\label{eq:Tt}
& T_t^2 = (t_{1} + t_{4})^2 = (t_{2} + t_{3})^2 = \ 
2C_{27} \ \ 
\tikeq{\begin{scope}[scale=0.7]
\draw[thick,dashed] (-1,.5) -- ++(2,0);
\draw[thick,dashed] (-1,1) -- ++(2,0);
\draw[thick,dashed] (-1,0) -- ++(2,0);
\draw[thick,dashed] (-1,-.5) -- ++(2,0);
\end{scope}}
\ - 2 \ \ 
\tikeq{\begin{scope}[scale=0.7]
\draw[thick,dashed] (-1,.5) -- ++(2,0);
\draw[thick,dashed] (-1,1) -- ++(2,0);
\draw[thick,dashed] (-1,0) -- ++(2,0);
\draw[thick,dashed] (-1,-.5) -- ++(2,0);
\draw[glu] (0,.5) -- (0,0);
\end{scope}} \ \ , \\
\notag \\
\label{eq:Tu}
& T_u^2 = (t_{1} + t_{3})^2 = (t_{2} + t_{4})^2 = 
\ 2C_{27} \  \ 
\tikeq{\begin{scope}[scale=0.7]
\draw[thick,dashed] (-1,.5) -- ++(2,0);
\draw[thick,dashed] (-1,1) -- ++(2,0);
\draw[thick,dashed] (-1,0) -- ++(2,0);
\draw[thick,dashed] (-1,-.5) -- ++(2,0);
\end{scope}}
\ - 2 \ \ 
\tikeq{\begin{scope}[scale=0.7]
\draw[thick,dashed] (-1,.5) -- ++(2,0);
\draw[thick,dashed] (-1,1) -- ++(2,0);
\draw[thick,dashed] (-1,0) -- ++(2,0);
\draw[thick,dashed] (-1,-.5) -- ++(2,0);
\draw[glu] (0,1) -- (0,0);
\end{scope}} \ \ , \\
\notag \\
\label{eq:Ts}
& T_s^2 = (t_{1} + t_{2})^2 = (t_{3} + t_{4})^2 = 
\ 2C_{27} \ \ 
\tikeq{\begin{scope}[scale=0.7]
\draw[thick,dashed] (-1,.5) -- ++(2,0);
\draw[thick,dashed] (-1,1) -- ++(2,0);
\draw[thick,dashed] (-1,0) -- ++(2,0);
\draw[thick,dashed] (-1,-.5) -- ++(2,0);
\end{scope}}
\ - 2 \ \ 
\tikeq{\begin{scope}[scale=0.7]
\draw[thick,dashed] (-1,.5) -- ++(2,0);
\draw[thick,dashed] (-1,1) -- ++(2,0);
\draw[thick,dashed] (-1,0) -- ++(2,0);
\draw[thick,dashed] (-1,-.5) -- ++(2,0);
\draw[glu] (0,1) -- (0,.5);
\end{scope}} \ \ ,
\end{align}
\end{subequations}
where the $\sun$ generators of the generalized parton $i$ ($i=1 \ldots 4$) are denoted by $t_i^a$ ($a=1 \ldots \Nc^2 -1$), satisfying $t_i^2 = t_i^a t_i^a = C_{27}$ (with $C_{27} = 2\Nc + 2$), and we used color conservation $t_1+t_2+t_3+t_4 = 0$. 
Note that the Casimir operator in the $s$-channel used previously and defined in \eq{C2hat-def} coincides with $T_s^2$ when viewed as an operator acting on kets, and $T_t^2$ and $T_u^2$ are therefore duly interpreted as the $t$-channel and $u$-channel Casimir operators acting on this space.

It is now easy to show that the operator $\mathcal{Q}$ defined by \eq{eq:Qop} is invariant under the action of the operators $\{\sigma_{\rm cc},\sigma_y,\sigma_x\}$ used in the previous section to define quantum numbers. From the birdtrack expressions of $T_t^2$ and $T_u^2$, and the fact that $\Irrep{27}{}$ is self-conjugate, $T_t^2$ and $T_u^2$ are clearly invariant under $\sigma_{\rm cc}$ (complex conjugation). 
Now consider the action of $\sigma_x = \sigma_{12} \sigma_{34}$ on the only non-trivial parts of $T_t^2$ and $T_u^2$ in \eq{eq:TtTuTs}. We have
\begin{subequations}
\begin{align} 
\label{Ttsquared-sigx}
\sigma_x(-t_2\cdot t_3)\sigma_x \ &=\ 
\tikeq{\begin{scope}[scale=0.6]
\draw[thick,dashed] (-1,.5) -- ++(2,0);
\draw[thick,dashed] (-1,1) -- ++(2,0);
\draw[thick,dashed] (-1,0) -- ++(2,0);
\draw[thick,dashed] (-1,-.5) -- ++(2,0);
\draw[glu] (0,.5) -- (0,0);
\draw[thick,dashed] (-1.2,1) to[out=180,in=0] (-3.2,.5);
\draw[thick,dashed] (-1.2,.5) to[out=180,in=0] (-3.2,1);
\draw[thick,dashed] (1.2,1) to[out=0,in=180] (3.2,.5);
\draw[thick,dashed] (1.2,.5) to[out=0,in=180] (3.2,1);
\draw[thick,dashed] (-1.2,0) to[out=180,in=0] (-3.2,-.5);
\draw[thick,dashed] (-1.2,-.5) to[out=180,in=0] (-3.2,0);
\draw[thick,dashed] (1.2,0) to[out=0,in=180] (3.2,-.5);
\draw[thick,dashed] (1.2,-.5) to[out=0,in=180] (3.2,0);
\end{scope}}
\ = \ 
\tikeq{\begin{scope}[scale=0.6]
\draw[thick,dashed] (-1,.5) -- ++(2,0);
\draw[thick,dashed] (-1,1) -- ++(2,0);
\draw[thick,dashed] (-1,0) -- ++(2,0);
\draw[thick,dashed] (-1,-.5) -- ++(2,0);
\draw[glu] (0,1) -- (0,-.5);
\end{scope}}
\ = \ - t_1\cdot t_4 \ = \ - t_2\cdot t_3 \ , \\[.3cm]
\sigma_x(-t_1\cdot t_3)\sigma_x \ &= \ 
\tikeq{\begin{scope}[scale=0.6]
\draw[thick,dashed] (-1,.5) -- ++(2,0);
\draw[thick,dashed] (-1,1) -- ++(2,0);
\draw[thick,dashed] (-1,0) -- ++(2,0);
\draw[thick,dashed] (-1,-.5) -- ++(2,0);
\draw[glu] (0,1) -- (0,0);
\draw[thick,dashed] (-1.2,1) to[out=180,in=0] (-3.2,.5);
\draw[thick,dashed] (-1.2,.5) to[out=180,in=0] (-3.2,1);
\draw[thick,dashed] (1.2,1) to[out=0,in=180] (3.2,.5);
\draw[thick,dashed] (1.2,.5) to[out=0,in=180] (3.2,1);
\draw[thick,dashed] (-1.2,0) to[out=180,in=0] (-3.2,-.5);
\draw[thick,dashed] (-1.2,-.5) to[out=180,in=0] (-3.2,0);
\draw[thick,dashed] (1.2,0) to[out=0,in=180] (3.2,-.5);
\draw[thick,dashed] (1.2,-.5) to[out=0,in=180] (3.2,0);
\end{scope}}
\ =\ 
\tikeq{\begin{scope}[scale=0.6]
\draw[thick,dashed] (-1,.5) -- ++(2,0);
\draw[thick,dashed] (-1,1) -- ++(2,0);
\draw[thick,dashed] (-1,0) -- ++(2,0);
\draw[thick,dashed] (-1,-.5) -- ++(2,0);
\draw[glu] (0,.5) -- (0,-.5);
\end{scope}}
\ = \  - t_2\cdot t_4 \ = \ - t_1\cdot t_3  \ , 
\end{align}
\end{subequations} 
where we used again color conservation, and the fact that $t_1^2 = t_2^2 = t_3^2 = t_4^2 = C_{27}$. Both $T_t^2$ and $T_u^2$ are thus invariant under $\sigma_x$. 
The invariance of $T_t^2$ and $T_u^2$ under $\sigma_y = \sigma_{14} \sigma_{23}$ is obtained in a similar way, and follows from the identities:  
\begin{subequations}
\begin{align}
\sigma_y(-t_2\cdot t_3)\sigma_y \ &= \ 
\tikeq{\begin{scope}[scale=0.6]
\draw[thick,dashed] (-1,.5) -- ++(2,0);
\draw[thick,dashed] (-1,1) -- ++(2,0);
\draw[thick,dashed] (-1,0) -- ++(2,0);
\draw[thick,dashed] (-1,-.5) -- ++(2,0);
\draw[glu] (0,.5) -- (0,0);
\draw[thick,dashed] (-1.2,1) to[out=180,in=0] (-3.2,-.5);
\draw[thick,dashed] (-1.2,-.5) to[out=180,in=0] (-3.2,1);
\draw[thick,dashed] (1.2,1) to[out=0,in=180] (3.2,-.5);
\draw[thick,dashed] (1.2,-.5) to[out=0,in=180] (3.2,1);
\draw[thick,dashed] (-1.2,0) to[out=180,in=0] (-3.2,.5);
\draw[thick,dashed] (-1.2,.5) to[out=180,in=0] (-3.2,0);
\draw[thick,dashed] (1.2,0) to[out=0,in=180] (3.2,.5);
\draw[thick,dashed] (1.2,.5) to[out=0,in=180] (3.2,0);
\end{scope}}
\ = \ 
\tikeq{\begin{scope}[scale=0.6]
\draw[thick,dashed] (-1,.5) -- ++(2,0);
\draw[thick,dashed] (-1,1) -- ++(2,0);
\draw[thick,dashed] (-1,0) -- ++(2,0);
\draw[thick,dashed] (-1,-.5) -- ++(2,0);
\draw[glu] (0,0) -- (0,.5);
\end{scope}}
\ = \ - t_2\cdot t_3 \ , \\[.3cm]
\sigma_y(-t_1\cdot t_3)\sigma_y \ &= \ 
\tikeq{\begin{scope}[scale=0.6]
\draw[thick,dashed] (-1,.5) -- ++(2,0);
\draw[thick,dashed] (-1,1) -- ++(2,0);
\draw[thick,dashed] (-1,0) -- ++(2,0);
\draw[thick,dashed] (-1,-.5) -- ++(2,0);
\draw[glu] (0,1) -- (0,0);
\draw[thick,dashed] (-1.2,1) to[out=180,in=0] (-3.2,-.5);
\draw[thick,dashed] (-1.2,-.5) to[out=180,in=0] (-3.2,1);
\draw[thick,dashed] (1.2,1) to[out=0,in=180] (3.2,-.5);
\draw[thick,dashed] (1.2,-.5) to[out=0,in=180] (3.2,1);
\draw[thick,dashed] (-1.2,0) to[out=180,in=0] (-3.2,.5);
\draw[thick,dashed] (-1.2,.5) to[out=180,in=0] (-3.2,0);
\draw[thick,dashed] (1.2,0) to[out=0,in=180] (3.2,.5);
\draw[thick,dashed] (1.2,.5) to[out=0,in=180] (3.2,0);
\end{scope}}
\ = \ 
\tikeq{\begin{scope}[scale=0.6]
\draw[thick,dashed] (-1,.5) -- ++(2,0);
\draw[thick,dashed] (-1,1) -- ++(2,0);
\draw[thick,dashed] (-1,0) -- ++(2,0);
\draw[thick,dashed] (-1,-.5) -- ++(2,0);
\draw[glu] (0,.5) -- (0,-.5);
\end{scope}}
\ = \ - t_2\cdot t_4 \ = \ - t_1\cdot t_3 \ . 
\end{align}
\end{subequations}

In summary, $T_t^2$ and $T_u^2$ are invariant under the action of the operators $\{ \sigma_{\rm cc},\sigma_y,\sigma_x \}$ defining quantum numbers, and so is the operator $\mathcal{Q}$. The simplicity of the proof makes obvious the $(21,1,1,1,6,6,6)$ block-diagonal structure of the sADM (\ie, the matrix representation of $\mathcal{Q}$) found in the basis $\mathcal{V}^{(s)}$ (see~\eq{eq:Qmat}). 

\subsection{Characteristic polynomial}
\label{sec:sADM-char-pol}

As a consequence of section~\ref{sec:sADM-structure}, the sADM characteristic polynomial factorizes as a product of seven factors, each factor being the characteristic polynomial of the block defining the sADM restricted to the subspace $\mathcal{V}_\lambda^{(s)}$ of quantum numbers $\lambda$. Namely, 
\begin{align}
\label{char-pol}
& {\rm Det} \left( x \, \unit - \mathcal{Q}^{(s)}(b) \right) \ = \ \prod_{\lambda} \ f_\lambda(x, b) \ ,
\end{align}
where the polynomials $f_\lambda(x, b)$, ordered according to their quantum number label $\lambda$ as in Table~\ref{table:q-numbers} (first column), are of degree $21,\,1,\,1,\,1,\,6,\,6,\,6$, respectively.  

The explicit calculation shows that each factor $f_\lambda(x, b)$ of degree $> 1$ can actually be factorized into polynomials of smaller degree. We obtain: 
\begin{subequations}
\label{char-pol-fact}
\begin{align}
& f_{_{(+,+,+)}}(x, b) \ = \ (x - r_1) (x - r_2) (x - r_3) p_1(x,b) p_2(x,b) p_2(x,-b)  p_3(x,b) \label{char-pol-a}  \\ 
& f_{_{(+,+,-)}}(x, b) \ = \ x - r_1 \label{char-pol-b}  \\ 
& f_{_{(+,-,+)}}(x, b) \ = \ x - r_2 \label{char-pol-c}  \\ 
& f_{_{(+,-,-)}}(x, b) \ = \ x - r_3 \label{char-pol-d}  \\ 
& f_{_{(-,+,-)}}(x, b) \ = \ (x - r_2) (x - r_3) p_2(x,b)  \label{char-pol-e} \\ 
& f_{_{(-,-,+)}}(x, b) \ = \ (x - r_1) (x - r_3) p_1(x,b)  \label{char-pol-f} \\ 
& f_{_{(-,-,-)}}(x, b) \ = \ (x - r_1) (x - r_2)  p_2(x,-b)  \label{char-pol-g} 
\end{align}
\end{subequations}
where the roots of the monomials appearing in the above equations read
\begin{align}
& r_1 = \frac{(5-b)(\Nc+1)}{2\Nc}   \ \ ; \ \ \ \  r_2 = 3 \frac{\Nc+1}{\Nc} \ \ ; \ \ \ \ r_3 = \frac{(5+b)(\Nc+1)}{2\Nc}  \ ,
\end{align}
and the polynomials $p_1(x,b), \, p_2(x,b), \, p_3(x,b)$ (of degree 4, 4, 6 respectively) are given explicitly in Appendix~\ref{app:D}. 

\vspace{2mm}
A few observations on the sADM characteristic polynomial and spectrum are in order: 
\bi
\item{} Since $p_1(x,b)$ and $p_3(x,b)$ are even in $b$, and $r_1(-b) = r_3(b)$, the polynomials $f_\lambda(x, b)$ are simply related to each other under $b \leftrightarrow -b$. Namely, under this symmetry \eq{char-pol-b} and \eq{char-pol-d} are exchanged, as well as \eq{char-pol-e} and \eq{char-pol-g}, and \eq{char-pol-a}, \eq{char-pol-c}, \eq{char-pol-f} are invariant. This can be easily understood from quantum numbers. Indeed, from \eq{eq:Qop} we see that $b \leftrightarrow -b$ is formally equivalent to $T_t^2 \leftrightarrow T_u^2$, \ie\ to the action of the permutation $\sigma_{34}$ (see \eq{eq:Tt}--\eq{eq:Tu}). We easily verify that under $\sigma_{34}$, the quantum numbers $\lambda=(\lambda_1,\lambda_2,\lambda_3)$ are matched to $(\lambda_1,\lambda_2 \lambda_3,\lambda_3)$.\footnote{This can be seen using $\sigma_{34}=\sigma_{13}\sigma_{14}\sigma_{13}$ and the successive transformations of quantum numbers under $\sigma_{13}$ and $\sigma_{14}$ given in section~\ref{sec:quantum-numbers} (see paragraph after \eq{KtsKus}).} Thus, under $\sigma_{34}$, the expression \eq{char-pol-e} becomes \eq{char-pol-g}. But the same result must be obtained under $b \leftrightarrow -b$. Alternatively, one can use the relation
\begin{align}
\label{eq:Q-rel}
\sigma_{34} \mathcal{Q}(b) \sigma_{34} = \mathcal{Q}(-b) \ ,
\end{align}
which is quickly proven by using $T_t^2 = \sigma_{13} T_s^2 \sigma_{13}$, $T_u^2 = \sigma_{14} T_s^2 \sigma_{14}$, $\sigma_{34} \sigma_{14} \sigma_{34} = \sigma_{13}$, and noticing that $\sigma_{34}T_s^2\sigma_{34} = T_s^2$. 
\item{} The blocks (of size $>1$) of the sADM corresponding to a given $\lambda$ can be further block-diagonalized (as can be seen from \eq{char-pol-a} and \eq{char-pol-e}--\eq{char-pol-g}). 
\item{} The sADM spectrum has a remarkable structure. 
Let's denote by $\mathcal{E}_{+}$ the subspace of basis $\mathcal{V}_{^{(+,+,+)}}$, and by 
$\mathcal{E}_{-}$ the supplementary subspace (characterized by quantum numbers different from $(+,+,+)$). We observe that the eigenvalues of the sADM restricted to $\mathcal{E}_{-}$, namely $r_1$, $r_2$, $r_3$ (each of multiplicity 3) and the 12 eigenvalues given by the roots of the polynomials $p_1(x,b)$, $p_2(x,b)$ and $p_2(x,-b)$, are also eigenvalues of the sADM restricted to $\mathcal{E}_{+}$. Since $r_1$, $r_2$, $r_3$ have multiplicity 1 in 
$\mathcal{E}_{+}$, there are six sADM eigenvalues which are `endemic' to this subspace, given by the roots of $p_3(x,b)$. 
\item{} An inspection of the polynomials $p_1(x,b)$, $p_2(x,b)$ and $p_3(x,b)$ (given in Appendix~\ref{app:D}) shows that the set of sADM eigenvalues for the $\Irrep{27}{} \otimes \Irrep{27}{} \to \Irrep{27}{} \otimes \Irrep{27}{}$ process does not exhibit any $b \leftrightarrow 1/\Nc$ symmetry, contrary to the processes $gg \rightarrow gg$ (as was observed in~\cite{Dokshitzer:2005ig} for some subset of the sADM eigenvalues), and $qq \rightarrow qq$ (as can be easily verified for the two eigenvalues of the associated sADM, a $2 \times 2$ matrix in that case). 
\ei 

\section{Summary and discussion}
\label{sec:conclusion}

In this manuscript, we have studied the color space of $\Irrep{27}{} \otimes \Irrep{27}{}$, by explicitly calculating all Hermitian projectors and transition operators associated with $\Irrep{27}{} \otimes \Irrep{27}{} \to \Irrep{27}{} \otimes \Irrep{27}{}$. As mentioned in the introduction, the study of this `process' has several motivations (both phenomenological and theoretical) and, to our knowledge, these operators have never been calculated before. Our study shows that with the help of the birdtrack pictorial technique, even non-trivial color calculations can be performed using only elementary tools (\eg\ the Fierz identity, color conservation, Schur's lemma, clebsches).
In particular, our calculations highlight the effectiveness of the quadratic Casimir operator $\hat{C}_2$, which acts as a powerful filter of the color space. This is because a given eigenvalue of $\hat{C}_2$ is shared by only a limited number of irreps of the decomposition of $\Irrep{27}{} \otimes \Irrep{27}{}$ (see Table~\ref{table:irreps}).   

When the $\hat{C}_2$ eigenspace is of dimension 1, the projector on the corresponding irrep is directly obtained from the eigenvector (after proper normalization). As for eigenspaces of dimension $> 1$, we have seen that they can be treated simply as follows.

When the eigenspace contains two equivalent irreps, the associated Hermitian projectors and transition operators are given by the $\hat{C}_2$ eigenvectors having specific symmetries under the $\psi \leftrightarrow \chi$ permutation of generalized partons, realized by the operators $\sigma_{12}$ and $\sigma_{34}$ acting on the initial or final pair. (As mentioned after \eq{irreps-3traces}, the fact that any irrep has a specific symmetry under $\psi \leftrightarrow \chi$ follows from the requirement that the pair satisfies either Bose-Einstein or Fermi-Dirac statistics.)

In general, an eigenspace of $\hat{C}_2$ can contain more than two equivalent irreps. In our study, this situation was encountered for the eigenvalue $C_R = C_{27} = 2(\Nc+1)$, corresponding to three equivalent irreps, one of which is antisymmetric and the other two symmetric. Since the two symmetric irreps cannot be distinguished by the $\psi \leftrightarrow \chi$ symmetry, their associated projectors are not uniquely defined -- only their sum is. To handle the trickier sector of the $\Irrep{27}{}$, we used clebsches, which simplify the derivation of the three projectors and six transition operators in this sector. 

Finally, when the $\hat{C}_2$ eigenspace contains some irreps which are complex conjugate to each other, those can be distinguished by selecting the eigenvectors which are even and odd under complex conjugation, and finding the precise linear combination of them having the proper symmetry (under permutation of quark/antiquark indices) characterizing the irrep and its complex conjugate.

The particular process we have considered, $\Irrep{27}{} \otimes \Irrep{27}{} \to \Irrep{27}{} \otimes \Irrep{27}{}$, exhausts all possible reasons for having a degenerate $\hat{C}_2$ eigenvalue, namely the presence of two complex conjugate irreps and/or two (or more) equivalent irreps. (Note that a $\hat{C}_2$ eigenspace can contain equivalent irreps as well as their complex conjugates, as is the case for the sector of the $\Irrep{35}{}$ discussed in section~\ref{sec:eigenspaces-dim8}.)    
So it seems clear that the procedure we have used to derive the complete color space, based on knowledge of the quadratic Casimir operator (which can be obtained in an elementary way using the Fierz identity), could be automated and applied to more complicated cases. Given its remarkable filtering power, the quadratic Casimir operator may also be a useful tool to implement in advanced algorithms for calculating Hermitian projectors~\cite{Keppeler:2012ih,Chargeishvili:2024pnq}. 

In the decomposition of $\Irrep{27}{} \otimes \Irrep{27}{}$, we found that the presence of two irreps $\Irrep{27}{}$ having the same symmetry under $\psi \leftrightarrow \chi$
(named $\Irrep{27}{+}$ and $\Irrep{27'}{\!\!+}$),
has an interesting consequence. Let us denote by $\mathcal{E}_P^{27\otimes 27}$ the vector space of all $20$ self-conjugate projectors and by $\mathcal{E}_T^{27 \otimes 27}$ the supplementary space, consisting of all transition operators and the (non-zero) differences $\Proj{R}-\Proj{\overline{R}}$ between a projector and its complex conjugate. 
We observed that the sum of the two transition operators between $\Irrep{27}{+}$ and $\Irrep{27'}{\!\!+}$, namely, $\TpTdag{\Irrep{27}{+} \to \Irrep{27'}{\!\!+}}$, is Hermitian and self-conjugate, and is moreover symmetric under left ($\sigma_{12}$) and right ($\sigma_{34}$) permutations. 
Thus, it has the same ($s$-channel) quantum numbers $(\sigma_{\rm cc}, \sigma_y, \sigma_x) =(+,+,+)$ 
as self-conjugate 
projectors.\footnote{The quantum numbers were defined with respect to the complex conjugation operator $\sigma_{\rm cc}$ and the permutation operators $\sigma_x=\sigma_{12 }\sigma_{34}$ and $\sigma_y=\sigma_{14} \sigma_{23}$, the generalized partons being labeled by $1 \cdots 4$ counterclockwise.} 
From this, we deduced that the self-conjugate projectors mix with $\TpTdag{\Irrep{27}{+} \to \Irrep{27'}{\!\!+}}$ under crossing symmetry. 
In other words, if, for a physics problem involving the color structure of $\Irrep{27}{} \otimes \Irrep{27}{} \to \Irrep{27}{} \otimes \Irrep{27}{}$, we need to know the (self-conjugate) projectors in the $t$-channel, it is not enough to have previously derived those in the $s$-channel, and one should consider the subspace 
$\mathcal{E}_{+} = \mathcal{E}_P^{27\otimes 27} \oplus \ave{\TpTdag{\Irrep{27}{+} \to \Irrep{27'}{\!\!+}}}$ instead of simply $\mathcal{E}_P^{27\otimes 27}$.
It should be noted that this situation differs from that encountered for the process $\Irrep{8}{} \otimes \Irrep{8}{} \rightarrow \Irrep{8}{} \otimes \Irrep{8}{}$~\cite{Dokshitzer:2005ig}. In the latter case, the subspace of $s$-channel self-conjugate projectors $\mathcal{E}_{P}^{8 \otimes 8} = \ave{ \{ \Proj{\Irrep{1}{}}, \Proj{\Irrep{8}{+}}, \Proj{\Irrep{8}{-}}, \Proj{\Irrep{10}{}+\Irrep{\overline{10}}{}}, \Proj{\Irrep{27}{}}, \Proj{\Irrep{0}{}} \} }$, and the supplementary subspace $\mathcal{E}_{T}^{8 \otimes 8} = \ave{ \{ \Trans{\Irrep{8}{+} \to \Irrep{8}{-}}, \Trans{\Irrep{8}{-} \to \Irrep{8}{+}},\Proj{\Irrep{10}{}} - \Proj{\Irrep{\overline{10}}{}} \} }$, are separately invariant under crossing symmetry.

To give an example of a simple observable where the aforementioned mixing between self-conjugate projectors and $\TpTdag{\Irrep{27}{+} \to \Irrep{27'}{\!\!+}}$ would be at work, let's consider the $p_\perp$-broadening of an energetic $\psi \chi$ pair of generalized partons (each being a two-gluon state in the $\Irrep{27}{}$) passing through a nucleus. Similarly to Ref.~\cite{Cougoulic:2017ust}, this observable involves (time-ordered) products of diagrams of this type: 
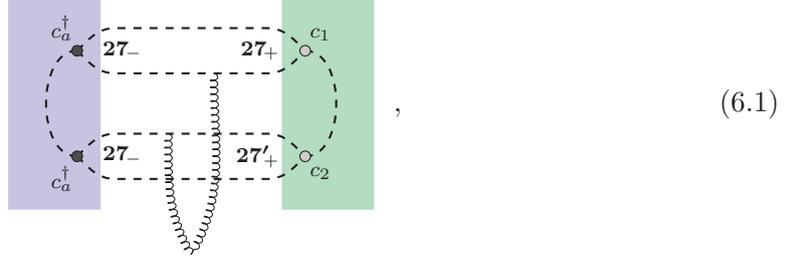
\begin{align}
\label{picture1}
\begin{tikzpicture}[scale=1]
\draw[fill=Blue!30!white,Blue!20!white] (-2.4,-1.4) rectangle (-1.2,1.4);
\draw[fill=Green!30!white,Green!30!white] (2.4,-1.4) rectangle (1.2,1.4);
\draw[thick,dashed] (-1.5,.7) to[out=45,in=180] (-1,1) -- (1,1) to [out=0,in=135] (1.5,.7);
\draw[thick,dashed] (-1.5,.7) to[out=-45,in=180] (-1,.4) -- (1,.4) to [out=0,in=-135]  (1.5,.7);
\draw[thick,dashed] (-1.5,-.7) to[out=-45,in=180] (-1,-1) -- (1,-1) to [out=0,in=-135] (1.5,-.7);
\draw[thick,dashed] (-1.5,-.7) to[out=45,in=180] (-1,-.4) -- (1,-.4) to [out=0,in=135]  (1.5,-.7);
\draw[thick,dashed] (-1.5,.7) to[out=180,in=180] (-1.5,-.7);
\draw[thick,dashed] (1.5,.7) to[out=0,in=0] (1.5,-.7);
\draw[fill=black!70!white] (-1.5,.7) circle (0.07);
\draw[fill=black!70!white] (-1.5,-.7) circle (0.07);
\draw[fill=black!20!white] (1.5,.7) circle (0.07);
\draw[fill=black!20!white] (1.5,-.7) circle (0.07);
\node[above] at (-1.7,.7) {\scriptsize $c_{a}^\dagger$};
\node[right] at (-1.3,.7) {\scriptsize $\Irrep{27}{\!-}$}; 
\node[below] at (-1.7,-.7) {\scriptsize $c_{a}^\dagger$};
\node[right] at (-1.3,-.7) {\scriptsize $\Irrep{27}{\!-}$}; 
\node[above] at (1.7,.7) {\scriptsize $c_1$};
\node[left] at (1.3,.7)  {\scriptsize $\Irrep{27}{\!+}$};
\node[below] at (1.7,-.7) {\scriptsize $c_2$};
\node[left] at (1.3,-.7)  {\scriptsize $\Irrep{27'}{\!\!+}$};
\draw[glu] (-.3,-.4) to[out=-90,in=120] (0,-2);
\draw[glu] (.3,.4) to[out=-90,in=60] (0,-2);
\end{tikzpicture} 
\ \ , 
\end{align} 
where {a dashed line denotes the irrep $\Irrep{27}{}$ (see \eq{dashproj-def})}, the upper (resp.~lower) half of the diagram belongs to the rescattering amplitude (resp.~conjugate amplitude), and at each time of the evolution the overall $\psi_1 \chi_2  \bar{\chi}_3 \bar{\psi}_4$ system is in a color-singlet (quadrupole) state.\footnote{This follows from the standard assumption that the rescattering is instantaneous. Here, for the sake of simplicity, the rescattering in \eq{picture1} is also treated in the `two-gluon approximation', where the exchange can be either a {\it real} (one gluon is exchanged with a target nucleon in both the amplitude and its conjugate), or {\it virtual} (two gluons are exchanged in the amplitude and none in its conjugate, or vice versa) contribution to the scattering cross section. The specific diagram in \eq{picture1} is a `real' contribution.}  
The generic diagram \eq{picture1} corresponds to a contribution to $p_\perp$-broadening where the $\psi \chi$ pair is in the irrep $\Irrep{27}{-}$ before rescattering (\ie, projected on $\Irrep{27}{-}$ in both the amplitude and conjugate amplitude), and in a {\it mixed state} after rescattering ($\Irrep{27}{+}$ in the amplitude, and $\Irrep{27'}{\!\!+}$ in the conjugate amplitude). 
The particular contribution \eq{picture1} is thus expressed in terms of the antisymmetric clebsch $c_a$ and symmetric clebsches $c_1$ and $c_2$ (defined explicitly in section~\ref{sec:proj27}), using the pictorial notation 
\eq{eq:CG-def}. 
%

To evaluate \eq{picture1}, we first express the $s$-channel projector $\Proj{\alpha}$ (with $\alpha = \Irrep{27}{-}$) appearing in the left rectangle in terms of $t$-channel operators $\mathcal{V}^{(t)}_\gamma$ using the matrix $K_{st} = K_{ts}^{-1}$, see \eq{Kts}. (The index $\gamma$ runs from 1 to 21, due to the block structure of $K_{st}$.) The gluon exchange, being equivalent to the action of $\frac{1}{2} T_t^2$ (see \eq{eq:Tt}),
then provides a factor $\frac{1}{2} C_\gamma$. Finally, the operator $\mathcal{V}^{(t)}_\gamma$ is expressed in terms of $s$-channel operators $\mathcal{V}^{(s)}_\delta$ using the matrix $K_{ts}$. The remaining birdtrack involves only $s$-channel operators, and is non-zero only if $\mathcal{V}^{(s)}_\delta =  \TpTdag{\Irrep{27}{+} \to \Irrep{27'}{\!\!+}}$ (\ie, if $\delta =21$, see the ordering of the components of $\mathcal{V}^{(s)}$ in \eq{Vsppp}). The transition operator $\Trans{\Irrep{27}{+} \rightarrow \Irrep{27'}{+}} = c_2^\dagger c_1$ (see \eq{eq:trans27-ppp-CG}) appearing in the right rectangle of \eq{picture1} is normalized according to \eq{eq:trans-norm}, and the result for \eq{picture1} reads: 
\begin{align}
\label{pocket-form}
\K{\Irrep{27}{}} \sum_{\gamma = 1}^{21}  (K_{st})_{\alpha \gamma} \, \frac{C_\gamma}{2} \, (K_{ts})_{\gamma \delta} \ 
\neq 0 \ . 
\end{align}
The latter expression shows that \eq{picture1} can only be non-zero if $K_{st}$ mixes $\TpTdag{\Irrep{27}{+} \to \Irrep{27'}{\!\!+}}$ with self-conjugate projectors, and the fact that it is indeed non-zero can be checked explicitly from knowledge of the $21 \times 21$ block $B$ in \eq{KtsKus-block}. This illustrates that $p_\perp$-broadening can be sensitive to the mixing of $\mathcal{E}_P^{27 \otimes 27}$ and $\Trans{\Irrep{27}{+} \rightarrow \Irrep{27'}{+}}$ under crossing symmetry.

Finally, let us emphasize that in principle, the elementary method used in our study makes it possible to trace the $\Nc$ dependence in the calculation (contained, for example, in Casimir charges and irreps dimensions), as well as the rather complex color structure of certain results. In particular, we could obtain the soft anomalous dimension matrix (sADM) of $\Irrep{27}{} \otimes \Irrep{27}{} \to \Irrep{27}{} \otimes \Irrep{27}{}$ almost effortlessly, due to the simple transformation of the chosen basis tensors under crossing symmetry. Indeed, the 42 tensors of the basis ${\cal T}$ form a representation of the symmetric group ${\rm S}_4$, as can be seen from the simple action of all permutation operators $\sigma_{ij}$ in this basis (see Table~\ref{tab:CrossingR}). We have seen that the organization of basis tensors in subsets of different `quantum numbers', defined as eigenvalues of $\sigma_x = \sigma_{12} \sigma_{34}$ and $\sigma_y = \sigma_{14}  \sigma_{23}$ (which operators form, together with $\sigma_x \sigma_y  = \sigma_{13} \sigma_{24}$ and the identity, an abelian subgroup of ${\rm S}_4$, the so-called Klein group), explains the main $(21,1,1,1,6,6,6)$ block structure of the sADM. 
We also observed that these blocks have several eigenvalues in common. We don't know if there is a simple reason for this, but it is plausible that the $S_4$ symmetry of the problem has something to do with it.

\acknowledgments 
S.~P. was funded by the Agence Nationale de la Recherche (ANR) under grant No.~ANR-18-CE31-0024 (COLDLOSS). 

\noindent 
F.~C. acknowledges support from 
the Polish National Science Center (NCN) grant No. 2022/46/E/ST2/00346.
F.~C. acknowledges support from 
María de Maeztu CEX2023-001318-M financed by MCIN/AEI/10.13039/501100011033. 
Xunta de Galicia (Centro singular de investigación de Galicia accreditation 2019-2022, ref. ED421G-2019/05, project ref. ED431C 2021/22, and CIGUS Network of Research Centres), by European Union ERDF, and by the Spanish Research State Agency under project PID2020119632GBI00. This work has been performed in the framework of the European Research Council project ERC-2018-ADG-835105 YoctoLHC and the MSCA RISE 823947 “Heavy ion collisions: collectivity and precision in saturation physics” (HIEIC), and has received funding from the European Union’s Horizon 2020 research and innovation programme under grant agreement No. 824093.

\eject

\appendix

\section{$\sun$ irreps of $\Irrep{27}{} \otimes \Irrep{27}{}$}
\label{app:A}

\vspace{-5mm}
\begin{table}[hb]
\renewcommand{\arraystretch}{1.15} 
\setlength\tabcolsep{8pt}
\hskip 0mm
\begin{center}
\begin{tabular}{|c|c|c|c|c|c|c|}
\hline
$\ba{c} {\rm Number} \\[-2mm] {\rm of \ traces} \ea$ & $\ba{c} {\rm SU}(3) \\[-2mm] {\rm name} \ea$ & $\sigma_{12}$ &$\ba{c} {\rm Young} \\[-2mm] {\rm diagram} \ea$ & Dimension $K_\al$ & Casimir & $\ba{c} {\rm Casimir} \\[-2mm] {\rm for} \ \Nc=3 \ea$ \\
\hline
4 & $\Irrep{1}{}$ & {($+$)} & $[0000]$ & $1$ & 0 & 0 \\
\hline
3 & $\Irrep{8}{+}$ & {($+$)} & $[1001]$ & $\Nc^2-1$ & $\Nc$ & $3$ \\
  & $\Irrep{8}{-}$ & {($-$)} & $[1001]$ & $\prime \prime$ & $\Nc$ & $3$ \\
\hline
2 & $\Irrep{0}{a}$  & {($+$)} & $[0110]$ & $\frac{1}{4} d_3 \Nc^2 u_1$ & $2\Nc-2$ & - \\
  & $\Irrep{10}{}$ & {($-$)}   & $[2010]$ & $\frac{1}{4} d_2 d_1 u_1 u_2$ & $2\Nc$ & $6$ \\
  & $\Irrep{\overline{10}}{}$ & {($-$)}   & $[0102]$ & $\prime \prime$ & $2\Nc$ & $6$ \\
  & $\Irrep{27}{+}$ & {($+$)} & $[2002]$ & $\frac{1}{4} d_1 \Nc^2 u_3$ & $2\Nc+2$ & $8$ \\
  & $\Irrep{27'}{\!\!+}$ & {($+$)} & $[2002]$ & $\prime \prime$ & $2\Nc+2$ & $8$ \\
  & $\Irrep{27}{-}$ & {($-$)} & $[2002]$ & $\prime \prime$ & $2\Nc+2$ & $8$ \\
\hline
1 & $\Irrep{0}{b+}$ & {($+$)} & $[1111]$ & $\frac{1}{9} d_3 d_1^2 u_1^2 u_3$ & $3\Nc$ & - \\
  & $\Irrep{0}{b-}$ & {($-$)} & $[1111]$ & $\prime \prime$ & $3\Nc$ & - \\
  & $\Irrep{35}{+}$ & {($+$)} & $[3011]$ & $\frac{1}{18} d_2 d_1 \Nc^2 u_2 u_4$ & $3\Nc + 3$ & 12 \\
  & $\Irrep{35}{-}$ & {($-$)} & $[3011]$ & $\prime \prime$ & $3\Nc + 3$ & 12 \\
  & $\Irrep{\overline{35}}{+}$ & {($+$)} & $[1103]$ & $\prime \prime$ & $3\Nc + 3$ & 12 \\
  & $\Irrep{\overline{35}}{-}$ & {($-$)} & $[1103]$ & $\prime \prime$ & $3\Nc + 3$ & 12 \\
  & $\Irrep{64}{+}$ & {($+$)} & $[3003]$ & $\frac{1}{36} d_1 \Nc^2 u_1^2 u_5$ & $3\Nc + 6$ & 15 \\
  & $\Irrep{64}{-}$ & {($-$)} &$[3003]$ & $\prime \prime$ & $3\Nc + 6$ & 15 \\
\hline 
0 & $\Irrep{0}{c}$  & {($+$)} & $[0220]$ & $\frac{1}{144} d_3 d_2^2 d_1 u_1 u_2^2 u_3$ & $4\Nc$ & - \\
  & $\Irrep{0}{d}$  & {($-$)}  & $[2120]$ & $\frac{1}{96} d_3 d_2 d_1 \Nc^2 u_1 u_3 u_4$ & $4\Nc+2$ & - \\
  & $\Irrep{\overline{0}}{d}$ & {($-$)} & $[0212]$ & $\prime \prime$ & $4\Nc+2$ & - \\
  & $\Irrep{0}{e}$ & {($+$)} & $[2112]$ & $\frac{1}{64} d_3 d_1^2 \Nc^2 u_2^2 u_5 $ & $4\Nc+4$ & - \\
  & $\Irrep{28}{}$ & {($+$)} & $[4020]$ & $ \frac{1}{288} d_2 d_1^2 \Nc^2 u_1 u_4 u_5$ & $4\Nc+6$ & $18$ \\
  & $\Irrep{\overline{28}}{}$ & {($+$)} & $[0204]$ & $\prime \prime$ & $4\Nc+6$ & $18$ \\
  & $\Irrep{81}{}$ & {($-$)}  & $[4012]$ & $\frac{1}{192} d_2 d_1 \Nc^2 u_1^2 u_3 u_6$ & $4\Nc+8$ & $20$ \\
  & $\Irrep{\overline{81}}{}$ & {($-$)}  & $[2104]$ & $\prime \prime$ & $4\Nc+8$ & $20$ \\
  & $\Irrep{125}{}$  & {($+$)}& $[4004]$ & $\frac{1}{576} d_1 \Nc^2 u_1^2 u_2^2 u_7$ & $4\Nc+12$ & $24$ \\
\hline
\end{tabular}
\end{center}
\caption{\label{table:irreps} 
$\sun$ irreps of the decomposition~\eq{27times27-deco} of $\Irrep{27}{} \otimes \Irrep{27}{}$. 
First column: number of traces taken in~\eq{psi-chi} giving rise to the irrep; Second column: ${\rm SU}(3)$ name; Third column: symmetry of the irrep under $\psi \leftrightarrow \chi$ (corresponding to the eigenvalue of the irrep's projector w.r.t.~the permutation operator $\sigma_{12}$, see section~\ref{sec:filtering}); Fourth column: ${\rm SU}(5)$ Dynkin name (see section~\ref{sec:sumup} for details); Fifth column: dimension $K_\al$ of the irreps $\al$, using the shorthand notation $d_k \equiv \Nc-k$ and $u_k \equiv \Nc+k$; Sixth column: $\hat{C}_2$ eigenvalue; Last column: $\hat{C}_2$ eigenvalues of existing irreps for $\Nc=3$.}
\end{table}

\section{Main properties of basis of tensors ${\cal T}$}
\label{app:B}

\subsection{List of tensors}
\label{app:B1}
\begin{table}[ht]
\renewcommand{\arraystretch}{.52} 
\setlength\tabcolsep{5pt}
\begin{tabular}{c|cccc}
  $m$ &  &  \quad \quad \quad \quad \quad tensors $\tau$ & &  
   \\[2mm] \hline \hline & & & & \\[-2mm]
4  & \makebox[0pt][l]{$~S$}\phantom{$M_8$} $\equiv\ \TS $ & & & 
\\[6mm] \hline \\[-2mm]  %
3  & \makebox[0pt][l]{$B_1$}\phantom{$M_8$} $\equiv\ \TBP$ & \makebox[0pt][l]{$B_2$}\phantom{$M_8$} $\equiv\ \TBM$ &  \makebox[0pt][l]{$B_3$}\phantom{$M_8$} $\equiv\ \TYP$ & \makebox[0pt][l]{$B_4$}\phantom{$M_8$} $\equiv\ \TYM$ 
\\[6mm] \hline \\[-2mm] %
& \makebox[0pt][l]{$C_1$}\phantom{$M_8$} $\equiv\ \TCP$ & \makebox[0pt][l]{$C_2$}\phantom{$M_8$} $\equiv\ \TCM$ & \makebox[0pt][l]{$C_3$}\phantom{$M_8$} $\equiv\ \TDP$ & \makebox[0pt][l]{$C_4$}\phantom{$M_8$} $\equiv\ \TDM$ \\[7mm]
2 & \makebox[0pt][l]{$Z_1$}\phantom{$M_8$} $\equiv\ \TEP$ & \makebox[0pt][l]{$Z_2$}\phantom{$M_8$} $\equiv\ \TEM$ & \makebox[0pt][l]{$Z_3$}\phantom{$M_8$} $\equiv\ \TFP$ & \makebox[0pt][l]{$Z_4$}\phantom{$M_8$} $\equiv\ \TFM$ \\[7mm]
& \makebox[0pt][l]{$H_1$}\phantom{$M_8$} $\equiv\ \THa$ & \makebox[0pt][l]{$H_2$}\phantom{$M_8$} $\equiv\ \THb$ & \makebox[0pt][l]{$H_3$}\phantom{$M_8$} $\equiv\ \THc$ & \makebox[0pt][l]{$H_4$}\phantom{$M_8$} $\equiv\ \THd$
\\[6mm] \hline \\[-2mm] %
& \makebox[0pt][l]{$J_1$}\phantom{$M_8$} $\equiv\ \TJa$ & \makebox[0pt][l]{$J_3$}\phantom{$M_8$} $\equiv\ \TJc$ & \makebox[0pt][l]{$J_5$}\phantom{$M_8$} $\equiv\ \TJe$ & \makebox[0pt][l]{$J_7$}\phantom{$M_8$} $\equiv\ \TJg$ \\[7mm]
& \makebox[0pt][l]{$J_2$}\phantom{$M_8$} $\equiv\ \TJb$ & \makebox[0pt][l]{$J_4$}\phantom{$M_8$} $\equiv\ \TJf$ & \makebox[0pt][l]{$J_6$}\phantom{$M_8$} $\equiv\ \TJd$ & \makebox[0pt][l]{$J_8$}\phantom{$M_8$} $\equiv\ \TJh$ \\
1 &&&& \\ 
& \makebox[0pt][l]{$K_1$}\phantom{$M_8$} $\equiv\ \TKg$ & \makebox[0pt][l]{$K_3$}\phantom{$M_8$} $\equiv\ \TKc$ & \makebox[0pt][l]{$K_5$}\phantom{$M_8$} $\equiv\ \TKe$ & \makebox[0pt][l]{$K_7$}\phantom{$M_8$} $\equiv\ \TKa$ \\[7mm]
& \makebox[0pt][l]{$K_2$}\phantom{$M_8$} $\equiv\ \TKh$ & \makebox[0pt][l]{$K_4$}\phantom{$M_8$} $\equiv\ \TKd$ & \makebox[0pt][l]{$K_6$}\phantom{$M_8$} $\equiv\ \TKf$ & \makebox[0pt][l]{$K_8$}\phantom{$M_8$} $\equiv\ \TKb$ 
\\[6mm] \hline \\[-2mm] %
& \makebox[0pt][l]{$L_1$}\phantom{$M_8$} $\equiv\ \TLa$ & \makebox[0pt][l]{$L_4$}\phantom{$M_8$} $\equiv\ \TLd$ & \makebox[0pt][l]{$L_6$}\phantom{$M_8$} $\equiv\ \TLh$ &  \makebox[0pt][l]{$L_8$}\phantom{$M_8$} $\equiv\ \TLf$ \\[7mm]
0 & \makebox[0pt][l]{$L_2$}\phantom{$M_8$} $\equiv\ \TLb$ & \makebox[0pt][l]{$L_5$}\phantom{$M_8$} $\equiv\ \TLe$ & \makebox[0pt][l]{$L_7$}\phantom{$M_8$} $\equiv\ \TLi$  & \makebox[0pt][l]{$L_9$}\phantom{$M_8$} $\equiv\ \TLg$  \\[7mm]
& \makebox[0pt][l]{$L_3$}\phantom{$M_8$} $\equiv\ \TLc$ & & &  \\[6mm]
\hline
\end{tabular}
\caption{\label{table:TensorBasis} 
The tensors $\tau$ of the basis ${\cal T}$ are classified according to the number $m$ of `traces' taken on each side of the corresponding birdtrack. To avoid any confusion that might result from deleting the quarks' arrows in the notation \eq{eq:tensor-abbrev}, we highlight the quarks (resp.~antiquarks) of the $4-m$ $q \bar{q}$ pairs appearing in the middle intermediate state of each birdtrack by orange (resp.~green) lines.}
\end{table}
\eject

\subsection{Action of simple operators in basis ${\cal T}$} 
\label{app:B2}

\begin{table}[h]
\renewcommand{\arraystretch}{0.84}
\setlength\tabcolsep{10pt}
\hskip 0mm
\begin{center}
\begin{tabular}{|c||c|c|c|c|c|c||c|c|}
\hline
$\tau$ & $\sigma_{12}\cdot\tau$ & $\sigma_{34}\cdot\tau$ & $\sigma_{14}\cdot\tau$ & $\sigma_{23}\cdot\tau$ & $\sigma_{13}\cdot\tau$ & $\sigma_{24}\cdot\tau$ & $\tau^*$ & $\tau^\dagger$\\[1mm]
\hline
\hline
$S$ & $S$ & $S$ & $L_2$ & $L_2$ & $L_1$ & $L_1$ & $S$ & $S$ \\
$L_1$ & $L_2$ & $L_2$ & $L_1$ & $L_1$ & $S$ & $S$ & $L_1$ & $L_1$ \\
$L_2$ & $L_1$ & $L_1$ & $S$ & $S$ &  $L_2$ & $L_2$ & $L_2$ & $L_2$ \\
\hline
$B_1$ & $B_4$ & $B_3$ & $L_7$ & $L_6$ & $J_2$ & $J_2$ & $B_2$ & $B_1$ \\
$B_2$ & $B_3$ & $B_4$ & $L_6$ & $L_7$ & $J_1$ & $J_1$ & $B_1$ & $B_2$ \\
$B_3$ & $B_2$ & $B_1$ & $K_2$ & $K_2$ & $L_5$ & $L_4$ & $B_4$ & $B_4$ \\
$B_4$ & $B_1$ & $B_2$ & $K_1$ & $K_1$ & $L_4$ & $L_5$ & $B_3$ & $B_3$ \\
$L_4$ & $L_6$ & $L_6$ & $J_1$ & $J_2$ & $B_4$ & $B_3$ & $L_5$ & $L_4$ \\
$L_5$ & $L_7$ & $L_7$ & $J_2$ & $J_1$ & $B_3$ & $B_4$ & $L_4$ & $L_5$ \\
$L_6$ & $L_4$ & $L_4$ & $B_2$ & $B_1$ & $K_1$ & $K_2$ & $L_7$ & $L_6$ \\
$L_7$ & $L_5$ & $L_5$ & $B_1$ & $B_2$ & $K_2$ & $K_1$ & $L_6$ & $L_7$ \\
$J_1$ & $K_2$ & $K_1$ & $L_4$ & $L_5$ & $B_2$ & $B_2$ & $J_2$ & $J_1$ \\
$J_2$ & $K_1$ & $K_2$ & $L_5$ & $L_4$ & $B_1$ & $B_1$ & $J_1$ & $J_2$ \\
$K_1$ & $J_2$ & $J_1$ & $B_4$ & $B_4$ & $L_6$ & $L_7$ & $K_2$ & $K_2$ \\
$K_2$ & $J_1$ & $J_2$ & $B_3$ & $B_3$ & $L_7$ & $L_6$ & $K_1$ & $K_1$ \\
\hline
$C_1$ & $C_4$ & $C_3$ & $L_8$ & $L_9$ & $C_2$ & $C_2$ & $C_2$ & $C_1$ \\
$C_2$ & $C_3$ & $C_4$ & $L_9$ & $L_8$ & $C_1$ & $C_1$ & $C_1$ & $C_2$ \\
$C_3$ & $C_2$ & $C_1$ & $C_4$ & $C_4$ & $L_9$ & $L_8$ & $C_4$ & $C_4$ \\
$C_4$ & $C_1$ & $C_2$ & $C_3$ & $C_3$ & $L_8$ & $L_9$ & $C_3$ & $C_3$ \\
$L_8$ & $L_9$ & $L_9$ & $C_1$ & $C_2$ & $C_4$ & $C_3$ & $L_9$ & $L_8$ \\
$L_9$ & $L_8$ & $L_8$ & $C_2$ & $C_1$ & $C_3$ & $C_4$ & $L_8$ & $L_9$ \\
\hline
$H_1$ & $H_2$ & $H_2$ & $L_3$ & $L_3$ & $H_1$ & $H_1$ & $H_1$ & $H_1$ \\
$H_2$ & $H_1$ & $H_1$ & $H_2$ & $H_2$ & $L_3$ & $L_3$ & $H_2$ & $H_2$ \\
$L_3$ & $L_3$ & $L_3$ & $H_1$ & $H_1$ & $H_2$ & $H_2$ & $L_3$ & $L_3$ \\
\hline
$H_3$ & $H_4$ & $H_4$ & $J_7$ & $J_8$ & $K_8$ & $K_7$ & $H_4$ & $H_3$ \\
$H_4$ & $H_3$ & $H_3$ & $J_8$ & $J_7$ & $K_7$ & $K_8$ & $H_3$ & $H_4$ \\
$J_7$ & $K_8$ & $K_7$ & $H_3$ & $H_4$ & $J_8$ & $J_8$ & $J_8$ & $J_7$ \\
$J_8$ & $K_7$ & $K_8$ & $H_4$ & $H_3$ & $J_7$ & $J_7$ & $J_7$ & $J_8$ \\
$K_7$ & $J_8$ & $J_7$ & $K_8$ & $K_8$ & $H_4$ & $H_3$ & $K_8$ & $K_8$ \\
$K_8$ & $J_7$ & $J_8$ & $K_7$ & $K_7$ & $H_3$ & $H_4$ & $K_7$ & $K_7$ \\
\hline
$Z_1$ & $Z_2$ & $Z_1$ & $K_4$ & $K_6$ & $J_4$ & $J_6$ & $Z_2$ & $Z_3$ \\
$Z_2$ & $Z_1$ & $Z_2$ & $K_5$ & $K_3$ & $J_5$ & $J_3$ & $Z_1$ & $Z_4$ \\
$Z_3$ & $Z_3$ & $Z_4$ & $K_3$ & $K_5$ & $J_6$ & $J_4$ & $Z_4$ & $Z_1$ \\
$Z_4$ & $Z_4$ & $Z_3$ & $K_6$ & $K_4$ & $J_3$ & $J_5$ & $Z_3$ & $Z_2$ \\
$J_3$ & $K_4$ & $K_3$ & $J_6$ & $J_3$ & $Z_4$ & $Z_2$ & $J_6$ & $J_3$ \\
$J_4$ & $K_3$ & $K_4$ & $J_4$ & $J_5$ & $Z_1$ & $Z_3$ & $J_5$ & $J_4$ \\
$J_5$ & $K_6$ & $K_5$ & $J_5$ & $J_4$ & $Z_2$ & $Z_4$ & $J_4$ & $J_5$ \\
$J_6$ & $K_5$ & $K_6$ & $J_3$ & $J_6$ & $Z_3$ & $Z_1$ & $J_3$ & $J_6$ \\
$K_3$ & $J_4$ & $J_3$ & $Z_3$ & $Z_2$ & $K_6$ & $K_3$ & $K_6$ & $K_4$ \\
$K_4$ & $J_3$ & $J_4$ & $Z_1$ & $Z_4$ & $K_4$ & $K_5$ & $K_5$ & $K_3$ \\
$K_5$ & $J_6$ & $J_5$ & $Z_2$ & $Z_3$ & $K_5$ & $K_4$ & $K_4$ & $K_6$ \\
$K_6$ & $J_5$ & $J_6$ & $Z_4$ & $Z_1$ & $K_3$ & $K_6$ & $K_3$ & $K_5$ \\
\hline
\end{tabular}
\end{center}
\caption{Action of permutation operators $\sigma_{ij}$, complex conjugation, and hermitian conjugation (in $s$-channel), on the 42 tensors $\tau$ of the basis ${\cal T}$.}
\label{tab:CrossingR}
\end{table}

\subsection{Traces of tensors}
\label{app:B3}

In order to normalize the projectors, it is necessary to calculate the trace of each element $\tau \in \mathcal{T}$, viewed as a map of ${\cal E} \equiv  \Irrep{27}{} \otimes \Irrep{27}{} \to {\cal E}$. The trace is expressed pictorially as\footnote{Let us recall that when applied to a projector $\Proj{R}$ on an irrep $R$ of $\Irrep{27}{} \otimes \Irrep{27}{}$, the trace defined in~\eq{eq:trace-def} gives the dimension of the irrep, $\tr (\Proj{R}) = K_R$.}
\begin{align}
\label{eq:trace-def}
\tr (\tau) = \tr \left\{  \TT{\tau} \right\} = \ \trTT{\tau}  \ \ .
\end{align}
We list all $\tr (\tau)$'s below, factoring out for convenience the dimension of the  
irrep $\Irrep{27}{}$ given in~\eq{27-dim} (see also Table~\ref{table:irreps}). 
\begin{subequations}
\begin{align}
&\tr{S} =  \tr{L_2} = \K{\Irrep{27}{}} \\[2mm]
&\tr{L_1} = (\K{\Irrep{27}{}})^2 \\[1mm]
&\tr{B_1} = \tr{B_2} = \tr{L_6} = \tr{L_7} = \frac{\Nc(\Nc+3)}{2(\Nc +2)} \, \K{\Irrep{27}{}} \\[1mm]
&\tr{B_3} = \tr{B_4} = \tr{K_1} =\tr{K_2} = - \frac{1}{(\Nc +2)} \, \K{\Irrep{27}{}}\\[1mm]
&\tr{J_1} = \tr{J_2} = \tr{L_4} = \tr{L_5} =  \frac{(\Nc-1)\Nc(\Nc+3)}{4} \, \K{\Irrep{27}{}} \\[1mm]
&\tr{C_1} = \tr{C_2} = \tr{L_8} = \tr{L_9} = \frac{(\Nc-1)\Nc(\Nc+3)}{2(\Nc +1)} \, \K{\Irrep{27}{}} \\[1mm]
&\tr{C_3} = \tr{C_4} = \frac{2}{(\Nc+1)(\Nc+2)}\, \K{\Irrep{27}{}} \\[1mm]
&\tr{H_1} = \tr{L_3} = \frac{\Nc^2(\Nc+3)}{4(\Nc+1)} \, \K{\Irrep{27}{}} \\[1mm]
&\tr{H_2} = \frac{\Nc^2+\Nc+2}{4(\Nc+1)(\Nc+2)} \, \K{\Irrep{27}{}} \\[1mm]
&\tr{H_3} = \tr{H_4} = \tr{J_7} = \tr{J_8} = \frac{\Nc^2(\Nc+3)}{4(\Nc+1)(\Nc+2)} \, \K{\Irrep{27}{}} \\[1mm]
&\tr{K_7} = \tr{K_8} = - \frac{\Nc(\Nc+3)}{4(\Nc+1)} \, \K{\Irrep{27}{}} \\[1mm]
&\tr{J_3} = \tr{J_6} = \tr{J_5} = \tr{J_4} = \frac{(\Nc-1)\Nc(\Nc+3)}{4(\Nc+1)} \, \K{\Irrep{27}{}} \\[1mm]
&\tr{Z_1} = \tr{Z_2} = \tr{Z_3} = \tr{Z_4} = \tr{K_3} = \tr{K_4} = \tr{K_5} = \tr{K_6} = - \frac{\Nc(\Nc+3) \K{\Irrep{27}{}}}{2(\Nc+1)(\Nc+2)} 
\end{align}
\end{subequations}

\section{Non-unicity of $(\Proj{\Irrep{27}{+}}$, $\Proj{\Irrep{27'}{\!\!+}})$}
\label{app:C}

\subsection{Set of solutions}
\label{app:C1}

Here we solve the system of equations \eq{syst-ab} and \eq{syst-ab-ortho} for the coefficients $a, \, b, \, a', \, b'$ entering the expressions \eq{eq:P27ppp-basis-tau} of the projectors $\Proj{\Irrep{27}{+}}$ and $\Proj{\Irrep{27'}{\!\!+}}$.

Using the change of variables:
\begin{align}
\label{eq:CV}
\left( \begin{matrix} a \\[2mm] b  \end{matrix}  \right) 
= U_{\theta} \cdot \left( \begin{matrix} X \\[2mm] Y \end{matrix}  \right) \ \ ; \ \ 
\left( \begin{matrix} a' \\[2mm] b'  \end{matrix}  \right) 
= U_{\theta} \cdot \left( \begin{matrix} X' \\[2mm] Y' \end{matrix}  \right)  \ \ ; \ \ 
U_{\theta} \equiv  \frac{1}{\sin{2\theta}} \left( \begin{matrix}\ \cos{\theta} \ & \ \sin{\theta} \ \\[2mm]
\ \cos{\theta} \ & \ -\sin{\theta} \ \end{matrix}  \right)  \ , 
\end{align}
where the angle $\theta \in \ ] 0,  \frac{\pi}{4} [$ is a fixed parameter defined by 
\begin{align}
\label{theta-def}
\cos{2\theta} = \frac{1}{\sqrt{\rho}} \ \ ; \ \ \sin{2\theta} = \sqrt{\frac{\rho-1}{\rho}} \ ,
\end{align}
the system of equations \eq{syst-ab}-\eq{syst-ab-ortho} becomes:
\begin{align}
& X^2 + Y^2  \ = \ X'^{\,2} + Y'^{\,2} \ = \ 1 \ \ ; \ \  X X' + Y Y'  \ = \ 0  \ .
\end{align}
This has an obvious solution:\footnote{The other solution obtained by multiplying $X'$ and $Y'$ by $-1$ changes the sign of $a'$ and $b'$ (see \eq{eq:CV}). This leads to the same projector \eq{eq:P27pp-basis-tau}, and to an irrelevant sign change in the transition operators involving the irrep $\Irrep{27'}{\!\!+}$. We can thus fix the sign of $X'$ and $Y'$ as in \eq{eq:XYXY}.}
\begin{align}
\label{eq:XYXY}
&X = \cos{t} \ \ ;  \ \ Y = -\sin{t}  \ \ ;  \ \ X' = \sin{t} \ \ ;  \ \ Y' = \cos{t}  \ ,
\end{align}
with $t \in [ 0,  2 \pi [$ an arbitrary parameter. 
Inserting \eq{eq:XYXY} into \eq{eq:CV} we get:
\begin{align}
\label{eq:abab}
& a(t) = \frac{\cos{(t+\theta)}}{\sin{2\theta}} \, ; \ b(t) = \frac{\cos{(t-\theta)}}{\sin{2\theta}} 
\, ; \ a'(t) = \frac{ \sin{(t+\theta)}}{\sin{2\theta}} \, ; \ b'(t) = \frac{\sin{(t-\theta)}}{\sin{2\theta}} \ , 
\end{align}
where the $t$-dependence of $a, \, b, \, a', \, b'$ is made explicit. Using \eq{eq:abab} in \eq{eq:P27ppp-basis-tau} leads to the expressions \eq{eq:P27ppp-param} of the projectors, parametrized by the continuous parameter $t$. 

Let us emphasize that the solution~\eq{eq:abab} for a parameter $t$ is related to the solution for $t=0$ as:
\begin{subequations}
\label{abab-rot}
\begin{align}
& \left( \begin{matrix} a(t) \\[2mm] a'(t)  \end{matrix}  \right) 
= R(t)  \cdot \left( \begin{matrix} a(0) \\[2mm] a'(0)  \end{matrix}  \right) \, ; \ 
\left( \begin{matrix} b(t) \\[2mm] b'(t)  \end{matrix}  \right) 
=  R(t) \cdot \left( \begin{matrix} b(0) \\[2mm] b'(0)  \end{matrix}  \right) \ , \\[2mm]
\label{Rt-def}
& \hskip 25mm R(t)  \equiv \left( \begin{matrix} \cos{t} \ & \ -\sin{t}\ \\[2mm] 
\sin{t} \ & \ \ \cos{t} \end{matrix}  \right) \, .
\end{align} 
\end{subequations}
It follows directly that the clebsches~\eq{eq:c1c2} for a parameter $t$ and for $t=0$ are related in the same way:
\begin{align}
\label{c1c2-rot}
\left( \begin{matrix} c_1(t) \\[2mm] c_2(t)  \end{matrix}  \right) 
=  R(t) \cdot \left( \begin{matrix} c_1(0) \\[2mm] c_2(0)  \end{matrix}  \right)  \ . \\[2mm] \nn
\end{align}

\subsection{A few possible choices for $(\Proj{\Irrep{27}{+}},\Proj{\Irrep{27'}{\!\!+}})$} 
\label{app:C2}

We list below some possible choices for the pair of projectors \eq{eq:P27ppp-param}, corresponding to simple values of the parameter $t$. 
Recall that the sum of projectors $\Proj{{\rm sum}}{}$ is given by \eq{eq:Psum}. 
\subsubsection*{(a) \bf{$t=0$}}
\vspace{-3mm}
\begin{subequations}
\label{P27pair-ex1}
\begin{align}
\label{eq:P27plus-ex1}
\Proj{\Irrep{27}{+}} \, =& \ \frac{1}{2(\sqrt{\rho}-1)} \left[ \sqrt{\rho} \left(\frac{\tilde{H}}{n_{11}} + \frac{\tilde{C}}{2(n_{22}+n_{23})} \right)  - \frac{\tilde{Z}}{2 n_{12}} \right] \ ,  \\[2mm]
\label{eq:P27plusp-ex1}
\Proj{\Irrep{27'}{\!\!+}} \, =& \ \frac{1}{2(\sqrt{\rho}+1)} \left[ \sqrt{\rho} \left(\frac{\tilde{H}}{n_{11}} + \frac{\tilde{C}}{2(n_{22}+n_{23})} \right) + \frac{\tilde{Z}}{2 n_{12}} \right]  \ .
\end{align}
\end{subequations}

\subsubsection*{(b) \bf{$t = \theta$}}
\vspace{-3mm}
\begin{subequations}
\label{P27pair-ex3}
\begin{align}
\label{eq:P27plusp-ex3}
\Proj{\Irrep{27}{+}} \, =& \ \frac{1}{\rho-1} \, \frac{\tilde{H}}{n_{11}} + \frac{\rho}{\rho-1} \, \frac{\tilde{C}}{2(n_{22}+n_{23})} + \frac{1}{1-\rho} \, \frac{\tilde{Z}}{2 n_{12}}  \ ,  \\[2mm]
\label{eq:P27plus-ex3}
\Proj{\Irrep{27'}{\!\!+}} \, =&  \ \frac{\tilde{H}}{n_{11}} \ .
\end{align}
\end{subequations}

\subsubsection*{(c) \bf{$t = \frac{\pi}{2}-\theta$ }}
\vspace{-3mm}
\begin{subequations}
\label{P27pair-ex2}
\begin{align}
\label{eq:P27plusp-ex2}
\Proj{\Irrep{27}{+}} \, =& \ \frac{\tilde{C}}{2(n_{22}+n_{23})} \ ,  \\[2mm]
\label{eq:P27plus-ex2}
\Proj{\Irrep{27'}{\!\!+}} \, =& \ \frac{\rho}{\rho-1} \, \frac{\tilde{H}}{n_{11}} + \frac{1}{\rho-1} \, \frac{\tilde{C}}{2(n_{22}+n_{23})} + \frac{1}{1-\rho} \, \frac{\tilde{Z}}{2 n_{12}} \ .
\end{align}
\end{subequations}

\subsubsection*{(d) \bf{ $t=\frac{\pi}{4}$}}
\vspace{-3mm}
\begin{subequations}
\label{P27pair-ex4}
\begin{align}
\label{eq:P27plus-ex4}
\Proj{\Irrep{27}{+}} \, =& \ \frac{1}{2} \, \Proj{{\rm sum}}{} - \frac{1}{2}\sqrt{ \frac{\rho}{\rho-1}} \left(\frac{\tilde{H}}{n_{11}} - \frac{\tilde{C}}{2(n_{22}+n_{23})} \right) \ ,  \\[2mm]
\label{eq:P27plusp-ex4}
\Proj{\Irrep{27'}{\!\!+}} \, =& \ \frac{1}{2} \, \Proj{{\rm sum}}{} + \frac{1}{2}\sqrt{ \frac{\rho}{\rho-1}} \left(\frac{\tilde{H}}{n_{11}} - \frac{\tilde{C}}{2(n_{22}+n_{23})} \right) \ . \\[3mm] \nn
\end{align}
\end{subequations}

\newpage
\section{Some factors of the sADM characteristic polynomial}
\label{app:D}
 
Here we give the explicit expressions of the polynomials $p_1(x,b)$, $p_2(x,b)$ and $ p_3(x,b)$ appearing as factors in the full characteristic polynomial of the sADM, see \eq{char-pol}--\eq{char-pol-fact}. 

\begin{align}
p_1(x,b) \ &= \ x^4 - \frac{x^3}{{\Nc}}(9 {\Nc}+8) -\frac{x^2}{{\Nc}^2}(b^2 {\Nc}^2+2 b^2 {\Nc}+7 b^2-30 {\Nc}^2-52 {\Nc}-19) \notag \\
&+\frac{x}{{\Nc}^3}(5 b^2 {\Nc}^3+16 b^2 {\Nc}^2+31 b^2 {\Nc}+12 b^2-44 {\Nc}^3-112 {\Nc}^2-79 {\Nc}-12) \notag \\
&-\frac{2}{{\Nc}^3}(3 b^2 {\Nc}^3+13 b^2 {\Nc}^2+23 b^2 {\Nc}+12 b^2-12 {\Nc}^3-40 {\Nc}^2-41 {\Nc}-12)  \ , \nn \\ \\[2mm]
p_2(x,b) \ &= \ x^4 
-\frac{x^3}{2 {\Nc}}(5 b \Nc+8 b+23 {\Nc}+24) \notag \\
&+ \frac{x^2}{4 {\Nc}^2}(5 b^2 {\Nc}^2+10 b^2 {\Nc}-28 b^2+86 b {\Nc}^2+220 b {\Nc}+128 b+197 {\Nc}^2+410 {\Nc}+204) \notag \\
&+ \frac{x}{8 {\Nc}^3}(5 b^3 {\Nc}^3+48 b^3 {\Nc}^2+216 b^3 {\Nc}+224 b^3-51 b^2 {\Nc}^3-112 b^2 {\Nc}^2+272 b^2 {\Nc}+320 b^2 \notag \\
&\phantom{+ \frac{x}{8 {\Nc}^3}(}-489 b {\Nc}^3-1712 b {\Nc}^2-1848 b {\Nc}-608 b-745 {\Nc}^3-2320 {\Nc}^2-2288 {\Nc}-704) \notag \\
&+\frac{1}{8 {\Nc}^4}(-3 b^4 {\Nc}^4-33 b^4 {\Nc}^3-150 b^4 {\Nc}^2-192 b^4 {\Nc}-20 b^3 {\Nc}^4-204 b^3 {\Nc}^3-832 b^3 {\Nc}^2 \notag \\
&\phantom{+\frac{1}{8 {\Nc}^4}(}-1152 b^3 {\Nc}-384 b^3+62 b^2 {\Nc}^4+122 b^2 {\Nc}^3-460 b^2 {\Nc}^2-960 b^2 {\Nc}-384 b^2 \notag \\
&\phantom{+\frac{1}{8 {\Nc}^4}(} +460 b {\Nc}^4+2036 b {\Nc}^3 +3120 b {\Nc}^2+1920 b {\Nc}+384 b+525 {\Nc}^4+2175 {\Nc}^3 \notag \\
&\phantom{+\frac{1}{8 {\Nc}^4}(}+3186 {\Nc}^2+1920 {\Nc}+384) \ , \nn \\ \\[2mm]
p_3(x,b) \ &= \ x^6 - x^5 \frac{16({\Nc}+1)}{{\Nc}} - \frac{x^4 }{{\Nc}^2}(5 b^2 {\Nc}^2+16 b^2 {\Nc}+47 b^2-105 {\Nc}^2-208 {\Nc}-91) \notag \\
&+\frac{2 x^3 }{{\Nc}^3}(29 b^2 {\Nc}^3+116 b^2 {\Nc}^2+283 b^2 {\Nc}+124 b^2-181 {\Nc}^3-532 {\Nc}^2-451 {\Nc}-92) \notag \\
&+\frac{4({\Nc}+1) x^2 }{{\Nc}^4} (b^4 {\Nc}^3+9 b^4 {\Nc}^2+33 b^4 {\Nc}-11 b^4-62 b^2 {\Nc}^3-234 b^2 {\Nc}^2 \notag \\
&\phantom{-\frac{4({\Nc}+1) x^2 }{{\Nc}^4} (} -438 b^2 {\Nc} -2 b^2+173 {\Nc}^3+497 {\Nc}^2+325 {\Nc}-35) \notag \\
&-\frac{8 ({\Nc}+1)^2 x}{{\Nc}^5} (3 b^4 {\Nc}^3+26 b^4 {\Nc}^2+79 b^4 {\Nc}-76 b^4-58 b^2 {\Nc}^3-204 b^2 {\Nc}^2 \notag \\
&\phantom{+\frac{8 ({\Nc}+1)^2 x}{{\Nc}^5}(}-258 b^2 {\Nc}+200 b^2+87 {\Nc}^3+242 {\Nc}^2+83 {\Nc}-124) \notag \\
&+\frac{32 (b^2-1) ({\Nc}^2-1) ({\Nc}+1)^2 }{{\Nc}^6} \left(b^2 {\Nc}^2+9 b^2 {\Nc}+30 b^2-9 {\Nc}^2-33 {\Nc}-30\right) \ . \nn \\ 
\end{align}

\bibliography{mybib}
\bibliographystyle{JHEP}

\end{document}

%% file: macro.tex
\def\ed{\end{document}}
\def\bc{\begin{center}}
\def\ec{\end{center}}
\def\bi{\begin{itemize}}
\def\ei{\end{itemize}}
\def\ben{\begin{enumerate}}
\def\een{\end{enumerate}}
\def\be{\begin{equation}}
\def\ee{\end{equation}}
\def\bea{\begin{eqnarray}}
\def\eea{\end{eqnarray}}
\newcommand{\nn}{\nonumber} 
\newcommand{\ie}{{i.e.}}  
\newcommand{\eg}{{e.g.}}
\def\bex{\begin{exercise}}
\def\eex{\end{exercise}}
\def\ba{\begin{array}}
\def\ea{\end{array}}
\def\bs{\begin{scope}}
\def\es{\end{scope}}

\def\sun{{\rm SU}(N)}
\newcommand{\unit}{{\mathds{1}}} 
\newcommand{\Nc}{N}

\newcommand{\eq}[1]{(\ref{#1})}

\newcommand{\ave}[1]{\langle{#1}\rangle}

\newcommand{\al}{\alpha}

\def\bm#1{\mbox{\boldmath$#1$}}
\newcommand{\tr}{\mathrm{Tr}\,}

\newcommand{\Bbar}{{\overline{B}}}
\newcommand{\Cbar}{{\overline{C}}}
\newcommand{\Lbar}{{\overline{L}}}
\newcommand{\Hbar}{{\overline{H}}}
\newcommand{\Zbar}{{\overline{Z}}}
\newcommand{\Jbar}{{\overline{J}}}
\newcommand{\Kbar}{{\overline{K}}}

\newcommand{\mathsize}[2]{\mbox{\fontsize{#1}{2}\selectfont $#2$}}

\newcommand{\Proj}[1]{{\mathds{P}}_{\mathsize{6}{\! #1}}} 
\newcommand{\K}[1]{{K}_{\mathsize{6}{\! #1}}} 
\newcommand{\E}[1]{{\cal E}_{\mathsize{6}{\! #1}}} 

\newcommand{\Irrep}[2]{{\bm{#1}}_{#2}}
\newcommand{\Trans}[1]{{\mathds{T}}_{\mathsize{6}{#1}}} 
\newcommand{\TpTdag}[1]{({\mathds{T}+\mathds{T}^\dagger})_{\mathsize{6}{#1}}} 
\newcommand{\TmTdag}[1]{({\mathds{T}-\mathds{T}^\dagger})_{\mathsize{6}{#1}}}

\newcommand{\Pair}[2]{(#1)_{_{\mathsize{6}{#2}}}}
\newcommand{\Syst}[3]{(#1)_{_{\mathsize{6}{#2}}}{#3}}

\newenvironment{psmallmatrix}
{\left(\begin{smallmatrix}}
	{\end{smallmatrix}\right)}

\def \drawqua (#1) {
\draw[thick,postaction={decorate},decoration={markings,mark=at position #1 with {\large\arrow{>}}}] 
}
\def \drawanti (#1) {
\draw[thick,postaction={decorate},decoration={markings,mark=at position #1 with {\large\arrow{<}}}] 
}

\usepackage{tikz}
\usetikzlibrary{patterns}
\usetikzlibrary{decorations.pathmorphing,decorations.markings,trees,calc,shapes.geometric}
\tikzset{
  qua/.style={postaction={decorate},decoration={markings,mark=at position .55 with {\large\arrow{>}}}},
  anti/.style={postaction={decorate},decoration={markings,mark=at position .55 with {\large\arrow{<}}}},
  glu/.style={decorate,decoration={coil,amplitude=1.4pt, segment length=2.6pt}} ,
  glu2/.style={draw = blue} ,
  vertical align/.style={baseline=-.5*(height("$+$")-depth("$+$"))},
  every picture/.style={vertical align}
}

\newcommand{\tikeq}[1]{\vcenter{\hbox{\begin{tikzpicture}#1\end{tikzpicture}}}}
\newcommand{\tikeqbis}[1]{\hbox{\begin{tikzpicture}#1\end{tikzpicture}}}

\newcommand{\Idqqalq}[3]{
\tikeqbis{
\begin{scope}[scale=#3,yscale=-1]
\draw[thick] (0,0.3) -- (1,0.3); 
\drawqua(0.6) (0.2,0.3) -- (0.3,0.3); 
\drawqua(0.6) (0.8,0.3) -- (0.85,0.3); 
\draw[thick] (0,0) -- (1,0); 
\drawqua(0.6) (0.2,0) -- (0.3,0); 
\drawqua(0.6) (0.8,0) -- (0.85,0); 
\draw[thick,#2] (0,-0.3) -- (1,-0.3); 
\draw[fill=#1,opacity=1] (0.4,-0.1) rectangle (0.6,0.4); 
\end{scope}
}}

\newcommand{\Aqqq}[1]{
\tikeqbis{
\begin{scope}[scale=#1,yscale=-1]
\draw[thick] (0,0.3) -- (1,0.3); 
\draw[thick,qua] (0.2,0.3) -- (0.3,0.3); 
\draw[thick] (0,0) -- (0.8,0);
\draw[thick,qua] (0.2,0) -- (0.3,0); 
\draw[thick,qua] (0,-0.3) -- (0.8,-0.3); 
\draw[fill=white,opacity=1] (0.4,-0.1) rectangle (0.6,0.4); 
\begin{scope}[xshift=10mm]
\draw[thick] (0,0.3) -- (1,0.3); \draw[thick,qua] (0.8,0.3) -- (0.85,0.3); 
\draw[thick] (0.2,0) -- (1,0); 
\draw[thick,qua] (0.8,0) -- (0.85,0); 
\draw[thick,qua] (0.2,-0.3) -- (1,-0.3); 
\draw[fill=white,opacity=1] (0.4,-0.1) rectangle (0.6,0.4); 
\end{scope}
\draw[thick,qua] (0.9,0.3) -- (1.1,0.3); 
\draw[thick] (0.8,0) .. controls +(0.2,0) and +(-0.2,0) .. (1.2,-0.3); 
\draw[thick] (0.8,-0.3) .. controls +(0.2,0) and +(-0.2,0) .. (1.2,0); 
\end{scope}
}}

\newcommand{\PqqqSym}[1]{\tikeqbis{\begin{scope}[scale=#1]
\drawqua(0.6) (0,0.3) -- (0.4,0.3); \drawqua(0.6) (0.6,0.3) -- (1,0.3); 
\drawqua(0.6) (0,0) -- (0.4,0); \drawqua(0.6) (0.6,0) -- (1,0);
\drawqua(0.6) (0,-0.3) -- (0.4,-0.3); \drawqua(0.6) (0.6,-0.3) -- (1,-0.3);
\draw[fill=white,opacity=1] (0.4,-0.4) rectangle (0.6,0.4); 
\end{scope}
}}

\newcommand{\Pqqqplus}[1]{\frac{4}{3} \ 
\tikeqbis{\begin{scope}[scale=#1,yscale=-1]
\draw[thick] (-1,0.3) -- (1,0.3); \draw[thick,qua] (-1,0.3) -- (-0.6,0.3); \drawqua(0.6) (0.6,0.3) -- (1,0.3); 
\draw[thick] (-1,0) -- (1,0); 
\draw[thick,qua] (-1,0) -- (-0.6,0); 
\drawqua(0.6) (0.6,0) -- (1,0);
\draw[thick] (-1,-0.3) -- (1,-0.3);
\draw[thick,qua] (-1,-0.3) -- (-0.6,-0.3);  
\drawqua(0.6) (0.6,-0.3) -- (1,-0.3);
\draw[thick,qua] (-0.2,0.3) -- (0.2,0.3);
\draw[thick,qua] (-0.4,0) -- (-0.1,0); 
\draw[thick,qua] (0.2,0) -- (0.4,0); 
\draw[fill=white,opacity=1] (-0.6,-0.1) rectangle (-0.4,0.4); 
\draw[fill=white,opacity=1] (0.4,-0.1) rectangle (0.6,0.4); 
\draw[fill=white!50!black,opacity=1] (-0.1,-0.4) rectangle (0.1,0.1); 
\end{scope}
}}

\def \xibird (#1,#2) {
\tikeqbis{  \begin{scope}[scale=#1]
	        \coordinate (O) at (0,0); 
	        \coordinate (S) at   ($(O)+(.5,0)$); 
	        \coordinate (A2) at  ($(O)+(0,.5)$) ;  
	        \coordinate (A1) at  ($(A2)+(0,.6)$) ; 
	        \coordinate (A3) at  ($(A2)+(1.,0)$); 
	        \coordinate (A4) at  ($(A1)+(1.,0)$);  
	        \coordinate (B2) at  ($(O)-(0,.5)$) ;  
	        \coordinate (B1) at  ($(B2)-(0,.6)$) ;  
	        \coordinate (B3) at  ($(B2)+(1.,0)$);  
	        \coordinate (B4) at  ($(B1)+(1.,0)$);          
    \draw ($(A1)+(-.2,.15)$) node {$1$};
    \draw ($(A2)+(-.2,-.15)$) node {$2$};
    \draw ($(A3)+(.2,-.15)$) node {$3$};
    \draw ($(A4)+(.2,.15)$) node {$4$};
	           
	        \draw[red]  (A1) -- ++ (-1,0) arc (90:180:.3) -- ++ (0,-1.6) arc (-180:-90:.3) -- (B1);
	        \draw[blue]  (A2) -- ++ (-.7,0) arc (90:180:.2) -- ++ (0,-0.6) arc (-180:-90:.2) -- (B2);
	        \draw[green]  (A4) -- ++ (1,0) arc (90:0:.3) -- ++ (0,-1.6) arc (0:-90:.3) -- (B4);
	        \draw[brown]  (A3) -- ++ (.7,0) arc (90:0:.2) -- ++ (0,-0.6) arc (0:-90:.2) -- (B3);

  \fill[lightgray,draw=black,opacity=0.6] ($(A1)+(0,.1)$) rectangle ($(A3)+(0,-.1)$);  
  \draw ($(A1)+(.5,-.3)$) node {${\cal M}$};
  \fill[lightgray,draw=black,opacity=0.6] ($(B2)+(0,.1)$) rectangle ($(B4)+(0,-.1)$); 
  \draw ($(B2)+(.5,-.3)$) node {${\cal M}^\star$};
 
 \draw[glu] ($(A1)+(-0.6,0)$) -- ++ (0,.2) arc (180:90:.3) -- ++ (1.6,0) arc (90:0:.3) -- ++ (0,#2); 
  \end{scope}
 }}
\def \xibirdstraight (#1,#2) {
\tikeqbis{  \begin{scope}[scale=#1]
	        \coordinate (O) at (0,0); 
	        \coordinate (S) at   ($(O)+(.5,0)$); 
	        \coordinate (A2) at  ($(O)+(0,.5)$) ;  
	        \coordinate (A1) at  ($(A2)+(0,.6)$) ; 
	        \coordinate (A3) at  ($(A2)+(1.,0)$); 
	        \coordinate (A4) at  ($(A1)+(1.,0)$);  
	        \coordinate (B2) at  ($(O)-(0,.5)$) ;  
	        \coordinate (B1) at  ($(B2)-(0,.6)$) ;  
	        \coordinate (B3) at  ($(B2)+(1.,0)$);  
	        \coordinate (B4) at  ($(B1)+(1.,0)$);          
    \draw ($(A1)+(-.2,.15)$) node {$1$};
    \draw ($(A2)+(-.2,-.15)$) node {$2$};
    \draw ($(A3)+(.2,-.15)$) node {$3$};
    \draw ($(A4)+(.2,.15)$) node {$4$};
	           
	        \draw[red]  (A1) -- ++ (-1,0) arc (90:180:.3) -- ++ (0,-1.6) arc (-180:-90:.3) -- (B1);
	        \draw[blue]  (A2) -- ++ (-.7,0) arc (90:180:.2) -- ++ (0,-0.6) arc (-180:-90:.2) -- (B2);
	        \draw[green]  (A4) -- ++ (1,0) arc (90:0:.3) -- ++ (0,-1.6) arc (0:-90:.3) -- (B4);
	        \draw[brown]  (A3) -- ++ (.7,0) arc (90:0:.2) -- ++ (0,-0.6) arc (0:-90:.2) -- (B3);

  \fill[lightgray,draw=black,opacity=0.6] ($(A1)+(0,.1)$) rectangle ($(A3)+(0,-.1)$);  
  \draw ($(A1)+(.5,-.3)$) node {${\cal M}$};
  \fill[lightgray,draw=black,opacity=0.6] ($(B2)+(0,.1)$) rectangle ($(B4)+(0,-.1)$); 
  \draw ($(B2)+(.5,-.3)$) node {${\cal M}^\star$};
 
 \draw[glu] ($(O)+(-1.3,0)$) -- ($(O)+(#2,0)$) ; 
  \end{scope}
 }}

\def \xibirdalpha (#1,#2,#3,#4) {
\tikeqbis{  \begin{scope}[scale=#1]
	        \coordinate (O) at (0,0); 
	        \coordinate (S) at   ($(O)+(.5,0)$); 
	        \coordinate (A2) at  ($(O)+(0,.5)$) ;  
	        \coordinate (A1) at  ($(A2)+(0,.6)$) ; 
	        \coordinate (A3) at  ($(A2)+(1.,0)$); 
	        \coordinate (A4) at  ($(A1)+(1.,0)$);  
	        \coordinate (B2) at  ($(O)-(0,.5)$) ;  
	        \coordinate (B1) at  ($(B2)-(0,.6)$) ;  
	        \coordinate (B3) at  ($(B2)+(1.,0)$);  
	        \coordinate (B4) at  ($(B1)+(1.,0)$);          
    \draw ($(A1)+(-.2,.15)$) node {$1$};
    \draw ($(A2)+(-.2,-.15)$) node {$2$};
    \draw ($(A3)+(.2,-.15)$) node {$3$};
    \draw ($(A4)+(.2,.15)$) node {$4$};       
	        \draw[red]  ($(A1)+(.2,0)$) -- (A1) -- ++ (-1.,0) arc (90:180:.3) -- ++ (0,-1.6) arc (-180:-90:.3) -- ($(B1)+(.2,0)$) ;
	        \draw[blue]  ($(A2)+(.2,0)$) -- (A2) -- ++ (-.7,0) arc (90:180:.2) -- ++ (0,-0.6) arc (-180:-90:.2) -- ($(B2)+(.2,0)$);
	        \draw[green]  ($(A4)+(-.2,0)$) -- (A4) -- ++ (1,0) arc (90:0:.3) -- ++ (0,-1.6) arc (0:-90:.3) -- ($(B4)+(-.2,0)$);
	        \draw[brown] ($(A3)+(-.2,0)$) -- (A3) -- ++ (.7,0) arc (90:0:.2) -- ++ (0,-0.6) arc (0:-90:.2) -- ($(B3)+(-.2,0)$);
   \fill[white,draw=black] ($(O)+(.5,.8)$) circle (.42);  
  \draw ($(O)+(.5,.8)$) node {$#3$};
    \fill[white,draw=black] ($(O)+(.5,-.8)$) circle (.42);  
  \draw ($(O)+(.5,-.8)$) node {$#4$};
 \draw[glu] ($(O)+(-1.3,0)$) -- ($(O)+(#2,0)$) ; 
  \end{scope}
 }}

\def \IdentitySinglet (#1,#2) {
\tikeq{  \begin{scope}[scale=#1]
	        \coordinate (O) at (0,0); 
	        \coordinate (A) at  ($(O)+(0,-1.1)$) ;  
	        \coordinate (B) at  ($(O)+(0,-.7)$) ; 
	        \coordinate (C) at  ($(O)+(0,.7)$); 
	        \coordinate (D) at  ($(O)+(0,1.1)$);              
	        \draw[red]  (A) -- ++ (#2,0);
	        \draw[blue] (B) -- ++ (#2,0);
	        \draw[brown] (C) -- ++ (#2,0);
	        \draw[green] (D) -- ++ (#2,0);
  \end{scope}
 }}

\def \OrthoKet (#1,#2) {
\tikeq{  \begin{scope}[scale=#1]
	        \coordinate (O) at (0,0); 
	        \coordinate (A2) at (.5,-.3) ;  
	        \coordinate (A1) at (1.1,-.3) ; 
	        \coordinate (A3) at (.5,.3) ;  
	        \coordinate (A4) at  (1.1,.3) ;   
    \draw (.2,-.9) node {$\bar{1}$};
    \draw (.2,-0.4) node {$\bar{2}$};
    \draw (.2,0.4) node {$3$};
    \draw (.2,.9) node {$4$};
	        \draw[red]  ($(A1)+(0,.2)$) -- (A1) -- ++ (0,-.5) arc (360:270:.3) -- ++ (-0.8,0) ;
	        \draw[blue]  ($(A2)+(0,.2)$) -- (A2) -- ++ (0,-.2) arc (360:270:.2) -- ++ (-.3,0.) ;
	        \draw[green]  ($(A4)+(0,-.2)$) -- (A4) -- ++ (0,.5) arc (0:90:.3) -- ++ (-0.8,0);
	        \draw[brown] ($(A3)+(0,-.2)$) -- (A3) -- ++ (0,.2) arc (0:90:.2) -- ++ (-.3,0);
   \fill[white,draw=black] ($(O)+(.8,0)$) circle (.42);  
  \draw ($(O)+(.8,0)$) node {$#2$};
  \end{scope}
 }}

\def \OrthoBra (#1,#2) {
\tikeq{  \begin{scope}[scale=#1]
 \coordinate (O) at (0,0); 
	        \coordinate (A2) at (-.5,-.3) ;  
	        \coordinate (A1) at (-1.1,-.3) ; 
	        \coordinate (A3) at (-.5,.3) ;  
	        \coordinate (A4) at  (-1.1,.3) ;  
    \draw (-.2,-.9) node {$\bar{1}$};
    \draw (-.2,-0.4) node {$\bar{2}$};
    \draw (-.2,0.4) node {$3$};
    \draw (-.2,.9) node {$4$}; 
	        \draw[red]  ($(A1)+(0,.2)$) -- (A1) -- ++ (0,-.5) arc (180:270:.3) -- ++ (0.8,0) ;
	        \draw[blue]  ($(A2)+(0,.2)$) -- (A2) -- ++ (0,-.2) arc (180:270:.2) -- ++ (.3,0.) ;
	        \draw[green]  ($(A4)+(0,-.2)$) -- (A4) -- ++ (0,.5) arc (180:90:.3) -- ++ (0.8,0);
	        \draw[brown] ($(A3)+(0,-.2)$) -- (A3) -- ++ (0,.2) arc (180:90:.2) -- ++ (.3,0);
   \fill[white,draw=black] ($(O)+(-.8,0)$) circle (.42);  
  \draw ($(O)+(-.8,0)$) node {$#2$};
  \end{scope}
 }}

\def \Balphabeta(#1,#2) {
\tikeq{  \begin{scope}[scale=#1]
	        \coordinate (O) at (0,0); 
	        \coordinate (A2) at  ($(O)+(-.5,0)$) ;  
	        \coordinate (A1) at  ($(A2)+(-.6,0)$) ; 
	        \coordinate (A3) at  ($(A2)+(0,1)$); 
	        \coordinate (A4) at  ($(A1)+(0,1)$);  
	        \coordinate (B2) at  ($(O)-(-.5,0)$) ;  
	        \coordinate (B1) at  ($(B2)-(-.6,0)$) ;  
	        \coordinate (B3) at  ($(B2)+(0,1)$);  
	        \coordinate (B4) at  ($(B1)+(0,1)$);          
    \draw ($(A2)+(.2,2)$) node {$4$};
    \draw ($(A2)+(.2,1.6)$) node {$3$};
    \draw ($(A2)+(.2,-.7)$) node {$\bar{1}$};
    \draw ($(A2)+(.2,-.1)$) node {$\bar{2}$};       
	        \draw[red]  ($(A1)+(0,.2)$) -- (A1) -- ++ (0,-.5) arc (180:270:.3) -- ++ (1.6,0) arc (-90:0:.3) -- ($(B1)+(0,.2)$) ;
	        \draw[blue]  ($(A2)+(0,.2)$) -- (A2) -- ++ (0,-.2) arc (180:270:.2) -- ++ (.6,0.) arc (-90:-0:.2) -- ($(B2)+(0,.2)$);
	        \draw[green]  ($(A4)+(0,-.2)$) -- (A4) -- ++ (0,.5) arc (180:90:.3) -- ++ (1.6,0) arc (90:0:.3) -- ($(B4)+(0,-.2)$);
	        \draw[brown] ($(A3)+(0,-.2)$) -- (A3) -- ++ (0,.2) arc (180:90:.2) -- ++ (.6,0) arc (90:0:.2) -- ($(B3)+(0,-.2)$);
   \fill[white,draw=black] ($(O)+(-.8,.5)$) circle (.42);  
  \draw ($(O)+(-.8,.5)$) node {$\al$};
    \fill[white,draw=black] ($(O)+(.8,.5)$) circle (.42);  
  \draw ($(O)+(.8,.5)$) node {$\beta$};
 \draw[glu] ($(O)+(0,-0.8)$) -- ($(O)+(0,#2)$) ; 
  \end{scope}
 }}

\def \Bbarone (#1) {
\tikeq{  \begin{scope}[scale=#1]
	        \coordinate (O) at (0,0); 
	        \coordinate (A2) at  ($(O)+(-.5,0)$) ;  
	        \coordinate (A1) at  ($(A2)+(-.6,0)$) ; 
	        \coordinate (A3) at  ($(A2)+(0,1)$); 
	        \coordinate (A4) at  ($(A1)+(0,1)$);  
	        \coordinate (B2) at  ($(O)-(-.5,0)$) ;  
	        \coordinate (B1) at  ($(B2)-(-.6,0)$) ;  
	        \coordinate (B3) at  ($(B2)+(0,1)$);  
	        \coordinate (B4) at  ($(B1)+(0,1)$);          
	        \draw[blue]  ($(A2)+(0,.2)$) -- (A2) -- ++ (0,-.2) arc (180:270:.2) -- ++ (.6,0.) arc (-90:-0:.2) -- ($(B2)+(0,.2)$);
	           \draw[brown] ($(A3)+(0,-.2)$) -- (A3) -- ++ (0,.2) arc (180:90:.2) -- ++ (.6,0) arc (90:0:.2) -- ($(B3)+(0,-.2)$);
	        \draw[qua]  ($(A4)+(0,-.2)$) -- (A4) -- ++ (0,.2) arc (0:90:.2) -- ++ (-.2,0) arc (90:180:.2) -- ++ (0,-1.4) arc (180:270:.2)  -- ++ (.2,0) arc (-90:0:.2) -- ($(A1)+(0,.2)$);
	         \draw[anti]  ($(B4)+(0,-.2)$) -- (B4) -- ++ (0,.2) arc (180:90:.2) -- ++ (.2,0) arc (90:0:.2) -- ++ (0,-1.4) arc (0:-90:.2)  -- ++ (-.2,0) arc (-90:-180:.2) -- ($(B1)+(0,.2)$);	        
   \fill[white,draw=black] ($(O)+(-.8,.5)$) circle (.42);  
  \draw ($(O)+(-.8,.5)$) node {$\al$};
    \fill[white,draw=black] ($(O)+(.8,.5)$) circle (.42);  
  \draw ($(O)+(.8,.5)$) node {$\beta$};
  \end{scope}
 }}
\def \Bbartwo (#1) {
\tikeq{  \begin{scope}[scale=#1]
	        \coordinate (O) at (0,0); 
	        \coordinate (A2) at  ($(O)+(-.5,0)$) ;  
	        \coordinate (A1) at  ($(A2)+(-.6,0)$) ; 
	        \coordinate (A3) at  ($(A2)+(0,1)$); 
	        \coordinate (A4) at  ($(A1)+(0,1)$);  
	        \coordinate (B2) at  ($(O)-(-.5,0)$) ;  
	        \coordinate (B1) at  ($(B2)-(-.6,0)$) ;  
	        \coordinate (B3) at  ($(B2)+(0,1)$);  
	        \coordinate (B4) at  ($(B1)+(0,1)$);          
	        \draw[anti]  ($(A1)+(0,.2)$) -- (A1) -- ++ (0,-.5) arc (180:270:.3) -- ++ (1.6,0) arc (-90:0:.3) -- ($(B1)+(0,.2)$) ;
	        \draw[blue]  ($(A2)+(0,.2)$) -- (A2) -- ++ (0,-.2) arc (180:270:.2) -- ++ (.6,0.) arc (-90:-0:.2) -- ($(B2)+(0,.2)$);
	        \draw[qua]  ($(A4)+(0,-.2)$) -- (A4) -- ++ (0,.5) arc (180:90:.3) -- ++ (1.6,0) arc (90:0:.3) -- ($(B4)+(0,-.2)$);
	        \draw[brown] ($(A3)+(0,-.2)$) -- (A3) -- ++ (0,.2) arc (180:90:.2) -- ++ (.6,0) arc (90:0:.2) -- ($(B3)+(0,-.2)$);
   \fill[white,draw=black] ($(O)+(-.8,.5)$) circle (.42);  
  \draw ($(O)+(-.8,.5)$) node {$\al$};
    \fill[white,draw=black] ($(O)+(.8,.5)$) circle (.42);  
  \draw ($(O)+(.8,.5)$) node {$\beta$};
  \end{scope}
 }}
\def \birdKtwo (#1) {
\tikeq{  \begin{scope}[scale=#1]
	        \coordinate (O) at (0,0); 
	        \coordinate (A2) at  ($(O)+(-.5,0)$) ;  
	        \coordinate (A1) at  ($(A2)+(-.6,0)$) ; 
	        \coordinate (A3) at  ($(A2)+(0,1)$); 
	        \coordinate (A4) at  ($(A1)+(0,1)$);               
	        \draw[blue]  ($(A2)+(0,.2)$) -- (A2) -- ++ (0,-.2) arc (180:270:.2) -- ++ (.3,0.);
	         \draw[brown] ($(A3)+(0,-.2)$) -- (A3) -- ++ (0,.2) arc (180:90:.2) -- ++ (.3,0) ;
	        \draw[qua]  ($(A4)+(0,-.2)$) -- (A4) -- ++ (0,.2) arc (0:90:.2) -- ++ (-.2,0) arc (90:180:.2) -- ++ (0,-1.4) arc (180:270:.2)  -- ++ (.2,0) arc (-90:0:.2) -- ($(A1)+(0,.2)$);              
   \fill[white,draw=black] ($(O)+(-.8,.5)$) circle (.42);  
  \draw ($(O)+(-.8,.5)$) node {$\al$};
  \end{scope}
 }
 = \frac{K_\al}{K_2} \ \ 
 \tikeq{  \begin{scope}[scale=#1]
	        \coordinate (O) at (0,0); 
	        \coordinate (A2) at  ($(O)+(-.5,0)$) ;  
	        \coordinate (A3) at  ($(A2)+(0,1)$);      
	        \draw[blue]  ($(A2)+(0,.5)$) -- (A2) -- ++ (0,-.2) arc (180:270:.2) -- ++ (.3,0.);
	         \draw[brown] ($(A3)+(0,-.5)$) -- (A3) -- ++ (0,.2) arc (180:90:.2) -- ++ (.3,0) ;
  \end{scope}
 }}

%% file: MacroFIG.tex
\newcommand{\subProjB}{
\draw[thick] (-5,-1) -- (5,-1);
\draw[thick] (-5,-.5) -- (5,-.5);
\draw[thick] (-5,.5) -- (5,.5);
\draw[thick] (-5,1) -- (5,1);
\draw[thick,->] (-3,-1) -- ++(-1,0);
\draw[thick,->] (-3,-.5) -- ++(-1,0);
\draw[thick,-<] (-3,1) -- ++(-1,0);
\draw[thick,-<] (-3,.5) -- ++(-1,0);
\draw[thick,->] (3,1) -- ++(1,0);
\draw[thick,->] (3,.5) -- ++(1,0);
\draw[thick,-<] (3,-.5) -- ++(1,0);
\draw[thick,-<] (3,-1) -- ++(1,0);
\draw[fill=white] (-3,1.2) rectangle (-2.6,.3);
\draw[fill=white] (-3,-1.2) rectangle (-2.6,-.3);
\draw[fill=white] (3,1.2) rectangle (2.6,.3);
\draw[fill=white] (3,-1.2) rectangle (2.6,-.3);
\draw[fill=white] (0,0) ellipse (1.2cm and 1.2cm);
\node at (0,0) {\scriptsize $27$};
}

\newcommand{\ProjB}[1]{
\begin{tikzpicture}[scale=#1]
\subProjB
\end{tikzpicture}}

\newcommand{\projdash}[2]{\tikeq{\begin{scope}[scale=#2]
\draw[dashed,thick] (-3,1) -- (3,1);
\draw[dashed,thick] (-3,-1) -- (3,-1);
\draw[fill=white] (0,0) ellipse (1.2cm and 2cm);
\node at (0,0) {$#1$};
\end{scope}}}

\newcommand{\schanprocess}[2]{\tikeq{\begin{scope}[scale=#2]
\draw[dashed,thick]  (-3,1) -- (3,1);
\node[left] at (-3,1) {$1$};  \node[right] at (3,1) {$4$}; 
\draw[dashed,thick] (-3,-1) -- (3,-1);
\node[left] at (-3,-1) {$2$};  \node[right] at (3,-1) {$3$}; 
\draw[fill=white] (0,0) ellipse (1.2cm and 2cm);
\node at (0,0) {$#1$};
\end{scope}}}

\newcommand{\irrepdash}[2]{\tikeqbis{\begin{scope}[scale=#2]
\draw[dashed,thick] (-2,0) -- (2,0);
\node at (0,0.6) {$#1$};
\end{scope}}}

\newcommand{\projdashG}[3]{\tikeqbis{\begin{scope}[scale=#3]
\draw[dashed,thick] (-2,0) -- (2,0);
\node at (0,0.6) {$#2$};
\bs[xshift=-1cm]
\draw[dashed,thick] (-5,1) -- (-2,1);
\draw[dashed,thick] (-5,-1) -- (-2,-1);
\draw[fill=white] (-2,0) ellipse (1.2cm and 1.2cm);
\node at (-2,0) {$#1$};
\es
\bs[xshift=1cm]
\draw[dashed,thick] (2,1) -- (5,1);
\draw[dashed,thick] (2,-1) -- (5,-1);
\draw[fill=white] (2,0) ellipse (1.2cm and 1.2cm);
\node at (2,0) {$#1^\dagger$};
\es
\end{scope}}}

\newcommand{\projdashCG}[5]{\tikeqbis{\begin{scope}[scale=#4,xscale=#5]
\draw[dashed,thick] (-2,0) -- (2,0);
\node at (0,0.6) {$#3$};
\bs[xshift=-0.3cm]
\draw[dashed,thick] (-4,1) -- (-2,0);
\draw[dashed,thick] (-4,-1) -- (-2,0);
\draw[fill=black] (-2,0) ellipse (0.2cm and 0.2cm);
\node at (-2,0.9) {$#1$};
\es
\bs[xshift=0.3cm]
\draw[dashed,thick] (2,0) -- (4,1);
\draw[dashed,thick] (2,0) -- (4,-1);
\draw[fill=black] (2,0) ellipse (0.2cm and 0.2cm);
\node at (2,0.9) {$#2$};
\es
\end{scope}}}

\newcommand{\CGdashG}[3]{\tikeqbis{\begin{scope}[scale=#3]
\draw[dashed,thick] (-2,0) -- (2,0);
\node at (0,0.6) {$#2$};
\bs[xshift=-1cm]
\draw[dashed,thick] (-5,1) -- (-2,1);
\draw[dashed,thick] (-5,-1) -- (-2,-1);
\draw[fill=white] (-2,0) ellipse (1.2cm and 1.2cm);
\node at (-2,0) {$#1$};
\es
\end{scope}}}

\newcommand{\CGdash}[4]{\tikeqbis{\begin{scope}[scale=#3,xscale=#4]
\draw[dashed,thick] (-1.7,0) -- (1.5,0);
\node at (0,0.6) {$#2$};
\bs[xshift=0.3cm]
\draw[dashed,thick] (-4,1) -- (-2,0);
\draw[dashed,thick] (-4,-1) -- (-2,0);
\draw[fill=black] (-2,0) ellipse (0.2cm and 0.2cm);
\node at (-2,0.9) {$#1$};
\es
\end{scope}}}

\newcommand{\CGdashdag}[4]{\tikeqbis{\begin{scope}[scale=#3,xscale=-1,xscale=#4]
\draw[dashed,thick] (-1.5,0) -- (1.7,0);
\node at (0,0.6) {$#2$};
\bs[xshift=0.3cm]
\draw[dashed,thick] (-4,1) -- (-2,0);
\draw[dashed,thick] (-4,-1) -- (-2,0);
\draw[fill=black] (-2,0) ellipse (0.2cm and 0.2cm);
\node at (-2,0.9) {$#1$};
\es
\end{scope}}}

\newcommand{\transdash}[3]{\tikeq{\bs[scale=#3]
\draw[dashed,thick] (-3,1) -- (3,1);
\draw[dashed,thick] (-3,-1) -- (3,-1);
\draw[fill=white] (0,0) ellipse (1.2cm and 2cm);
\node at (0,0) {$#1$};
\bs[xshift=8cm]
\draw[dashed,thick] (-3.1,1) -- (3,1);
\draw[dashed,thick] (-3.1,-1) -- (3,-1);
\draw[fill=white] (0,0) ellipse (1.2cm and 2cm);
\node at (0,0) {$#2$};
\es
\draw[fill=white] (4,0) ellipse (1.5cm and 2.5cm);
\es}}

\newcommand{\identitydash}[1]{\tikeq{\begin{scope}[scale=#1]
\draw[dashed,thick] (-2,1) -- (2,1);
\draw[dashed,thick] (-2,-1) -- (2,-1);
\end{scope}}}

\newcommand{\dashproj}[1]{\tikeqbis{\begin{scope}[scale=#1]
\draw[dashed,thick] (-2,0) -- (2,0);
\end{scope}}}

\newcommand{\fullidentity}[2]{\tikeq{\begin{scope}[scale=#1]
\subProjB \begin{scope}[yshift=-4cm] \subProjB \end{scope}
\end{scope}}}

\newcommand{\tensors}[2]{\tikeq{\begin{scope}[scale=#2]
\subProjB \begin{scope}[yshift=-4cm] \subProjB \end{scope}
\begin{scope}[xshift=15cm] 
\subProjB  \begin{scope}[yshift=-4cm] \subProjB \end{scope}
\end{scope}
\draw[fill=white] (7.5,-2) node{$#1$} ellipse (3cm and 5cm);
\end{scope}}}

\newcommand{\simpletensors}[1]{\tikeq{\begin{scope}[scale=\setscaletikz]
\node at (0,0) {\TLa};
\draw[fill=white] (0,0) ellipse (1.8cm and 3cm);
\node at (0,0) {$#1$};
\end{scope}}}

\newcommand{\Casimirdef}[1]{\tikeq{\begin{scope}[scale=\setscaletikz]
\node at (0,0) {\TLa};
\draw[fill=white] (0,0) ellipse (1.8cm and 3cm);
\node at (0,0) {$#1$};
\draw[glu] (-3,-2.8) .. controls +(0,-2) and +(-1,0) .. (0,-5.5) 
.. controls +(1,0) and +(0,-2) .. (3,-2.8); 
\draw[fill=white!70!black] (3,0) ellipse (.4 and 3.2);
\draw[fill=white!70!black] (-3,0) ellipse (.4 and 3.2);
\end{scope}}}

\newcommand{\IDqqqq}{\tikeq{\bs[scale=\setscaletikz,yscale=1.2]
\draw[thick,\qbcolor] (0,1) -- (3,1);
\draw[thick,\qcolor] (0,-1) -- (3,-1); 
\draw[thick,\qbcolor] (0,-2) -- (3,-2);
\draw[thick,\qcolor] (0,2) -- (3,2); 
\es
}}

\newcommand{\potential}[2]{\tikeq{\bs[scale=\setscaletikz,yscale=1.2]
\draw[thick,\qbcolor] (0,1) -- (3,1);
\draw[thick,\qcolor] (0,-1) -- (3,-1); 
\draw[thick,\qbcolor] (0,-2) -- (3,-2);
\draw[thick,\qcolor] (0,2) -- (3,2); 
\draw[glu] (1.5,#1) -- (1.5,#2) ; 
\es
}}

\newcommand{\Vonethree}[1]{\tikeq{\bs[scale=#1]
\draw[thick,qua] (0,-0.25) -- (0.4,-0.25);
\draw[thick,qua] (0.4,-0.25) -- (0.8,-0.25);
\draw[thick,qua] (0,0.25) -- (0.4,0.25);
\draw[thick,qua] (0.4,0.25) -- (0.8,0.25);
\draw[glu] (0.4,-0.25) -- (0.4,0.25);
\es
}}

\newcommand{\X}[1]{\tikeq{\bs[scale=#1]
\drawqua(0.25) (0,-0.25) -- (0.7,0.25);  
\drawqua(0.25) (0,0.25) -- (0.7,-0.25);
\es
}}

\newcommand{\Iqq}[1]{\tikeq{\bs[scale=#1]
\draw[thick,qua] (0,-0.25) -- (0.7,-0.25);  
\draw[thick,qua] (0,0.25) -- (0.7,0.25);	
\es
}}

\newcommand{\Vonetwo}[1]{\tikeq{\bs[scale=#1]
\draw[thick,anti] (0,-0.25) -- (0.4,-0.25);
\draw[thick,anti] (0.4,-0.25) -- (0.8,-0.25);
\draw[thick,qua] (0,0.25) -- (0.4,0.25);
\draw[thick,qua] (0.4,0.25) -- (0.8,0.25);
\draw[glu] (0.4,-0.25) -- (0.4,0.25);
\es
}}

\newcommand{\Sqq}[1]{\tikeq{\bs[scale=#1]
\draw[thick,qua] (-0.1,0.25) -- (0,0.25) -- (0,-0.25) -- (-0.1,-0.25);  
\draw[thick,qua] (0.5,-0.25) -- (0.4,-0.25) -- (0.4,0.25) -- (0.5,0.25);
\es
}}

\newcommand{\Iqqbar}[1]{\tikeq{\bs[scale=#1]
\draw[thick,anti] (0,-0.25) -- (0.7,-0.25);  
\draw[thick,qua] (0,0.25) -- (0.7,0.25);	
\es
}}

\newcommand{\Id}[1]{\tikeq{\bs[scale=#1]
\drawqua(0.6) (0,-1) -- ++(-2,0);
\drawqua(0.6) (0,-.5) -- ++(-2,0);
\drawanti(0.6) (0,.5) -- ++(-2,0);
\drawanti(0.6) (0,1) -- ++(-2,0);
\drawanti(0.6) (0,-1) -- ++(2,0);
\drawanti(0.6) (0,-.5) -- ++(2,0);
\drawqua(0.6) (0,.5) -- ++(2,0);
\drawqua(0.6) (0,1) -- ++(2,0);
\draw[fill=white] (-.2,1.2) rectangle (.2,.3);
\draw[fill=white] (-.2,-1.2) rectangle (.2,-.3);
\es
}}

\newcommand{\Q}[1]{\tikeq{\bs[scale=#1]
\drawqua(0.8) (-1,-1) -- ++(-1.5,0);
\drawqua(0.8) (-1,-.5) -- ++(-1.5,0);
\drawanti(0.8) (-1,.5) -- ++(-1.5,0);
\drawanti(0.8) (-1,1) -- ++(-1.5,0);
\drawanti(0.8) (1,-1) -- ++(1.5,0);
\drawanti(0.8) (1,-.5) -- ++(1.5,0);
\drawqua(0.8) (1,.5) -- ++(1.5,0);
\drawqua(0.8) (1,1) -- ++(1.5,0);
\draw[thick] (-1,1) -- (1,1);
\draw[thick] (-1,-1) -- (1,-1);
\draw[thick] (-1,.5) -- (-.5,.5) -- (-.5,-.5) -- (-1,-.5);
\draw[thick] (+1,.5) -- (+.5,.5) -- (+.5,-.5) -- (+1,-.5);
\draw[fill=white] (-1.5,1.2) rectangle ++(.4,-.9);
\draw[fill=white] (-1.5,-1.2) rectangle ++(.4,+.9);
\draw[fill=white] (+1.5,1.2) rectangle ++(-.4,-.9);
\draw[fill=white] (+1.5,-1.2) rectangle ++(-.4,+.9);
\es
}}

\renewcommand{\S}[1]{\tikeq{\bs[scale=#1]
\drawqua(0.8) (-1,-1) -- ++(-1.5,0);
\drawqua(0.8) (-1,-.5) -- ++(-1.5,0);
\drawanti(0.8) (-1,.5) -- ++(-1.5,0);
\drawanti(0.8) (-1,1) -- ++(-1.5,0);
\drawanti(0.8) (1,-1) -- ++(1.5,0);
\drawanti(0.8) (1,-.5) -- ++(1.5,0);
\drawqua(0.8) (1,.5) -- ++(1.5,0);
\drawqua(0.8) (1,1) -- ++(1.5,0);
\draw[thick] (-1,.5) -- (-1,-.5);
\draw[thick] (-1,1) -- (-.5,1) -- (-.5,-1) -- (-1,-1);
\draw[fill=white] (-1.2,-.2) rectangle (-.3,.2);
\draw[thick] (1,.5) -- (1,-.5);
\draw[thick] (1,1) -- (.5,1) -- (.5,-1) -- (1,-1);
\draw[fill=white] (1.2,-.2) rectangle (.3,.2);
\es
}}

\newcommand{\Squ}{
\draw[thick] (-3,2.5) -- ++(-1,0);
\draw[thick] (-3,2) -- ++(-1,0);
\draw[thick] (-3,1) -- ++(-1,0);
\draw[thick] (-3,.5) -- ++(-1,0);
\draw[thick] (-3,-.5) -- ++(-1,0);
\draw[thick] (-3,-1) -- ++(-1,0);
\draw[thick] (-3,-2) -- ++(-1,0);
\draw[thick] (-3,-2.5) -- ++(-1,0);
\draw[thick] (3,2.5) -- ++(1,0);
\draw[thick] (3,2) -- ++(1,0);
\draw[thick] (3,1) -- ++(1,0);
\draw[thick] (3,.5) -- ++(1,0);
\draw[thick] (3,-.5) -- ++(1,0);
\draw[thick] (3,-1) -- ++(1,0);
\draw[thick] (3,-2) -- ++(1,0);
\draw[thick] (3,-2.5) -- ++(1,0);
\draw[fill=white] (-4,2.7) rectangle ++(-.4,-.9);
\draw[fill=white] (-4,1.2) rectangle ++(-.4,-.9);
\draw[fill=white] (4,2.7) rectangle ++(.4,-.9);
\draw[fill=white] (4,1.2) rectangle ++(.4,-.9);
\draw[fill=white] (-4,-2.7) rectangle ++(-.4,.9);
\draw[fill=white] (-4,-1.2) rectangle ++(-.4,.9);
\draw[fill=white] (4,-2.7) rectangle ++(.4,.9);
\draw[fill=white] (4,-1.2) rectangle ++(.4,.9);
}

\newcommand{\TS}{
\begin{tikzpicture}[scale=\setscaletikz]
\Squ
\draw[thick] (-3,1) -- ++(.5,0)-- ++(0,-2) -- ++(-.5,0) -- (-3,-1);
\draw[thick] (-3,2) -- ++(1,0)-- ++(0,-4) -- ++(-1,0) -- (-3,-2);
\draw[thick] (-3,2.5) -- ++(1.5,0)-- ++(0,-5) -- ++(-1.5,0) -- (-3,-2.5);
\draw[thick] (3,1) -- ++(-.5,0)-- ++(0,-2) -- ++(.5,0) -- (3,-1);
\draw[thick] (3,2) -- ++(-1,0)-- ++(0,-4) -- ++(1,0) -- (3,-2);
\draw[thick] (3,2.5) -- ++(-1.5,0)-- ++(0,-5) -- ++(1.5,0) -- (3,-2.5);
\draw[thick] (-3,.5) -- (-3,-.5);
\draw[thick] (3,.5) -- (3,-.5);
\end{tikzpicture}}

\newcommand{\tbp}{{\ensuremath{B_1}}}
\newcommand{\TBP}{
\begin{tikzpicture}[scale=\setscaletikz]
\draw[thick] (-3,2.5) -- ++(-1,0);
\draw[thick] (-3,2) -- ++(-1,0);
\draw[thick] (-3,1) -- ++(-1,0);
\draw[thick] (-3,.5) -- ++(-1,0);
\draw[thick] (-3,-.5) -- ++(-1,0);
\draw[thick] (-3,-1) -- ++(-1,0);
\draw[thick] (-3,-2) -- ++(-1,0);
\draw[thick] (-3,-2.5) -- ++(-1,0);
\draw[thick] (3,2.5) -- ++(1,0);
\draw[thick] (3,2) -- ++(1,0);
\draw[thick] (3,1) -- ++(1,0);
\draw[thick] (3,.5) -- ++(1,0);
\draw[thick] (3,-.5) -- ++(1,0);
\draw[thick] (3,-1) -- ++(1,0);
\draw[thick] (3,-2) -- ++(1,0);
\draw[thick] (3,-2.5) -- ++(1,0);
\draw[fill=white] (-4,2.7) rectangle ++(-.4,-.9);
\draw[fill=white] (-4,1.2) rectangle ++(-.4,-.9);
\draw[fill=white] (4,2.7) rectangle ++(.4,-.9);
\draw[fill=white] (4,1.2) rectangle ++(.4,-.9);
\draw[fill=white] (-4,-2.7) rectangle ++(-.4,.9);
\draw[fill=white] (-4,-1.2) rectangle ++(-.4,.9);
\draw[fill=white] (4,-2.7) rectangle ++(.4,.9);
\draw[fill=white] (4,-1.2) rectangle ++(.4,.9);
\draw[thick] (-3,1) -- ++(.5,0)-- ++(0,-2) -- ++(-.5,0) -- (-3,-1);
\draw[thick] (-3,2) -- ++(1,0)-- ++(0,-4) -- ++(-1,0) -- (-3,-2);
\draw[thick] (-3,2.5) -- ++(1.5,0)-- ++(0,-5) -- ++(-1.5,0) -- (-3,-2.5);
\draw[thick] (3,1) -- ++(-.5,0)-- ++(0,-2) -- ++(.5,0) -- (3,-1);
\draw[thick] (3,2) -- ++(-1,0)-- ++(0,-4) -- ++(1,0) -- (3,-2);
\draw[thick] (3,2.5) -- ++(-1.5,0)-- ++(0,-5) -- ++(1.5,0) -- (3,-2.5);
\draw[thick,\qbcolor] (-3,.5) -- (3,.5);
\draw[thick,\qcolor] (-3,-.5) -- (3,-.5);
\end{tikzpicture}}

\newcommand{\TBM}{
\begin{tikzpicture}[scale=\setscaletikz]
\draw[thick] (-3,2.5) -- ++(-1,0);
\draw[thick] (-3,2) -- ++(-1,0);
\draw[thick] (-3,1) -- ++(-1,0);
\draw[thick] (-3,.5) -- ++(-1,0);
\draw[thick] (-3,-.5) -- ++(-1,0);
\draw[thick] (-3,-1) -- ++(-1,0);
\draw[thick] (-3,-2) -- ++(-1,0);
\draw[thick] (-3,-2.5) -- ++(-1,0);
\draw[thick] (3,2.5) -- ++(1,0);
\draw[thick] (3,2) -- ++(1,0);
\draw[thick] (3,1) -- ++(1,0);
\draw[thick] (3,.5) -- ++(1,0);
\draw[thick] (3,-.5) -- ++(1,0);
\draw[thick] (3,-1) -- ++(1,0);
\draw[thick] (3,-2) -- ++(1,0);
\draw[thick] (3,-2.5) -- ++(1,0);
\draw[fill=white] (-4,2.7) rectangle ++(-.4,-.9);
\draw[fill=white] (-4,1.2) rectangle ++(-.4,-.9);
\draw[fill=white] (4,2.7) rectangle ++(.4,-.9);
\draw[fill=white] (4,1.2) rectangle ++(.4,-.9);
\draw[fill=white] (-4,-2.7) rectangle ++(-.4,.9);
\draw[fill=white] (-4,-1.2) rectangle ++(-.4,.9);
\draw[fill=white] (4,-2.7) rectangle ++(.4,.9);
\draw[fill=white] (4,-1.2) rectangle ++(.4,.9);
\draw[thick] (-3,1) -- ++(.5,0)-- ++(0,-2) -- ++(-.5,0) -- (-3,-1);
\draw[thick] (-3,2) -- ++(1,0)-- ++(0,-4) -- ++(-1,0) -- (-3,-2);
\draw[thick] (3,1) -- ++(-.5,0)-- ++(0,-2) -- ++(.5,0) -- (3,-1);
\draw[thick] (3,2) -- ++(-1,0)-- ++(0,-4) -- ++(1,0) -- (3,-2);
\draw[thick] (-3,.5) -- (-3,-.5);
\draw[thick] (3,.5) -- (3,-.5);
\draw[thick,\qbcolor] (-3,-2.5) -- (3,-2.5);
\draw[thick,\qcolor] (-3,2.5) -- (3,2.5);
\end{tikzpicture}}

\newcommand{\TYP}{
\begin{tikzpicture}[scale=\setscaletikz],
\draw[thick] (-3,2.5) -- ++(-1,0);
\draw[thick] (-3,2) -- ++(-1,0);
\draw[thick] (-3,1) -- ++(-1,0);
\draw[thick] (-3,.5) -- ++(-1,0);
\draw[thick] (-3,-.5) -- ++(-1,0);
\draw[thick] (-3,-1) -- ++(-1,0);
\draw[thick] (-3,-2) -- ++(-1,0);
\draw[thick] (-3,-2.5) -- ++(-1,0);
\draw[thick] (3,2.5) -- ++(1,0);
\draw[thick] (3,2) -- ++(1,0);
\draw[thick] (3,1) -- ++(1,0);
\draw[thick] (3,.5) -- ++(1,0);
\draw[thick] (3,-.5) -- ++(1,0);
\draw[thick] (3,-1) -- ++(1,0);
\draw[thick] (3,-2) -- ++(1,0);
\draw[thick] (3,-2.5) -- ++(1,0);
\draw[fill=white] (-4,2.7) rectangle ++(-.4,-.9);
\draw[fill=white] (-4,1.2) rectangle ++(-.4,-.9);
\draw[fill=white] (4,2.7) rectangle ++(.4,-.9);
\draw[fill=white] (4,1.2) rectangle ++(.4,-.9);
\draw[fill=white] (-4,-2.7) rectangle ++(-.4,.9);
\draw[fill=white] (-4,-1.2) rectangle ++(-.4,.9);
\draw[fill=white] (4,-2.7) rectangle ++(.4,.9);
\draw[fill=white] (4,-1.2) rectangle ++(.4,.9);
\draw[thick] (-3,1) -- ++(.5,0)-- ++(0,-2) -- ++(-.5,0) -- (-3,-1);
\draw[thick] (-3,2) -- ++(1,0)-- ++(0,-4) -- ++(-1,0) -- (-3,-2);
\draw[thick] (-3,2.5) -- ++(1.5,0)-- ++(0,-5) -- ++(-1.5,0) -- (-3,-2.5);
\draw[thick] (3,1) -- ++(-.5,0)-- ++(0,-2) -- ++(.5,0) -- (3,-1);
\draw[thick] (3,2) -- ++(-1,0)-- ++(0,-4) -- ++(1,0) -- (3,-2);
\draw[thick] (3,.5) -- (3,-.5);
\draw[thick,\qbcolor] (-3,.5) --(-1.5,.5) to[out=0, in =180] (1.5,-2.5) -- (3,-2.5);
\draw[thick,\qcolor] (-3,-.5)-- (-1.5,-.5) to[out=0, in =180] (1.5,2.5) -- (3,2.5);
\end{tikzpicture}}

\newcommand{\TYM}{
\begin{tikzpicture}[scale=\setscaletikz]
\draw[thick] (-3,2.5) -- ++(-1,0);
\draw[thick] (-3,2) -- ++(-1,0);
\draw[thick] (-3,1) -- ++(-1,0);
\draw[thick] (-3,.5) -- ++(-1,0);
\draw[thick] (-3,-.5) -- ++(-1,0);
\draw[thick] (-3,-1) -- ++(-1,0);
\draw[thick] (-3,-2) -- ++(-1,0);
\draw[thick] (-3,-2.5) -- ++(-1,0);
\draw[thick] (3,2.5) -- ++(1,0);
\draw[thick] (3,2) -- ++(1,0);
\draw[thick] (3,1) -- ++(1,0);
\draw[thick] (3,.5) -- ++(1,0);
\draw[thick] (3,-.5) -- ++(1,0);
\draw[thick] (3,-1) -- ++(1,0);
\draw[thick] (3,-2) -- ++(1,0);
\draw[thick] (3,-2.5) -- ++(1,0);
\draw[fill=white] (-4,2.7) rectangle ++(-.4,-.9);
\draw[fill=white] (-4,1.2) rectangle ++(-.4,-.9);
\draw[fill=white] (4,2.7) rectangle ++(.4,-.9);
\draw[fill=white] (4,1.2) rectangle ++(.4,-.9);
\draw[fill=white] (-4,-2.7) rectangle ++(-.4,.9);
\draw[fill=white] (-4,-1.2) rectangle ++(-.4,.9);
\draw[fill=white] (4,-2.7) rectangle ++(.4,.9);
\draw[fill=white] (4,-1.2) rectangle ++(.4,.9);
\draw[thick] (-3,1) -- ++(.5,0)-- ++(0,-2) -- ++(-.5,0) -- (-3,-1);
\draw[thick] (-3,2) -- ++(1,0)-- ++(0,-4) -- ++(-1,0) -- (-3,-2);
\draw[thick] (3,1) -- ++(-.5,0)-- ++(0,-2) -- ++(.5,0) -- (3,-1);
\draw[thick] (3,2) -- ++(-1,0)-- ++(0,-4) -- ++(1,0) -- (3,-2);
\draw[thick] (3,2.5) -- ++(-1.5,0)-- ++(0,-5) -- ++(1.5,0) -- (3,-2.5);
\draw[thick] (-3,.5) -- (-3,-.5);
\draw[thick,\qbcolor] (-3,-2.5) --(-1.5,-2.5) to[out=0, in =180] (1.5,.5) -- (3,.5);
\draw[thick,\qcolor] (-3,2.5)-- (-1.5,2.5) to[out=0, in =180] (1.5,-.5) -- (3,-.5);
\end{tikzpicture}}

\newcommand{\TCP}{
\begin{tikzpicture}[scale=\setscaletikz]
\draw[thick] (-3,2.5) -- ++(-1,0);
\draw[thick] (-3,2) -- ++(-1,0);
\draw[thick] (-3,1) -- ++(-1,0);
\draw[thick] (-3,.5) -- ++(-1,0);
\draw[thick] (-3,-.5) -- ++(-1,0);
\draw[thick] (-3,-1) -- ++(-1,0);
\draw[thick] (-3,-2) -- ++(-1,0);
\draw[thick] (-3,-2.5) -- ++(-1,0);
\draw[thick] (3,2.5) -- ++(1,0);
\draw[thick] (3,2) -- ++(1,0);
\draw[thick] (3,1) -- ++(1,0);
\draw[thick] (3,.5) -- ++(1,0);
\draw[thick] (3,-.5) -- ++(1,0);
\draw[thick] (3,-1) -- ++(1,0);
\draw[thick] (3,-2) -- ++(1,0);
\draw[thick] (3,-2.5) -- ++(1,0);
\draw[fill=white] (-4,2.7) rectangle ++(-.4,-.9);
\draw[fill=white] (-4,1.2) rectangle ++(-.4,-.9);
\draw[fill=white] (4,2.7) rectangle ++(.4,-.9);
\draw[fill=white] (4,1.2) rectangle ++(.4,-.9);
\draw[fill=white] (-4,-2.7) rectangle ++(-.4,.9);
\draw[fill=white] (-4,-1.2) rectangle ++(-.4,.9);
\draw[fill=white] (4,-2.7) rectangle ++(.4,.9);
\draw[fill=white] (4,-1.2) rectangle ++(.4,.9);
\draw[thick] (-3,2) -- ++(1,0)-- ++(0,-4) -- ++(-1,0) -- (-3,-2);
\draw[thick] (-3,2.5) -- ++(1.5,0)-- ++(0,-5) -- ++(-1.5,0) -- (-3,-2.5);
\draw[thick] (3,2) -- ++(-1,0)-- ++(0,-4) -- ++(1,0) -- (3,-2);
\draw[thick] (3,2.5) -- ++(-1.5,0)-- ++(0,-5) -- ++(1.5,0) -- (3,-2.5);
\draw[thick,\qbcolor] (-3,1) -- (3,1);
\draw[thick,\qbcolor] (-3,.5) -- (3,.5);
\draw[thick,\qcolor] (-3,-1) -- (3,-1);
\draw[thick,\qcolor] (-3,-.5) -- (3,-.5);
\end{tikzpicture}}

\newcommand{\TCM}{
\begin{tikzpicture}[scale=\setscaletikz]
\draw[thick] (-3,2.5) -- ++(-1,0);
\draw[thick] (-3,2) -- ++(-1,0);
\draw[thick] (-3,1) -- ++(-1,0);
\draw[thick] (-3,.5) -- ++(-1,0);
\draw[thick] (-3,-.5) -- ++(-1,0);
\draw[thick] (-3,-1) -- ++(-1,0);
\draw[thick] (-3,-2) -- ++(-1,0);
\draw[thick] (-3,-2.5) -- ++(-1,0);
\draw[thick] (3,2.5) -- ++(1,0);
\draw[thick] (3,2) -- ++(1,0);
\draw[thick] (3,1) -- ++(1,0);
\draw[thick] (3,.5) -- ++(1,0);
\draw[thick] (3,-.5) -- ++(1,0);
\draw[thick] (3,-1) -- ++(1,0);
\draw[thick] (3,-2) -- ++(1,0);
\draw[thick] (3,-2.5) -- ++(1,0);
\draw[fill=white] (-4,2.7) rectangle ++(-.4,-.9);
\draw[fill=white] (-4,1.2) rectangle ++(-.4,-.9);
\draw[fill=white] (4,2.7) rectangle ++(.4,-.9);
\draw[fill=white] (4,1.2) rectangle ++(.4,-.9);
\draw[fill=white] (-4,-2.7) rectangle ++(-.4,.9);
\draw[fill=white] (-4,-1.2) rectangle ++(-.4,.9);
\draw[fill=white] (4,-2.7) rectangle ++(.4,.9);
\draw[fill=white] (4,-1.2) rectangle ++(.4,.9);
\draw[thick] (-3,1) -- ++(.5,0)-- ++(0,-2) -- ++(-.5,0) -- (-3,-1);
\draw[thick] (3,1) -- ++(-.5,0)-- ++(0,-2) -- ++(.5,0) -- (3,-1);
\draw[thick] (-3,.5) -- (-3,-.5);
\draw[thick] (3,.5) -- (3,-.5);
\draw[thick,\qcolor] (-3,2.5) -- (3,2.5);
\draw[thick,\qcolor] (-3,2) -- (3,2);
\draw[thick,\qbcolor] (-3,-2.5) -- (3,-2.5);
\draw[thick,\qbcolor] (-3,-2) -- (3,-2);
\end{tikzpicture}}

\newcommand{\TCMblob}{
\begin{tikzpicture}[scale=\setscaletikz]
\draw[thick] (-3,2.5) -- ++(-1,0);
\draw[thick] (-3,2) -- ++(-1,0);
\draw[thick] (-3,1) -- ++(-1,0);
\draw[thick] (-3,.5) -- ++(-1,0);
\draw[thick] (-3,-.5) -- ++(-1,0);
\draw[thick] (-3,-1) -- ++(-1,0);
\draw[thick] (-3,-2) -- ++(-1,0);
\draw[thick] (-3,-2.5) -- ++(-1,0);
\draw[thick] (3,2.5) -- ++(1,0);
\draw[thick] (3,2) -- ++(1,0);
\draw[thick] (3,1) -- ++(1,0);
\draw[thick] (3,.5) -- ++(1,0);
\draw[thick] (3,-.5) -- ++(1,0);
\draw[thick] (3,-1) -- ++(1,0);
\draw[thick] (3,-2) -- ++(1,0);
\draw[thick] (3,-2.5) -- ++(1,0);
\draw[fill=white] (-4,2.7) rectangle ++(-.4,-.9);
\draw[fill=white] (-4,1.2) rectangle ++(-.4,-.9);
\draw[fill=white] (4,2.7) rectangle ++(.4,-.9);
\draw[fill=white] (4,1.2) rectangle ++(.4,-.9);
\draw[fill=white] (-4,-2.7) rectangle ++(-.4,.9);
\draw[fill=white] (-4,-1.2) rectangle ++(-.4,.9);
\draw[fill=white] (4,-2.7) rectangle ++(.4,.9);
\draw[fill=white] (4,-1.2) rectangle ++(.4,.9);
\draw[thick] (-3,1) -- ++(.5,0)-- ++(0,-2) -- ++(-.5,0) -- (-3,-1);
\draw[thick] (3,1) -- ++(-.5,0)-- ++(0,-2) -- ++(.5,0) -- (3,-1);
\draw[thick] (-3,.5) -- (-3,-.5);
\draw[thick] (3,.5) -- (3,-.5);
\draw[thick,\qcolor] (-3,2.5) -- (3,2.5);
\draw[thick,\qcolor] (-3,2) -- (3,2);
\draw[thick,\qbcolor] (-3,-2.5) -- (3,-2.5);
\draw[thick,\qbcolor] (-3,-2) -- (3,-2);
\draw[fill=white] (0,0) ellipse (1.3cm and 3cm);
\node at (0,0) {\scriptsize $27$};
\end{tikzpicture}}

\newcommand{\TDP}{
\begin{tikzpicture}[scale=\setscaletikz]
\draw[thick] (-3,2.5) -- ++(-1,0);
\draw[thick] (-3,2) -- ++(-1,0);
\draw[thick] (-3,1) -- ++(-1,0);
\draw[thick] (-3,.5) -- ++(-1,0);
\draw[thick] (-3,-.5) -- ++(-1,0);
\draw[thick] (-3,-1) -- ++(-1,0);
\draw[thick] (-3,-2) -- ++(-1,0);
\draw[thick] (-3,-2.5) -- ++(-1,0);
\draw[thick] (3,2.5) -- ++(1,0);
\draw[thick] (3,2) -- ++(1,0);
\draw[thick] (3,1) -- ++(1,0);
\draw[thick] (3,.5) -- ++(1,0);
\draw[thick] (3,-.5) -- ++(1,0);
\draw[thick] (3,-1) -- ++(1,0);
\draw[thick] (3,-2) -- ++(1,0);
\draw[thick] (3,-2.5) -- ++(1,0);
\draw[fill=white] (-4,2.7) rectangle ++(-.4,-.9);
\draw[fill=white] (-4,1.2) rectangle ++(-.4,-.9);
\draw[fill=white] (4,2.7) rectangle ++(.4,-.9);
\draw[fill=white] (4,1.2) rectangle ++(.4,-.9);
\draw[fill=white] (-4,-2.7) rectangle ++(-.4,.9);
\draw[fill=white] (-4,-1.2) rectangle ++(-.4,.9);
\draw[fill=white] (4,-2.7) rectangle ++(.4,.9);
\draw[fill=white] (4,-1.2) rectangle ++(.4,.9);
\draw[thick] (-3,2) -- ++(1,0)-- ++(0,-4) -- ++(-1,0) -- (-3,-2);
\draw[thick] (-3,2.5) -- ++(1.5,0)-- ++(0,-5) -- ++(-1.5,0) -- (-3,-2.5);
\draw[thick] (3,1) -- ++(-.5,0)-- ++(0,-2) -- ++(.5,0) -- (3,-1);
\draw[thick] (3,.5) -- (3,-.5);
\draw[thick,\qbcolor] (-3,1) -- (-1.5,1) to[out=0,in=180] (1.5,-2) -- (3,-2);
\draw[thick,\qbcolor] (-3,.5) -- (-1.5,.5) to[out=0,in=180] (1.5,-2.5) -- (3,-2.5);
\draw[thick,\qcolor] (-3,-1) -- (-1.5,-1) to[out=0,in=180] (1.5,2) -- (3,2);
\draw[thick,\qcolor] (-3,-.5) -- (-1.5,-.5) to[out=0,in=180] (1.5,2.5) -- (3,2.5);
\end{tikzpicture}}

\newcommand{\TDPblob}{
\begin{tikzpicture}[scale=\setscaletikz]
\draw[thick] (-3,2.5) -- ++(-1,0);
\draw[thick] (-3,2) -- ++(-1,0);
\draw[thick] (-3,1) -- ++(-1,0);
\draw[thick] (-3,.5) -- ++(-1,0);
\draw[thick] (-3,-.5) -- ++(-1,0);
\draw[thick] (-3,-1) -- ++(-1,0);
\draw[thick] (-3,-2) -- ++(-1,0);
\draw[thick] (-3,-2.5) -- ++(-1,0);
\draw[thick] (-3,2) -- ++(1,0)-- ++(0,-4) -- ++(-1,0) -- (-3,-2);
\draw[thick] (-3,2.5) -- ++(1.5,0)-- ++(0,-5) -- ++(-1.5,0) -- (-3,-2.5);
\draw[fill=white] (-4,2.7) rectangle ++(-.4,-.9);
\draw[fill=white] (-4,1.2) rectangle ++(-.4,-.9);
\draw[fill=white] (-4,-2.7) rectangle ++(-.4,.9);
\draw[fill=white] (-4,-1.2) rectangle ++(-.4,.9);
\begin{scope}[xshift=2cm]
\draw[thick] (3,2.5) -- ++(1,0);
\draw[thick] (3,2) -- ++(1,0);
\draw[thick] (3,1) -- ++(1,0);
\draw[thick] (3,.5) -- ++(1,0);
\draw[thick] (3,-.5) -- ++(1,0);
\draw[thick] (3,-1) -- ++(1,0);
\draw[thick] (3,-2) -- ++(1,0);
\draw[thick] (3,-2.5) -- ++(1,0);
\draw[thick] (3,1) -- ++(-.5,0)-- ++(0,-2) -- ++(.5,0) -- (3,-1);
\draw[thick] (3,.5) -- (3,-.5);
\draw[fill=white] (4,2.7) rectangle ++(.4,-.9);
\draw[fill=white] (4,1.2) rectangle ++(.4,-.9);
\draw[fill=white] (4,-2.7) rectangle ++(.4,.9);
\draw[fill=white] (4,-1.2) rectangle ++(.4,.9);
\end{scope}
\draw[thick,\qbcolor] (-3,1) -- (-1.5,1) to[out=0,in=180] (1.5,-2) -- (5,-2);
\draw[thick,\qbcolor] (-3,.5) -- (-1.5,.5) to[out=0,in=180] (1.5,-2.5) -- (5,-2.5);
\draw[thick,\qcolor] (-3,-1) -- (-1.5,-1) to[out=0,in=180] (1.5,2) -- (5,2);
\draw[thick,\qcolor] (-3,-.5) -- (-1.5,-.5) to[out=0,in=180] (1.5,2.5) -- (5,2.5);
\draw[fill=white] (2.5,0) ellipse (1.3cm and 3cm);
\node at (2.5,0) {\scriptsize $27$};
\end{tikzpicture}}

\newcommand{\TDM}{
\begin{tikzpicture}[scale=\setscaletikz]
\draw[thick] (-3,2.5) -- ++(-1,0);
\draw[thick] (-3,2) -- ++(-1,0);
\draw[thick] (-3,1) -- ++(-1,0);
\draw[thick] (-3,.5) -- ++(-1,0);
\draw[thick] (-3,-.5) -- ++(-1,0);
\draw[thick] (-3,-1) -- ++(-1,0);
\draw[thick] (-3,-2) -- ++(-1,0);
\draw[thick] (-3,-2.5) -- ++(-1,0);
\draw[thick] (3,2.5) -- ++(1,0);
\draw[thick] (3,2) -- ++(1,0);
\draw[thick] (3,1) -- ++(1,0);
\draw[thick] (3,.5) -- ++(1,0);
\draw[thick] (3,-.5) -- ++(1,0);
\draw[thick] (3,-1) -- ++(1,0);
\draw[thick] (3,-2) -- ++(1,0);
\draw[thick] (3,-2.5) -- ++(1,0);
\draw[fill=white] (-4,2.7) rectangle ++(-.4,-.9);
\draw[fill=white] (-4,1.2) rectangle ++(-.4,-.9);
\draw[fill=white] (4,2.7) rectangle ++(.4,-.9);
\draw[fill=white] (4,1.2) rectangle ++(.4,-.9);
\draw[fill=white] (-4,-2.7) rectangle ++(-.4,.9);
\draw[fill=white] (-4,-1.2) rectangle ++(-.4,.9);
\draw[fill=white] (4,-2.7) rectangle ++(.4,.9);
\draw[fill=white] (4,-1.2) rectangle ++(.4,.9);
\draw[thick] (-3,1) -- ++(.5,0)-- ++(0,-2) -- ++(-.5,0) -- (-3,-1);
\draw[thick] (3,2) -- ++(-1,0)-- ++(0,-4) -- ++(1,0) -- (3,-2);
\draw[thick] (3,2.5) -- ++(-1.5,0)-- ++(0,-5) -- ++(1.5,0) -- (3,-2.5);
\draw[thick] (-3,.5) -- (-3,-.5);
\draw[thick,\qbcolor] (-3,-2) -- (-1.5,-2) to[out=0,in=180] (1.5,1) -- (3,1);
\draw[thick,\qbcolor] (-3,-2.5) -- (-1.5,-2.5) to[out=0,in=180] (1.5,.5) -- (3,.5);
\draw[thick,\qcolor] (-3,2.5) -- (-1.5,2.5) to[out=0,in=180] (1.5,-.5) -- (3,-.5);
\draw[thick,\qcolor] (-3,2) -- (-1.5,2) to[out=0,in=180] (1.5,-1) -- (3,-1);
\end{tikzpicture}}

\newcommand{\TEP}{
\begin{tikzpicture}[scale=\setscaletikz]
\draw[thick] (-3,2.5) -- ++(-1,0);
\draw[thick] (-3,2) -- ++(-1,0);
\draw[thick] (-3,1) -- ++(-1,0);
\draw[thick] (-3,.5) -- ++(-1,0);
\draw[thick] (-3,-.5) -- ++(-1,0);
\draw[thick] (-3,-1) -- ++(-1,0);
\draw[thick] (-3,-2) -- ++(-1,0);
\draw[thick] (-3,-2.5) -- ++(-1,0);
\draw[thick] (3,2.5) -- ++(1,0);
\draw[thick] (3,2) -- ++(1,0);
\draw[thick] (3,1) -- ++(1,0);
\draw[thick] (3,.5) -- ++(1,0);
\draw[thick] (3,-.5) -- ++(1,0);
\draw[thick] (3,-1) -- ++(1,0);
\draw[thick] (3,-2) -- ++(1,0);
\draw[thick] (3,-2.5) -- ++(1,0);
\draw[fill=white] (-4,2.7) rectangle ++(-.4,-.9);
\draw[fill=white] (-4,1.2) rectangle ++(-.4,-.9);
\draw[fill=white] (4,2.7) rectangle ++(.4,-.9);
\draw[fill=white] (4,1.2) rectangle ++(.4,-.9);
\draw[fill=white] (-4,-2.7) rectangle ++(-.4,.9);
\draw[fill=white] (-4,-1.2) rectangle ++(-.4,.9);
\draw[fill=white] (4,-2.7) rectangle ++(.4,.9);
\draw[fill=white] (4,-1.2) rectangle ++(.4,.9);
\draw[thick] (-3,2) -- ++(1,0)-- ++(0,-4) -- ++(-1,0) -- (-3,-2);
\draw[thick] (-3,2.5) -- ++(1.5,0)-- ++(0,-5) -- ++(-1.5,0) -- (-3,-2.5);
\draw[thick] (3,2.5) -- ++(-1.5,0)-- ++(0,-5) -- ++(1.5,0) -- (3,-2.5);
\draw[thick] (3,.5) -- (3,-.5);
\draw[thick,\qbcolor] (-3,1) -- ++(1.5,0) to[out=0,in=180] (1.5,1) -- ++(1.5,0);
\draw[thick,\qbcolor] (-3,.5) -- ++(1.5,0) to[out=0,in=180] (1.5,-2) -- ++(1.5,0);
\draw[thick,\qcolor] (-3,-.5) -- ++(1.5,0) to[out=0,in=180] (1.5,2) -- ++(1.5,0);
\draw[thick,\qcolor] (-3,-1) -- ++(1.5,0) to[out=0,in=180] (1.5,-1) -- ++(1.5,0);
\end{tikzpicture}}

\newcommand{\TEM}{
\begin{tikzpicture}[scale=\setscaletikz]
\draw[thick] (-3,2.5) -- ++(-1,0);
\draw[thick] (-3,2) -- ++(-1,0);
\draw[thick] (-3,1) -- ++(-1,0);
\draw[thick] (-3,.5) -- ++(-1,0);
\draw[thick] (-3,-.5) -- ++(-1,0);
\draw[thick] (-3,-1) -- ++(-1,0);
\draw[thick] (-3,-2) -- ++(-1,0);
\draw[thick] (-3,-2.5) -- ++(-1,0);
\draw[thick] (3,2.5) -- ++(1,0);
\draw[thick] (3,2) -- ++(1,0);
\draw[thick] (3,1) -- ++(1,0);
\draw[thick] (3,.5) -- ++(1,0);
\draw[thick] (3,-.5) -- ++(1,0);
\draw[thick] (3,-1) -- ++(1,0);
\draw[thick] (3,-2) -- ++(1,0);
\draw[thick] (3,-2.5) -- ++(1,0);
\draw[fill=white] (-4,2.7) rectangle ++(-.4,-.9);
\draw[fill=white] (-4,1.2) rectangle ++(-.4,-.9);
\draw[fill=white] (4,2.7) rectangle ++(.4,-.9);
\draw[fill=white] (4,1.2) rectangle ++(.4,-.9);
\draw[fill=white] (-4,-2.7) rectangle ++(-.4,.9);
\draw[fill=white] (-4,-1.2) rectangle ++(-.4,.9);
\draw[fill=white] (4,-2.7) rectangle ++(.4,.9);
\draw[fill=white] (4,-1.2) rectangle ++(.4,.9);
\draw[thick] (-3,1) -- ++(.5,0)-- ++(0,-2) -- ++(-.5,0) -- (-3,-1);
\draw[thick] (3,2.5) -- ++(-1.5,0)-- ++(0,-5) -- ++(1.5,0) -- (3,-2.5);
\draw[thick] (-3,.5) -- (-3,-.5);
\draw[thick] (3,.5) -- (3,-.5);
\draw[thick,\qbcolor] (-3,-2) -- ++(1.5,0) to[out=0,in=180] (1.5,1) -- ++(1.5,0);
\draw[thick,\qbcolor] (-3,-2.5) -- ++(1.5,0) to[out=0,in=180] (1.5,-2) -- ++(1.5,0);
\draw[thick,\qcolor] (-3,2.5) -- ++(1.5,0) to[out=0,in=180] (1.5,2) -- ++(1.5,0);
\draw[thick,\qcolor] (-3,2) -- ++(1.5,0) to[out=0,in=180] (1.5,-1) -- ++(1.5,0);
\end{tikzpicture}}

\newcommand{\TFP}{
\begin{tikzpicture}[scale=\setscaletikz]
\draw[thick] (-3,2.5) -- ++(-1,0);
\draw[thick] (-3,2) -- ++(-1,0);
\draw[thick] (-3,1) -- ++(-1,0);
\draw[thick] (-3,.5) -- ++(-1,0);
\draw[thick] (-3,-.5) -- ++(-1,0);
\draw[thick] (-3,-1) -- ++(-1,0);
\draw[thick] (-3,-2) -- ++(-1,0);
\draw[thick] (-3,-2.5) -- ++(-1,0);
\draw[thick] (3,2.5) -- ++(1,0);
\draw[thick] (3,2) -- ++(1,0);
\draw[thick] (3,1) -- ++(1,0);
\draw[thick] (3,.5) -- ++(1,0);
\draw[thick] (3,-.5) -- ++(1,0);
\draw[thick] (3,-1) -- ++(1,0);
\draw[thick] (3,-2) -- ++(1,0);
\draw[thick] (3,-2.5) -- ++(1,0);
\draw[fill=white] (-4,2.7) rectangle ++(-.4,-.9);
\draw[fill=white] (-4,1.2) rectangle ++(-.4,-.9);
\draw[fill=white] (4,2.7) rectangle ++(.4,-.9);
\draw[fill=white] (4,1.2) rectangle ++(.4,-.9);
\draw[fill=white] (-4,-2.7) rectangle ++(-.4,.9);
\draw[fill=white] (-4,-1.2) rectangle ++(-.4,.9);
\draw[fill=white] (4,-2.7) rectangle ++(.4,.9);
\draw[fill=white] (4,-1.2) rectangle ++(.4,.9);
\draw[thick] (-3,2.5) -- ++(1.5,0)-- ++(0,-5) -- ++(-1.5,0) -- (-3,-2.5);
\draw[thick] (3,2) -- ++(-1,0)-- ++(0,-4) -- ++(1,0) -- (3,-2);
\draw[thick] (3,2.5) -- ++(-1.5,0)-- ++(0,-5) -- ++(1.5,0) -- (3,-2.5);
\draw[thick] (-3,.5) -- (-3,-.5);
\draw[thick,\qbcolor] (-3,1) -- ++(1.5,0) to[out=0,in=180] (1.5,1) -- ++(1.5,0);
\draw[thick,\qbcolor] (-3,-2) -- ++(1.5,0) to[out=0,in=180] (1.5,.5) -- ++(1.5,0);
\draw[thick,\qcolor] (-3,2) -- ++(1.5,0) to[out=0,in=180] (1.5,-.5) -- ++(1.5,0);
\draw[thick,\qcolor] (-3,-1) -- ++(1.5,0) to[out=0,in=180] (1.5,-1) -- ++(1.5,0);
\end{tikzpicture}}

\newcommand{\TFPscaled}[3]{\tikeq{\bs[scale=\setscaletikz]
\bs[xshift=-2cm]
\draw[thick] (-1,2.5) -- ++(-3,0);
\draw[thick] (-1,2) -- ++(-3,0);
\draw[thick] (-1,1) -- ++(-3,0);
\draw[thick] (-1.5,.5) -- ++(-2.5,0);
\draw[thick] (-1.5,-.5) -- ++(-2.5,0);
\draw[thick] (-1,-1) -- ++(-3,0);
\draw[thick] (-1,-2) -- ++(-3,0);
\draw[thick] (-1,-2.5) -- ++(-3,0);
\draw[thick] (-1,2.5) -- (-1,-2.5);
\draw[thick] (-1.5,.5) -- (-1.5,-.5);
\draw[fill=white] (-4,2.7) rectangle ++(-.4,-.9);
\draw[fill=white] (-4,1.2) rectangle ++(-.4,-.9);
\draw[fill=white] (-4,-2.7) rectangle ++(-.4,.9);
\draw[fill=white] (-4,-1.2) rectangle ++(-.4,.9);
\es
\bs[xshift=2cm]
\draw[thick] (1,2.5) -- ++(3,0);
\draw[thick] (1.5,2) -- ++(2.5,0);
\draw[thick] (1,1) -- ++(3,0);
\draw[thick] (1,.5) -- ++(3,0);
\draw[thick] (1,-.5) -- ++(3,0);
\draw[thick] (1,-1) -- ++(3,0);
\draw[thick] (1.5,-2) -- ++(2.5,0);
\draw[thick] (1,-2.5) -- ++(3,0);
\draw[thick] (1.5,2) -- (1.5,-2);
\draw[thick] (1,2.5) -- (1,-2.5);
\draw[fill=white] (4,2.7) rectangle ++(.4,-.9);
\draw[fill=white] (4,1.2) rectangle ++(.4,-.9);
\draw[fill=white] (4,-2.7) rectangle ++(.4,.9);
\draw[fill=white] (4,-1.2) rectangle ++(.4,.9);
\es
\draw[thick,\qbcolor] (-3,1) -- ++(1.5,0) to[out=0,in=180] (1.5,1) -- ++(1.5,0);
\draw[thick,\qcolor] (-3,-1) -- ++(1.5,0) to[out=0,in=180] (1.5,-1) -- ++(1.5,0);
\draw[thick,\qbcolor] (-3,-2) -- ++(3,0) to[out=0,in=180] (3,.5) -- ++(0,0);
\draw[thick,\qcolor] (-3,2) -- ++(3,0) to[out=0,in=180] (3,-.5) -- ++(0,0);
\es
\draw[glu] (#1,-.6) .. controls +(0,-0.5) and +(-0.25,0) .. (#3,-1.3) 
.. controls +(0.25,0) and +(0,-0.5) .. (#2,-.6); 
\draw[fill=white!70!black] (#1,0) ellipse (.1 and 0.8);
\draw[fill=white!70!black] (#2,0) ellipse (.1 and 0.8);
}}

\newcommand{\TFM}{
\begin{tikzpicture}[scale=\setscaletikz]
\draw[thick] (-3,2.5) -- ++(-1,0);
\draw[thick] (-3,2) -- ++(-1,0);
\draw[thick] (-3,1) -- ++(-1,0);
\draw[thick] (-3,.5) -- ++(-1,0);
\draw[thick] (-3,-.5) -- ++(-1,0);
\draw[thick] (-3,-1) -- ++(-1,0);
\draw[thick] (-3,-2) -- ++(-1,0);
\draw[thick] (-3,-2.5) -- ++(-1,0);
\draw[thick] (3,2.5) -- ++(1,0);
\draw[thick] (3,2) -- ++(1,0);
\draw[thick] (3,1) -- ++(1,0);
\draw[thick] (3,.5) -- ++(1,0);
\draw[thick] (3,-.5) -- ++(1,0);
\draw[thick] (3,-1) -- ++(1,0);
\draw[thick] (3,-2) -- ++(1,0);
\draw[thick] (3,-2.5) -- ++(1,0);
\draw[fill=white] (-4,2.7) rectangle ++(-.4,-.9);
\draw[fill=white] (-4,1.2) rectangle ++(-.4,-.9);
\draw[fill=white] (4,2.7) rectangle ++(.4,-.9);
\draw[fill=white] (4,1.2) rectangle ++(.4,-.9);
\draw[fill=white] (-4,-2.7) rectangle ++(-.4,.9);
\draw[fill=white] (-4,-1.2) rectangle ++(-.4,.9);
\draw[fill=white] (4,-2.7) rectangle ++(.4,.9);
\draw[fill=white] (4,-1.2) rectangle ++(.4,.9);
\draw[thick] (-3,.5) -- (-3,-.5);
\draw[thick] (-3,2.5) -- ++(1.5,0)-- ++(0,-5) -- ++(-1.5,0) -- (-3,-2.5);
\draw[thick] (3,.5) -- (3,-.5);
\draw[thick] (3,1) -- ++(-.5,0)-- ++(0,-2) -- ++(.5,0) -- (3,-1);
\draw[thick,\qbcolor] (-3,1) -- ++(1.5,0) to[out=0,in=180] (1.5,-2) -- ++(1.5,0);
\draw[thick,\qbcolor] (-3,-2) -- ++(1.5,0) to[out=0,in=180] (1.5,-2.5) -- ++(1.5,0);
\draw[thick,\qcolor] (-3,2) -- ++(1.5,0) to[out=0,in=180] (1.5,2.5) -- ++(1.5,0);
\draw[thick,\qcolor] (-3,-1) -- ++(1.5,0) to[out=0,in=180] (1.5,2) -- ++(1.5,0);
\end{tikzpicture}}

\newcommand{\TFMblob}{
\begin{tikzpicture}[scale=\setscaletikz]
\draw[thick] (-3,2.5) -- ++(-1,0);
\draw[thick] (-3,2) -- ++(-1,0);
\draw[thick] (-3,1) -- ++(-1,0);
\draw[thick] (-3,.5) -- ++(-1,0);
\draw[thick] (-3,-.5) -- ++(-1,0);
\draw[thick] (-3,-1) -- ++(-1,0);
\draw[thick] (-3,-2) -- ++(-1,0);
\draw[thick] (-3,-2.5) -- ++(-1,0);
\draw[thick] (3,2.5) -- ++(1,0);
\begin{scope}[xshift=2cm]
\draw[thick] (3,2) -- ++(1,0);
\draw[thick] (3,1) -- ++(1,0);
\draw[thick] (3,.5) -- ++(1,0);
\draw[thick] (3,-.5) -- ++(1,0);
\draw[thick] (3,-1) -- ++(1,0);
\draw[thick] (3,-2) -- ++(1,0);
\draw[thick] (3,-2.5) -- ++(1,0);
\draw[thick] (3,.5) -- (3,-.5);
\draw[thick] (3,1) -- ++(-.5,0)-- ++(0,-2) -- ++(.5,0) -- (3,-1);
\end{scope}
\draw[fill=white] (-4,2.7) rectangle ++(-.4,-.9);
\draw[fill=white] (-4,1.2) rectangle ++(-.4,-.9);
\draw[fill=white] (6,2.7) rectangle ++(.4,-.9);
\draw[fill=white] (6,1.2) rectangle ++(.4,-.9);
\draw[fill=white] (-4,-2.7) rectangle ++(-.4,.9);
\draw[fill=white] (-4,-1.2) rectangle ++(-.4,.9);
\draw[fill=white] (6,-2.7) rectangle ++(.4,.9);
\draw[fill=white] (6,-1.2) rectangle ++(.4,.9);
\draw[thick] (-3,.5) -- (-3,-.5);
\draw[thick] (-3,2.5) -- ++(1.5,0)-- ++(0,-5) -- ++(-1.5,0) -- (-3,-2.5);
\draw[thick,\qbcolor] (-3,1) -- ++(1.5,0) to[out=0,in=180] (1.5,-2) -- ++(4.5,0);
\draw[thick,\qbcolor] (-3,-2) -- ++(1.5,0) to[out=0,in=180] (1.5,-2.5) -- ++(4.5,0);
\draw[thick,\qcolor] (-3,2) -- ++(1.5,0) to[out=0,in=180] (1.5,2.5) -- ++(4.5,0);
\draw[thick,\qcolor] (-3,-1) -- ++(1.5,0) to[out=0,in=180] (1.5,2) -- ++(4.5,0);
\draw[fill=white] (2.5,0) ellipse (1.3cm and 3cm);
\node at (2.5,0) {\scriptsize $27$};
\end{tikzpicture}}

\newcommand{\VextexOctetONE}{
\begin{tikzpicture}[scale=\setscaletikz]
\draw[thick] (-3,2.5) -- ++(-1,0);
\draw[thick] (-3,2) -- ++(-1,0);
\draw[thick] (-3,1) -- ++(-1,0);
\draw[thick] (-3,.5) -- ++(-1,0);
\draw[thick] (-3,-.5) -- ++(-1,0);
\draw[thick] (-3,-1) -- ++(-1,0);
\draw[thick] (-3,-2) -- ++(-1,0);
\draw[thick] (-3,-2.5) -- ++(-1,0);
\draw[thick] (-3,2) -- ++(1,0)-- ++(0,-4) -- ++(-1,0) -- (-3,-2);
\draw[thick] (-3,2.5) -- ++(1.5,0)-- ++(0,-5) -- ++(-1.5,0) -- (-3,-2.5);
\draw[fill=white] (-4,2.7) rectangle ++(-.4,-.9);
\draw[fill=white] (-4,1.2) rectangle ++(-.4,-.9);
\draw[fill=white] (-4,-2.7) rectangle ++(-.4,.9);
\draw[fill=white] (-4,-1.2) rectangle ++(-.4,.9);
\draw[thick,\qbcolor] (-3,.5) -- (-1.5,.5) to[out=0,in=180] (1.5,-2.5) -- (4,-2.5);
\draw[thick] (-3,1) -- ++(0.5,0) --++(0,-2)  -- (-3,-1);
\draw[thick,\qcolor] (-3,-.5) -- (-1.5,-.5) to[out=0,in=180] (1.5,2.5) -- (4,2.5);
\draw[fill=white] (2.5,0) ellipse (1.3cm and 3cm);
\node at (2.5,0) {\scriptsize $\Irrep{8}{}$};
\end{tikzpicture}
}

\newcommand{\VextexOctetTWO}{
\begin{tikzpicture}[scale=\setscaletikz]
\draw[thick] (-3,2.5) -- ++(-1,0);
\draw[thick] (-3,2) -- ++(-1,0);
\draw[thick] (-3,1) -- ++(-1,0);
\draw[thick] (-3,.5) -- ++(-1,0);
\draw[thick] (-3,-.5) -- ++(-1,0);
\draw[thick] (-3,-1) -- ++(-1,0);
\draw[thick] (-3,-2) -- ++(-1,0);
\draw[thick] (-3,-2.5) -- ++(-1,0);
\draw[thick] (-3,2) -- ++(1,0)-- ++(0,-4) -- ++(-1,0) -- (-3,-2);
\draw[fill=white] (-4,2.7) rectangle ++(-.4,-.9);
\draw[fill=white] (-4,1.2) rectangle ++(-.4,-.9);
\draw[fill=white] (-4,-2.7) rectangle ++(-.4,.9);
\draw[fill=white] (-4,-1.2) rectangle ++(-.4,.9);
\draw[thick,\qbcolor] (-3,-2.5) -- ++(1.5,0) to[out=0,in=180] (1.5,-2) -- (4,-2);
\draw[thick,\qcolor] (-3,2.5) -- ++(1.5,0) to[out=0,in=180] (1.5,2) -- (4,2);
\draw[thick] (-3,1) -- ++(0.5,0) --++(0,-2)  -- (-3,-1);
\draw[fill=white] (2.5,0) ellipse (1.3cm and 3cm);
\draw[thick] (-3,.5) -- (-3,-.5);
\node at (2.5,0) {\scriptsize $\Irrep{8}{}$};
\end{tikzpicture}
}

\newcommand{\VextexZerobONE}{
\begin{tikzpicture}[scale=\setscaletikz]
\draw[thick] (-3,2.5) -- ++(-1,0);
\draw[thick] (-3,2) -- ++(-1,0);
\draw[thick] (-3,1) -- ++(-1,0);
\draw[thick] (-3,.5) -- ++(-1,0);
\draw[thick] (-3,-.5) -- ++(-1,0);
\draw[thick] (-3,-1) -- ++(-1,0);
\draw[thick] (-3,-2) -- ++(-1,0);
\draw[thick] (-3,-2.5) -- ++(-1,0);
\draw[thick] (3,2.5) -- ++(1,0);
\draw[fill=white] (-4,2.7) rectangle ++(-.4,-.9);
\draw[fill=white] (-4,1.2) rectangle ++(-.4,-.9);
\draw[fill=white] (-4,-2.7) rectangle ++(-.4,.9);
\draw[fill=white] (-4,-1.2) rectangle ++(-.4,.9);
\draw[thick] (-3,2.5) -- ++(1,0)-- ++(0,-5) -- ++(-1,0) -- (-3,-2.5);
\draw[thick,\qbcolor] (-3,1) -- ++(1.5,0) to[out=0,in=180] (1.5,-1.5) -- ++(2.5,0);
\draw[thick,\qbcolor] (-3,.5) -- ++(1.5,0) to[out=0,in=180] (1.5,-2.0) -- ++(2.5,0);
\draw[thick,\qbcolor] (-3,-2) -- ++(1.5,0) to[out=0,in=180] (1.5,-2.5) -- ++(2.5,0);
\draw[thick,\qcolor] (-3,2) -- ++(1.5,0) to[out=0,in=180] (1.5,2.5) -- ++(2.5,0);
\draw[thick,\qcolor] (-3,-.5) -- ++(1.5,0) to[out=0,in=180] (1.5,2) -- ++(2.5,0);
\draw[thick,\qcolor] (-3,-1) -- ++(1.5,0) to[out=0,in=180] (1.5,1.5) -- ++(2.5,0);
\draw[fill=white] (2.5,0) ellipse (1.3cm and 3cm);
\node at (2.5,0) {\scriptsize $\Irrep{0}{b}$};
\end{tikzpicture}
}

\newcommand{\VextexZerobTWO}{
\begin{tikzpicture}[scale=\setscaletikz]
\draw[thick] (-3,2.5) -- ++(-1,0);
\draw[thick] (-3,2) -- ++(-1,0);
\draw[thick] (-3,1) -- ++(-1,0);
\draw[thick] (-3,.5) -- ++(-1,0);
\draw[thick] (-3,-.5) -- ++(-1,0);
\draw[thick] (-3,-1) -- ++(-1,0);
\draw[thick] (-3,-2) -- ++(-1,0);
\draw[thick] (-3,-2.5) -- ++(-1,0);
\draw[thick] (3,2.5) -- ++(1,0);
\draw[fill=white] (-4,2.7) rectangle ++(-.4,-.9);
\draw[fill=white] (-4,1.2) rectangle ++(-.4,-.9);
\draw[fill=white] (-4,-2.7) rectangle ++(-.4,.9);
\draw[fill=white] (-4,-1.2) rectangle ++(-.4,.9);
\draw[thick] (-3,.5) -- (-3,-.5);
\draw[thick,\qbcolor] (-3,1) -- ++(1.5,0) to[out=0,in=180] (1.5,-2.5) -- ++(2.5,0);
\draw[thick,\qbcolor] (-3,-2) -- ++(1.5,0) to[out=0,in=180] (1.5,-1.5) -- ++(2.5,0);
\draw[thick,\qbcolor] (-3,-2.5) -- ++(1.5,0) to[out=0,in=180] (1.5,-2.0) -- ++(2.5,0);
\draw[thick,\qcolor] (-3,2.5) -- ++(1.5,0) to[out=0,in=180] (1.5,2.0) -- ++(2.5,0);
\draw[thick,\qcolor] (-3,2) -- ++(1.5,0) to[out=0,in=180] (1.5,1.5) -- ++(2.5,0);
\draw[thick,\qcolor] (-3,-1) -- ++(1.5,0) to[out=0,in=180] (1.5,2.5) -- ++(2.5,0);
\draw[fill=white] (2.5,0) ellipse (1.3cm and 3cm);
\node at (2.5,0) {\scriptsize $\Irrep{0}{b}$};
\end{tikzpicture}
}

\newcommand{\VoneDagVone}{
\begin{tikzpicture}[scale=\setscaletikz]
\draw[thick] (-3,2.5) -- ++(-1,0);
\draw[thick] (-3,2) -- ++(-1,0);
\draw[thick] (-3,1) -- ++(-1,0);
\draw[thick] (-3,.5) -- ++(-1,0);
\draw[thick] (-3,-.5) -- ++(-1,0);
\draw[thick] (-3,-1) -- ++(-1,0);
\draw[thick] (-3,-2) -- ++(-1,0);
\draw[thick] (-3,-2.5) -- ++(-1,0);
\draw[thick] (8,2.5) -- ++(1,0);
\draw[thick] (8,2) -- ++(1,0);
\draw[thick] (8,1) -- ++(1,0);
\draw[thick] (8,.5) -- ++(1,0);
\draw[thick] (8,-.5) -- ++(1,0);
\draw[thick] (8,-1) -- ++(1,0);
\draw[thick] (8,-2) -- ++(1,0);
\draw[thick] (8,-2.5) -- ++(1,0);
\draw[fill=white] (-4,2.7) rectangle ++(-.4,-.9);
\draw[fill=white] (-4,1.2) rectangle ++(-.4,-.9);
\draw[fill=white] (9,2.7) rectangle ++(.4,-.9);
\draw[fill=white] (9,1.2) rectangle ++(.4,-.9);
\draw[fill=white] (-4,-2.7) rectangle ++(-.4,.9);
\draw[fill=white] (-4,-1.2) rectangle ++(-.4,.9);
\draw[fill=white] (9,-2.7) rectangle ++(.4,.9);
\draw[fill=white] (9,-1.2) rectangle ++(.4,.9);
\draw[thick] (-3,2.5) -- ++(1,0)-- ++(0,-5) -- ++(-1,0) -- (-3,-2.5);
\draw[thick] (8,2.5) -- ++(-1,0)-- ++(0,-5) -- ++(1,0) -- (8,-2.5);
\draw[thick,\qbcolor] (-3,1) -- ++(1.5,0) to[out=0,in=180] (1.5,-1.5) -- ++(2,0) to[out=0,in=180] (6.5,1) -- ++(1.5,0);
\draw[thick,\qbcolor] (-3,.5) -- ++(1.5,0) to[out=0,in=180] (1.5,-2.0) -- ++(2,0) to[out=0,in=180] (6.5,.5) -- ++(1.5,0);
\draw[thick,\qbcolor] (-3,-2) -- ++(1.5,0) to[out=0,in=180] (1.5,-2.5) -- ++(2.5,0);
\draw[thick,\qbcolor] (-3,-2) -- ++(1.5,0) to[out=0,in=180] (1.5,-2.5) -- ++(2,0) to[out=0,in=180] (6.5,-2) -- ++(1.5,0);
\draw[thick,\qcolor] (-3,2) -- ++(1.5,0) to[out=0,in=180] (1.5,2.5) -- ++(2,0) to[out=0,in=180] (6.5,2) -- ++(1.5,0);
\draw[thick,\qcolor] (-3,-.5) -- ++(1.5,0) to[out=0,in=180] (1.5,2) -- ++(2,0) to[out=0,in=180] (6.5,-.5) -- ++(1.5,0);
\draw[thick,\qcolor] (-3,-1) -- ++(1.5,0) to[out=0,in=180] (1.5,1.5) -- ++(2,0) to[out=0,in=180] (6.5,-1) -- ++(1.5,0);
\draw[fill=white] (2.5,0) ellipse (1.3cm and 3cm);
\node at (2.5,0) {\scriptsize $\Irrep{0}{b}$};
\end{tikzpicture}
}

\newcommand{\PZerob}{
\begin{tikzpicture}[scale=\setscaletikz]
\draw[thick,->] (-5,2.5) -- ++(2.5,0);
\draw[thick,->] (-5,1.7) -- ++(2.5,0);
\draw[thick,->] (-5,0.9) -- ++(2.5,0);
\draw[thick,-<] (5,2.5) -- ++(-2.5,0);
\draw[thick,-<] (5,1.7) -- ++(-2.5,0);
\draw[thick,-<] (5,0.9) -- ++(-2.5,0);
\draw[thick,-<] (-5,-2.5) -- ++(2.5,0);
\draw[thick,-<] (-5,-1.7) -- ++(2.5,0);
\draw[thick,-<] (-5,-0.9) -- ++(2.5,0);
\draw[thick,->] (5,-2.5) -- ++(-2.5,0);
\draw[thick,->] (5,-1.7) -- ++(-2.5,0);
\draw[thick,->] (5,-0.9) -- ++(-2.5,0);
\draw[thick] (-5,2.5) -- ++(10,0);
\draw[thick] (-5,1.7) -- ++(10,0);
\draw[thick] (-5,0.9) -- ++(10,0);
\draw[thick] (-5,-2.5) -- ++(10,0);
\draw[thick] (-5,-1.7) -- ++(10,0);
\draw[thick] (-5,-0.9) -- ++(10,0);
\draw[fill=white] (-3.8,.7) rectangle (-4.2,2);
\draw[fill=white] (3.8,.7) rectangle (4.2,2);
\draw[fill=white] (-3.8,-.7) rectangle (-4.2,-2);
\draw[fill=white] (3.8,-.7) rectangle (4.2,-2);
\draw[fill=white] (0,1.7) ellipse (1.3cm and 1.6cm);
\node at (0,1.7) {\scriptsize $\Irrep{8}{}$};
\draw[fill=white] (0,-1.7) ellipse (1.3cm and 1.6cm);
\node at (0,-1.7) {\scriptsize ${\Irrep{8}{}}$};
\end{tikzpicture}
}

\newcommand{\ClebschONE}{
\begin{tikzpicture}[scale=\setscaletikz]
\draw[thick] (-3,2.5) -- ++(-1,0);
\draw[thick] (-3,2) -- ++(-1,0);
\draw[thick] (-3,1) -- ++(-1,0);
\draw[thick] (-3,.5) -- ++(-1,0);
\draw[thick] (-3,-.5) -- ++(-1,0);
\draw[thick] (-3,-1) -- ++(-1,0);
\draw[thick] (-3,-2) -- ++(-1,0);
\draw[thick] (-3,-2.5) -- ++(-1,0);
\draw[thick] (3,2.5) -- ++(1,0);
\draw[fill=white] (-4,2.7) rectangle ++(-.4,-.9);
\draw[fill=white] (-4,1.2) rectangle ++(-.4,-.9);
\draw[fill=white] (-4,-2.7) rectangle ++(-.4,.9);
\draw[fill=white] (-4,-1.2) rectangle ++(-.4,.9);
\draw[thick] (-3,.5) -- (-3,-.5);
\draw[thick] (-3,2.5) -- ++(1.5,0)-- ++(0,-5) -- ++(-1.5,0) -- (-3,-2.5);
\draw[thick,\qbcolor] (-3,1) -- ++(1.5,0) to[out=0,in=180] (1.5,-2) -- ++(2.5,0);
\draw[thick,\qbcolor] (-3,-2) -- ++(1.5,0) to[out=0,in=180] (1.5,-2.5) -- ++(2.5,0);
\draw[thick,\qcolor] (-3,2) -- ++(1.5,0) to[out=0,in=180] (1.5,2.5) -- ++(2.5,0);
\draw[thick,\qcolor] (-3,-1) -- ++(1.5,0) to[out=0,in=180] (1.5,2) -- ++(2.5,0);
\draw[fill=white] (2.5,0) ellipse (1.3cm and 3cm);
\node at (2.5,0) {\scriptsize $27$};
\end{tikzpicture}}

\newcommand{\ClebschTWO}{
\begin{tikzpicture}[scale=\setscaletikz]
\draw[thick] (-3,2.5) -- ++(-1,0);
\draw[thick] (-3,2) -- ++(-1,0);
\draw[thick] (-3,1) -- ++(-1,0);
\draw[thick] (-3,.5) -- ++(-1,0);
\draw[thick] (-3,-.5) -- ++(-1,0);
\draw[thick] (-3,-1) -- ++(-1,0);
\draw[thick] (-3,-2) -- ++(-1,0);
\draw[thick] (-3,-2.5) -- ++(-1,0);
\draw[thick] (-3,2) -- ++(1,0)-- ++(0,-4) -- ++(-1,0) -- (-3,-2);
\draw[thick] (-3,2.5) -- ++(1.5,0)-- ++(0,-5) -- ++(-1.5,0) -- (-3,-2.5);
\draw[fill=white] (-4,2.7) rectangle ++(-.4,-.9);
\draw[fill=white] (-4,1.2) rectangle ++(-.4,-.9);
\draw[fill=white] (-4,-2.7) rectangle ++(-.4,.9);
\draw[fill=white] (-4,-1.2) rectangle ++(-.4,.9);
\draw[thick,\qbcolor] (-3,1) -- (-1.5,1) to[out=0,in=180] (1.5,-2) -- (4,-2);
\draw[thick,\qbcolor] (-3,.5) -- (-1.5,.5) to[out=0,in=180] (1.5,-2.5) -- (4,-2.5);
\draw[thick,\qcolor] (-3,-1) -- (-1.5,-1) to[out=0,in=180] (1.5,2) -- (4,2);
\draw[thick,\qcolor] (-3,-.5) -- (-1.5,-.5) to[out=0,in=180] (1.5,2.5) -- (4,2.5);
\draw[fill=white] (2.5,0) ellipse (1.3cm and 3cm);
\node at (2.5,0) {\scriptsize $27$};
\end{tikzpicture}}

\newcommand{\ClebschTHREE}{
\begin{tikzpicture}[scale=\setscaletikz]
\draw[thick] (-3,2.5) -- ++(-1,0);
\draw[thick] (-3,2) -- ++(-1,0);
\draw[thick] (-3,1) -- ++(-1,0);
\draw[thick] (-3,.5) -- ++(-1,0);
\draw[thick] (-3,-.5) -- ++(-1,0);
\draw[thick] (-3,-1) -- ++(-1,0);
\draw[thick] (-3,-2) -- ++(-1,0);
\draw[thick] (-3,-2.5) -- ++(-1,0);
\draw[fill=white] (-4,2.7) rectangle ++(-.4,-.9);
\draw[fill=white] (-4,1.2) rectangle ++(-.4,-.9);
\draw[fill=white] (-4,-2.7) rectangle ++(-.4,.9);
\draw[fill=white] (-4,-1.2) rectangle ++(-.4,.9);
\draw[thick] (-3,1) -- ++(.5,0)-- ++(0,-2) -- ++(-.5,0) -- (-3,-1);
\draw[thick] (-3,.5) -- (-3,-.5);
\draw[thick,\qcolor] (-3,2.5) -- (1.5,2.5);
\draw[thick,\qcolor] (-3,2) -- (1.5,2);
\draw[thick,\qbcolor] (-3,-2.5) -- (1.5,-2.5);
\draw[thick,\qbcolor] (-3,-2) -- (1.5,-2);
\draw[fill=white] (0,0) ellipse (1.3cm and 3cm);
\node at (0,0) {\scriptsize $27$};
\end{tikzpicture}}

\newcommand{\THb}{
\begin{tikzpicture}[scale=\setscaletikz]
\draw[thick] (-3,2.5) -- ++(-1,0);
\draw[thick] (-3,2) -- ++(-1,0);
\draw[thick] (-3,1) -- ++(-1,0);
\draw[thick] (-3,.5) -- ++(-1,0);
\draw[thick] (-3,-.5) -- ++(-1,0);
\draw[thick] (-3,-1) -- ++(-1,0);
\draw[thick] (-3,-2) -- ++(-1,0);
\draw[thick] (-3,-2.5) -- ++(-1,0);
\draw[thick] (3,2.5) -- ++(1,0);
\draw[thick] (3,2) -- ++(1,0);
\draw[thick] (3,1) -- ++(1,0);
\draw[thick] (3,.5) -- ++(1,0);
\draw[thick] (3,-.5) -- ++(1,0);
\draw[thick] (3,-1) -- ++(1,0);
\draw[thick] (3,-2) -- ++(1,0);
\draw[thick] (3,-2.5) -- ++(1,0);
\draw[fill=white] (-4,2.7) rectangle ++(-.4,-.9);
\draw[fill=white] (-4,1.2) rectangle ++(-.4,-.9);
\draw[fill=white] (4,2.7) rectangle ++(.4,-.9);
\draw[fill=white] (4,1.2) rectangle ++(.4,-.9);
\draw[fill=white] (-4,-2.7) rectangle ++(-.4,.9);
\draw[fill=white] (-4,-1.2) rectangle ++(-.4,.9);
\draw[fill=white] (4,-2.7) rectangle ++(.4,.9);
\draw[fill=white] (4,-1.2) rectangle ++(.4,.9);
\draw[thick] (-3,.5) -- (-3,-.5);
\draw[thick] (-3,2.5) -- ++(1.5,0)-- ++(0,-5) -- ++(-1.5,0) -- (-3,-2.5);
\draw[thick] (3,.5) -- (3,-.5);
\draw[thick] (3,2.5) -- ++(-1.5,0)-- ++(0,-5) -- ++(1.5,0) -- (3,-2.5);
\draw[thick,\qbcolor] (-3,1) -- ++(1.5,0) to[out=0,in=180] (1.5,-2) -- ++(1.5,0);
\draw[thick,\qbcolor] (-3,-2) -- ++(1.5,0) to[out=0,in=180] (1.5,1) -- ++(1.5,0);
\draw[thick,\qcolor] (-3,2) -- ++(1.5,0) to[out=0,in=180] (1.5,-1) -- ++(1.5,0);
\draw[thick,\qcolor] (-3,-1) -- ++(1.5,0) to[out=0,in=180] (1.5,2) -- ++(1.5,0);
\end{tikzpicture}}

\newcommand{\THa}{
\begin{tikzpicture}[scale=\setscaletikz]
\draw[thick] (-3,2.5) -- ++(-1,0);
\draw[thick] (-3,2) -- ++(-1,0);
\draw[thick] (-3,1) -- ++(-1,0);
\draw[thick] (-3,.5) -- ++(-1,0);
\draw[thick] (-3,-.5) -- ++(-1,0);
\draw[thick] (-3,-1) -- ++(-1,0);
\draw[thick] (-3,-2) -- ++(-1,0);
\draw[thick] (-3,-2.5) -- ++(-1,0);
\draw[thick] (3,2.5) -- ++(1,0);
\draw[thick] (3,2) -- ++(1,0);
\draw[thick] (3,1) -- ++(1,0);
\draw[thick] (3,.5) -- ++(1,0);
\draw[thick] (3,-.5) -- ++(1,0);
\draw[thick] (3,-1) -- ++(1,0);
\draw[thick] (3,-2) -- ++(1,0);
\draw[thick] (3,-2.5) -- ++(1,0);
\draw[fill=white] (-4,2.7) rectangle ++(-.4,-.9);
\draw[fill=white] (-4,1.2) rectangle ++(-.4,-.9);
\draw[fill=white] (4,2.7) rectangle ++(.4,-.9);
\draw[fill=white] (4,1.2) rectangle ++(.4,-.9);
\draw[fill=white] (-4,-2.7) rectangle ++(-.4,.9);
\draw[fill=white] (-4,-1.2) rectangle ++(-.4,.9);
\draw[fill=white] (4,-2.7) rectangle ++(.4,.9);
\draw[fill=white] (4,-1.2) rectangle ++(.4,.9);
\draw[thick] (-3,.5) -- (-3,-.5);
\draw[thick] (-3,2.5) -- ++(1.5,0)-- ++(0,-5) -- ++(-1.5,0) -- (-3,-2.5);
\draw[thick] (3,.5) -- (3,-.5);
\draw[thick] (3,2.5) -- ++(-1.5,0)-- ++(0,-5) -- ++(1.5,0) -- (3,-2.5);
\draw[thick,\qbcolor] (-3,1) -- ++(1.5,0) to[out=0,in=180] (1.5,1) -- ++(1.5,0);
\draw[thick,\qbcolor] (-3,-2) -- ++(1.5,0) to[out=0,in=180] (1.5,-2) -- ++(1.5,0);
\draw[thick,\qcolor] (-3,2) -- ++(1.5,0) to[out=0,in=180] (1.5,2) -- ++(1.5,0);
\draw[thick,\qcolor] (-3,-1) -- ++(1.5,0) to[out=0,in=180] (1.5,-1) -- ++(1.5,0);
\end{tikzpicture}}

\newcommand{\THc}{
\begin{tikzpicture}[scale=\setscaletikz]
\draw[thick] (-3,2.5) -- ++(-1,0);
\draw[thick] (-3,2) -- ++(-1,0);
\draw[thick] (-3,1) -- ++(-1,0);
\draw[thick] (-3,.5) -- ++(-1,0);
\draw[thick] (-3,-.5) -- ++(-1,0);
\draw[thick] (-3,-1) -- ++(-1,0);
\draw[thick] (-3,-2) -- ++(-1,0);
\draw[thick] (-3,-2.5) -- ++(-1,0);
\draw[thick] (3,2.5) -- ++(1,0);
\draw[thick] (3,2) -- ++(1,0);
\draw[thick] (3,1) -- ++(1,0);
\draw[thick] (3,.5) -- ++(1,0);
\draw[thick] (3,-.5) -- ++(1,0);
\draw[thick] (3,-1) -- ++(1,0);
\draw[thick] (3,-2) -- ++(1,0);
\draw[thick] (3,-2.5) -- ++(1,0);
\draw[fill=white] (-4,2.7) rectangle ++(-.4,-.9);
\draw[fill=white] (-4,1.2) rectangle ++(-.4,-.9);
\draw[fill=white] (4,2.7) rectangle ++(.4,-.9);
\draw[fill=white] (4,1.2) rectangle ++(.4,-.9);
\draw[fill=white] (-4,-2.7) rectangle ++(-.4,.9);
\draw[fill=white] (-4,-1.2) rectangle ++(-.4,.9);
\draw[fill=white] (4,-2.7) rectangle ++(.4,.9);
\draw[fill=white] (4,-1.2) rectangle ++(.4,.9);
\draw[thick] (-3,.5) -- (-3,-.5);
\draw[thick] (-3,2.5) -- ++(1.5,0)-- ++(0,-5) -- ++(-1.5,0) -- (-3,-2.5);
\draw[thick] (3,.5) -- (3,-.5);
\draw[thick] (3,2.5) -- ++(-1.5,0)-- ++(0,-5) -- ++(1.5,0) -- (3,-2.5);
\draw[thick,\qbcolor] (-3,1) -- ++(1.5,0) to[out=0,in=180] (1.5,-2) -- ++(1.5,0);
\draw[thick,\qbcolor] (-3,-2) -- ++(1.5,0) to[out=0,in=180] (1.5,1) -- ++(1.5,0);
\draw[thick,\qcolor] (-3,2) -- ++(1.5,0) to[out=0,in=180] (1.5,2) -- ++(1.5,0);
\draw[thick,\qcolor] (-3,-1) -- ++(1.5,0) to[out=0,in=180] (1.5,-1) -- ++(1.5,0);
\end{tikzpicture}}

\newcommand{\THd}{
\begin{tikzpicture}[scale=\setscaletikz]
\draw[thick] (-3,2.5) -- ++(-1,0);
\draw[thick] (-3,2) -- ++(-1,0);
\draw[thick] (-3,1) -- ++(-1,0);
\draw[thick] (-3,.5) -- ++(-1,0);
\draw[thick] (-3,-.5) -- ++(-1,0);
\draw[thick] (-3,-1) -- ++(-1,0);
\draw[thick] (-3,-2) -- ++(-1,0);
\draw[thick] (-3,-2.5) -- ++(-1,0);
\draw[thick] (3,2.5) -- ++(1,0);
\draw[thick] (3,2) -- ++(1,0);
\draw[thick] (3,1) -- ++(1,0);
\draw[thick] (3,.5) -- ++(1,0);
\draw[thick] (3,-.5) -- ++(1,0);
\draw[thick] (3,-1) -- ++(1,0);
\draw[thick] (3,-2) -- ++(1,0);
\draw[thick] (3,-2.5) -- ++(1,0);
\draw[fill=white] (-4,2.7) rectangle ++(-.4,-.9);
\draw[fill=white] (-4,1.2) rectangle ++(-.4,-.9);
\draw[fill=white] (4,2.7) rectangle ++(.4,-.9);
\draw[fill=white] (4,1.2) rectangle ++(.4,-.9);
\draw[fill=white] (-4,-2.7) rectangle ++(-.4,.9);
\draw[fill=white] (-4,-1.2) rectangle ++(-.4,.9);
\draw[fill=white] (4,-2.7) rectangle ++(.4,.9);
\draw[fill=white] (4,-1.2) rectangle ++(.4,.9);
\draw[thick] (-3,.5) -- (-3,-.5);
\draw[thick] (-3,2.5) -- ++(1.5,0)-- ++(0,-5) -- ++(-1.5,0) -- (-3,-2.5);
\draw[thick] (3,.5) -- (3,-.5);
\draw[thick] (3,2.5) -- ++(-1.5,0)-- ++(0,-5) -- ++(1.5,0) -- (3,-2.5);
\draw[thick,\qbcolor] (-3,1) -- ++(1.5,0) to[out=0,in=180] (1.5,1) -- ++(1.5,0);
\draw[thick,\qbcolor] (-3,-2) -- ++(1.5,0) to[out=0,in=180] (1.5,-2) -- ++(1.5,0);
\draw[thick,\qcolor] (-3,2) -- ++(1.5,0) to[out=0,in=180] (1.5,-1) -- ++(1.5,0);
\draw[thick,\qcolor] (-3,-1) -- ++(1.5,0) to[out=0,in=180] (1.5,2) -- ++(1.5,0);
\end{tikzpicture}}

\newcommand{\TJa}{
\begin{tikzpicture}[scale=\setscaletikz]
\draw[thick] (-3,2.5) -- ++(-1,0);
\draw[thick] (-3,2) -- ++(-1,0);
\draw[thick] (-3,1) -- ++(-1,0);
\draw[thick] (-3,.5) -- ++(-1,0);
\draw[thick] (-3,-.5) -- ++(-1,0);
\draw[thick] (-3,-1) -- ++(-1,0);
\draw[thick] (-3,-2) -- ++(-1,0);
\draw[thick] (-3,-2.5) -- ++(-1,0);
\draw[thick] (3,2.5) -- ++(1,0);
\draw[thick] (3,2) -- ++(1,0);
\draw[thick] (3,1) -- ++(1,0);
\draw[thick] (3,.5) -- ++(1,0);
\draw[thick] (3,-.5) -- ++(1,0);
\draw[thick] (3,-1) -- ++(1,0);
\draw[thick] (3,-2) -- ++(1,0);
\draw[thick] (3,-2.5) -- ++(1,0);
\draw[fill=white] (-4,2.7) rectangle ++(-.4,-.9);
\draw[fill=white] (-4,1.2) rectangle ++(-.4,-.9);
\draw[fill=white] (4,2.7) rectangle ++(.4,-.9);
\draw[fill=white] (4,1.2) rectangle ++(.4,-.9);
\draw[fill=white] (-4,-2.7) rectangle ++(-.4,.9);
\draw[fill=white] (-4,-1.2) rectangle ++(-.4,.9);
\draw[fill=white] (4,-2.7) rectangle ++(.4,.9);
\draw[fill=white] (4,-1.2) rectangle ++(.4,.9);
\draw[thick] (-3,2.5) -- ++(1.5,0)-- ++(0,-5) -- ++(-1.5,0) -- (-3,-2.5);
\draw[thick] (3,2.5) -- ++(-1.5,0)-- ++(0,-5) -- ++(1.5,0) -- (3,-2.5);
\draw[thick,\qbcolor] (-3,1) -- ++(1.5,0) to[out=0,in=180] (1.5,1) -- ++(1.5,0);
\draw[thick,\qbcolor] (-3,.5) -- ++(1.5,0) to[out=0,in=180] (1.5,.5) -- ++(1.5,0);
\draw[thick,\qbcolor] (-3,-2) -- ++(1.5,0) to[out=0,in=180] (1.5,-2) -- ++(1.5,0);
\draw[thick,\qcolor] (-3,2) -- ++(1.5,0) to[out=0,in=180] (1.5,2) -- ++(1.5,0);
\draw[thick,\qcolor] (-3,-.5) -- ++(1.5,0) to[out=0,in=180] (1.5,-.5) -- ++(1.5,0);
\draw[thick,\qcolor] (-3,-1) -- ++(1.5,0) to[out=0,in=180] (1.5,-1) -- ++(1.5,0);
\end{tikzpicture}}

\newcommand{\TJc}{
\begin{tikzpicture}[scale=\setscaletikz]
\draw[thick] (-3,2.5) -- ++(-1,0);
\draw[thick] (-3,2) -- ++(-1,0);
\draw[thick] (-3,1) -- ++(-1,0);
\draw[thick] (-3,.5) -- ++(-1,0);
\draw[thick] (-3,-.5) -- ++(-1,0);
\draw[thick] (-3,-1) -- ++(-1,0);
\draw[thick] (-3,-2) -- ++(-1,0);
\draw[thick] (-3,-2.5) -- ++(-1,0);
\draw[thick] (3,2.5) -- ++(1,0);
\draw[thick] (3,2) -- ++(1,0);
\draw[thick] (3,1) -- ++(1,0);
\draw[thick] (3,.5) -- ++(1,0);
\draw[thick] (3,-.5) -- ++(1,0);
\draw[thick] (3,-1) -- ++(1,0);
\draw[thick] (3,-2) -- ++(1,0);
\draw[thick] (3,-2.5) -- ++(1,0);
\draw[fill=white] (-4,2.7) rectangle ++(-.4,-.9);
\draw[fill=white] (-4,1.2) rectangle ++(-.4,-.9);
\draw[fill=white] (4,2.7) rectangle ++(.4,-.9);
\draw[fill=white] (4,1.2) rectangle ++(.4,-.9);
\draw[fill=white] (-4,-2.7) rectangle ++(-.4,.9);
\draw[fill=white] (-4,-1.2) rectangle ++(-.4,.9);
\draw[fill=white] (4,-2.7) rectangle ++(.4,.9);
\draw[fill=white] (4,-1.2) rectangle ++(.4,.9);
\draw[thick] (-3,2.5) -- ++(1.5,0)-- ++(0,-5) -- ++(-1.5,0) -- (-3,-2.5);
\draw[thick] (3,2.5) -- ++(-1.5,0)-- ++(0,-5) -- ++(1.5,0) -- (3,-2.5);
\draw[thick,\qbcolor] (-3,1) -- ++(1.5,0) to[out=0,in=180] (1.5,1) -- ++(1.5,0);
\draw[thick,\qbcolor] (-3,.5) -- ++(1.5,0) to[out=0,in=180] (1.5,.5) -- ++(1.5,0);
\draw[thick,\qbcolor] (-3,-2) -- ++(1.5,0) to[out=0,in=180] (1.5,-2) -- ++(1.5,0);
\draw[thick,\qcolor] (-3,2) -- ++(1.5,0) to[out=0,in=180] (1.5,-.5) -- ++(1.5,0);
\draw[thick,\qcolor] (-3,-.5) -- ++(1.5,0) to[out=0,in=180] (1.5,2) -- ++(1.5,0);
\draw[thick,\qcolor] (-3,-1) -- ++(1.5,0) to[out=0,in=180] (1.5,-1) -- ++(1.5,0);
\end{tikzpicture}}

\newcommand{\TJe}{
\begin{tikzpicture}[scale=\setscaletikz]
\draw[thick] (-3,2.5) -- ++(-1,0);
\draw[thick] (-3,2) -- ++(-1,0);
\draw[thick] (-3,1) -- ++(-1,0);
\draw[thick] (-3,.5) -- ++(-1,0);
\draw[thick] (-3,-.5) -- ++(-1,0);
\draw[thick] (-3,-1) -- ++(-1,0);
\draw[thick] (-3,-2) -- ++(-1,0);
\draw[thick] (-3,-2.5) -- ++(-1,0);
\draw[thick] (3,2.5) -- ++(1,0);
\draw[thick] (3,2) -- ++(1,0);
\draw[thick] (3,1) -- ++(1,0);
\draw[thick] (3,.5) -- ++(1,0);
\draw[thick] (3,-.5) -- ++(1,0);
\draw[thick] (3,-1) -- ++(1,0);
\draw[thick] (3,-2) -- ++(1,0);
\draw[thick] (3,-2.5) -- ++(1,0);
\draw[fill=white] (-4,2.7) rectangle ++(-.4,-.9);
\draw[fill=white] (-4,1.2) rectangle ++(-.4,-.9);
\draw[fill=white] (4,2.7) rectangle ++(.4,-.9);
\draw[fill=white] (4,1.2) rectangle ++(.4,-.9);
\draw[fill=white] (-4,-2.7) rectangle ++(-.4,.9);
\draw[fill=white] (-4,-1.2) rectangle ++(-.4,.9);
\draw[fill=white] (4,-2.7) rectangle ++(.4,.9);
\draw[fill=white] (4,-1.2) rectangle ++(.4,.9);
\draw[thick] (-3,2.5) -- ++(1.5,0)-- ++(0,-5) -- ++(-1.5,0) -- (-3,-2.5);
\draw[thick] (3,2.5) -- ++(-1.5,0)-- ++(0,-5) -- ++(1.5,0) -- (3,-2.5);
\draw[thick,\qbcolor] (-3,1) -- ++(1.5,0) to[out=0,in=180] (1.5,1) -- ++(1.5,0);
\draw[thick,\qbcolor] (-3,.5) -- ++(1.5,0) to[out=0,in=180] (1.5,-2) -- ++(1.5,0);
\draw[thick,\qbcolor] (-3,-2) -- ++(1.5,0) to[out=0,in=180] (1.5,.5) -- ++(1.5,0);
\draw[thick,\qcolor] (-3,2) -- ++(1.5,0) to[out=0,in=180] (1.5,2) -- ++(1.5,0);
\draw[thick,\qcolor] (-3,-.5) -- ++(1.5,0) to[out=0,in=180] (1.5,-.5) -- ++(1.5,0);
\draw[thick,\qcolor] (-3,-1) -- ++(1.5,0) to[out=0,in=180] (1.5,-1) -- ++(1.5,0);
\end{tikzpicture}}

\newcommand{\TJg}{
\begin{tikzpicture}[scale=\setscaletikz]
\draw[thick] (-3,2.5) -- ++(-1,0);
\draw[thick] (-3,2) -- ++(-1,0);
\draw[thick] (-3,1) -- ++(-1,0);
\draw[thick] (-3,.5) -- ++(-1,0);
\draw[thick] (-3,-.5) -- ++(-1,0);
\draw[thick] (-3,-1) -- ++(-1,0);
\draw[thick] (-3,-2) -- ++(-1,0);
\draw[thick] (-3,-2.5) -- ++(-1,0);
\draw[thick] (3,2.5) -- ++(1,0);
\draw[thick] (3,2) -- ++(1,0);
\draw[thick] (3,1) -- ++(1,0);
\draw[thick] (3,.5) -- ++(1,0);
\draw[thick] (3,-.5) -- ++(1,0);
\draw[thick] (3,-1) -- ++(1,0);
\draw[thick] (3,-2) -- ++(1,0);
\draw[thick] (3,-2.5) -- ++(1,0);
\draw[fill=white] (-4,2.7) rectangle ++(-.4,-.9);
\draw[fill=white] (-4,1.2) rectangle ++(-.4,-.9);
\draw[fill=white] (4,2.7) rectangle ++(.4,-.9);
\draw[fill=white] (4,1.2) rectangle ++(.4,-.9);
\draw[fill=white] (-4,-2.7) rectangle ++(-.4,.9);
\draw[fill=white] (-4,-1.2) rectangle ++(-.4,.9);
\draw[fill=white] (4,-2.7) rectangle ++(.4,.9);
\draw[fill=white] (4,-1.2) rectangle ++(.4,.9);
\draw[thick] (-3,2.5) -- ++(1.5,0)-- ++(0,-5) -- ++(-1.5,0) -- (-3,-2.5);
\draw[thick] (3,2.5) -- ++(-1.5,0)-- ++(0,-5) -- ++(1.5,0) -- (3,-2.5);
\draw[thick,\qbcolor] (-3,1) -- ++(1.5,0) to[out=0,in=180] (1.5,1) -- ++(1.5,0);
\draw[thick,\qbcolor] (-3,.5) -- ++(1.5,0) to[out=0,in=180] (1.5,-2) -- ++(1.5,0);
\draw[thick,\qbcolor] (-3,-2) -- ++(1.5,0) to[out=0,in=180] (1.5,.5) -- ++(1.5,0);
\draw[thick,\qcolor] (-3,2) -- ++(1.5,0) to[out=0,in=180] (1.5,-.5) -- ++(1.5,0);
\draw[thick,\qcolor] (-3,-.5) -- ++(1.5,0) to[out=0,in=180] (1.5,2) -- ++(1.5,0);
\draw[thick,\qcolor] (-3,-1) -- ++(1.5,0) to[out=0,in=180] (1.5,-1) -- ++(1.5,0);
\end{tikzpicture}}

\newcommand{\TJb}{
\begin{tikzpicture}[scale=\setscaletikz]
\draw[thick] (-3,2.5) -- ++(-1,0);
\draw[thick] (-3,2) -- ++(-1,0);
\draw[thick] (-3,1) -- ++(-1,0);
\draw[thick] (-3,.5) -- ++(-1,0);
\draw[thick] (-3,-.5) -- ++(-1,0);
\draw[thick] (-3,-1) -- ++(-1,0);
\draw[thick] (-3,-2) -- ++(-1,0);
\draw[thick] (-3,-2.5) -- ++(-1,0);
\draw[thick] (3,2.5) -- ++(1,0);
\draw[thick] (3,2) -- ++(1,0);
\draw[thick] (3,1) -- ++(1,0);
\draw[thick] (3,.5) -- ++(1,0);
\draw[thick] (3,-.5) -- ++(1,0);
\draw[thick] (3,-1) -- ++(1,0);
\draw[thick] (3,-2) -- ++(1,0);
\draw[thick] (3,-2.5) -- ++(1,0);
\draw[fill=white] (-4,2.7) rectangle ++(-.4,-.9);
\draw[fill=white] (-4,1.2) rectangle ++(-.4,-.9);
\draw[fill=white] (4,2.7) rectangle ++(.4,-.9);
\draw[fill=white] (4,1.2) rectangle ++(.4,-.9);
\draw[fill=white] (-4,-2.7) rectangle ++(-.4,.9);
\draw[fill=white] (-4,-1.2) rectangle ++(-.4,.9);
\draw[fill=white] (4,-2.7) rectangle ++(.4,.9);
\draw[fill=white] (4,-1.2) rectangle ++(.4,.9);
\draw[thick] (-3,.5) -- (-3,-.5);
\draw[thick] (3,.5) -- (3,-.5);
\draw[thick,\qbcolor] (-3,1) -- ++(1.5,0) to[out=0,in=180] (1.5,1) -- ++(1.5,0);
\draw[thick,\qbcolor] (-3,-2) -- ++(1.5,0) to[out=0,in=180] (1.5,-2) -- ++(1.5,0);
\draw[thick,\qbcolor] (-3,-2.5) -- ++(1.5,0) to[out=0,in=180] (1.5,-2.5) -- ++(1.5,0);
\draw[thick,\qcolor] (-3,2.5) -- ++(1.5,0) to[out=0,in=180] (1.5,2.5) -- ++(1.5,0);
\draw[thick,\qcolor] (-3,2) -- ++(1.5,0) to[out=0,in=180] (1.5,2) -- ++(1.5,0);
\draw[thick,\qcolor] (-3,-1) -- ++(1.5,0) to[out=0,in=180] (1.5,-1) -- ++(1.5,0);
\end{tikzpicture}}

\newcommand{\TJd}{
\begin{tikzpicture}[scale=\setscaletikz]
\draw[thick] (-3,2.5) -- ++(-1,0);
\draw[thick] (-3,2) -- ++(-1,0);
\draw[thick] (-3,1) -- ++(-1,0);
\draw[thick] (-3,.5) -- ++(-1,0);
\draw[thick] (-3,-.5) -- ++(-1,0);
\draw[thick] (-3,-1) -- ++(-1,0);
\draw[thick] (-3,-2) -- ++(-1,0);
\draw[thick] (-3,-2.5) -- ++(-1,0);
\draw[thick] (3,2.5) -- ++(1,0);
\draw[thick] (3,2) -- ++(1,0);
\draw[thick] (3,1) -- ++(1,0);
\draw[thick] (3,.5) -- ++(1,0);
\draw[thick] (3,-.5) -- ++(1,0);
\draw[thick] (3,-1) -- ++(1,0);
\draw[thick] (3,-2) -- ++(1,0);
\draw[thick] (3,-2.5) -- ++(1,0);
\draw[fill=white] (-4,2.7) rectangle ++(-.4,-.9);
\draw[fill=white] (-4,1.2) rectangle ++(-.4,-.9);
\draw[fill=white] (4,2.7) rectangle ++(.4,-.9);
\draw[fill=white] (4,1.2) rectangle ++(.4,-.9);
\draw[fill=white] (-4,-2.7) rectangle ++(-.4,.9);
\draw[fill=white] (-4,-1.2) rectangle ++(-.4,.9);
\draw[fill=white] (4,-2.7) rectangle ++(.4,.9);
\draw[fill=white] (4,-1.2) rectangle ++(.4,.9);
\draw[thick] (-3,.5) -- (-3,-.5);
\draw[thick] (3,.5) -- (3,-.5);
\draw[thick,\qbcolor] (-3,1) -- ++(1.5,0) to[out=0,in=180] (1.5,-2) -- ++(1.5,0);
\draw[thick,\qbcolor] (-3,-2) -- ++(1.5,0) to[out=0,in=180] (1.5,1) -- ++(1.5,0);
\draw[thick,\qbcolor] (-3,-2.5) -- ++(1.5,0) to[out=0,in=180] (1.5,-2.5) -- ++(1.5,0);
\draw[thick,\qcolor] (-3,2.5) -- ++(1.5,0) to[out=0,in=180] (1.5,2.5) -- ++(1.5,0);
\draw[thick,\qcolor] (-3,2) -- ++(1.5,0) to[out=0,in=180] (1.5,2) -- ++(1.5,0);
\draw[thick,\qcolor] (-3,-1) -- ++(1.5,0) to[out=0,in=180] (1.5,-1) -- ++(1.5,0);
\end{tikzpicture}}

\newcommand{\TJf}{
\begin{tikzpicture}[scale=\setscaletikz]
\draw[thick] (-3,2.5) -- ++(-1,0);
\draw[thick] (-3,2) -- ++(-1,0);
\draw[thick] (-3,1) -- ++(-1,0);
\draw[thick] (-3,.5) -- ++(-1,0);
\draw[thick] (-3,-.5) -- ++(-1,0);
\draw[thick] (-3,-1) -- ++(-1,0);
\draw[thick] (-3,-2) -- ++(-1,0);
\draw[thick] (-3,-2.5) -- ++(-1,0);
\draw[thick] (3,2.5) -- ++(1,0);
\draw[thick] (3,2) -- ++(1,0);
\draw[thick] (3,1) -- ++(1,0);
\draw[thick] (3,.5) -- ++(1,0);
\draw[thick] (3,-.5) -- ++(1,0);
\draw[thick] (3,-1) -- ++(1,0);
\draw[thick] (3,-2) -- ++(1,0);
\draw[thick] (3,-2.5) -- ++(1,0);
\draw[fill=white] (-4,2.7) rectangle ++(-.4,-.9);
\draw[fill=white] (-4,1.2) rectangle ++(-.4,-.9);
\draw[fill=white] (4,2.7) rectangle ++(.4,-.9);
\draw[fill=white] (4,1.2) rectangle ++(.4,-.9);
\draw[fill=white] (-4,-2.7) rectangle ++(-.4,.9);
\draw[fill=white] (-4,-1.2) rectangle ++(-.4,.9);
\draw[fill=white] (4,-2.7) rectangle ++(.4,.9);
\draw[fill=white] (4,-1.2) rectangle ++(.4,.9);
\draw[thick] (-3,.5) -- (-3,-.5);
\draw[thick] (3,.5) -- (3,-.5);
\draw[thick,\qbcolor] (-3,1) -- ++(1.5,0) to[out=0,in=180] (1.5,1) -- ++(1.5,0);
\draw[thick,\qbcolor] (-3,-2) -- ++(1.5,0) to[out=0,in=180] (1.5,-2) -- ++(1.5,0);
\draw[thick,\qbcolor] (-3,-2.5) -- ++(1.5,0) to[out=0,in=180] (1.5,-2.5) -- ++(1.5,0);
\draw[thick,\qcolor] (-3,2.5) -- ++(1.5,0) to[out=0,in=180] (1.5,2.5) -- ++(1.5,0);
\draw[thick,\qcolor] (-3,2) -- ++(1.5,0) to[out=0,in=180] (1.5,-1) -- ++(1.5,0);
\draw[thick,\qcolor] (-3,-1) -- ++(1.5,0) to[out=0,in=180] (1.5,2) -- ++(1.5,0);
\end{tikzpicture}}

\newcommand{\TJh}{
\begin{tikzpicture}[scale=\setscaletikz]
\draw[thick] (-3,2.5) -- ++(-1,0);
\draw[thick] (-3,2) -- ++(-1,0);
\draw[thick] (-3,1) -- ++(-1,0);
\draw[thick] (-3,.5) -- ++(-1,0);
\draw[thick] (-3,-.5) -- ++(-1,0);
\draw[thick] (-3,-1) -- ++(-1,0);
\draw[thick] (-3,-2) -- ++(-1,0);
\draw[thick] (-3,-2.5) -- ++(-1,0);
\draw[thick] (3,2.5) -- ++(1,0);
\draw[thick] (3,2) -- ++(1,0);
\draw[thick] (3,1) -- ++(1,0);
\draw[thick] (3,.5) -- ++(1,0);
\draw[thick] (3,-.5) -- ++(1,0);
\draw[thick] (3,-1) -- ++(1,0);
\draw[thick] (3,-2) -- ++(1,0);
\draw[thick] (3,-2.5) -- ++(1,0);
\draw[fill=white] (-4,2.7) rectangle ++(-.4,-.9);
\draw[fill=white] (-4,1.2) rectangle ++(-.4,-.9);
\draw[fill=white] (4,2.7) rectangle ++(.4,-.9);
\draw[fill=white] (4,1.2) rectangle ++(.4,-.9);
\draw[fill=white] (-4,-2.7) rectangle ++(-.4,.9);
\draw[fill=white] (-4,-1.2) rectangle ++(-.4,.9);
\draw[fill=white] (4,-2.7) rectangle ++(.4,.9);
\draw[fill=white] (4,-1.2) rectangle ++(.4,.9);
\draw[thick] (-3,.5) -- (-3,-.5);
\draw[thick] (3,.5) -- (3,-.5);
\draw[thick,\qbcolor] (-3,1) -- ++(1.5,0) to[out=0,in=180] (1.5,-2) -- ++(1.5,0);
\draw[thick,\qbcolor] (-3,-2) -- ++(1.5,0) to[out=0,in=180] (1.5,1) -- ++(1.5,0);
\draw[thick,\qbcolor] (-3,-2.5) -- ++(1.5,0) to[out=0,in=180] (1.5,-2.5) -- ++(1.5,0);
\draw[thick,\qcolor] (-3,2.5) -- ++(1.5,0) to[out=0,in=180] (1.5,2.5) -- ++(1.5,0);
\draw[thick,\qcolor] (-3,2) -- ++(1.5,0) to[out=0,in=180] (1.5,-1) -- ++(1.5,0);
\draw[thick,\qcolor] (-3,-1) -- ++(1.5,0) to[out=0,in=180] (1.5,2) -- ++(1.5,0);
\end{tikzpicture}}

\newcommand{\TKa}{
\begin{tikzpicture}[scale=\setscaletikz]
\draw[thick] (-3,2.5) -- ++(-1,0);
\draw[thick] (-3,2) -- ++(-1,0);
\draw[thick] (-3,1) -- ++(-1,0);
\draw[thick] (-3,.5) -- ++(-1,0);
\draw[thick] (-3,-.5) -- ++(-1,0);
\draw[thick] (-3,-1) -- ++(-1,0);
\draw[thick] (-3,-2) -- ++(-1,0);
\draw[thick] (-3,-2.5) -- ++(-1,0);
\draw[thick] (3,2.5) -- ++(1,0);
\draw[thick] (3,2) -- ++(1,0);
\draw[thick] (3,1) -- ++(1,0);
\draw[thick] (3,.5) -- ++(1,0);
\draw[thick] (3,-.5) -- ++(1,0);
\draw[thick] (3,-1) -- ++(1,0);
\draw[thick] (3,-2) -- ++(1,0);
\draw[thick] (3,-2.5) -- ++(1,0);
\draw[fill=white] (-4,2.7) rectangle ++(-.4,-.9);
\draw[fill=white] (-4,1.2) rectangle ++(-.4,-.9);
\draw[fill=white] (4,2.7) rectangle ++(.4,-.9);
\draw[fill=white] (4,1.2) rectangle ++(.4,-.9);
\draw[fill=white] (-4,-2.7) rectangle ++(-.4,.9);
\draw[fill=white] (-4,-1.2) rectangle ++(-.4,.9);
\draw[fill=white] (4,-2.7) rectangle ++(.4,.9);
\draw[fill=white] (4,-1.2) rectangle ++(.4,.9);
\draw[thick] (-3,2.5) -- ++(1.5,0)-- ++(0,-5) -- ++(-1.5,0) -- (-3,-2.5);
\draw[thick] (3,.5) -- (3,-.5);
\draw[thick,\qbcolor] (-3,1) -- ++(1.5,0) to[out=0,in=180] (1.5,1) -- ++(1.5,0);
\draw[thick,\qbcolor] (-3,.5) -- ++(1.5,0) to[out=0,in=180] (1.5,-2.5) -- ++(1.5,0);
\draw[thick,\qbcolor] (-3,-2) -- ++(1.5,0) to[out=0,in=180] (1.5,-2) -- ++(1.5,0);
\draw[thick,\qcolor] (-3,2) -- ++(1.5,0) to[out=0,in=180] (1.5,2) -- ++(1.5,0);
\draw[thick,\qcolor] (-3,-.5) -- ++(1.5,0) to[out=0,in=180] (1.5,2.5) -- ++(1.5,0);
\draw[thick,\qcolor] (-3,-1) -- ++(1.5,0) to[out=0,in=180] (1.5,-1) -- ++(1.5,0);
\end{tikzpicture}}

\newcommand{\TKc}{
\begin{tikzpicture}[scale=\setscaletikz]
\draw[thick] (-3,2.5) -- ++(-1,0);
\draw[thick] (-3,2) -- ++(-1,0);
\draw[thick] (-3,1) -- ++(-1,0);
\draw[thick] (-3,.5) -- ++(-1,0);
\draw[thick] (-3,-.5) -- ++(-1,0);
\draw[thick] (-3,-1) -- ++(-1,0);
\draw[thick] (-3,-2) -- ++(-1,0);
\draw[thick] (-3,-2.5) -- ++(-1,0);
\draw[thick] (3,2.5) -- ++(1,0);
\draw[thick] (3,2) -- ++(1,0);
\draw[thick] (3,1) -- ++(1,0);
\draw[thick] (3,.5) -- ++(1,0);
\draw[thick] (3,-.5) -- ++(1,0);
\draw[thick] (3,-1) -- ++(1,0);
\draw[thick] (3,-2) -- ++(1,0);
\draw[thick] (3,-2.5) -- ++(1,0);
\draw[fill=white] (-4,2.7) rectangle ++(-.4,-.9);
\draw[fill=white] (-4,1.2) rectangle ++(-.4,-.9);
\draw[fill=white] (4,2.7) rectangle ++(.4,-.9);
\draw[fill=white] (4,1.2) rectangle ++(.4,-.9);
\draw[fill=white] (-4,-2.7) rectangle ++(-.4,.9);
\draw[fill=white] (-4,-1.2) rectangle ++(-.4,.9);
\draw[fill=white] (4,-2.7) rectangle ++(.4,.9);
\draw[fill=white] (4,-1.2) rectangle ++(.4,.9);
\draw[thick] (-3,2.5) -- ++(1.5,0)-- ++(0,-5) -- ++(-1.5,0) -- (-3,-2.5);
\draw[thick] (3,.5) -- (3,-.5);
\draw[thick,\qbcolor] (-3,1) -- ++(1.5,0) to[out=0,in=180] (1.5,-2) -- ++(1.5,0);
\draw[thick,\qbcolor] (-3,.5) -- ++(1.5,0) to[out=0,in=180] (1.5,-2.5) -- ++(1.5,0);
\draw[thick,\qbcolor] (-3,-2) -- ++(1.5,0) to[out=0,in=180] (1.5,1) -- ++(1.5,0);
\draw[thick,\qcolor] (-3,2) -- ++(1.5,0) to[out=0,in=180] (1.5,2) -- ++(1.5,0);
\draw[thick,\qcolor] (-3,-.5) -- ++(1.5,0) to[out=0,in=180] (1.5,2.5) -- ++(1.5,0);
\draw[thick,\qcolor] (-3,-1) -- ++(1.5,0) to[out=0,in=180] (1.5,-1) -- ++(1.5,0);
\end{tikzpicture}}

\newcommand{\TKe}{
\begin{tikzpicture}[scale=\setscaletikz]
\draw[thick] (-3,2.5) -- ++(-1,0);
\draw[thick] (-3,2) -- ++(-1,0);
\draw[thick] (-3,1) -- ++(-1,0);
\draw[thick] (-3,.5) -- ++(-1,0);
\draw[thick] (-3,-.5) -- ++(-1,0);
\draw[thick] (-3,-1) -- ++(-1,0);
\draw[thick] (-3,-2) -- ++(-1,0);
\draw[thick] (-3,-2.5) -- ++(-1,0);
\draw[thick] (3,2.5) -- ++(1,0);
\draw[thick] (3,2) -- ++(1,0);
\draw[thick] (3,1) -- ++(1,0);
\draw[thick] (3,.5) -- ++(1,0);
\draw[thick] (3,-.5) -- ++(1,0);
\draw[thick] (3,-1) -- ++(1,0);
\draw[thick] (3,-2) -- ++(1,0);
\draw[thick] (3,-2.5) -- ++(1,0);
\draw[fill=white] (-4,2.7) rectangle ++(-.4,-.9);
\draw[fill=white] (-4,1.2) rectangle ++(-.4,-.9);
\draw[fill=white] (4,2.7) rectangle ++(.4,-.9);
\draw[fill=white] (4,1.2) rectangle ++(.4,-.9);
\draw[fill=white] (-4,-2.7) rectangle ++(-.4,.9);
\draw[fill=white] (-4,-1.2) rectangle ++(-.4,.9);
\draw[fill=white] (4,-2.7) rectangle ++(.4,.9);
\draw[fill=white] (4,-1.2) rectangle ++(.4,.9);
\draw[thick] (-3,2.5) -- ++(1.5,0)-- ++(0,-5) -- ++(-1.5,0) -- (-3,-2.5);
\draw[thick] (3,.5) -- (3,-.5);
\draw[thick,\qbcolor] (-3,1) -- ++(1.5,0) to[out=0,in=180] (1.5,1) -- ++(1.5,0);
\draw[thick,\qbcolor] (-3,.5) -- ++(1.5,0) to[out=0,in=180] (1.5,-2.5) -- ++(1.5,0);
\draw[thick,\qbcolor] (-3,-2) -- ++(1.5,0) to[out=0,in=180] (1.5,-2) -- ++(1.5,0);
\draw[thick,\qcolor] (-3,2) -- ++(1.5,0) to[out=0,in=180] (1.5,-1) -- ++(1.5,0);
\draw[thick,\qcolor] (-3,-.5) -- ++(1.5,0) to[out=0,in=180] (1.5,2.5) -- ++(1.5,0);
\draw[thick,\qcolor] (-3,-1) -- ++(1.5,0) to[out=0,in=180] (1.5,2) -- ++(1.5,0);
\end{tikzpicture}}

\newcommand{\TKg}{
\begin{tikzpicture}[scale=\setscaletikz]
\draw[thick] (-3,2.5) -- ++(-1,0);
\draw[thick] (-3,2) -- ++(-1,0);
\draw[thick] (-3,1) -- ++(-1,0);
\draw[thick] (-3,.5) -- ++(-1,0);
\draw[thick] (-3,-.5) -- ++(-1,0);
\draw[thick] (-3,-1) -- ++(-1,0);
\draw[thick] (-3,-2) -- ++(-1,0);
\draw[thick] (-3,-2.5) -- ++(-1,0);
\draw[thick] (3,2.5) -- ++(1,0);
\draw[thick] (3,2) -- ++(1,0);
\draw[thick] (3,1) -- ++(1,0);
\draw[thick] (3,.5) -- ++(1,0);
\draw[thick] (3,-.5) -- ++(1,0);
\draw[thick] (3,-1) -- ++(1,0);
\draw[thick] (3,-2) -- ++(1,0);
\draw[thick] (3,-2.5) -- ++(1,0);
\draw[fill=white] (-4,2.7) rectangle ++(-.4,-.9);
\draw[fill=white] (-4,1.2) rectangle ++(-.4,-.9);
\draw[fill=white] (4,2.7) rectangle ++(.4,-.9);
\draw[fill=white] (4,1.2) rectangle ++(.4,-.9);
\draw[fill=white] (-4,-2.7) rectangle ++(-.4,.9);
\draw[fill=white] (-4,-1.2) rectangle ++(-.4,.9);
\draw[fill=white] (4,-2.7) rectangle ++(.4,.9);
\draw[fill=white] (4,-1.2) rectangle ++(.4,.9);
\draw[thick] (-3,2.5) -- ++(1.5,0)-- ++(0,-5) -- ++(-1.5,0) -- (-3,-2.5);
\draw[thick] (3,.5) -- (3,-.5);
\draw[thick,\qbcolor] (-3,1) -- ++(1.5,0) to[out=0,in=180] (1.5,-2) -- ++(1.5,0);
\draw[thick,\qbcolor] (-3,.5) -- ++(1.5,0) to[out=0,in=180] (1.5,-2.5) -- ++(1.5,0);
\draw[thick,\qbcolor] (-3,-2) -- ++(1.5,0) to[out=0,in=180] (1.5,1) -- ++(1.5,0);
\draw[thick,\qcolor] (-3,2) -- ++(1.5,0) to[out=0,in=180] (1.5,-1) -- ++(1.5,0);
\draw[thick,\qcolor] (-3,-.5) -- ++(1.5,0) to[out=0,in=180] (1.5,2.5) -- ++(1.5,0);
\draw[thick,\qcolor] (-3,-1) -- ++(1.5,0) to[out=0,in=180] (1.5,2) -- ++(1.5,0);
\end{tikzpicture}}

\newcommand{\TKb}{
\begin{tikzpicture}[scale=\setscaletikz]
\draw[thick] (-3,2.5) -- ++(-1,0);
\draw[thick] (-3,2) -- ++(-1,0);
\draw[thick] (-3,1) -- ++(-1,0);
\draw[thick] (-3,.5) -- ++(-1,0);
\draw[thick] (-3,-.5) -- ++(-1,0);
\draw[thick] (-3,-1) -- ++(-1,0);
\draw[thick] (-3,-2) -- ++(-1,0);
\draw[thick] (-3,-2.5) -- ++(-1,0);
\draw[thick] (3,2.5) -- ++(1,0);
\draw[thick] (3,2) -- ++(1,0);
\draw[thick] (3,1) -- ++(1,0);
\draw[thick] (3,.5) -- ++(1,0);
\draw[thick] (3,-.5) -- ++(1,0);
\draw[thick] (3,-1) -- ++(1,0);
\draw[thick] (3,-2) -- ++(1,0);
\draw[thick] (3,-2.5) -- ++(1,0);
\draw[fill=white] (-4,2.7) rectangle ++(-.4,-.9);
\draw[fill=white] (-4,1.2) rectangle ++(-.4,-.9);
\draw[fill=white] (4,2.7) rectangle ++(.4,-.9);
\draw[fill=white] (4,1.2) rectangle ++(.4,-.9);
\draw[fill=white] (-4,-2.7) rectangle ++(-.4,.9);
\draw[fill=white] (-4,-1.2) rectangle ++(-.4,.9);
\draw[fill=white] (4,-2.7) rectangle ++(.4,.9);
\draw[fill=white] (4,-1.2) rectangle ++(.4,.9);
\draw[thick] (-3,.5) -- (-3,-.5);
\draw[thick] (3,2.5) -- ++(-1.5,0)-- ++(0,-5) -- ++(1.5,0) -- (3,-2.5);
\draw[thick,\qcolor] (-3,2.5) -- ++(1.5,0) to[out=0,in=180] (1.5,-.5) -- ++(1.5,0);
\draw[thick,\qcolor] (-3,2) -- ++(1.5,0) to[out=0,in=180] (1.5,2) -- ++(1.5,0);
\draw[thick,\qcolor] (-3,-1) -- ++(1.5,0) to[out=0,in=180] (1.5,-1) -- ++(1.5,0);
\draw[thick,\qbcolor] (-3,1) -- ++(1.5,0) to[out=0,in=180] (1.5,1) -- ++(1.5,0);
\draw[thick,\qbcolor] (-3,-2.5) -- ++(1.5,0) to[out=0,in=180] (1.5,.5) -- ++(1.5,0);
\draw[thick,\qbcolor] (-3,-2) -- ++(1.5,0) to[out=0,in=180] (1.5,-2) -- ++(1.5,0);
\end{tikzpicture}}

\newcommand{\TKd}{
\begin{tikzpicture}[scale=\setscaletikz]
\draw[thick] (-3,2.5) -- ++(-1,0);
\draw[thick] (-3,2) -- ++(-1,0);
\draw[thick] (-3,1) -- ++(-1,0);
\draw[thick] (-3,.5) -- ++(-1,0);
\draw[thick] (-3,-.5) -- ++(-1,0);
\draw[thick] (-3,-1) -- ++(-1,0);
\draw[thick] (-3,-2) -- ++(-1,0);
\draw[thick] (-3,-2.5) -- ++(-1,0);
\draw[thick] (3,2.5) -- ++(1,0);
\draw[thick] (3,2) -- ++(1,0);
\draw[thick] (3,1) -- ++(1,0);
\draw[thick] (3,.5) -- ++(1,0);
\draw[thick] (3,-.5) -- ++(1,0);
\draw[thick] (3,-1) -- ++(1,0);
\draw[thick] (3,-2) -- ++(1,0);
\draw[thick] (3,-2.5) -- ++(1,0);
\draw[fill=white] (-4,2.7) rectangle ++(-.4,-.9);
\draw[fill=white] (-4,1.2) rectangle ++(-.4,-.9);
\draw[fill=white] (4,2.7) rectangle ++(.4,-.9);
\draw[fill=white] (4,1.2) rectangle ++(.4,-.9);
\draw[fill=white] (-4,-2.7) rectangle ++(-.4,.9);
\draw[fill=white] (-4,-1.2) rectangle ++(-.4,.9);
\draw[fill=white] (4,-2.7) rectangle ++(.4,.9);
\draw[fill=white] (4,-1.2) rectangle ++(.4,.9);
\draw[thick] (-3,.5) -- (-3,-.5);
\draw[thick] (3,2.5) -- ++(-1.5,0)-- ++(0,-5) -- ++(1.5,0) -- (3,-2.5);
\draw[thick,\qcolor] (-3,2.5) -- ++(1.5,0) to[out=0,in=180] (1.5,-.5) -- ++(1.5,0);
\draw[thick,\qcolor] (-3,2) -- ++(1.5,0) to[out=0,in=180] (1.5,2) -- ++(1.5,0);
\draw[thick,\qcolor] (-3,-1) -- ++(1.5,0) to[out=0,in=180] (1.5,-1) -- ++(1.5,0);
\draw[thick,\qbcolor] (-3,1) -- ++(1.5,0) to[out=0,in=180] (1.5,-2) -- ++(1.5,0);
\draw[thick,\qbcolor] (-3,-2.5) -- ++(1.5,0) to[out=0,in=180] (1.5,.5) -- ++(1.5,0);
\draw[thick,\qbcolor] (-3,-2) -- ++(1.5,0) to[out=0,in=180] (1.5,1) -- ++(1.5,0);
\end{tikzpicture}}

\newcommand{\TKf}{
\begin{tikzpicture}[scale=\setscaletikz]
\draw[thick] (-3,2.5) -- ++(-1,0);
\draw[thick] (-3,2) -- ++(-1,0);
\draw[thick] (-3,1) -- ++(-1,0);
\draw[thick] (-3,.5) -- ++(-1,0);
\draw[thick] (-3,-.5) -- ++(-1,0);
\draw[thick] (-3,-1) -- ++(-1,0);
\draw[thick] (-3,-2) -- ++(-1,0);
\draw[thick] (-3,-2.5) -- ++(-1,0);
\draw[thick] (3,2.5) -- ++(1,0);
\draw[thick] (3,2) -- ++(1,0);
\draw[thick] (3,1) -- ++(1,0);
\draw[thick] (3,.5) -- ++(1,0);
\draw[thick] (3,-.5) -- ++(1,0);
\draw[thick] (3,-1) -- ++(1,0);
\draw[thick] (3,-2) -- ++(1,0);
\draw[thick] (3,-2.5) -- ++(1,0);
\draw[fill=white] (-4,2.7) rectangle ++(-.4,-.9);
\draw[fill=white] (-4,1.2) rectangle ++(-.4,-.9);
\draw[fill=white] (4,2.7) rectangle ++(.4,-.9);
\draw[fill=white] (4,1.2) rectangle ++(.4,-.9);
\draw[fill=white] (-4,-2.7) rectangle ++(-.4,.9);
\draw[fill=white] (-4,-1.2) rectangle ++(-.4,.9);
\draw[fill=white] (4,-2.7) rectangle ++(.4,.9);
\draw[fill=white] (4,-1.2) rectangle ++(.4,.9);
\draw[thick] (-3,.5) -- (-3,-.5);
\draw[thick] (3,2.5) -- ++(-1.5,0)-- ++(0,-5) -- ++(1.5,0) -- (3,-2.5);
\draw[thick,\qcolor] (-3,2.5) -- ++(1.5,0) to[out=0,in=180] (1.5,-.5) -- ++(1.5,0);
\draw[thick,\qcolor] (-3,2) -- ++(1.5,0) to[out=0,in=180] (1.5,-1) -- ++(1.5,0);
\draw[thick,\qcolor] (-3,-1) -- ++(1.5,0) to[out=0,in=180] (1.5,2) -- ++(1.5,0);
\draw[thick,\qbcolor] (-3,1) -- ++(1.5,0) to[out=0,in=180] (1.5,1) -- ++(1.5,0);
\draw[thick,\qbcolor] (-3,-2.5) -- ++(1.5,0) to[out=0,in=180] (1.5,.5) -- ++(1.5,0);
\draw[thick,\qbcolor] (-3,-2) -- ++(1.5,0) to[out=0,in=180] (1.5,-2) -- ++(1.5,0);
\end{tikzpicture}}

\newcommand{\TKh}{
\begin{tikzpicture}[scale=\setscaletikz]
\draw[thick] (-3,2.5) -- ++(-1,0);
\draw[thick] (-3,2) -- ++(-1,0);
\draw[thick] (-3,1) -- ++(-1,0);
\draw[thick] (-3,.5) -- ++(-1,0);
\draw[thick] (-3,-.5) -- ++(-1,0);
\draw[thick] (-3,-1) -- ++(-1,0);
\draw[thick] (-3,-2) -- ++(-1,0);
\draw[thick] (-3,-2.5) -- ++(-1,0);
\draw[thick] (3,2.5) -- ++(1,0);
\draw[thick] (3,2) -- ++(1,0);
\draw[thick] (3,1) -- ++(1,0);
\draw[thick] (3,.5) -- ++(1,0);
\draw[thick] (3,-.5) -- ++(1,0);
\draw[thick] (3,-1) -- ++(1,0);
\draw[thick] (3,-2) -- ++(1,0);
\draw[thick] (3,-2.5) -- ++(1,0);
\draw[fill=white] (-4,2.7) rectangle ++(-.4,-.9);
\draw[fill=white] (-4,1.2) rectangle ++(-.4,-.9);
\draw[fill=white] (4,2.7) rectangle ++(.4,-.9);
\draw[fill=white] (4,1.2) rectangle ++(.4,-.9);
\draw[fill=white] (-4,-2.7) rectangle ++(-.4,.9);
\draw[fill=white] (-4,-1.2) rectangle ++(-.4,.9);
\draw[fill=white] (4,-2.7) rectangle ++(.4,.9);
\draw[fill=white] (4,-1.2) rectangle ++(.4,.9);
\draw[thick] (-3,.5) -- (-3,-.5);
\draw[thick] (3,2.5) -- ++(-1.5,0)-- ++(0,-5) -- ++(1.5,0) -- (3,-2.5);
\draw[thick,\qcolor] (-3,2.5) -- ++(1.5,0) to[out=0,in=180] (1.5,-.5) -- ++(1.5,0);
\draw[thick,\qcolor] (-3,2) -- ++(1.5,0) to[out=0,in=180] (1.5,-1) -- ++(1.5,0);
\draw[thick,\qcolor] (-3,-1) -- ++(1.5,0) to[out=0,in=180] (1.5,2) -- ++(1.5,0);
\draw[thick,\qbcolor] (-3,1) -- ++(1.5,0) to[out=0,in=180] (1.5,-2) -- ++(1.5,0);
\draw[thick,\qbcolor] (-3,-2.5) -- ++(1.5,0) to[out=0,in=180] (1.5,.5) -- ++(1.5,0);
\draw[thick,\qbcolor] (-3,-2) -- ++(1.5,0) to[out=0,in=180] (1.5,1) -- ++(1.5,0);
\end{tikzpicture}}

\newcommand{\TLa}{
\begin{tikzpicture}[scale=\setscaletikz]
\draw[thick] (-3,2.5) -- ++(-1,0);
\draw[thick] (-3,2) -- ++(-1,0);
\draw[thick] (-3,1) -- ++(-1,0);
\draw[thick] (-3,.5) -- ++(-1,0);
\draw[thick] (-3,-.5) -- ++(-1,0);
\draw[thick] (-3,-1) -- ++(-1,0);
\draw[thick] (-3,-2) -- ++(-1,0);
\draw[thick] (-3,-2.5) -- ++(-1,0);
\draw[thick] (3,2.5) -- ++(1,0);
\draw[thick] (3,2) -- ++(1,0);
\draw[thick] (3,1) -- ++(1,0);
\draw[thick] (3,.5) -- ++(1,0);
\draw[thick] (3,-.5) -- ++(1,0);
\draw[thick] (3,-1) -- ++(1,0);
\draw[thick] (3,-2) -- ++(1,0);
\draw[thick] (3,-2.5) -- ++(1,0);
\draw[fill=white] (-4,2.7) rectangle ++(-.4,-.9);
\draw[fill=white] (-4,1.2) rectangle ++(-.4,-.9);
\draw[fill=white] (4,2.7) rectangle ++(.4,-.9);
\draw[fill=white] (4,1.2) rectangle ++(.4,-.9);
\draw[fill=white] (-4,-2.7) rectangle ++(-.4,.9);
\draw[fill=white] (-4,-1.2) rectangle ++(-.4,.9);
\draw[fill=white] (4,-2.7) rectangle ++(.4,.9);
\draw[fill=white] (4,-1.2) rectangle ++(.4,.9);
\draw[thick,\qcolor] (-3,2.5) to[out=0,in=180] (3,2.5);
\draw[thick,\qcolor] (-3,2) to[out=0,in=180] (3,2);
\draw[thick,\qcolor] (-3,-.5) to[out=0,in=180] (3,-.5);
\draw[thick,\qcolor] (-3,-1) to[out=0,in=180] (3,-1);
\draw[thick,\qbcolor] (-3,1) to[out=0,in=180] (3,1);
\draw[thick,\qbcolor] (-3,.5) to[out=0,in=180] (3,.5);
\draw[thick,\qbcolor] (-3,-2) to[out=0,in=180] (3,-2);
\draw[thick,\qbcolor] (-3,-2.5) to[out=0,in=180] (3,-2.5);
\end{tikzpicture}}

\newcommand{\TLb}{
\begin{tikzpicture}[scale=\setscaletikz]
\draw[thick] (-3,2.5) -- ++(-1,0);
\draw[thick] (-3,2) -- ++(-1,0);
\draw[thick] (-3,1) -- ++(-1,0);
\draw[thick] (-3,.5) -- ++(-1,0);
\draw[thick] (-3,-.5) -- ++(-1,0);
\draw[thick] (-3,-1) -- ++(-1,0);
\draw[thick] (-3,-2) -- ++(-1,0);
\draw[thick] (-3,-2.5) -- ++(-1,0);
\draw[thick] (3,2.5) -- ++(1,0);
\draw[thick] (3,2) -- ++(1,0);
\draw[thick] (3,1) -- ++(1,0);
\draw[thick] (3,.5) -- ++(1,0);
\draw[thick] (3,-.5) -- ++(1,0);
\draw[thick] (3,-1) -- ++(1,0);
\draw[thick] (3,-2) -- ++(1,0);
\draw[thick] (3,-2.5) -- ++(1,0);
\draw[fill=white] (-4,2.7) rectangle ++(-.4,-.9);
\draw[fill=white] (-4,1.2) rectangle ++(-.4,-.9);
\draw[fill=white] (4,2.7) rectangle ++(.4,-.9);
\draw[fill=white] (4,1.2) rectangle ++(.4,-.9);
\draw[fill=white] (-4,-2.7) rectangle ++(-.4,.9);
\draw[fill=white] (-4,-1.2) rectangle ++(-.4,.9);
\draw[fill=white] (4,-2.7) rectangle ++(.4,.9);
\draw[fill=white] (4,-1.2) rectangle ++(.4,.9);
\draw[thick,\qcolor] (-3,2.5) to[out=0,in=180] (3,-.5);
\draw[thick,\qcolor] (-3,2) to[out=0,in=180] (3,-1);
\draw[thick,\qcolor] (-3,-.5) to[out=0,in=180] (3,2.5);
\draw[thick,\qcolor] (-3,-1) to[out=0,in=180] (3,2);
\draw[thick,\qbcolor] (-3,1) to[out=0,in=180] (3,-2);
\draw[thick,\qbcolor] (-3,.5) to[out=0,in=180] (3,-2.5);
\draw[thick,\qbcolor] (-3,-2) to[out=0,in=180] (3,1);
\draw[thick,\qbcolor] (-3,-2.5) to[out=0,in=180] (3,.5);
\end{tikzpicture}}

\newcommand{\TLc}{
\begin{tikzpicture}[scale=\setscaletikz]
\draw[thick] (-3,2.5) -- ++(-1,0);
\draw[thick] (-3,2) -- ++(-1,0);
\draw[thick] (-3,1) -- ++(-1,0);
\draw[thick] (-3,.5) -- ++(-1,0);
\draw[thick] (-3,-.5) -- ++(-1,0);
\draw[thick] (-3,-1) -- ++(-1,0);
\draw[thick] (-3,-2) -- ++(-1,0);
\draw[thick] (-3,-2.5) -- ++(-1,0);
\draw[thick] (3,2.5) -- ++(1,0);
\draw[thick] (3,2) -- ++(1,0);
\draw[thick] (3,1) -- ++(1,0);
\draw[thick] (3,.5) -- ++(1,0);
\draw[thick] (3,-.5) -- ++(1,0);
\draw[thick] (3,-1) -- ++(1,0);
\draw[thick] (3,-2) -- ++(1,0);
\draw[thick] (3,-2.5) -- ++(1,0);
\draw[fill=white] (-4,2.7) rectangle ++(-.4,-.9);
\draw[fill=white] (-4,1.2) rectangle ++(-.4,-.9);
\draw[fill=white] (4,2.7) rectangle ++(.4,-.9);
\draw[fill=white] (4,1.2) rectangle ++(.4,-.9);
\draw[fill=white] (-4,-2.7) rectangle ++(-.4,.9);
\draw[fill=white] (-4,-1.2) rectangle ++(-.4,.9);
\draw[fill=white] (4,-2.7) rectangle ++(.4,.9);
\draw[fill=white] (4,-1.2) rectangle ++(.4,.9);
\draw[thick,\qcolor] (-3,2.5) to[out=0,in=180] (3,2.5);
\draw[thick,\qcolor] (-3,2) to[out=0,in=180] (3,-.5);
\draw[thick,\qcolor] (-3,-.5) to[out=0,in=180] (3,2);
\draw[thick,\qcolor] (-3,-1) to[out=0,in=180] (3,-1);
\draw[thick,\qbcolor] (-3,1) to[out=0,in=180] (3,1);
\draw[thick,\qbcolor] (-3,.5) to[out=0,in=180] (3,-2);
\draw[thick,\qbcolor] (-3,-2) to[out=0,in=180] (3,.5);
\draw[thick,\qbcolor] (-3,-2.5) to[out=0,in=180] (3,-2.5);
\end{tikzpicture}}

\newcommand{\TLf}{
\begin{tikzpicture}[scale=\setscaletikz]
\draw[thick] (-3,2.5) -- ++(-1,0);
\draw[thick] (-3,2) -- ++(-1,0);
\draw[thick] (-3,1) -- ++(-1,0);
\draw[thick] (-3,.5) -- ++(-1,0);
\draw[thick] (-3,-.5) -- ++(-1,0);
\draw[thick] (-3,-1) -- ++(-1,0);
\draw[thick] (-3,-2) -- ++(-1,0);
\draw[thick] (-3,-2.5) -- ++(-1,0);
\draw[thick] (3,2.5) -- ++(1,0);
\draw[thick] (3,2) -- ++(1,0);
\draw[thick] (3,1) -- ++(1,0);
\draw[thick] (3,.5) -- ++(1,0);
\draw[thick] (3,-.5) -- ++(1,0);
\draw[thick] (3,-1) -- ++(1,0);
\draw[thick] (3,-2) -- ++(1,0);
\draw[thick] (3,-2.5) -- ++(1,0);
\draw[fill=white] (-4,2.7) rectangle ++(-.4,-.9);
\draw[fill=white] (-4,1.2) rectangle ++(-.4,-.9);
\draw[fill=white] (4,2.7) rectangle ++(.4,-.9);
\draw[fill=white] (4,1.2) rectangle ++(.4,-.9);
\draw[fill=white] (-4,-2.7) rectangle ++(-.4,.9);
\draw[fill=white] (-4,-1.2) rectangle ++(-.4,.9);
\draw[fill=white] (4,-2.7) rectangle ++(.4,.9);
\draw[fill=white] (4,-1.2) rectangle ++(.4,.9);
\draw[thick,\qcolor] (-3,2.5) to[out=0,in=180] (3,2.5);
\draw[thick,\qcolor] (-3,2) to[out=0,in=180] (3,2);
\draw[thick,\qcolor] (-3,-.5) to[out=0,in=180] (3,-.5);
\draw[thick,\qcolor] (-3,-1) to[out=0,in=180] (3,-1);
\draw[thick,\qbcolor] (-3,1) to[out=0,in=180] (3,-2);
\draw[thick,\qbcolor] (-3,.5) to[out=0,in=180] (3,-2.5);
\draw[thick,\qbcolor] (-3,-2) to[out=0,in=180] (3,1);
\draw[thick,\qbcolor] (-3,-2.5) to[out=0,in=180] (3,.5);
\end{tikzpicture}}

\newcommand{\TLg}{
\begin{tikzpicture}[scale=\setscaletikz]
\draw[thick] (-3,2.5) -- ++(-1,0);
\draw[thick] (-3,2) -- ++(-1,0);
\draw[thick] (-3,1) -- ++(-1,0);
\draw[thick] (-3,.5) -- ++(-1,0);
\draw[thick] (-3,-.5) -- ++(-1,0);
\draw[thick] (-3,-1) -- ++(-1,0);
\draw[thick] (-3,-2) -- ++(-1,0);
\draw[thick] (-3,-2.5) -- ++(-1,0);
\draw[thick] (3,2.5) -- ++(1,0);
\draw[thick] (3,2) -- ++(1,0);
\draw[thick] (3,1) -- ++(1,0);
\draw[thick] (3,.5) -- ++(1,0);
\draw[thick] (3,-.5) -- ++(1,0);
\draw[thick] (3,-1) -- ++(1,0);
\draw[thick] (3,-2) -- ++(1,0);
\draw[thick] (3,-2.5) -- ++(1,0);
\draw[fill=white] (-4,2.7) rectangle ++(-.4,-.9);
\draw[fill=white] (-4,1.2) rectangle ++(-.4,-.9);
\draw[fill=white] (4,2.7) rectangle ++(.4,-.9);
\draw[fill=white] (4,1.2) rectangle ++(.4,-.9);
\draw[fill=white] (-4,-2.7) rectangle ++(-.4,.9);
\draw[fill=white] (-4,-1.2) rectangle ++(-.4,.9);
\draw[fill=white] (4,-2.7) rectangle ++(.4,.9);
\draw[fill=white] (4,-1.2) rectangle ++(.4,.9);
\draw[thick,\qcolor] (-3,2.5) to[out=0,in=180] (3,-.5);
\draw[thick,\qcolor] (-3,2) to[out=0,in=180] (3,-1);
\draw[thick,\qcolor] (-3,-.5) to[out=0,in=180] (3,2.5);
\draw[thick,\qcolor] (-3,-1) to[out=0,in=180] (3,2);
\draw[thick,\qbcolor] (-3,1) to[out=0,in=180] (3,1);
\draw[thick,\qbcolor] (-3,.5) to[out=0,in=180] (3,.5);
\draw[thick,\qbcolor] (-3,-2) to[out=0,in=180] (3,-2);
\draw[thick,\qbcolor] (-3,-2.5) to[out=0,in=180] (3,-2.5);
\end{tikzpicture}}

\newcommand{\TLd}{
\begin{tikzpicture}[scale=\setscaletikz]
\draw[thick] (-3,2.5) -- ++(-1,0);
\draw[thick] (-3,2) -- ++(-1,0);
\draw[thick] (-3,1) -- ++(-1,0);
\draw[thick] (-3,.5) -- ++(-1,0);
\draw[thick] (-3,-.5) -- ++(-1,0);
\draw[thick] (-3,-1) -- ++(-1,0);
\draw[thick] (-3,-2) -- ++(-1,0);
\draw[thick] (-3,-2.5) -- ++(-1,0);
\draw[thick] (3,2.5) -- ++(1,0);
\draw[thick] (3,2) -- ++(1,0);
\draw[thick] (3,1) -- ++(1,0);
\draw[thick] (3,.5) -- ++(1,0);
\draw[thick] (3,-.5) -- ++(1,0);
\draw[thick] (3,-1) -- ++(1,0);
\draw[thick] (3,-2) -- ++(1,0);
\draw[thick] (3,-2.5) -- ++(1,0);
\draw[fill=white] (-4,2.7) rectangle ++(-.4,-.9);
\draw[fill=white] (-4,1.2) rectangle ++(-.4,-.9);
\draw[fill=white] (4,2.7) rectangle ++(.4,-.9);
\draw[fill=white] (4,1.2) rectangle ++(.4,-.9);
\draw[fill=white] (-4,-2.7) rectangle ++(-.4,.9);
\draw[fill=white] (-4,-1.2) rectangle ++(-.4,.9);
\draw[fill=white] (4,-2.7) rectangle ++(.4,.9);
\draw[fill=white] (4,-1.2) rectangle ++(.4,.9);
\draw[thick,\qcolor] (-3,2.5) to[out=0,in=180] (3,2.5);
\draw[thick,\qcolor] (-3,2) to[out=0,in=180] (3,2);
\draw[thick,\qcolor] (-3,-.5) to[out=0,in=180] (3,-.5);
\draw[thick,\qcolor] (-3,-1) to[out=0,in=180] (3,-1);
\draw[thick,\qbcolor] (-3,1) to[out=0,in=180] (3,1);
\draw[thick,\qbcolor] (-3,.5) to[out=0,in=180] (3,-2);
\draw[thick,\qbcolor] (-3,-2) to[out=0,in=180] (3,.5);
\draw[thick,\qbcolor] (-3,-2.5) to[out=0,in=180] (3,-2.5);
\end{tikzpicture}}

\newcommand{\TLe}{
\begin{tikzpicture}[scale=\setscaletikz]
\draw[thick] (-3,2.5) -- ++(-1,0);
\draw[thick] (-3,2) -- ++(-1,0);
\draw[thick] (-3,1) -- ++(-1,0);
\draw[thick] (-3,.5) -- ++(-1,0);
\draw[thick] (-3,-.5) -- ++(-1,0);
\draw[thick] (-3,-1) -- ++(-1,0);
\draw[thick] (-3,-2) -- ++(-1,0);
\draw[thick] (-3,-2.5) -- ++(-1,0);
\draw[thick] (3,2.5) -- ++(1,0);
\draw[thick] (3,2) -- ++(1,0);
\draw[thick] (3,1) -- ++(1,0);
\draw[thick] (3,.5) -- ++(1,0);
\draw[thick] (3,-.5) -- ++(1,0);
\draw[thick] (3,-1) -- ++(1,0);
\draw[thick] (3,-2) -- ++(1,0);
\draw[thick] (3,-2.5) -- ++(1,0);
\draw[fill=white] (-4,2.7) rectangle ++(-.4,-.9);
\draw[fill=white] (-4,1.2) rectangle ++(-.4,-.9);
\draw[fill=white] (4,2.7) rectangle ++(.4,-.9);
\draw[fill=white] (4,1.2) rectangle ++(.4,-.9);
\draw[fill=white] (-4,-2.7) rectangle ++(-.4,.9);
\draw[fill=white] (-4,-1.2) rectangle ++(-.4,.9);
\draw[fill=white] (4,-2.7) rectangle ++(.4,.9);
\draw[fill=white] (4,-1.2) rectangle ++(.4,.9);
\draw[thick,\qcolor] (-3,2.5) to[out=0,in=180] (3,2.5);
\draw[thick,\qcolor] (-3,2) to[out=0,in=180] (3,-.5);
\draw[thick,\qcolor] (-3,-.5) to[out=0,in=180] (3,2);
\draw[thick,\qcolor] (-3,-1) to[out=0,in=180] (3,-1);
\draw[thick,\qbcolor] (-3,1) to[out=0,in=180] (3,1);
\draw[thick,\qbcolor] (-3,.5) to[out=0,in=180] (3,.5);
\draw[thick,\qbcolor] (-3,-2) to[out=0,in=180] (3,-2);
\draw[thick,\qbcolor] (-3,-2.5) to[out=0,in=180] (3,-2.5);
\end{tikzpicture}}

\newcommand{\TLh}{
\begin{tikzpicture}[scale=\setscaletikz]
\draw[thick] (-3,2.5) -- ++(-1,0);
\draw[thick] (-3,2) -- ++(-1,0);
\draw[thick] (-3,1) -- ++(-1,0);
\draw[thick] (-3,.5) -- ++(-1,0);
\draw[thick] (-3,-.5) -- ++(-1,0);
\draw[thick] (-3,-1) -- ++(-1,0);
\draw[thick] (-3,-2) -- ++(-1,0);
\draw[thick] (-3,-2.5) -- ++(-1,0);
\draw[thick] (3,2.5) -- ++(1,0);
\draw[thick] (3,2) -- ++(1,0);
\draw[thick] (3,1) -- ++(1,0);
\draw[thick] (3,.5) -- ++(1,0);
\draw[thick] (3,-.5) -- ++(1,0);
\draw[thick] (3,-1) -- ++(1,0);
\draw[thick] (3,-2) -- ++(1,0);
\draw[thick] (3,-2.5) -- ++(1,0);
\draw[fill=white] (-4,2.7) rectangle ++(-.4,-.9);
\draw[fill=white] (-4,1.2) rectangle ++(-.4,-.9);
\draw[fill=white] (4,2.7) rectangle ++(.4,-.9);
\draw[fill=white] (4,1.2) rectangle ++(.4,-.9);
\draw[fill=white] (-4,-2.7) rectangle ++(-.4,.9);
\draw[fill=white] (-4,-1.2) rectangle ++(-.4,.9);
\draw[fill=white] (4,-2.7) rectangle ++(.4,.9);
\draw[fill=white] (4,-1.2) rectangle ++(.4,.9);
\draw[thick,\qcolor] (-3,2.5) to[out=0,in=180] (3,-.5);
\draw[thick,\qcolor] (-3,2) to[out=0,in=180] (3,-1);
\draw[thick,\qcolor] (-3,-.5) to[out=0,in=180] (3,2.5);
\draw[thick,\qcolor] (-3,-1) to[out=0,in=180] (3,2);
\draw[thick,\qbcolor] (-3,1) to[out=0,in=180] (3,1);
\draw[thick,\qbcolor] (-3,.5) to[out=0,in=180] (3,-2);
\draw[thick,\qbcolor] (-3,-2.5) to[out=0,in=180] (3,-2.5);
\draw[thick,\qbcolor] (-3,-2) to[out=0,in=180] (3,.5);
\end{tikzpicture}}

\newcommand{\TLi}{
\begin{tikzpicture}[scale=\setscaletikz]
\draw[thick] (-3,2.5) -- ++(-1,0);
\draw[thick] (-3,2) -- ++(-1,0);
\draw[thick] (-3,1) -- ++(-1,0);
\draw[thick] (-3,.5) -- ++(-1,0);
\draw[thick] (-3,-.5) -- ++(-1,0);
\draw[thick] (-3,-1) -- ++(-1,0);
\draw[thick] (-3,-2) -- ++(-1,0);
\draw[thick] (-3,-2.5) -- ++(-1,0);
\draw[thick] (3,2.5) -- ++(1,0);
\draw[thick] (3,2) -- ++(1,0);
\draw[thick] (3,1) -- ++(1,0);
\draw[thick] (3,.5) -- ++(1,0);
\draw[thick] (3,-.5) -- ++(1,0);
\draw[thick] (3,-1) -- ++(1,0);
\draw[thick] (3,-2) -- ++(1,0);
\draw[thick] (3,-2.5) -- ++(1,0);
\draw[fill=white] (-4,2.7) rectangle ++(-.4,-.9);
\draw[fill=white] (-4,1.2) rectangle ++(-.4,-.9);
\draw[fill=white] (4,2.7) rectangle ++(.4,-.9);
\draw[fill=white] (4,1.2) rectangle ++(.4,-.9);
\draw[fill=white] (-4,-2.7) rectangle ++(-.4,.9);
\draw[fill=white] (-4,-1.2) rectangle ++(-.4,.9);
\draw[fill=white] (4,-2.7) rectangle ++(.4,.9);
\draw[fill=white] (4,-1.2) rectangle ++(.4,.9);
\draw[thick,\qcolor] (-3,2.5) to[out=0,in=180] (3,2.5);
\draw[thick,\qcolor] (-3,2) to[out=0,in=180] (3,-.5);
\draw[thick,\qcolor] (-3,-1) to[out=0,in=180] (3,-1);
\draw[thick,\qcolor] (-3,-.5) to[out=0,in=180] (3,2);
\draw[thick,\qbcolor] (-3,1) to[out=0,in=180] (3,-2);
\draw[thick,\qbcolor] (-3,.5) to[out=0,in=180] (3,-2.5);
\draw[thick,\qbcolor] (-3,-2) to[out=0,in=180] (3,1);
\draw[thick,\qbcolor] (-3,-2.5) to[out=0,in=180] (3,.5);
\end{tikzpicture}}

\newcommand{\TT}[1]{
\begin{tikzpicture}[scale=\setscaletikz]
\draw[thick] (-3,2.5) -- ++(-1,0);
\draw[thick] (-3,2) -- ++(-1,0);
\draw[thick] (-3,1) -- ++(-1,0);
\draw[thick] (-3,.5) -- ++(-1,0);
\draw[thick] (-3,-.5) -- ++(-1,0);
\draw[thick] (-3,-1) -- ++(-1,0);
\draw[thick] (-3,-2) -- ++(-1,0);
\draw[thick] (-3,-2.5) -- ++(-1,0);
\draw[thick] (3,2.5) -- ++(1,0);
\draw[thick] (3,2) -- ++(1,0);
\draw[thick] (3,1) -- ++(1,0);
\draw[thick] (3,.5) -- ++(1,0);
\draw[thick] (3,-.5) -- ++(1,0);
\draw[thick] (3,-1) -- ++(1,0);
\draw[thick] (3,-2) -- ++(1,0);
\draw[thick] (3,-2.5) -- ++(1,0);
\draw[fill=white] (-4,2.7) rectangle ++(-.4,-.9);
\draw[fill=white] (-4,1.2) rectangle ++(-.4,-.9);
\draw[fill=white] (4,2.7) rectangle ++(.4,-.9);
\draw[fill=white] (4,1.2) rectangle ++(.4,-.9);
\draw[fill=white] (-4,-2.7) rectangle ++(-.4,.9);
\draw[fill=white] (-4,-1.2) rectangle ++(-.4,.9);
\draw[fill=white] (4,-2.7) rectangle ++(.4,.9);
\draw[fill=white] (4,-1.2) rectangle ++(.4,.9);
\draw[thick,\qcolor] (-3,2.5) to[out=0,in=180] (3,2.5);
\draw[thick,\qcolor] (-3,2) to[out=0,in=180] (3,2);
\draw[thick,\qcolor] (-3,-.5) to[out=0,in=180] (3,-.5);
\draw[thick,\qcolor] (-3,-1) to[out=0,in=180] (3,-1);
\draw[thick,\qbcolor] (-3,1) to[out=0,in=180] (3,1);
\draw[thick,\qbcolor] (-3,.5) to[out=0,in=180] (3,.5);
\draw[thick,\qbcolor] (-3,-2) to[out=0,in=180] (3,-2);
\draw[thick,\qbcolor] (-3,-2.5) to[out=0,in=180] (3,-2.5);
\draw[fill=white] (0,0) ellipse (2.4cm and 4cm);
\node at (0,0) {$\mathsize{14}{#1}$};
\end{tikzpicture}}

\newcommand{\trTT}[1]{
\begin{tikzpicture}[scale=\setscaletikz]
\draw[thick] (-3,2.5) -- ++(-1,0);
\draw[thick] (-3,2) -- ++(-1,0);
\draw[thick] (-3,1) -- ++(-1,0);
\draw[thick] (-3,.5) -- ++(-1,0);
\draw[thick] (-3,-.5) -- ++(-1,0);
\draw[thick] (-3,-1) -- ++(-1,0);
\draw[thick] (-3,-2) -- ++(-1,0);
\draw[thick] (-3,-2.5) -- ++(-1,0);
\draw[thick] (3,2.5) -- ++(1,0);
\draw[thick] (3,2) -- ++(1,0);
\draw[thick] (3,1) -- ++(1,0);
\draw[thick] (3,.5) -- ++(1,0);
\draw[thick] (3,-.5) -- ++(1,0);
\draw[thick] (3,-1) -- ++(1,0);
\draw[thick] (3,-2) -- ++(1,0);
\draw[thick] (3,-2.5) -- ++(1,0);
\draw[thick,\qcolor] (-4,2.5) -- (-5,2.5) -- (-5,4.5) -- (5,4.5) -- (5,2.5) -- (4,2.5);
\draw[thick,\qcolor] (-4,2) -- (-5.5,2) -- (-5.5,5) -- (5.5,5) -- (5.5,2) -- (4,2);
\draw[thick,\qbcolor] (-4,1) -- (-6.5,1) -- (-6.5,6) -- (6.5,6) -- (6.5,1) -- (4,1);
\draw[thick,\qbcolor] (-4,.5) -- (-7,.5) -- (-7,6.5) -- (7,6.5) -- (7,.5) -- (4,.5);
\draw[thick,\qbcolor] (-4,-2.5) -- (-5,-2.5) -- (-5,-4.5) -- (5,-4.5) -- (5,-2.5) -- (4,-2.5);
\draw[thick,\qbcolor] (-4,-2) -- (-5.5,-2) -- (-5.5,-5) -- (5.5,-5) -- (5.5,-2) -- (4,-2);
\draw[thick,\qcolor] (-4,-1) -- (-6.5,-1) -- (-6.5,-6) -- (6.5,-6) -- (6.5,-1) -- (4,-1);
\draw[thick,\qcolor] (-4,-.5) -- (-7,-.5) -- (-7,-6.5) -- (7,-6.5) -- (7,-.5) -- (4,-.5);
\draw[fill=white] (-4,2.7) rectangle ++(-.4,-.9);
\draw[fill=white] (-4,1.2) rectangle ++(-.4,-.9);
\draw[fill=white] (4,2.7) rectangle ++(.4,-.9);
\draw[fill=white] (4,1.2) rectangle ++(.4,-.9);
\draw[fill=white] (-4,-2.7) rectangle ++(-.4,.9);
\draw[fill=white] (-4,-1.2) rectangle ++(-.4,.9);
\draw[fill=white] (4,-2.7) rectangle ++(.4,.9);
\draw[fill=white] (4,-1.2) rectangle ++(.4,.9);
\draw[thick,\qcolor] (-3,2.5) to[out=0,in=180] (3,2.5);
\draw[thick,\qcolor] (-3,2) to[out=0,in=180] (3,2);
\draw[thick,\qcolor] (-3,-.5) to[out=0,in=180] (3,-.5);
\draw[thick,\qcolor] (-3,-1) to[out=0,in=180] (3,-1);
\draw[thick,\qbcolor] (-3,1) to[out=0,in=180] (3,1);
\draw[thick,\qbcolor] (-3,.5) to[out=0,in=180] (3,.5);
\draw[thick,\qbcolor] (-3,-2) to[out=0,in=180] (3,-2);
\draw[thick,\qbcolor] (-3,-2.5) to[out=0,in=180] (3,-2.5);
\draw[fill=white] (0,0) ellipse (2.4cm and 4cm);
\node at (0,0) {$\mathsize{14}{#1}$};
\end{tikzpicture}}

\newcommand{\TfaBT}[1]{\tikeqbis{\bs[scale=#1,yshift=15mm]
\drawqua(0.6) (-4,0) --(0,0); 
\drawqua(0.6) (0,0) -- (4,0);
\draw[glu] (0,0) -- (0,-3.5) node[below] {\scriptsize $a$};
\es}}

\newcommand{\trtt}[1]{\tikeq{\bs[scale=#1]
 \drawanti(0.25) (0,0) circle (2);
 \draw[glu] (2,0) -- (5,0); \draw[glu] (-5,0) -- (-2,0);
\es
}}

\newcommand{\BTtf}[1]{\tikeq{\bs[scale=#1]
 \draw[thick] (-2,-2) -- (0,0) -- (-2,2); 
 \draw[thick,-<] (-2,-2) -- (-1,-1);
 \draw[glu] (0,0) -- (3,0);
\es
}}

\newcommand{\BTtfb}[1]{\tikeq{\bs[scale=#1]
 \draw[thick] (-4,2) to[out=-45,in=180] (-2,-2) to[out=0,in=-90] (-1,0) to[out=90,in=0] (-2,2) to[out=180,in=45] (-4,-2); 
 \draw[thick,-<] (-2,-2) -- (-2.001,-2);
 \draw[glu] (-1,0) -- (2,0);
\es
}}

\newcommand{\DeltaAdj}{
 \begin{tikzpicture}[scale=.9*\setscaletikz]
 \draw[glu] (-2,0) -- (2,0);
 \end{tikzpicture}
}

\newcommand{\ColorConservationA}[1]{\tikeqbis{\bs[scale=#1]
 \draw[thick,->] (3,2) -- ++(1,0); \draw[thick] (4,2) -- ++(1,0); 
 \draw[thick] (3,2) to[out=180,in=45] (0,1);
 \draw[thick,-<] (3,1) -- ++(1,0); \draw[thick] (3,1) -- ++(2,0);
 \draw[thick] (3,1) to[out=180,in=30] (0,.5);
 \draw[glu] (0,0) to (3,0) -- ++(2,0);
 \draw[glu] (0,-1) to[out=-45,in=180] (3,-2) -- ++(2,0);
 \draw[fill=white!70!black] (0,0) ellipse (1 and 2); 
 \draw[fill=black] (4.5,-.5) circle (0.05);
 \draw[fill=black] (4.5,-1) circle (0.05);
 \draw[fill=black] (4.5,-1.5) circle (0.05);
 \draw[glu,red] (3,2) to[out=-90,in=180] (5,-4);
\es
}}

\newcommand{\ColorConservationB}[1]{\tikeqbis{\bs[scale=#1]
 \draw[thick,->] (3,2) -- ++(1,0); \draw[thick] (4,2) -- ++(1,0); 
 \draw[thick] (3,2) to[out=180,in=45] (0,1);
 \draw[thick,-<] (3,1) -- ++(1,0); \draw[thick] (3,1) -- ++(2,0);
 \draw[thick] (3,1) to[out=180,in=30] (0,.5);
 \draw[glu] (0,0) to (3,0) -- ++(2,0);
 \draw[glu] (0,-1) to[out=-45,in=180] (3,-2) -- ++(2,0);
 \draw[fill=white!70!black] (0,0) ellipse (1 and 2); 
 \draw[fill=black] (4.5,-.5) circle (0.05);
 \draw[fill=black] (4.5,-1) circle (0.05);
 \draw[fill=black] (4.5,-1.5) circle (0.05);
 \draw[glu,red] (3,1) to[out=-90,in=180] (5,-4);
\es
}}

\newcommand{\ColorConservationC}[1]{\tikeqbis{\bs[scale=#1]
 \draw[thick,->] (3,2) -- ++(1,0); \draw[thick] (4,2) -- ++(1,0); 
 \draw[thick] (3,2) to[out=180,in=45] (0,1);
 \draw[thick,-<] (3,1) -- ++(1,0); \draw[thick]  (3,1) -- ++(2,0);
 \draw[thick] (3,1) to[out=180,in=30] (0,.5);
 \draw[glu] (0,0) to (3,0) -- ++(2,0);
 \draw[glu] (0,-1) to[out=-45,in=180] (3,-2) -- ++(2,0);
 \draw[fill=white!70!black] (0,0) ellipse (1 and 2); 
 \draw[fill=black] (4.5,-.5) circle (0.05);
 \draw[fill=black] (4.5,-1) circle (0.05);
 \draw[fill=black] (4.5,-1.5) circle (0.05);
 \draw[glu,red] (3,0) to[out=-90,in=180] (5,-4);
\es
}}

\newcommand{\ColorConservationD}[1]{\tikeqbis{\bs[scale=#1]
 \draw[thick,->] (3,2) -- ++(1,0); \draw[thick] (4,2) -- ++(1,0); 
 \draw[thick] (3,2) to[out=180,in=45] (0,1);
 \draw[thick,-<] (3,1) -- ++(1,0); \draw[thick] (3,1) -- ++(2,0);
 \draw[thick] (3,1) to[out=180,in=30] (0,.5);
 \draw[glu] (0,0) to (3,0) -- ++(2,0);
 \draw[glu] (0,-1) to[out=-45,in=180] (3,-2) -- ++(2,0);
 \draw[fill=white!70!black] (0,0) ellipse (1 and 2); 
 \draw[fill=black] (4.5,-.5) circle (0.05);
 \draw[fill=black] (4.5,-1) circle (0.05);
 \draw[fill=black] (4.5,-1.5) circle (0.05);
 \draw[glu,red] (3,-2) to[out=-90,in=180] (5,-4);
\es
}}

%% file: 27x27.bbl
\providecommand{\href}[2]{#2}\begingroup\raggedright\begin{thebibliography}{10}

\bibitem{Cvitanovic:1976am}
{Cvitanovi\'c, Predrag}, {\it {Group theory for Feynman diagrams in non-Abelian
  gauge theories}},  {\em Phys. Rev. D} {\bf 14} (1976) 1536--1553.

\bibitem{Cvitanovic:2008zz}
{Cvitanovi\'c, Predrag}, {\em {Group Theory: Birdtracks, Lie's, and Exceptional
  Groups}}.
\newblock Princeton University Press, (5, 2020).

\bibitem{Dokshitzer:1995fv}
Y.~L. Dokshitzer, {\it {Perturbative QCD (and beyond)}},  {\em Lect. Notes
  Phys.} {\bf 496} (1997) 87--135.

\bibitem{Sjodahl:2015}
M.~Sj\"odahl, {\em {The Magic of Color}}.
\newblock Spa, Belgium, (2015).

\bibitem{Keppeler:2017kwt}
S.~Keppeler, {\it {Birdtracks for SU(N)}},  {\em SciPost Phys. Lect. Notes}
  {\bf 3} (2018) 1, [\href{http://arxiv.org/abs/1707.07280}{{\tt
  arXiv:1707.07280}}].

\bibitem{Peigne:2023iwm}
S.~Peign\'e, {\it {Introduction to color in QCD: Initiation to the birdtrack
  pictorial technique}},  in {\em {6th Chilean School of High Energy Physics}},
  (2, 2023).
\newblock \href{http://arxiv.org/abs/2302.07574}{{\tt arXiv:2302.07574}}.

\bibitem{Peigne:2024srm}
S.~Peign\'e, {\em {Color in QCD: An Introduction Featuring the Birdtrack
  Pictorial Technique}}.
\newblock SpringerBriefs in Physics. Springer, (5, 2024).

\bibitem{Botts:1989kf}
J.~Botts and G.~F. Sterman, {\it {Hard Elastic Scattering in QCD: Leading
  Behavior}},  {\em Nucl. Phys. B} {\bf 325} (1989) 62--100.

\bibitem{Sotiropoulos:1993rd}
M.~G. Sotiropoulos and G.~F. Sterman, {\it {Color exchange in near forward hard
  elastic scattering}},  {\em Nucl. Phys. B} {\bf 419} (1994) 59--76,
  [\href{http://arxiv.org/abs/hep-ph/9310279}{{\tt hep-ph/9310279}}].

\bibitem{Contopanagos:1996nh}
H.~Contopanagos, E.~Laenen, and G.~F. Sterman, {\it {Sudakov factorization and
  resummation}},  {\em Nucl. Phys. B} {\bf 484} (1997) 303--330,
  [\href{http://arxiv.org/abs/hep-ph/9604313}{{\tt hep-ph/9604313}}].

\bibitem{Kidonakis:1998nf}
N.~Kidonakis, G.~Oderda, and G.~F. Sterman, {\it {Evolution of color exchange
  in QCD hard scattering}},  {\em Nucl. Phys. B} {\bf 531} (1998) 365--402,
  [\href{http://arxiv.org/abs/hep-ph/9803241}{{\tt hep-ph/9803241}}].

\bibitem{Oderda:1999kr}
G.~Oderda, {\it {Dijet rapidity gaps in photoproduction from perturbative
  QCD}},  {\em Phys. Rev. D} {\bf 61} (2000) 014004,
  [\href{http://arxiv.org/abs/hep-ph/9903240}{{\tt hep-ph/9903240}}].

\bibitem{Kyrieleis:2005dt}
A.~Kyrieleis and M.~H. Seymour, {\it {The Colour evolution of the process q q
  ---\ensuremath{>} q q g}},  {\em JHEP} {\bf 01} (2006) 085,
  [\href{http://arxiv.org/abs/hep-ph/0510089}{{\tt hep-ph/0510089}}].

\bibitem{Dokshitzer:2005ig}
Y.~L. Dokshitzer and G.~Marchesini, {\it {Soft gluons at large angles in hadron
  collisions}},  {\em JHEP} {\bf 01} (2006) 007,
  [\href{http://arxiv.org/abs/hep-ph/0509078}{{\tt hep-ph/0509078}}].

\bibitem{Kovner:2001vi}
A.~Kovner and U.~A. Wiedemann, {\it {Eikonal evolution and gluon radiation}},
  {\em Phys. Rev. D} {\bf 64} (2001) 114002,
  [\href{http://arxiv.org/abs/hep-ph/0106240}{{\tt hep-ph/0106240}}].

\bibitem{Angelopoulou:2023qdm}
A.-K. Angelopoulou, A.~D. Le, and S.~Munier, {\it {Scattering from an external
  field in quantum chromodynamics at high energies: from foundations to
  interdisciplinary connections}},  {\em SciPost Phys. Lect. Notes} {\bf 2025}
  (2025) 92, [\href{http://arxiv.org/abs/2311.14796}{{\tt arXiv:2311.14796}}].

\bibitem{Nikolaev:2003zf}
N.~N. Nikolaev, W.~Schafer, B.~G. Zakharov, and V.~R. Zoller, {\it {Nonlinear
  k-perpendicular factorization for forward dijets in DIS off nuclei in the
  saturation regime}},  {\em J. Exp. Theor. Phys.} {\bf 97} (2003) 441--465,
  [\href{http://arxiv.org/abs/hep-ph/0303024}{{\tt hep-ph/0303024}}].

\bibitem{Nikolaev:2005zj}
N.~N. Nikolaev, W.~Schafer, and B.~G. Zakharov, {\it {Nonlinear
  k(perpendicular)-factorization for gluon-gluon dijets produced off nuclear
  targets}},  {\em Phys. Rev. D} {\bf 72} (2005) 114018,
  [\href{http://arxiv.org/abs/hep-ph/0508310}{{\tt hep-ph/0508310}}].

\bibitem{Nikolaev:2005dd}
N.~N. Nikolaev, W.~Schafer, B.~G. Zakharov, and V.~R. Zoller, {\it {Nonlinear
  k-perpendicular-factorization for quark-gluon dijet production off nuclei}},
  {\em Phys. Rev. D} {\bf 72} (2005) 034033,
  [\href{http://arxiv.org/abs/hep-ph/0504057}{{\tt hep-ph/0504057}}].

\bibitem{Cougoulic:2017ust}
F.~Cougoulic and S.~Peign\'e, {\it {Nuclear $p_\perp$-broadening of an
  energetic parton pair}},  {\em JHEP} {\bf 05} (2018) 203,
  [\href{http://arxiv.org/abs/1712.01953}{{\tt arXiv:1712.01953}}].

\bibitem{Li:2023qkg}
M.~Li, T.~Lappi, X.~Zhao, and C.~A. Salgado, {\it {On the momentum broadening
  of in-medium jet evolution using a light-front Hamiltonian approach}},  {\em
  PoS} {\bf HardProbes2023} (2024) 151,
  [\href{http://arxiv.org/abs/2307.11251}{{\tt arXiv:2307.11251}}].

\bibitem{Aurenche:2011rd}
P.~Aurenche and B.~G. Zakharov, {\it {Jet color chemistry and anomalous baryon
  production in $AA$-collisions}},  {\em Eur. Phys. J. C} {\bf 71} (2011) 1829,
  [\href{http://arxiv.org/abs/1109.6819}{{\tt arXiv:1109.6819}}].

\bibitem{Zakharov:2018hfz}
B.~G. Zakharov, {\it {Color randomization of fast gluon-gluon pairs in the
  quark-gluon plasma}},  {\em J. Exp. Theor. Phys.} {\bf 128} (2019) 243--258,
  [\href{http://arxiv.org/abs/1806.04723}{{\tt arXiv:1806.04723}}].

\bibitem{Zakharov:2019fov}
B.~G. Zakharov, {\it {Radiative Contribution to p$_{\perp}$ Broadening of Fast
  Partons in a Quark\textendash{}Gluon Plasma}},  {\em J. Exp. Theor. Phys.}
  {\bf 129} (2019) 521--540, [\href{http://arxiv.org/abs/1912.04875}{{\tt
  arXiv:1912.04875}}].

\bibitem{Arleo:2010rb}
F.~Arleo, S.~Peign\'e, and T.~Sami, {\it {Revisiting scaling properties of
  medium-induced gluon radiation}},  {\em Phys. Rev. D} {\bf 83} (2011) 114036,
  [\href{http://arxiv.org/abs/1006.0818}{{\tt arXiv:1006.0818}}].

\bibitem{Arleo:2012rs}
F.~Arleo and S.~Peign\'e, {\it {Heavy-quarkonium suppression in p-A collisions
  from parton energy loss in cold QCD matter}},  {\em JHEP} {\bf 03} (2013)
  122, [\href{http://arxiv.org/abs/1212.0434}{{\tt arXiv:1212.0434}}].

\bibitem{Liou:2014rha}
T.~Liou and A.~H. Mueller, {\it {Parton energy loss in high energy hard forward
  processes in proton-nucleus collisions}},  {\em Phys. Rev. D} {\bf 89} (2014)
  074026, [\href{http://arxiv.org/abs/1402.1647}{{\tt arXiv:1402.1647}}].

\bibitem{Peigne:2014rka}
S.~Peign\'e and R.~Kolevatov, {\it {Medium-induced soft gluon radiation in
  forward dijet production in relativistic proton-nucleus collisions}},  {\em
  JHEP} {\bf 01} (2015) 141, [\href{http://arxiv.org/abs/1405.4241}{{\tt
  arXiv:1405.4241}}].

\bibitem{Peigne:2014uha}
S.~Peign\'e, F.~Arleo, and R.~Kolevatov, {\it {Coherent medium-induced gluon
  radiation in hard forward $1 \to 1$ partonic processes}},  {\em Phys. Rev. D}
  {\bf 93} (2016) 014006, [\href{http://arxiv.org/abs/1402.1671}{{\tt
  arXiv:1402.1671}}].

\bibitem{Munier:2016oih}
S.~Munier, S.~Peign\'e, and E.~Petreska, {\it {Medium-induced gluon radiation
  in hard forward parton scattering in the saturation formalism}},  {\em Phys.
  Rev. D} {\bf 95} (2017) 014014, [\href{http://arxiv.org/abs/1603.01028}{{\tt
  arXiv:1603.01028}}].

\bibitem{Arleo:2020hat}
F.~Arleo, F.~Cougoulic, and S.~Peign\'e, {\it {Fully coherent energy loss
  effects on light hadron production in pA collisions}},  {\em JHEP} {\bf 09}
  (2020) 190, [\href{http://arxiv.org/abs/2003.06337}{{\tt arXiv:2003.06337}}].

\bibitem{Arleo:2021bpv}
F.~Arleo, G.~Jackson, and S.~Peign\'e, {\it {Impact of fully coherent energy
  loss on heavy meson production in pA collisions}},  {\em JHEP} {\bf 01}
  (2022) 164, [\href{http://arxiv.org/abs/2107.05871}{{\tt arXiv:2107.05871}}].

\bibitem{Arleo:2021krm}
F.~Arleo, G.~Jackson, and S.~Peign\'e, {\it {Depletion of atmospheric neutrino
  fluxes from parton energy loss}},  {\em Phys. Lett. B} {\bf 835} (2022)
  137541, [\href{http://arxiv.org/abs/2112.10791}{{\tt arXiv:2112.10791}}].

\bibitem{Jackson:2023adv}
G.~Jackson, S.~Peign\'e, and K.~Watanabe, {\it {Coherent gluon radiation:
  beyond leading-log accuracy}},  {\em JHEP} {\bf 05} (2024) 207,
  [\href{http://arxiv.org/abs/2312.11650}{{\tt arXiv:2312.11650}}].

\bibitem{Keppeler:2012ih}
S.~Keppeler and M.~Sj\"odahl, {\it {Orthogonal multiplet bases in SU(Nc) color
  space}},  {\em JHEP} {\bf 09} (2012) 124,
  [\href{http://arxiv.org/abs/1207.0609}{{\tt arXiv:1207.0609}}].

\bibitem{Chargeishvili:2024pnq}
B.~Chargeishvili, {\it {Automatic generation of orthogonal multiplet bases in
  $\mathrm{SU}(N_c)$ color space}},
  \href{http://arxiv.org/abs/2404.02443}{{\tt arXiv:2404.02443}}.

\bibitem{Mathematica}
W.~R. Inc., ``Mathematica, {V}ersion 14.2.''
\newblock Champaign, IL, 2024.

\bibitem{Sjodahl:2012nk}
M.~Sj\"odahl, {\it {ColorMath - A package for color summed calculations in
  SU(Nc)}},  {\em Eur. Phys. J. C} {\bf 73} (2013) 2310,
  [\href{http://arxiv.org/abs/1211.2099}{{\tt arXiv:1211.2099}}].

\bibitem{Sjodahl:2014opa}
M.~Sj\"odahl, {\it {ColorFull -- a C++ library for calculations in SU(Nc) color
  space}},  {\em Eur. Phys. J. C} {\bf 75} (2015) 236,
  [\href{http://arxiv.org/abs/1412.3967}{{\tt arXiv:1412.3967}}].

\bibitem{Keppeler:2013yla}
S.~Keppeler and M.~Sj\"odahl, {\it {Hermitian Young Operators}},  {\em J. Math.
  Phys.} {\bf 55} (2014) 021702, [\href{http://arxiv.org/abs/1307.6147}{{\tt
  arXiv:1307.6147}}].

\bibitem{Alcock-Zeilinger:2016sxc}
J.~Alcock-Zeilinger and H.~Weigert, {\it {Compact Hermitian Young Projection
  Operators}},  {\em J. Math. Phys.} {\bf 58} (2017) 051702,
  [\href{http://arxiv.org/abs/1610.10088}{{\tt arXiv:1610.10088}}].

\bibitem{Alcock-Zeilinger:2016cva}
J.~Alcock-Zeilinger and H.~Weigert, {\it {Transition Operators}},  {\em J.
  Math. Phys.} {\bf 58} (2017) 051703,
  [\href{http://arxiv.org/abs/1610.08802}{{\tt arXiv:1610.08802}}].

\bibitem{Alcock-Zeilinger:2016xgf}
J.~Alcock-Zeilinger and H.~Weigert, {\it {Simplification Rules for Birdtrack
  Operators}},  {\em J. Math. Phys.} {\bf 58} (2017) 051701,
  [\href{http://arxiv.org/abs/1610.08801}{{\tt arXiv:1610.08801}}].

\bibitem{Alcock-Zeilinger:2017ija}
J.~M. Alcock-Zeilinger, {\em {Symmetry Implications for Wilson Line Correlators
  in QCD at High Energies}}.
\newblock PhD thesis, Cape Town U., (2017).

\bibitem{Alcock-Zeilinger:2018fha}
J.~Alcock-Zeilinger and H.~Weigert, {\it {Compact construction algorithms for
  the singlets of SU(N) over mixed tensor product spaces}},
  \href{http://arxiv.org/abs/1812.11223}{{\tt arXiv:1812.11223}}.

\bibitem{Maltoni:2002mq}
F.~Maltoni, K.~Paul, T.~Stelzer, and S.~Willenbrock, {\it {Color Flow
  Decomposition of QCD Amplitudes}},  {\em Phys. Rev. D} {\bf 67} (2003)
  014026, [\href{http://arxiv.org/abs/hep-ph/0209271}{{\tt hep-ph/0209271}}].

\bibitem{Sjodahl:2009wx}
M.~Sj\"odahl, {\it {Color structure for soft gluon resummation: A General
  recipe}},  {\em JHEP} {\bf 09} (2009) 087,
  [\href{http://arxiv.org/abs/0906.1121}{{\tt arXiv:0906.1121}}].

\bibitem{Kilian:2012pz}
W.~Kilian, T.~Ohl, J.~Reuter, and C.~Speckner, {\it {QCD in the Color-Flow
  Representation}},  {\em JHEP} {\bf 10} (2012) 022,
  [\href{http://arxiv.org/abs/1206.3700}{{\tt arXiv:1206.3700}}].

\bibitem{Platzer:2013fha}
S.~Pl\"atzer, {\it {Summing Large-$N$ Towers in Colour Flow Evolution}},  {\em
  Eur. Phys. J. C} {\bf 74} (2014) 2907,
  [\href{http://arxiv.org/abs/1312.2448}{{\tt arXiv:1312.2448}}].

\bibitem{AngelesMartinez:2018cfz}
R.~\'Angeles~Mart\'\i{}nez, M.~De~Angelis, J.~R. Forshaw, S.~Pl\"atzer, and
  M.~H. Seymour, {\it {Soft gluon evolution and non-global logarithms}},  {\em
  JHEP} {\bf 05} (2018) 044, [\href{http://arxiv.org/abs/1802.08531}{{\tt
  arXiv:1802.08531}}].

\bibitem{DeAngelis:2020rvq}
M.~De~Angelis, J.~R. Forshaw, and S.~Pl\"atzer, {\it {Resummation and
  Simulation of Soft Gluon Effects beyond Leading Color}},  {\em Phys. Rev.
  Lett.} {\bf 126} (2021) 112001, [\href{http://arxiv.org/abs/2007.09648}{{\tt
  arXiv:2007.09648}}].

\bibitem{Platzer:2020lbr}
S.~Pl\"atzer and I.~Ruffa, {\it {Towards Colour Flow Evolution at Two Loops}},
  {\em JHEP} {\bf 06} (2021) 007, [\href{http://arxiv.org/abs/2012.15215}{{\tt
  arXiv:2012.15215}}].

\bibitem{Alcock-Zeilinger:2022hrk}
J.~Alcock-Zeilinger, S.~Keppeler, S.~Pl\"atzer, and M.~Sj\"odahl, {\it {Wigner
  6j symbols for SU(N): Symbols with at least two quark-lines}},  {\em J. Math.
  Phys.} {\bf 64} (2023) 023504, [\href{http://arxiv.org/abs/2209.15013}{{\tt
  arXiv:2209.15013}}].

\bibitem{Keppeler:2023msu}
S.~Keppeler, S.~Pl\"atzer, and M.~Sj\"odahl, {\it {Wigner 6j symbols with gluon
  lines: completing the set of 6j symbols required for color decomposition}},
  {\em JHEP} {\bf 05} (2024) 051, [\href{http://arxiv.org/abs/2312.16688}{{\tt
  arXiv:2312.16688}}].

\bibitem{BandaGuzman:2020wrz}
V.~M. Banda~Guzm\'an, R.~Flores-Mendieta, J.~Hern\'andez, and F.~d.~J.
  Rosales-Aldape, {\it {Spin and flavor projection operators in the $SU(2N_f)$
  spin-flavor group}},  {\em Phys. Rev. D} {\bf 102} (2020) 036010,
  [\href{http://arxiv.org/abs/2006.14169}{{\tt arXiv:2006.14169}}].

\bibitem{Guzman:2023vpq}
V.~M.~B. Guzm\'an, R.~Flores-Mendieta, and J.~Hernandez, {\it {Baryon-meson
  scattering amplitude in the $1/N_c$ expansion}},
  \href{http://arxiv.org/abs/2305.00879}{{\tt arXiv:2305.00879}}.

\bibitem{Blok:2010ge}
B.~Blok, Y.~Dokshitzer, L.~Frankfurt, and M.~Strikman, {\it {The Four jet
  production at LHC and Tevatron in QCD}},  {\em Phys. Rev. D} {\bf 83} (2011)
  071501, [\href{http://arxiv.org/abs/1009.2714}{{\tt arXiv:1009.2714}}].

\bibitem{Georgi:1999wka}
H.~Georgi, {\em {Lie algebras in particle physics}}, vol.~54.
\newblock Perseus Books, Reading, MA, 2nd ed.~ed., (1999).

\bibitem{White:1992aa}
P.~L. White, {\it {Discrete symmetries from broken SU(N) and the MSSM}},  {\em
  Nucl. Phys. B} {\bf 403} (1993) 141--158,
  [\href{http://arxiv.org/abs/hep-ph/9207231}{{\tt hep-ph/9207231}}].

\bibitem{Gross:1993hu}
D.~J. Gross and W.~Taylor, {\it {Two-dimensional QCD is a string theory}},
  {\em Nucl. Phys. B} {\bf 400} (1993) 181--208,
  [\href{http://arxiv.org/abs/hep-th/9301068}{{\tt hep-th/9301068}}].

\bibitem{Dietrich:2006cm}
D.~D. Dietrich and F.~Sannino, {\it {Conformal window of SU(N) gauge theories
  with fermions in higher dimensional representations}},  {\em Phys. Rev. D}
  {\bf 75} (2007) 085018, [\href{http://arxiv.org/abs/hep-ph/0611341}{{\tt
  hep-ph/0611341}}].

\bibitem{Adams1986-ea}
D.~Adams, {\em {The Hitchhiker's Guide to the Galaxy: The Original Radio
  Scripts}}.
\newblock Pan, London, England, (1985).

\end{thebibliography}\endgroup
